\documentclass[12pt,a4paper,english]{article}
\pdfoutput=1

\usepackage[bindingoffset=0.3cm,textheight=22.5cm,hdivide={2.7cm,*,2.7cm}, 
vdivide={*,22cm,*}]{geometry}
\usepackage{cite}
\usepackage[bookmarksnumbered=true,breaklinks=true]{hyperref}

\usepackage{amsmath,amsfonts,amssymb,babel,slashed,graphicx,color,empheq}
\usepackage[font=small,labelfont=bf]{caption}
\usepackage[font=small,labelfont=bf]{subcaption}
\IfFileExists{dsfont.sty}
	{\usepackage{dsfont}
         \newcommand{\id}{\mathds{1}}}
	{\typeout{Package dsfont.sty was not found, using alternative macros.}
         \let\mathds=\mathbb
         \newcommand{\id}{\mbox{1 \kern-.59em {\rm l}}}}
\let\one=\id

\makeatletter

         \@addtoreset{equation}{section}
\makeatother
\usepackage{tikz}
\usetikzlibrary{shapes}
\usetikzlibrary{positioning}
\usetikzlibrary{chains}
\usetikzlibrary{arrows,fit,decorations.pathreplacing, shapes.misc}
\tikzstyle{every picture}+=[remember picture]
\tikzstyle{na} = [baseline=-.5ex]

\usepackage{paralist}
\def\nn{\nonumber}

\def\a{\alpha}          
\def\b{\beta}           
\def\d{\delta}

\def\g{\gamma}

\def\l{\lambda} \def\L{\Lambda}

\def\r{\rho}
  
\def\t{\tau}

\def\cA{{\cal A}}  \def\cC{{\cal C}}
\def\cD{{\cal D}}  
 \def\cH{{\cal H}} \def\cI{{\cal I}}
  \def\cL{{\cal L}}
 \def\cN{{\cal N}} \def\cO{{\cal O}}
  
\def\cS{{\cal S}}  
\def\cV{{\cal V}} \def\cW{{\cal W}}

\newcommand{\R}{\mathds{R}}
\newcommand{\C}{\mathds{C}}

\newcommand{\Z}{\mathds{Z}}

\newcommand{\msu}{\mathfrak{s}\mathfrak{u}}

\newcommand{\mmu}{\mathfrak{u}}

\def\diag{\mbox{diag}}
\newcommand{\End}{\operatorname{End}}
\newcommand{\Hom}{\operatorname{Hom}}

\def\del{\partial}

\newcommand{\im}{\mathrm{i}}

\newcommand{\tr}{\mbox{tr}}

\def\End{\mathrm{End}}
\def\Hom{\mathrm{Hom}}

\newcommand{\Di}{{\slashed{D}}}

 \allowdisplaybreaks[3]

\renewcommand{\title}[1]{\vspace{10mm}\noindent{\Large{\bf
#1}}\vspace{8mm}} \newcommand{\authors}[1]{\noindent{\large
#1}\vspace{5mm}} \newcommand{\address}[1]{{\itshape #1\vspace{2mm}}}

\begin{document}
\hypersetup{pageanchor=false}

\begin{titlepage}

\begin{flushright}
UWThPh-2018-13 
\end{flushright}

\begin{center}

\title{ \Large Intersecting branes,  Higgs sector, and chirality \\[1ex]
from ${\cal N}=4$ SYM with soft SUSY breaking
}

\vskip 3mm

\authors{Marcus Sperling{\footnote{marcus.sperling@univie.ac.at}}, 
Harold C.\ Steinacker{\footnote{harold.steinacker@univie.ac.at}}
}
 
\vskip 3mm

 \address{ 

{\it Faculty of Physics, University of Vienna\\
Boltzmanngasse 5, A-1090 Vienna, Austria  }  
  }
\end{center}
\vskip 1.4cm
\begin{abstract}
We consider $SU(N)$ $\cN=4$ super Yang-Mills with cubic and quadratic 
soft SUSY breaking potential, such that the global $SU(4)_R$ is broken to 
$SU(3)$ or further. 
As shown recently, this set-up supports a rich set of non-trivial vacua with 
the geometry of self-intersecting $SU(3)$ branes in 6 extra dimensions.
The zero modes on these branes can be interpreted as 3 generations of bosonic 
and chiral fermionic strings connecting the branes at their intersections. 
Here, we uncover a large class of exact solutions consisting of branes 
connected by Higgs condensates, leading to Yukawa couplings between the chiral 
fermionic zero modes.
Under certain decoupling conditions, the backreaction of the Higgs on the 
branes vanishes exactly. 
The resulting physics is that of a spontaneously broken chiral gauge theory
on branes with fluxes.
In particular, we identify combined \emph{brane plus Higgs} configurations 
which lead to gauge fields that couple to chiral fermions at low energy. This 
turns out to be quite close to the Standard Model and its constructions via 
branes in string theory. 
As a by-product, we construct a $G_2$-brane solution corresponding to a 
squashed fuzzy coadjoint orbit of $G_2$. 
\end{abstract}

% \vskip 1.4cm

%\textbf{keywords:}  fuzzy extra dimensions; $N=4$ super-Yang-Mills;
%self-intersecting branes; chiral gauge theory

\end{titlepage}
%%%%%%%%%%%%%%%%%%%%%%%%%%%%%%%%%%%%%%%%%%%%%%%%%%
\hypersetup{pageanchor=true} 
\tableofcontents

\section{Introduction}
%%%%%%%%%%%%%%%%%%%%%%%%%%%%%

Among all 4-dimensional supersymmetric gauge theories, $\cN=4$ super Yang-Mills 
(SYM) is the 
%%H simplest and 
most special and rigid model. Due to this rigidity, $\cN=4$ SYM has 
remarkable properties as a quantum field theory, such as conformal invariance 
and UV-finiteness \cite{Mandelstam:1982cb,Brink:1982wv}. It arises in various contexts including 
compactification of 10-dimensional SYM on $T^6$,
IIB string theory on D3 branes, and the AdS/CFT correspondence 
\cite{Maldacena:1997re}. 
Its amplitudes exhibit a rich  structure \cite{Brandhuber:2007yx,ArkaniHamed:2010kv}, 
and powerful tools such as integrability provided many remarkable insights \cite{Beisert:2010jr}.

On the other hand, the special structure of $\cN=4$ SYM  limits its 
applicability to real-world physics. 
The SUSY vacua of $\cN=4$ SYM with $SU(N)$ gauge group have a simple structure, parametrized by the 
three commuting vacuum expectation values (VEVs) of the $\cN=1$ chiral 
superfields.
However, deformed or softly broken versions of $\cN=4$ SYM, for example via 
extra terms in the potential, may lead to more interesting vacua and low-energy 
physics. 
For instance, it is well-known that certain mass-deformations to the 
superpotential, which preserve $\cN=1$, have a much richer vacuum 
structure \cite{Vafa:1994tf,Donagi:1995cf,Dorey:2000fc,Polchinski:2000uf,Berenstein:2000ux,Berenstein:2014isa}, 
including stacks of fuzzy spheres which partially or completely break the gauge symmetry.
The resulting low-energy effective theory around 
such vacua behaves like a 6-dimensional theory with two fuzzy 
extra dimensions and a finite tower of KK modes, cf.\ 
\cite{Aschieri:2006uw,Andrews:2006aw,Harland:2009kg}.
Adding suitable soft SUSY breaking terms to the potential can lead to vacua with even richer geometries, 
including $S^2 \times S^2$
 \cite{Behr:2005wp,CastroVillarreal:2005uu,Azuma:2005pm,Kurkcuoglu:2015eta}, $S^4$ \cite{Steinacker:2015dra}, and others. 
 This generation of fuzzy extra dimensions is particularly interesting from the 
matrix model point 
 of view. In fact, the IKKT matrix model on noncommutative D3-brane solutions 
 reduces to the $U(N)$ $\cN=4$ SYM \cite{Aoki:1999vr}, where space-time and 
fuzzy extra dimensions are treated on the samen footing. Here, we focus, 
however, on the familiar gauge theory setting on $\R^{3,1}$.

The above examples lead to two obvious questions: (i) Which other, more 
interesting fuzzy geometries can be 
realized as vacua of deformed or softly broken $\cN=4$ SYM? (ii) What is the 
low-energy physics around these vacua? In particular, is it possible to obtain a 
gauge theory with chiral low-energy physics in some sense? 
More specifically, we have in mind
a  gauge theory where fermions with different chirality couple 
differently to the gauge fields, such as the Standard Model (SM) of particle 
physics.

The issue of chirality is particularly challenging, and one might be led to quickly dismiss this possibility.
In the aforementioned examples, 
the low-energy effective theory exhibits the corresponding KK tower, but it 
typically results in a non-chiral theory \cite{Chatzistavrakidis:2009ix}.
On the other hand, according to \cite{Chatzistavrakidis:2011gs}, chiral fermions 
do appear on intersecting
non-commutative brane solutions in the IKKT matrix model, if the
intersecting branes span all six extra dimensions. 
This works nicely for flat branes, which cannot be realized in $SU(N)$ 
SYM, though. In contrast, intersecting configurations of curved fuzzy branes 
tend to be unstable, cf.\ \cite{Azuma:2007jr}. 

Remarkably, intersecting compact brane solutions of softly broken $\cN=4$ SYM without instabilities 
were found in \cite{Steinacker:2014lma,Steinacker:2015mia}, by projecting
coadjoint orbits $\cO [\mu]$ of $SU(3)$ along the Cartan directions. This results 
in squashed orbits $\cC[\mu]$ of dimension 4 and 6, but both types locally span all 
six internal directions at the intersections.
Moreover, chiral fermionic zero modes arise at the intersections of (the sheets 
of) the branes.  
While it was shown that there are no negative modes around the vacua, there 
do exist numerous zero modes, which lead to Yukawa couplings between the 
fermionic zero modes. 

These zero modes arising on such intersecting fuzzy branes $\cO [\mu]$ are particularly 
interesting. The bosonic ones are denoted as \emph{Higgs sector} here, and will 
be the main focus of the present paper.
It had been argued in \cite{Steinacker:2015mia} that some of these Higgs modes 
may acquire a VEV, which in turn may lead to a chiral low-energy theory on 
$\R^{1,3}\times \cC[\mu]$ in suitable configurations. More specifically, 
some of the fermionic zero modes (a \emph{mirror sector}) should become 
massive due to the Yukawa couplings, leaving a low-energy sector of fermions 
with index zero, but chiral coupling to the low-energy gauge fields. 
However, the existence of such non-trivial \emph{Higgs vacua} had not been 
established up to now.

In the present paper, we study this Higgs sector in detail, and show that there 
exist exact solutions with non-vanishing Higgs VEV linking the $\cC[\mu]$ 
branes at their intersections, 
leading to appropriate Yukawa couplings. We find a remarkably rich structure of 
such solutions, with many similarities, but also some differences to the case 
of supersymmetric gauge theories. In particular, the zero modes form a 
\emph{graded ring}. 
We establish a class of exact solutions with vanishing backreaction to the 
branes, which is discussed and illustrated in a number of examples,
including a brane with $G_2$ structure.
For the simplest 6-dimensional (locally space-filling) branes, 
we find suitable Higgs solutions which link the 3+3 left and right-handed 
sheets, as required for a chiral gauge theory with 3 generations.
This suggests that a low-energy behavior reminiscent of the Standard Model might 
be achieved for larger branes; however for generic branes we 
can only provide evidence for the existence of such Higgs solutions.

\paragraph{Summary.}
Due to the length of the article, we provide a brief summary of the key points 
already at this stage.
Our starting point is the $\cN=4$ SYM Lagrangian for gauge group $SU(N)$ with 
the following cubic and quadratic soft SUSY breaking terms
\begin{equation}
V_{\mathrm{soft}}[X]= -4\, \tr \left([X_1^+,X_2^+] X_3^+ +\mathrm{h.c.} 
\right)+ 4 \sum_{i=1}^3 M_{i}^2 \, \tr \left( X_i^- X_i^+ \right)\, 
\end{equation}
with $X_i^- \coloneqq (X_i^+)^\dagger , \, i=1,2,3$.
We will focus on vacuum solutions $X_i^\pm   \in \mathrm{Mat}(N,\C)$.  
Our first new insight is the following observation: The 
second order, i.e.\ double commutator, equations of motion (eom) for $X$ admit 
\emph{first integral equations}
\begin{align}
 \left[X_i^+,X_j^+\right] = \sum_{k=1}^3 \varepsilon_{ijk}X_k^- 
 \, ,\; \forall i,j
 \qquad \text{and} \qquad
 \sum_{i=1}^3 \left[X_i^+,X_i^-\right] =0 \; .
 \label{eq:1st-integral}
\end{align}
The full equations of motion with arbitrary mass term 
$M\equiv M_i \in[0,\frac{1}{2}]$ are solved by a rescaling
\begin{align}
 \widetilde{X}_i^+ = R(M) \cdot X_i^+ \, , 
 \label{eq:rescale-X}
\end{align}
with $R(M) = \frac{1}{2} (1+\sqrt{1-4M^2})$
and any $X_i^+$ that satisfy \eqref{eq:1st-integral}.
As demonstrated in 
\cite{Steinacker:2014lma}, this allows for vacua describing squashed 
fuzzy $SU(3)$ coadjoint orbits $\cC[\mu]$. The (bosonic and fermionic) fluctuations around such a background $X$ 
are governed by a matrix Laplacian $\cO_V^X$ and a Dirac operator $\Di^X$. For 
vanishing masses $M$, the operator $\cO_V^X$ on $\cC[\mu]$ is positive 
semi-definite, and the zero modes split into \emph{regular} and 
\emph{exceptional} modes. The regular modes are fully classified by the 
representation theory origins of $\cC[\mu]$. Moreover, regular zero modes are 
characterized by the \emph{decoupling conditions} 
\begin{equation}
 \left[X_i^+ , \phi_j^+\right]=0 \, , \; \forall i\neq j \,,
 \qquad \text{and} \qquad 
 \left[X_i^-, \phi_i^+\right]=0 \, \; \forall i \,,
 \label{eq:decoupling}
\end{equation}
between background $X$ and fluctuations $\phi$.
Our second new insight concerns properties of the regular zero modes: first, the set 
of regular zero modes, in the sense of \eqref{eq:decoupling}, form a 
ring which is graded by the integral weights of $\msu(3)$. 
Second, the decoupling conditions between $X$ and $\phi$ are sufficient for a 
decoupling of the  potential
\begin{subequations}
\begin{equation}
 V[X+\phi] = V[X]+V[\phi]
\end{equation}
and the full equations of motion also decouple,
\begin{equation}
 \mathrm{eom}(X+\phi) = \mathrm{eom}(X) + \mathrm{eom}(\phi)  \,,
 \label{eq:additivity_eom}
\end{equation}
\end{subequations}
for arbitrary mass values.

Our third new insight lies in the following construction: starting from a 
background $X$, we add regular zero modes $\phi$ which satisfy 
the equations of motion. Then $X+\phi$ is 
an exact solution due to  \eqref{eq:additivity_eom}. 
Examples include (i) zero modes $\phi$ with maximal rank that
reduce $X+ \phi$ to the fuzzy 2-sphere, and (ii)  
zero modes $\phi$ with rank one such that $X+\phi$ is interpreted as 
$\cC[\mu]$ brane plus string modes. The latter provide Yukawa couplings for 
the analogous fermionic zero modes at the brane intersections.

While the spectra of $\cO_V^{X+\phi}$ and $\Di^{X+\phi}$ 
are not understood analytically,  numerical studies show the existence of 
instabilities in the massless case. However, a surprising but simple observation shows 
that the full potential can be expressed in complete squares, such that it 
becomes positive semi-definite for $M\geq \frac{\sqrt{2}}{3}$. While for 
$M>\frac{\sqrt{2}}{3}$ the solution with $V=0$ are trivial, 
for the particular mass $M^*=\frac{\sqrt{2}}{3}$ the solutions with $V=0$ are 
given precisely by \eqref{eq:1st-integral}, upon 
rescaling \eqref{eq:rescale-X}. 
Thus $\cO_V^{X+\phi}$ has \emph{no instabilities},
and moreover numerical studies indicate that the number of zero modes is 
drastically reduced in comparison to $\cO_V^{X}$, and independent of $N$.
This implies that our non-trivial (brane+Higgs) solutions are locally stable up to a compact moduli space,
which is understood in several interesting cases.
%%H

These results are remarkable, because they point to the possibility to obtain 
interesting chiral low energy gauge theories from softly broken $\cN=4$ SYM. 
The $\cC[\mu]$ solutions behave as self-intersecting branes in extra dimensions,
with chiral fermionic zero modes located at the intersections. 
In \cite{Steinacker:2014lma,Steinacker:2015mia} it has been argued that 
some Higgs moduli on the $\cC[\mu]$ branes need to acquire VEVs 
in order to stabilize the system and to give masses to undesired 
fermions via Yukawa couplings. Here, we established the existence of such 
Higgs VEVs, which give masses not only to fermions, but also to most of 
the bosonic moduli, as shown by the spectrum of $\cO_V^{X+\phi}$. 

%%H
There are two types of branes $\cC[(N,0)]$ and $\cC[(N_1,N_2)]$ with different chirality properties. 
The space-filling 6-dimensional $\cC[(N_1,N_2)]$ have a built-in separation of chiral modes, 
as seen through a suitable gauge boson mode of $\chi$ \eqref{chi-def}.
We focus on the simplest $\cC[(1,1)]$ brane, for which we find an exact Higgs 
solution,
and the more general \emph{rigged} $\cC[(N,1)]$ branes. These are the basis 
for 
our approach to the Standard Model.
%%H
On the 4-dimensional counterparts $\cC[(N,0)]$, a chiral gauge theory would require a  
configuration of Higgs modes which is not supported by our results.

There is another interesting general message: the underlying models are
gauge theories with a gauge group of large rank $N \gg 1$. In the trivial 
vacuum, such large-$N$ gauge theories are governed by the t'Hooft coupling $\l = 
g^2 N$.
In contrast, the $\cC[(N_1,N_2)]$ vacua with large $N_i$ behave as 
semi-classical, large fuzzy branes. The fluctuation spectrum on these vacua 
consist of a tower of KK states with a finite mass gap (independent of $N$), as 
well as the Higgs sector of zero modes. 
As discussed in section \ref{sec:regular-zm}, this Higgs sector consists of 
\emph{string-like modes}, which link some sheets of these branes with coupling 
strength $g$, 
as well as \emph{semi-classical  modes}, which are almost free. 
Moreover, most of the latter disappear as the string-like modes acquire some 
VEV.
Hence, the large number of massless fluctuations in the original large-$N$ 
theory is reduced to a small sector of string-like modes, which acquire VEVs, and leave
few remaining zero modes. 
This mean that the original large-$N$ gauge theory reduces to an
effective low-energy theory with few modes and an interesting geometric structure. 
This certainly provides strong motivation to study such scenarios in more detail.

\paragraph{Outline.}
The paper is organized as follows:
We start  by recalling $\cN=4$ SYM in section \ref{sec:background}, which 
is modified by cubic and quadratic soft breaking terms. The
properties of the squashed coadjoint $SU(3)$ orbits and the classification of 
the bosonic zero modes are reviewed, and the potential is organized 
accordingly.
In section \ref{sec:4D-branes},
we focus on the exact (classical) $\cC[\mu]$ solutions of 4-dimensional type, 
and study their zero mode sector. 
We find various exact solutions in the massless case, and show that all of these are 
free of instabilities in the presence of certain mass parameters. 
Moving on to 6-dimensional squashed orbits and their zero modes, we show in 
section \ref{sec:6D-branes} that chiral settings are simpler and more naturally 
obtained here, and identify the chirality observable $\chi$, see 
\eqref{chi-def}.
We find again exact solutions for the simplest 6-dimensional case $\cC[(1,1)]$, 
and comment on generalizations for larger branes and the issue of three 
generations. 
The fermionic zero modes and their Yukawa couplings are discussed in section \ref{sec:fermions}.
In section \ref{sec:standardmodel}, we give a qualitative discussion how 
low-energy theories resembling the Standard Model might be obtained using the 
present framework. Finally, we conclude in section \ref{sec:conclusion}.

Two appendices complement this article, exemplifying solutions to the 
equations of motion in appendix \ref{app:explicit_sol}, and  spelling out 
details of 
the novel combined solutions and the spectra of the vector Laplacian and the 
Dirac operator  in appendix \ref{app:solutions_spectrum}.
% 
%%%%%%%%%%%%%%%%%%%%%%%%%%%%%%%%%%%%%%%%%%%%%%%%%%%%%%%%%%%%%%%%%%
%%%%%%%%%%%%%%%%%%%%%%%%%%%%%%%%%%%%%%%%%%%%%%%%%%%%%%%%%%%%%%%%%%
% 
\section{Background, zero modes and Higgs potential}
\label{sec:background}
First we recall the setting from 
\cite{Steinacker:2014lma,Steinacker:2014eua,Steinacker:2015mia}. 
We start with the action of $\cN=4$ $SU(N)$ SYM, which is organized most transparently 
in terms of 10-dimensional SYM reduced to 4 dimensions:
\begin{equation}
\begin{aligned}
 S_{\rm YM}
=  \int d^4 y \ \frac 1{4g^2} &\tr\Big(-F^{\mu\nu} F_{\mu\nu} 
 - 2  D^\mu \phi^a D_\mu \phi_a +  [\phi^a,\phi^b][\phi_a,\phi_b] \Big)\\
&+  \tr\Big(\bar\Psi\g^{\mu} \im D_\mu \Psi + \bar\Psi\Gamma^a 
[\phi_a,\Psi]\Big) \, .
\end{aligned} 
\label{N=4SYM}
\end{equation}
Here $F_{\mu\nu}$ is the field strength, $D_\mu = \del_\mu - \im 
[A_\mu,\cdot]$ the gauge covariant derivative,
$\phi^a,\ a \in \{1,2,4,5,6,7\}$ are 6 scalar fields, 
$\Psi$ is a matrix-valued Majorana-Weyl (MW) spinor of $SO(9,1)$ dimensionally  
reduced to 4-dimensions, and $\Gamma^a$ arise from the 10-dimensional gamma matrices.
 All fields transform take values in $\msu(N)$ and transform
 in the adjoint of the $SU(N)$ gauge symmetry. The global $SO(6)_R$ symmetry is manifest.

It will be useful to work with the following complex linear combinations of 
dimensionless scalar fields
\begin{equation}
\label{X-T-definition}
\begin{aligned}
m X_1^\pm &= \frac{1}{\sqrt{2}}\left(\phi_4\pm \im \phi_5 \right) 
, \\
m X_2^\pm &= \frac{1}{\sqrt{2}}\left(\phi_6\mp \im \phi_7 \right)
,\\ 
m X_3^\pm &= \frac{1}{\sqrt{2}}\left(\phi_1\mp \im \phi_2 \right)
\end{aligned}
\end{equation}
with $m$ having the dimension of a mass.
For later, we also introduce the notation $X_j^\pm \equiv X_{\pm \a_j}$ with a
normalization such that 
\begin{align}
 X^\a =  X_{-\a} \quad \forall  \a \in \cI = \left\{ \pm \a_j,\ j=1,2,3 
\right\} \,.
\end{align}
To introduce a scale and to allow non-trivial \emph{brane} solutions, we add  
soft terms to the potential,
\begin{align}
 \cV[\phi] &=  \frac {m^4}{g^2}  \big(V_4[X] + V_{\mathrm{soft}}[X]\big)
 \equiv  \frac {m^4}{g^2} V[X] 
 \label{eq:V_total}
 \end{align}
 where
 \begin{align}
 V_4[X] &= -\frac 14 \tr \sum_{\a,\b\in\cI}[X_\a,X_\b][X^\a,X^\b] \\
 &=- \frac{1}{2} \tr \sum_{i,j} [X_i^+,X_j^+][X_i^-,X_j^-] 
 - \frac{1}{2} \tr   \sum_{i,j} [X_i^+,X_j^-][X_i^-,X_j^+]  
 , \nn\\
 V_{\rm soft}[X] &= 4\, \tr \big(-[X_1^+,X_2^+] X_3^+ - [X_2^-,X_1^-] X_3^- + 
M_{i}^2 X_i^- X_i^+ \big) \equiv V_3[X]+V_2[X] \;,
 \label{V-soft}
\end{align}
thereby fixing the scale $m$.
The cubic potential can be written as  
 \begin{align}
 V_3[X] = - \frac {4}3\tr \left(\sum_{i,j,k}\varepsilon_{ijk} X_i^+ X_j^+ X_k^+ 
 + \text{h.c.}\right)\  \equiv \frac{\im}{3}  \tr \sum_{\a,\b,\g \in \cI} 
(g_{\a\b\g} X^\a X^\b X^\g)  ,
\label{cubic-flux}
\end{align}
corresponding to a holomorphic 3-form on $\C^3$.  
Rewritten in a  real basis, this is recognized as the structure constants of 
$\msu(3)$ projected to the root generators \cite{Steinacker:2014lma}.

The cubic term breaks the global $SU(4)_R$ symmetry to $SU(3)_R$, which is in a sense the minimal breaking possible.
The mass terms $M_i$ may break this further: for 
equal  $M_i \equiv M \geq0$, the $SU(3)_R$ is maintained. For $M_1=M_2 
\neq M_3 $ (or permutations thereof) one has a global $(SU(2)\times U(1))_R$, 
and if all masses are distinct there is only a $(U(1)\times 
U(1))_R$ left.
% 
%%%%%%%%%%%%%%%%%%%%%%%%%%%%%%%%%%%%%%%%%%%%%%%%%%%%%%%%%%%%%%%%%%%%
%%%%%%%%%%%%%%%%%%%%%%%%%%%%%%%%%%%%%%%%%%%%%%%%%%%%%%%%%%%%%%%%%%%%
%
\subsection{Preliminaries}
\label{sec:massive_potential}
In this section we perform some algebraic manipulations of the full potential, 
which allow to derive first integral equations. Also, we introduce notation for 
the treatment of  perturbations.
\paragraph{Rewriting the potential.}
We reconsider the full potential $V[X]= V_4[X]+V_3[X]+V_2[X]$. To begin with, 
we rewrite the quartic 
potential by using the Jacobi identity of the  commutator and cyclicity 
of the trace. This results in
\begin{subequations}
\begin{align}
 V_4 &= - \tr\sum_{i,j} [X_i^+,X_j^+][X_i^-,X_j^-] 
 + \frac{1}{2} \tr \sum_{i} [X_i^+,X_i^-] \sum_{j} [X_j^+,X_j^-]  \,.
\end{align}
The cubic potential \eqref{cubic-flux} can be expressed as
\begin{align}
 V_3 = -\frac{2}{3} \tr \left( \sum_{i,j,k} \varepsilon_{ijk} [ X_i^+, X_j^+] 
X_k^+ 
 - \sum_{i,j,k} \varepsilon_{ijk} [X_i^-, X_j^-] X_k^-  \right) \,.
\end{align}
\end{subequations}
By completing the square, we arrive at the following expression for 
the total potential
\begin{equation}
\begin{aligned}
 V[X] =\tr \bigg( &\sum_{i,j} 
 \Big( [X_i^+,X_j^+] -\frac{2}{3} \sum_k \varepsilon_{ijk} X_k^-\Big) 
 \Big(-[X_i^-, X_j^-] - \frac{2}{3} \sum_{k} \varepsilon_{ijk} X_k^+ \Big) \\
&+ \frac{1}{2}  \sum_{i} [X_i^+,X_i^-] \sum_{j} [X_j^+,X_j^-]
+4 \sum_i \Big(M_i^2 -\frac{2}{9}\Big)  X_i^- X_i^+ \bigg) \,.
\end{aligned}
\end{equation}
Next, we define the suggestive notation
\begin{subequations}
\label{eq:notation_FandD}
\begin{align}
F_l &=\frac{1}{2} \sum_{i,j} \varepsilon_{lij} B_{ij} \, , \quad  B_{ij}= 
[X_i^+,X_j^+] -\frac{2}{3} \sum_k \varepsilon_{ijk} X_k^-  \, , \\
D&= \sum_{i} [X_i^+,X_i^-] \,,
\end{align}
\end{subequations}
and arrive at
\begin{equation}
\label{eq:potential_rewritten}
 V[X]
%  &= \tr( B_{ij} B_{ij}^\dagger) + \frac{1}{2} \tr (D D^\dagger)   + 4 
% \sum_i(M_i^2 - \tfrac{2}{9}) \tr(X_i^- X_i^+)   \nn\\
= 2\tr( F_{l} F_{l}^\dagger) + \frac{1}{2} \tr (D D^\dagger)   + 4 
\sum_i(M_i^2 - \tfrac{2}{9}) \tr(X_i^- X_i^+)
\end{equation}
It is apparent that $\tr( F_{l} F_{l}^\dagger),\tr (D D^\dagger), \tr(X_i^- 
X_i^+)  \geq0$. Therefore, we conclude the following:
\begin{compactenum}[(i)]

  \item Solutions with $V<0$ can only exist if $0\leq 
M_i^2 < \frac{2}{9}$ holds at least for one index $i$.
 \item  If $M_i^2 > \frac{2}{9}$ for all $i$, then the only solution with $V=0$ 
 is the trivial one $X_i^\pm =0$ for all $i$.
\item For the special case $M_i = \frac{\sqrt{2}}{3}$ for all $i$, the vacuum 
configurations with $V=0$ are characterized by the two simultaneous conditions
\begin{equation}
\begin{aligned}
 F_l &=0\,  \quad \forall l  & &\Longleftrightarrow  \quad &
 [X_i^+,X_j^+] -\frac{2}{3} \sum_k \varepsilon_{ijk} X_k^- &=0\,  \quad 
\forall 
i,j \\
 D&=0  & &\Longleftrightarrow & \sum_{i} [X_i^+,X_i^-] &=0 \; . 
\end{aligned} 
\label{firstorder-eom}
\end{equation}
Note that although the potential simplifies to $V= 2\tr( F_{l} F_{l}^\dagger) + 
\frac{1}{2} \tr (D D^\dagger) $, the $F_l$ are not F-terms in the usual sense, 
because they cannot be integrated to a superpotential. In other words, the 
would-be F-terms $F_l$ contain both the holomorphic as well as anti-holomorphic 
multiplets; 
therefore, they cannot originate from derivatives of a holomorphic 
superpotential. Nonetheless, $D$ behaves like the D-term for $\mathcal{N}=4$ 
SYM.
\end{compactenum}
\paragraph{Relation to $\mathcal{N}=1^*$ SUSY.}
Although SUSY is generically broken by the addition of soft terms, there is a 
special case $M_1=M_2=0, M_3 = \sqrt{2}$ where the soft potential 
can be expressed as a mass deformation of an $\cN=1$ superpotential. 
Suppose the superpotential is of the form
\begin{align}
 W = \tr ([X_1^+,X_2^+]X_3^- 
 - m_1 X_1^+ X_1^+ 
 - m_2 X_2^+ X_2^+   
 - m_3 X_3^- X_3^- ) \; .
\end{align}
Then the F-term contributions to the scalar potential are as follows:
\begin{align}
 V_F = - \tr \bigg( &[X_1^+,X_2^+] [X_1^-,X_2^-]  +  [X_1^+,X_3^-] 
[X_1^-,X_3^+] +  [X_2^+,X_3^-] [X_2^-,X_3^+] \nn \\
&+ m_1 ([X_2^+,X_3^-]X_1^- - [X_2^-,X_3^+]X_1^+) \nn \\
&+ m_2 ([X_3^-,X_1^+]X_2^- - [X_3^+,X_1^-]X_2^-)\nn \\
&+ m_3 ([X_1^+,X_2^+]X_3^+ - [X_1^-,X_2^-]X_3^-)\nn \\
&+m_1^2 X_1^+ X_1^- +m_2^2 X_2^+ X_2^- +m_3^2 X_3^+ X_3^-\bigg)
\end{align}
to match the cubic potential \eqref{cubic-flux}, one is forced to set 
$m_1=m_2=0$ and 
$m_3=\sqrt{2}$. Hence, this has an unbroken $\cN=1$ SUSY and is known as 
$\cN=1^*$ SUSY, see for 
instance \cite{Vafa:1994tf,Donagi:1995cf,Dorey:2000fc,Polchinski:2000uf}.
% 
%%%%%%%%%%%%%%%%%%%%%%%%%%%%%%%%%%%%%%%%%%%%%%%%%%%%%%%%%%%%%%%%%%%%
%%%%%%%%%%%%%%%%%%%%%%%%%%%%%%%%%%%%%%%%%%%%%%%%%%%%%%%%%%%%%%%%%%%%
%
\paragraph{Perturbation of the background.}
Let us add a  perturbation $\phi^\a$ to a background $X^\a$,
\begin{align}
Y^\a  = X^\a + \phi^\a \, .
\label{eq:perturbe_background}
\end{align}
As discussed in \cite{Steinacker:2015mia}, the perturbations imply further 
symmetry breaking and lead to interesting low-energy physics
in the zero-mode sector of the background $X$ .
The complete potential can be worked out,
\begin{align}
  V[X+\phi] 
&= V[X] + V[\phi] +
\tr\Big(\phi^\a \Box_X X_\a + X^\a \Box_\phi \phi_\a 
        +\frac 12\phi^\a  \big(\Box_X + 2 \Di_{ad} \big) \phi_\a - \frac 12 f^2 \Big) \nn\\
     &\ \ + 4\tr \big(- \varepsilon_{ijk} \phi_i^+ X_j^+ X_k^+ 
      - \varepsilon_{ijk} \phi_i^+ \phi_j^+ X_k^+ + M_{i}^2 \phi_i^- X_i^+
      + \text{h.c.} \big)  \; .
 \label{V-full-expand}
\end{align}
Here 
\begin{align}
 f &= i[\phi_\a,X^\a]  
\end{align}
can be viewed as gauge-fixing function in extra dimensions, and, following 
\cite{Steinacker:2014eua}, we define following operators 
\begin{align}
 \Box_X &= \sum_{a\in\cI} [X_\a,[X^\a,\cdot]]
 = \sum_{j=1}^3 [X_j^+, [X_j^-,\cdot ]] + [X_j^-, [X_j^+,\cdot]] ,  \\
% \label{matrix-Box}  \\
(\Di_{ad} \phi)_\a &= \sum_\b [[X_\a,X^{\b}],\phi_{\b}] 
=  ((\slashed{D}_{\rm mix}  + \slashed{D}_{\rm diag}) \phi)_\a \nn\\
(\slashed{D}_{\rm mix} \phi)_\a &= \sum_{\b\neq \a} [[X_\a,X^{\b}],\phi_{\b}]  \nn\\
(\slashed{D}_{\rm diag} \phi)_\a &= [[X_\a,X_{-\a}],\phi_{\a}]  \qquad\mbox{(no sum)} \ .
 \label{higgs-decouple-1}
\end{align}
\paragraph{Equations of motion.}
The equations of motion (eom) for the background $X$ can be written as
\begin{align}
 0=\big(\Box_4  + m^2(\Box_X + 4 M_{i}^2)\big)X_i^+ + 4 m^2 \varepsilon_{ijk} X_j^- X_k^- 
 \label{eom-roots}
\end{align}
where $\Box_4 = -D_\mu D^\mu$ is the 4-dimensional covariant d'Alembertian.
For classical vacua, i.e.\ space-time independent $X$, the eom reduce and can 
be cast in the following form:
\begin{align}
 0 = 2\sum_j \left[ \widetilde{B}_{ij} ,X_j^- \right]
 +\left[ D,X_i^+ \right] + 4M_i^2 X_i^+  \quad
 \text{with } \quad \widetilde{B}_{ij} =[X_i^+,X_j^+] - \sum_k 
\varepsilon_{ijk} X_k^-
\label{eq:eom_new}
\end{align}
with $D$ as  in \eqref{eq:notation_FandD}. We observe that $B_{ij}$ of  
\eqref{eq:notation_FandD} and 
$\widetilde{B}_{ij}$ are equivalent upon rescaling $X$.
\paragraph{Homogeneity of potential.}
The full potential exhibits a certain homogeneity pattern, which implies  
the relation $4V_4 +3 V_3+2V_2=0 $ for solutions. Hence the potential value at a 
solution $X$ of the eom can be computed via
\begin{align}
 V[X]\big|_{\text{sol. eom}}= \frac{1}{4} V_3[X] +\frac{1}{2}V_2[X] \; .
 \label{energy-homogeneous}
\end{align}
% 
%%%%%%%%%%%%%%%%%%%%%%%%%%%%%%%%%%%%%%%%%%%%%%%%%%%%%%
%%%%%%%%%%%%%%%%%%%%%%%%%%%%%%%%%%%%%%%%%%%%%%%%%%%%%%
%
\subsection{Squashed \texorpdfstring{$SU(3)$}{SU(3)} brane solutions}
\label{sec:solutions}
It is well-known that the potential \eqref{eq:V_total} with 
\eqref{V-soft} has fuzzy sphere solutions $X_i^\pm \sim R_i^\pm J_i$ 
where $J_i$ are generators of $\msu(2)$
\cite{DuboisViolette:1989at,Vafa:1994tf,Myers:1999ps,Polchinski:2000uf,Iso:2001mg,Berenstein:2002jq,Andrews:2006aw,Aschieri:2006uw}.
However as shown in \cite{Steinacker:2014lma}, there are also solutions with much richer structure
corresponding to (stacks of) squashed fuzzy coadjoint $SU(3)$ orbits $\cC_N[\mu]$, obtained by 
the following ansatz 
\begin{align} \ 
 X_i^\pm = R_i \pi(T_i^\pm)  \ .
  \label{basic-branes}
\end{align}
We denote these as \emph{(squashed) $SU(3)$ branes}, and they are the focus of 
this paper. Here
\begin{align}
 T_1^\pm &\equiv T_{\pm \a_1}, \qquad T_2^\pm \equiv T_{\pm \a_2},  \qquad
 T_3^\pm \equiv T_{\pm \a_3} 
 \label{X-T-definition-roots}
\end{align}
 are \emph{root} generators of $\msu(3)_X \subset \msu(N)$,
$\pi$ is any representation on $\cH \cong \C^N$, and
$\a_{1}, \a_2$ are the simple roots with $\a_3 = -(\a_1+\a_2)$. 
In these conventions\footnote{We use field theory conventions, while in 
\cite{Steinacker:2014lma} group-theory friendly conventions were used. In 
particular, the $\a_i$ are related to  the standard 
basis $\tilde\a_i$ of positive roots of $\msu(3)$ used in group theory via 
$\a_1=\tilde\a_1,\a_2=\tilde\a_2,\a_3=-\tilde\a_3$, such that 
$\a_1+\a_2+\a_3=0$; this is more natural here.}, the Lie algebra relations 
are
\begin{align}
 [T_\a,T_\b] &= \pm T_{\a+\b}, \qquad 0 \neq \a+\b\in \cI  \nn\\
 [T_{\a_i},T_{-\a_i}] &= H_{i} \equiv H_{\a_i} \nn\\
 [H,T_\a] &= \a(H) T_\a \ , 
 \label{Cartan-Weyl}
\end{align}
and in particular
\begin{subequations}
\label{eq:Lie_algebra_eom}
\begin{align}
 [T_i^+,T_j^+] = \varepsilon_{ijk} T_k^- 
 \label{root-algebra}
\end{align}
and $\a_i(H_i) = (\a_i,\a_i) =2$ where 
$(\cdot,\cdot)$ denotes the Killing form of $\msu(3)$.
Note also that the choice of labeling of $\a_3 = -\a_1 -\a_2$ implies that 
\begin{align}
 \sum_j [T_j^+,T_j^-] = \sum_j H_j =0 \, .
\end{align}
\end{subequations}
Using the Lie algebra relations, the equations of motion \eqref{eom-roots} 
become 
\begin{align}
0   &=  R_1\big(2 R_1^2 +R_2^2 + R_3^2 - 4 \frac{R_2 R_3}{R_1} + 4M_{1}^2 \big) 
T_1^+ \nn\\
 0  &= R_2\big(R_1^2 + 2 R_2^2 + R_3^2 - 4 \frac{R_1 R_3}{R_2}  + 4M_{2}^2\big) 
T_2^+  \nn\\
 0 &= R_3\big(R_1^2 + R_2^2 + 2R_3^2 - 4 \frac{R_1 R_2}{R_3}  + 4M_{3}^2\big) 
T_3^+ .
  \label{eom-general}
\end{align}
Assuming  $M_i=0$ for simplicity, these equations have 
the non-trivial solution
\begin{align}
 R_i = 1 \equiv R \ .
 \label{R-solution-equal}
\end{align}
This can be seen from \eqref{eq:eom_new}, as for vanishing masses the eom are 
necessarily satisfied for any configuration with $\widetilde{B}_{ij}=0=D$. 
Inspecting the relations \eqref{eq:Lie_algebra_eom} reveals that the ansatz 
\eqref{basic-branes} solves the eom for \eqref{R-solution-equal}.
Based on investigations with \texttt{Mathematica} presented in 
appendix \ref{app:explicit_sol}, we conclude that all $R_i$ must be equal (up to 
an irrelevant sign) if all $M_i^2$ are equal. 
In addition, we observe that there are no solution if one of the mass 
parameters satisfies $M_i^2 \geq \frac 43$, in agreement  with earlier findings 
\cite{Steinacker:2015mia}.

\begin{figure}[t!]
\centering
\begin{subfigure}{0.48\textwidth}
\begin{center}
 \includegraphics[width=0.45\textwidth]{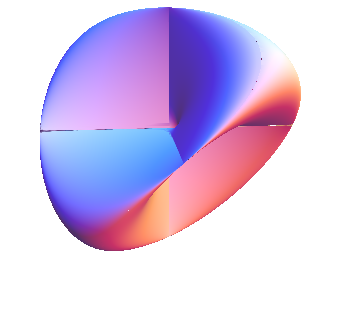}
 \end{center}
 \caption{}
 \label{fig:squashedCP2}
\end{subfigure}
\begin{subfigure}{0.48\textwidth}
\begin{center}
 \includegraphics[width=0.40\textwidth]{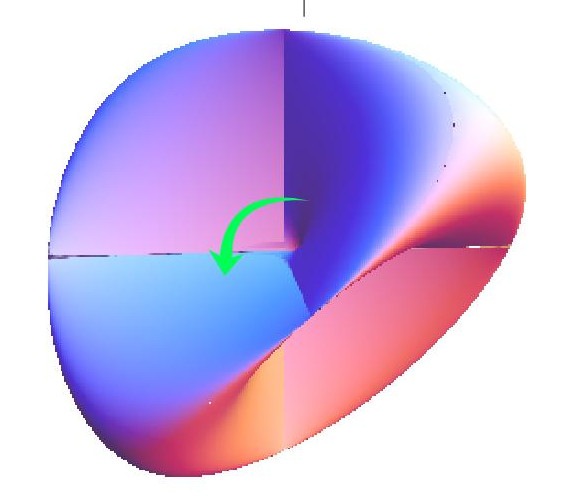}
 \end{center}
 \caption{}
 \label{fig:squashedCP2-Higgs}
\end{subfigure}
\caption{(\subref{fig:squashedCP2}) 3-dimensional section of squashed $\C P^2$.
(\subref{fig:squashedCP2-Higgs}) 3-dimensional section of squashed $\C P^2$, 
with zero mode connecting 2 
sheets. }
\end{figure}

Let $\mu$ be the highest weight vector of an $SU(3)$ irrep $\pi_\mu$, which 
enters the 
definition \eqref{basic-branes} of the squashed orbit $\cC[\mu]$. 
The highest weight $\mu$ has associated Dynkin 
labels $n_i =2\frac{( \mu ,\alpha_i )}{( \alpha_i ,\alpha_i 
)}\in \mathbb{N}$ ($\alpha_i$,  $i=1,2$ are the simple roots), and we 
simply write $\mu=(n_1,n_2)$ instead of $\mu = n_1 \Lambda_1 +n_2 \Lambda_2$, 
with $\Lambda_i$ the fundamental weights. 
Generically the $\cC_N[\mu]$ are 6-dimensional (fuzzy) varieties, while for 
$\mu =(n,0)$ and $\mu = (0,n)$ they are 
4-dimensional projections of (fuzzy) $\C P^2$.
Such a \emph{squashed} $\C P^2$ has a triple self-intersection at the origin, 
as visualized in figure 
\ref{fig:squashedCP2}.

The $\cC_N[\mu]$ backgrounds $X$ break $SU(3)_R$ to $(U(1) \times U(1))_K$  
generated by 
\begin{align}
 K_i = 2\t_i-[H_{\a_i},\cdot], \quad i=1,2,3\qquad  \text{with} \quad 
K_1+K_2+K_3 = 0 \; .
 \label{K-i-def}
\end{align}
%%H The remaining global symmetry
where $\t_i$ denotes the Cartan generators of $SU(3)_R$.
This is a combination of the global 
$(U(1)\times U(1))_R \subset SU(3)_R$ symmetry and the $SU(3)_X\subset 
SU(N)$ gauge symmetry.
We denote these $U(1)_{K_i}$-charges by 
\begin{align}
 K(\phi) \quad \in \ \msu(3)^*
\end{align}
for any\footnote{In particular, $K(X_\a) = 0$ expresses the invariance of the background.} scalar field $\phi$, 
and similarly also for fermionic fields. 
They define a $\msu(3)$ weight lattice.
The $\msu(3)_X\subset \msu(N)$ gauge charges 
will be denoted as usual by $\l(\phi) \in \msu(3)^*$, so that $K=2\t-\l$. They all live in the same $\msu(3)$ weight lattice.
Furthermore, we will need the \emph{$\t$-parity} generator $\t$ in $U(3)_R$ 
defined by
\begin{align}
 \t \phi_i^\pm = \pm \phi_i^\pm \ ,
 \label{tau-parity} 
\end{align}
which is broken by the cubic potential \eqref{cubic-flux}. 
\paragraph{Minimal branes.}
The simplest solutions are the minimal branes, which arise for $\mu=(1,0)$ and $\mu=(0,1)$. 
 To see e.g. $\cC[(1,0)]$ explicitly, denote the 3 extremal weight vectors of $(1,0)$ with 
$\{|1,\mu\rangle,|2,\mu\rangle,|3,\mu\rangle\}$.
Then the minimal squashed $\C P^2$ brane is given by
\begin{align} 
 X_i^+ &\coloneqq  T_i^+ = |i, \mu\rangle \langle i-1,\mu|
\end{align}
where $T$ is the fundamental representation of $\msu(3)$, and we dropped $r$ for simplicity.
The  energy for this minimal brane for $M_i=0$ is obtained using 
\eqref{energy-homogeneous} as 
\begin{align}
 V_4 + V_{3} = \frac 14 V_{3} = - 2 \tr(T_1^+[T_2^+,T_3^+] + \text{h.c.}) = -8 \ .
 \label{V-minimalCP2}
\end{align}
\subsection{Geometric significance: self-intersecting branes}
The qualitative features of the above solutions, and in particular the chirality 
properties of the 
fermionic zero modes discussed below, can be understood 
in terms of the semi-classical geometry of the solutions, interpreted as self-intersecting branes in $\R^6$.
As for all quantized coadjoint orbits, the semi-classical geometry of 
$\cC_N[\mu]$ can be extracted using coherent states.
These are precisely the $SU(3)$ orbits $\{g \cdot |x\rangle   \, ,\; g\in SU(3) 
\}$  of the 
extremal weight states $|x\rangle$ of the irreps $\cH_\L\in \End(\cH_\mu)$, 
where $\cH_\mu$ is the representation space for the $\msu(3)$ representation 
$\pi_\mu$.
The set of coherent states $|g x\rangle \coloneqq g \cdot |x\rangle$
forms a $U(1)$ bundle over the coadjoint orbit 
$\cO[\mu]$, and the semi-classical
base manifold embedded in $\R^6$ (the \emph{brane}) is recovered from the 
expectation values of these coherent states
\begin{align}
 \cC &= \left\{\langle g x| X^\a | g x\rangle, \  g \in SU(3) \right\} 
\subset \R^6 .
\end{align}
This would be the full coadjoint orbit $\cO[\mu]$ if the $X^\a$ were 
supplemented by the Cartan generators $H_3, H_8$, but 
here we obtain the projection $\cC[\mu]$ of $\cO[\mu]$ along the Cartan 
generators, cf.\ 
\cite{Steinacker:2014lma,Steinacker:2014eua,Steinacker:2015mia}.

The extremal weight states $\cW |x\rangle$, which lie on the discrete orbit of 
the Weyl group $\cW$ through $|x\rangle$, are projected to the origin of weight 
space. To see this, it is sufficient to note that $\langle x|X_\a|x\rangle = 0$ 
as any $X_j^\pm$ annihilates either $|x\rangle$ or $\langle x|$.
The tangent space of the sub-variety $\cC_N[\mu]$ is obtained by acting with 
the three $SU(2)_\a$ subgroups, which correspond to the roots $X_{\a}$ of 
$\msu(3)$, on the set of extremal weight states $\cW|x\rangle$. 
For $\mu=(N,0)$ and $\mu=(0,N)$, one of these actions vanishes, leading to the 
4-dimensional self-intersecting
$\C P^2$ branes depicted in figure \ref{fig:squashedCP2-Higgs}. For generic 
$\mu=(N_1,N_2)$, the $\cC_N[\mu]$ are 6-dimensional sub-varieties in $\R^6$,
i.e.\ they are locally space-filling branes. Moreover, the non-degenerate 
Poisson structure on $\cC_N[\mu]$ is recovered from the 
commutators $[X^\a,X^\b] \sim i \{x^\a,x^\b\}$, whose Pfaffian is measured by 
the operator $\chi$, see \eqref{chi-def}.
Therefore the generic 6-dimensional branes, which carry a rank 6 flux due to 
the symplectic form, consist of 3+3 locally space-filling sheets that cover 
the origin.
\subsection{Fluctuations and zero modes}
Consider the fluctuations of the scalar fields on a background brane $\cC[\mu]$ 
with representation space $\cH$.
To organize the degrees of freedom $\End(\cH)$, which denotes the algebra of 
all possible functions, we note that 
the solutions \eqref{basic-branes} define an embedding $SU(3)_X \subset SU(N)$, 
which acts via the adjoint on all the fields.  
Consequently, we decompose the $\msu(N)$-valued fields into harmonics, i.e.\ 
irreps of this $SU(3)_X$
\begin{align}
\msu(N) \cong \End(\cH)\big|_{SU(3)_X} = \oplus_\L  n_\L \cH_\L 
\label{eq:decomposition_End}
\end{align}
where $\cH_\L$ denotes the  irreps with highest weight $\L$ appearing with
multiplicity $n_\L$.
For the case of $\mu=(N,0)$ or $\mu=(0,N)$, this decomposition is given by
\begin{align}
 \End(\cH)\big|_{SU(3)_X} = \oplus_{n=0}^N  \cH_{(n,n)} \ . 
\label{eq:decomposition_End-CP2}
\end{align}
While the $SU(3)_X$ gauge transformations do \emph{not}
act on the indices $\a$ of the scalar fields $\phi_\a$, the $(U(1)\times U(1))_{K_i}$ symmetry 
\eqref{K-i-def} does act, and the $\a$ realize 3+3 of the 8 states in 
$(1,1)$ of the $\msu(3)$ weight lattice due to the origin of $\cC[\mu]$ as a $SU(3)$ coadjoint orbit. 
This allows to organize the various harmonics, which will be very useful.

Assume first $M_i=0$. Then the squashed brane backgrounds $X_\a$ admit a number 
of bosonic zero modes $\phi^{(0)}_\a$, as shown in \cite{Steinacker:2014lma}.
To see this, we note that the bilinear form defined by $\slashed{D}_{\rm mix}$ 
on a background \eqref{basic-branes} can be simplified, for example, as follows:
\begin{align}
 \tr \big(\phi_i^-(\slashed{D}_{\rm mix} \phi)^+_i\big) 
  &= \sum_{j\neq i} \tr (\phi_i^-[[X_i^+,X_{j}^+],\phi_{j}^-]) 
  =  - \varepsilon_{ikj}\frac{R_i R_j}{R_k} \tr(\phi_i^-[X_{k}^-,\phi_{j}^-])
  \label{Dmix-explicit}
\end{align}
using \eqref{root-algebra}, where the $\phi_j^\pm$ are 
perturbations as in \eqref{eq:perturbe_background}. For $R_i\equiv R$,
this has the form of the quadratic contribution from the cubic potential  
\eqref{V-full-expand}, and the quadratic terms in the potential can be combined 
as follows:
\begin{align}
 V_2[\phi] = \frac 12\tr \phi^\a \cO_V^X \phi_\a  \ ,   %\nn\\[1ex]
  \qquad \cO_V^X &\stackrel{R_i=R}{=} \Box_X + 2\slashed{D}_{\rm diag} + 2(1- 
\frac{2}{R})\slashed{D}_{\rm mix} \ \nn\\
              &\stackrel{R_i=1}{=} \Box_X + 2\slashed{D}_{\rm diag}  -2 
\slashed{D}_{\rm mix} \ .
 \label{fluct-equation}
\end{align}
It has been proven in \cite{Steinacker:2014lma} that, under these conditions, 
$\cO_V^X$ is positive semi-definite for all representations $\pi$.
The zero modes of $\cO_V^X$ fall into two classes, denoted as \emph{regular} 
and \emph{exceptional} zero modes. We will focus on the regular zero modes, and 
show that their classification is based on a \emph{decoupling condition}, as discussed in 
the next section.
\subsection{Regular zero modes and decoupling condition}
\label{sec:regular-zm}
Let $\phi_\a$ and $\psi_\a, \ \a\in \cI$ be two arbitrary matrix 
configurations, each consisting of three complex matrices or equivalently six 
hermitian matrices.
We will say that $\phi_\a$ and $\psi_\a$ satisfy the \emph{decoupling 
condition} if
\begin{align}
 [\psi_\a,\phi_\b]=0\qquad \mbox{whenever}\quad  \a+\b \in \cI \quad \mbox{or}\ 
\  \a+\b=0
  \label{extremal-char-1}
 \end{align}
This condition is symmetric under $\psi \leftrightarrow \phi$, and 
equivalent to
 \begin{align}
 [\psi_{\a},\phi_{-\b}]=0\qquad \mbox{whenever}\quad  \a-\b \in \cI \quad \mbox{or}\ \  \a-\b=0 \ .
   \label{extremal-char-2}
\end{align}
We define \emph{regular zero modes} $\phi_\a$ of $\cO_V^X$ to be modes that 
satisfy 
the decoupling condition w.r.t.\ $X_\a$.
As we will see momentarily, these modes are in fact zero modes of $\cO_V^X$ 
for $R_i=R$ and $M_i =0$, and this definition is then equivalent to 
the one given in \cite{Steinacker:2014lma,Steinacker:2014eua}.
The conditions \eqref{extremal-char-1} or \eqref{extremal-char-2} amount to the 
requirement that $\phi_\a$
is annihilated by three ladder operators out of the six $X_j^\pm$. For example, 
the condition for $\phi_i^+$ is 
\begin{align}
 [X_j^+,\phi_i^+] &= 0 \quad \text{for}\ \ j\neq i, \quad \text{and} \quad 
[X_i^-,\phi_i^+] = 0 \quad \text{(no sum)}.
 \label{extremal-3}
\end{align}
Before classifying them more explicitly, we state some important consequences.
For $M_i=0$, the regular zero modes $\phi_{\a}$ satisfy
\begin{equation}
\begin{aligned}
 (\slashed{D}_{\rm mix}^X \phi)_{i}^+ &= 0 \label{D-mix-vanish} \\
 \cO_{V,M_i\equiv0}^X \phi_{i}^+ &= [H_R,\phi_{i}^+] \ \stackrel{R_i=R}{=} \  0
\end{aligned} 
\end{equation}
where 
\begin{align}
  H_R &\coloneqq R_1^2 H_1+R_2^2 H_2 + R_3^2 H_3  \ \stackrel{R_i=R}{=} \  0 \ .
  \label{HR-def}
\end{align}
To prove the first statement of \eqref{D-mix-vanish}, consider
 \begin{align}
(\slashed{D}_{\rm mix}^X \phi)_{i}^+ &= \sum_j [[X_i^+,X_j^+],\phi_j^-] 
+ \sum_{j\neq i} [[X_i^+,X_j^+],\phi_j^+] \nn \\
&=\sum_{j,k} \frac{R_i R_j}{R_k} 
\varepsilon_{ijk}\underbrace{[X_k^-,\phi_j^-]}_{=0} 
-\sum_{j\neq i} [\underbrace{[X_j^-,\phi_j^+]}_{=0},X_i^+] 
-\sum_{j\neq i} [\underbrace{[\phi_j^+,X_i^+]}_{=0},X_j^-] 
=0
 \label{Dmix-zero-explicit}
\end{align}
where we have used the algebra relation \eqref{root-algebra} and the Jacobi 
identity. All indicated commutators vanish due to decoupling condition 
\eqref{extremal-3}.
For the second statement of \eqref{D-mix-vanish}, consider 
\begin{align}
 \Box_X \phi_{i}^+ &=
 \sum_j [X_j^+,[X_j^-,\phi_i^+]]
 +\sum_j [X_j^-,[X_j^+,\phi_i^+]] %\\
 =\sum_{j\neq i} [X_j^+,[X_j^-,\phi_i^+]]
 + [X_i^-,[X_i^+,\phi_i^+]] \nn \\
 &= \sum_{j\neq i} [[X_j^+,X_j^-],\phi_i^+] -[[X_i^+,X_i^-],\phi_i^+]
 =\sum_{j\neq i}R_j^2 [H_j,\phi_i^+] - R_i^2 [H_i,\phi_i^+] \\
  \slashed{D}_{\rm diag}\phi_{i}^+ &= R_i^2 [[T_{i},T_{-i}],\phi_{i}^+]
  = R_i^2 [H_i,\phi_{i}^+] \ 
\end{align}
using \eqref{extremal-3}. Therefore 
\begin{align}
 \cO_{V,M_i\equiv0}^X \phi_{i}^+ =  (\Box_X + 2\slashed{D}_{\rm diag})\phi_{i}^+
  = [H_R,\phi_{i}^+]  
  \stackrel{R_i=R}{=} 0 
 \label{bosonic-zeromodes-ri}
\end{align}
Consequently, the regular zero modes, as defined by the decoupling condition, 
are indeed zero modes of the $SU(3)$ branes if all $R_i$ are equal \emph{and} 
all mass parameters $M_i$ vanish. We shall keep this name also in the general 
case of non-vanishing masses and distinct $R_i$, which is discussed in section 
\ref{sec:masses}.

Now we relate this to the group-theoretical classification of 
the regular zero modes on a squashed $SU(3)$ brane $\cC[\mu]$ given in
\cite{Steinacker:2014lma,Steinacker:2015mia}.
The decoupling conditions imply that any regular zero mode $\phi_\a$ for any fixed $\a\in\cI$  is an extremal weight vector 
with $\msu(3)_X$ weight $\l$ in some irrep
$\cH_{\L} \subset \End(\cH)$ of the decomposition \eqref{eq:decomposition_End}. 
In view of \eqref{extremal-3}, the \emph{arrow}
$\l$ must be the extremal weight vector in the  Weyl chamber opposite to the 
\emph{polarization} $\a$ (or possibly on its wall).
Recalling the unbroken  $U(1)_{K_i}$ of the background, this means that it is 
one of the six extremal  $U(1)_{K_i}$ weights\footnote{These are the corners 
of the convex set of weights,
or equivalently of the maximal $\msu(3)$ irrep in $(1,1)\otimes \cH_{\L^+}$.} 
$\L'=\a-\l$ of any $\phi_\a \in \cH_{\L}$, and we denote it by
\begin{align}
 \phi_{\a,\L'} \ \in \cH_{\L} , \qquad \L'=\a-\l \equiv K(\phi_{\a,\L'})  \ . 
 \label{zeromodes-weights-Weyl}
\end{align}
Hence the regular zero modes $\phi_{\a,\L'}$ have charge 
$\L'= \a-\l$ under the $K_i$, 
corresponding to a point of the $\msu(3)$ weight lattice in 
(the interior of) the Weyl chamber of $\a$. The eigenvalue  
$\t=\pm 1$ determined by the parity of the Weyl chamber of $\a=\pm\a_i$.
Clearly there is only one (extremal) state in $\cH_\L$ for any such $\L'$. Since $\Di_{\rm mix}$ preserves $\L'$
but flips the $\t$-parity (recalling $\t \Di_{\rm mix} = - \Di_{\rm mix} \t$ 
from \cite{Steinacker:2015mia}), it follows again that 
$\slashed{D}_{\rm mix}^X \phi = 0$, i.e.\ \eqref{D-mix-vanish} holds. This 
provides another way to characterize the regular zero 
modes.

%%H
We now observe that the regular zero modes form a \emph{ring}, in the 
following sense: 
For each $\a$, let $V_\a=\{ \phi_{\a,\L'} \}$ be the vector space of regular zero modes with polarization $\a$. 
According to \eqref{zeromodes-weights-Weyl}, this vector space is graded by the 
integral weights in the Weyl chamber corresponding to $\a$.
Moreover, the decoupling 
condition \eqref{extremal-char-1} -- or equivalently the extremal weight property -- implies that 
if $\phi_\a$, $\phi'_\a$ are regular zero modes with the same polarization $\a$, 
then so is their (matrix) product $\phi_\a \phi'_\a$. Hence, each $V_\a$ is a 
graded ring. Moreover, the vector space $V = \oplus_\a V_\a$ 
forms a ring graded by the (integral) weights 
of $\msu(3)$ (or rather of $U(1)_{K_i}$),
where we define the product of zero modes in different Weyl chambers to vanish\footnote{This does not mean that 
their matrix product vanishes. However, this concept helps to organize the 
regular zero modes}.
This structure is respected by the Weyl group $\cW$, which relates the different $V_\a$.
In particular, all ring elements are nilpotent, due to the cutoff in $\L$.

The same analysis goes through for stacks of branes, where analogous zero modes 
arise as strings connecting different branes. 
As aforementioned, we will denote the space of zero modes as \emph{Higgs 
sector}.
Labeling the six-component vector $\phi = (\phi_\a)$ of zero modes by the 
dominant $U(1)_{K_i}$ weight, we learn that the Higgs sector forms a 
\emph{nilpotent ring graded by integral weights of $\msu(3)$}.

\paragraph{Examples.}
Starting from the simplest solution to the decoupling condition
\begin{align}
 \phi_{\a, \a} = T_{-\a}
\end{align}
one can use the ring multiplication of the regular zero modes to 
construct modes of the form
\begin{align}
 \phi_{\a,(n+1)\a} = (T_{-\a})^{n} ,
 \label{zeromodes-examples}
\end{align}
and any linear combinations of these.
A possible background with such a zero mode would then be 
\begin{align}
 Y_\a = X_\a + \varepsilon_\a (T_{-\a})^n .
\end{align}
On  a single squashed $\C P^2_N$ brane $\cC[(N,0)]$, these 
exhausts all regular zero modes. The ring structure is given by
$V_\a = \C[\phi_{\a, \a}]/\langle\phi_{\a, \a}^{N+1}\rangle$.
%%H
In particular,  the regular zero modes with maximal $n=N$
 on squashed $\C P^2_N$ link the 3 intersecting $\R^4$ sheets at the 
origin, with polarization along the common $\R^2$ \cite{Steinacker:2014lma}. 
These \emph{string-modes} are given by $|i-1, \mu\rangle \langle i,\mu|$, where
$\{|1,\mu\rangle,|2,\mu\rangle,|3,\mu\rangle\}$
denote the 3 extremal weight vectors of $\mu=(N,0)$.
These are coherent states located at the origin
on each of the three sheets.
An artist's rendering of such a string mode 
 underlying this solution 
is given in figure \ref{fig:squashedCP2-Higgs}.
More generally, the regular zero modes can be interpreted as strings linking these sheets,
shifted along their intersection.

\paragraph{Exceptional zero modes.}
In addition to the regular zero modes, there are certain exceptional zero modes
which are described in \cite{Steinacker:2014lma,Steinacker:2015mia}. For the 
squashed $\C P^2_N$ solutions, the only 
exceptional modes are the 
six Goldstone bosons corresponding to the spontaneously broken 
$SU(3)\slash (U(1)\times U(1))$. These are 
easily understood also in the deformed settings considered below. 
For the more general branes, the explicit description of the exceptional zero modes is not known.
This is one reason why we focus mostly on the $\C P^2$ branes in this paper.

\paragraph{$\Z_3$ and generations.} 
Note that the $U(3)_R$ label $\a$ for the scalar fields determines a 3-family structure, 
which reflects the Weyl group $\Z_3 \times \Z_2 = \cW$. 
This coincides with the  family and the $\tau$-parity as determined by  
the  $U(1)_i$ charges of the zero modes. 
 The $\Z_3 \times \Z_2$ structure is indicated by the field labels as in
 $\phi_{i\pm}$. This will be  useful for selection rules etc.
\paragraph{Stringy versus semi-classical modes.}
The above characterization of regular zero modes hides the fact that they come with very different characteristics.
We single out two extreme types: (i) the maximal or stringy zero modes, and 
(ii) the semi-classical or almost-commutative zero modes.
The distinction corresponds to the separation of functions on non-squashed $SU(3)$ coadjoint orbits $\cC[\mu]$ into UV and IR sector,
as discussed in \cite{Steinacker:2016nsc}, but they are also distinguished by their coupling strength:

All Higgs modes couple to all other (scalar, gauge and fermionic) fields through 
commutators $[\phi_\a,\cdot]$. To quantify this strength, we first need to 
proper normalize the Higgs modes. This
is dictated by the kinetic term\footnote{the canonical form is obtained by 
absorbing $g$ in the  scalar field $\phi_a$.}  
\begin{align}
 - \int d^4 x \ \frac 1{2g^2} \tr (D^\mu \phi^a D_\mu \phi_a)
 \label{kinetic-phi-g}
\end{align}
i.e.\ all modes should be normalized w.r.t.\ the trace $\tr()$.
Now the \emph{maximal Higgs} modes $\phi^{\mathrm{string}}_{\a,\L'}$ are the 
ones with extremal $\msu(3)_K$ weight $\L'$; 
these modes are given by rank one operators 
\begin{align}
 \phi^{\mathrm{string}}_{\a,\L'} = |x\rangle\langle x'|
 \label{string-modes}
\end{align}
linking  the extremal (i.e.\ coherent) weight states $|x\rangle$, $|x'\rangle$ 
of $\cH_\L\subset \End(\cH)$ whose weight difference is maximal.
From the brane point of view, these are the extreme UV modes with the maximal 
momentum \cite{Steinacker:2016nsc}.
Since the expectation values vanish, $\langle x|X_\a|x\rangle = 0$, these 
states are localized at the origin, and $\phi^{\rm string}$ should be 
interpreted as strings linking the sheets of squashed $\cC[\mu]$ at the origin. 
For large $N$, there are other \emph{almost-maximal} zero modes, for example 
with rank 2 such as the modes considered in section \ref{sec:6D-branes}.
These almost-maximal modes have a similar properties as the maximal modes; 
hence, this broader class will be called \emph{stringy} modes. 
The commutator of these zero modes is of order one, 
\begin{align}
 [\phi_\a^{\mathrm{string}},\phi_\b^{\mathrm{string}}]= O(1)
\end{align}
i.e.\ they are completely non--commutative \emph{functions} on the branes.
Putting back $g$ \eqref{kinetic-phi-g}, we see that
their interaction strength is characterized by $g$.

In contrast, the  zero modes $\phi^{\rm low}_{\a,\L'}$ with small weight $\L'$ 
are matrices with high, typically maximal rank. These modes correspond to slowly-varying, semi-classical functions 
$\phi^{\rm low}_{\a,\L'}(x)$  on the $\cC[\mu]$ brane, given by polynomials of small degree in the 
$X_\a$, the lowest mode being
\begin{align}
 \phi_\a = \frac 1{c_N} X_{-\a} \ \sim \frac 1{c_N} x_{-\a} \ .
\end{align}
The normalization $c_N$  for the minimal zero mode on an 
irreducible  $\cC[\mu]$ can be obtained from the quadratic Casimir on $\cH_\mu$,
\begin{align}
 1= \tr (\phi_a \phi_a) =  \frac {r^2}{c_N^2}\, \tr  (T_a T_a )
  = \frac {r^2}{c_N^2}\, \frac{\dim(\cH_\mu)}{4}(\mu,\mu+2\rho) \quad \text{(no 
sum)} \, .
\end{align}
%  (no sum). 
Thus
\begin{align}
 c_N^2 =  \frac{r^2\dim(\cH_\mu)}{4}(\mu,\mu+2\rho)  = O(N^2)\,,
\end{align}
which means that the $\phi_a$ are almost-commutative functions on 
$\cC[\mu]$,
\begin{align}
 [\phi_\a^{\mathrm{low}},\phi_\b^{\mathrm{low}}] = \frac{1}{c_N^2} c_{\a\b}^a 
T_a   \ 
  &\sim i \{\phi_\a^{\mathrm{low}},\phi_\b^{\mathrm{low}}\} \ \ll 1 
  \label{almost-comm}
\end{align}
where $\{\cdot,\cdot\}$ denotes the Poisson bracket, as usual for 
low-energy functions on fuzzy spaces. Hence their interaction strength is small, and 
they are become free fields as $N\to\infty$. 

To summarize, the vast number of scalar modes for large $N$ becomes gapped on an irreducible $\cC[\mu]$ vacuum, and the zero modes 
consist of stringy modes with interaction strength $g$, as well as weakly interacting semi-classical modes.
We will see below that the stringy modes form bound states which are stable at least for a special value of the mass parameter, 
in which case the zero modes are further reduced to a small number independent of $N$.

% 
%%%%%%%%%%%%%%%%%%%%%%%%%%%%%%%%%%%%%%%%%%%%%%%%%%%%%%%%%%%%%
%%%%%%%%%%%%%%%%%%%%%%%%%%%%%%%%%%%%%%%%%%%%%%%%%%%%%%%%%%%%%
% 
\subsection{Aspects of the Higgs potential}
\label{sec:interactions}
Now consider the interacting potential for the Higgs sector, i.e.\ the zero 
modes $\phi_\a$ on a background solution $X$. The linear term in $\phi$ 
vanishes due to the eom for $X$, so that the effective potential for $\phi$ 
obtained from \eqref{V-full-expand} is 
\begin{equation}
\begin{aligned}
 V_{\mathrm{eff}}[\phi] &\coloneqq V[X+\phi] -V[X] \\
 &\, = V[\phi] + \tr \left( \frac{1}{2} \phi^\a  \big(\Box_X + 2 \Di_{\rm diag} 
\big) \phi_\a 
 + (X^\a + \frac 14\phi^\a)\Box_\phi \phi_\a  - \frac{1}{2}f^2\right)   \; .
\end{aligned}
\end{equation}
For the regular zero modes,
the cubic interaction $V_{31}$ arising from the quartic term drops out. To see this, consider
\begin{align} 
 V_{31}\ \ni\ \tr X^\a\Box_\phi \phi_\a          
 &=  \tr[X_\a,\phi_\b][\phi^\a,\phi^\b]     \qquad\qquad\mbox{(no sum)}\nn\\
  &=  -\tr \phi_\b [[\phi^\a,\phi^\b],X_\a] 
  =  \tr \phi_\b \Big([[\phi^\b,X_\a],\phi^\a]  + [[X_\a,\phi^\a],\phi^\b] \Big)\nn\\ 
  &=  -\tr \phi_\b [[\phi_{-\b},X_\a],\phi^\a] 
  \label{V-phi-rewrite}
\end{align} 
using the Jacobi identity,  $\phi^\b = \phi_{-\b}$, and
the gauge-fixing condition $f = [X_\a,\phi^\a] =0$, which follows from the 
decoupling condition \eqref{extremal-char-1}.
Since either $\a \pm \b \in \cI$ or $\a\pm\b=0$ for any pair of roots $\a,\b$ of 
$\msu(3)$,
the $V_{31}$ term \eqref{V-phi-rewrite} vanishes for the regular zero-modes, 
again due to the decoupling condition.
Therefore $V_{\mathrm{eff}}[\phi]$ reduces for the regular zero mode sector to
\begin{align}
 V_{\mathrm{eff}}[\phi] 
 &=V[\phi]+ \tr \left( \frac 12 \phi^\a  \big(\Box_X + 2 \Di_{\rm diag}^X \big) 
\phi_\a  \right) \, .
 \label{V-phi-eff}
\end{align}
The second term is nothing else than $\frac{1}{2}\tr(\phi^\a 
\cO_{V,M_i\equiv0}^X \phi_\a) $ and vanishes since $\phi_\a$ are regular zero 
modes, see \eqref{D-mix-vanish}.
Hence, the effective potential for the regular zero modes is given by 
\begin{align}
  \cV[\phi] &= \frac {m^4}{g^2} V[\phi]  =\frac{m^4}{g^2}  \tr\left(
  V_4[\phi]+ V_{\rm soft}[\phi]\right) 
 \label{V-reg-zero-noM}
 \end{align}
 and we arrive at
\begin{align}
\cV[X+\phi] =  \cV[X] + \cV[\phi] \ .
\label{V-additivity}
\end{align}
We emphasize that the argument remains valid for any element of the ring of regular 
zero modes:  suppose $\phi$ and $\phi'$ are two six-component vectors of 
regular zero modes, then
\begin{align}
\cV[X+(\phi+\phi')] =  \cV[X] + \cV[\phi+\phi']
\quad \text{and} \quad
\cV[X+(\phi\cdot\phi')] =  \cV[X] + \cV[\phi\cdot\phi']
\end{align}
holds, with the component-wise multiplication defined in section \ref{sec:regular-zm}.
However, $\cV[\phi+\phi']$ or $\cV[\phi\cdot\phi']$ do not 
necessarily decompose in any linear fashion.
Similarly, one can extend to discussion to Higgs modes connecting stacks 
of branes and the argument remains the same. 

In particular, the potential for $\phi$ has the same structure as the original 
potential \eqref{V-soft} for the model.
Thus the quadratic potential $V_2[\phi]$ vanishes again in the absence of mass 
terms, but the cubic term $V_3[\phi]$ may entail some unstable directions.
Some of the $\phi_\a$ are then expected to take a non-trivial VEV, which is 
then stabilized by the quartic term. We will indeed find such non-trivial minima 
for $ \phi_\a$, which  will be denoted as \emph{Higgs vacua}.
Such a non-trivial Higgs linking different branes leads to a \emph{bound 
state} of the branes.

As a further remark, we note that the mixed term quadratic in both $X$ and $\phi$ 
can be written equivalently as
\begin{align}
 \tr \Big( \phi^\a  \big(\Box_X + 2 \Di_{\rm diag}^X \big) \phi_\a \Big)
  = \tr \Big(X^\a  \big(\Box_\phi + 2 \Di_{\rm diag}^\phi \big) X_\a  \Big)
  \label{X-phi-quad}
\end{align}
assuming the decoupling condition between $\phi$ and $X$.

The above argument for \eqref{V-phi-rewrite} to vanish does not apply to the 
exceptional zero modes.
For single branes of $\C P^2$ type, these are precisely 
the $SU(3)_R$ Goldstone bosons, which can be studied separately.
However there exist other exceptional zero modes, e.g.\
$\L\in\cW(1,0), \ \L'\in\cW(2,0)$ (or conjugate) connecting $\cC[(0,1)]$ with $\cC[(1,0)]$
which need to be studied separately.

We will see that 
in the presence of positive mass terms $M_i^2 > 0$,  the above instability can be stabilized.
However if the $M^2_i$ are not all equal, some of the $\phi^\a$ turn out to acquire a negative mass.
This will also lead to a non-trivial Higgs vacuum. 
Quantum corrections\footnote{Another conceivable mechanism 
is a rotation of the branes, see \cite{Steinacker:2014eua}.} might also
play an important role, however we will assume that these are subleading if the branes are sufficiently large,
so that the semi-classical description is valid.
\paragraph{Full equations of motion and decoupling.}

We have just seen that  $\cV(X+\phi) = \cV(X) + \cV(\phi)$ 
for a sum of backgrounds $X$, $\phi$ satisfying the decoupling condition 
\eqref{extremal-char-1}.
Now, we would like to know if such a composed background $ Y^\a = X^\a + 
\phi^\a $ can be an exact solution of the full action.
To address this, we cannot rely on the above reduced action for the Higgs sector, 
because we need $\d S = 0$ for arbitrary fluctuations of $Y^\a$.

Consider combined (static) configurations of the form $X^a + \phi^a$
which satisfies the equation of motion 
\begin{align}
  0=(\Box_{X+\phi} + 4 M_{i}^2)\big)(X+\phi)_i^+ + 4 \varepsilon_{ijk} 
(X+\phi)_j^- (X+\phi)_k^- .
 \label{eom-comb}
\end{align}
To simplify this, consider the matrix Laplacian $\Box_{X+\phi}$ acting on an 
arbitrary $\psi \in \End(\cH)$. We obtain
\begin{align}
  \Box_{X+\phi} \psi &=  (\Box_\phi + \Box_X) \psi
  + [X_\b,[\phi^\b, \psi]] + [\phi_\b,[X^\b, \psi]] \nn\\
 &=  (\Box_\phi + \Box_X) \psi 
  + 2 [\phi_\b,[X^\b, \psi]] - [\psi,[X_\b,\phi^\b]] \; .
\end{align}
Applying this on $X+\phi$ leads to
\begin{align}
  \Box_{X+\phi} (X_\a + \phi_\a) &=  (\Box_\phi + \Box_X) (X_\a+\phi_\a) 
  + 2 [\phi_\b,[X^\b, X_\a]] +  2 [X_\b,[\phi^\b, \phi_\a]] \nn\\
  & \quad - [\phi_\a,[\phi_\b,X^\b]]  - [X_\a,[X_\b,\phi^\b]]  \nn\\
  &=  (\Box_\phi + \Box_X) (X_\a+\phi_\a) + 2 (\Di_{ad}^X \phi)_\a + 2  
(\Di_{ad}^\phi X)_\a  \nn\\
  &\quad  - [\phi_\a,[\phi_\b,X^\b]]  - [X_\a,[X_\b,\phi^\b]]  \ .
  \label{box-x+y}
\end{align}
If $X_\a$ and $\phi_\a$ satisfy the decoupling condition 
\eqref{extremal-char-1}, we can replace $\Di_{\mathrm{ad}}^X \phi 
=\Di_{\mathrm{diag}}^X \phi$ etc., and also $[X_\b,Y^\b] = 0$ holds. 
This further simplifies the Laplacian to
\begin{align}
 \Box_{X+\phi} (X_\a + \phi_\a) &= (\Box_\phi + \Box_X) (X_\a+\phi_\a) + 2 
(\Di_{\mathrm{diag}}^X \phi)_\a + 2  (\Di_{\mathrm{diag}}^\phi X)_\a \ .
 \label{Box-Sum-zeromodes}
\end{align}
Furthermore, we note that the cubic term simplifies via \eqref{extremal-3} to
\begin{align}
 \varepsilon_{ijk} (X+\phi)_j^- (X+\phi)_k^-  
  &= \varepsilon_{ijk} X_j^- X_k^- + \varepsilon_{ijk} \phi_j^- \phi_k^-  + 
\varepsilon_{ijk} [X_j^-, \phi_k^-]  \nn\\
  &= \varepsilon_{ijk} X_j^- X_k^- + \varepsilon_{ijk} \phi_j^- \phi_k^- \; .
\end{align}
Collecting all the intermediate steps, the full eom reduces to
\begin{align}
 0 &= \big(  \Box_\phi + 2\Di_{\rm diag}^\phi\big) X_i^+ +
 \big(\Box_X  + 4 M_i^2 \big) X_i^+ 
 +2  \varepsilon_{ijk}[X_j^-,X_k^-]  \nn\\
  & + \big( \Box_X + 2\Di_{\rm diag}^X\big) \phi_\a
  + \big(\Box_\phi  + 4M_i^2 \big) \phi_i^+
  + 2 \varepsilon_{ijk}[\phi_j^-,\phi_k^-] 
  \label{eom-X+Y}
\end{align}
provided $X$ and $\phi$ satisfy the decoupling conditions 
\eqref{extremal-char-1}.
Now recall that
\begin{subequations}
\begin{align}
 (\Box_X + 2\Di_{ad}^X) \phi_i^+ = \cO_{V,M_i\equiv0}^X \phi_i^+ = 0 
 \quad \text{and} \quad
 (\Box_\phi + 2\Di_{ad}^\phi) X_i^+ = \cO_{V,M_i\equiv0}^\phi X_i^+ = 0
\end{align}
\end{subequations}
because the symmetry of the decoupling conditions implies that $X$ and $\phi$ 
are regular zero modes of $\cO_{V,M_i\equiv0}^\phi$ and $\cO_{V,M_i\equiv0}^X$, 
respectively.
 As a consequence, the full eom for $X+\phi$ reduce to the sum of the 
individual eom for $X$ and $\phi$ separately.
Thus, \emph{$X+\phi$ is an exact solution 
if both $X$ and $\phi$ satisfy their individual equations of motion}
\begin{align}
 \big(\Box_\phi  + 4M_i^2 \big) \phi_i^+
  + 2 \varepsilon_{ijk}[\phi_j^-,\phi_k^-] =0
 \label{eom-phi}
\end{align}
\emph{and similar for $X$, as well as the decoupling condition 
\eqref{extremal-3}.
}

In fact, \eqref{eom-X+Y} are precisely the eom obtain using the above reduced 
potential, 
i.e.\ dropping the mixed $V_{31}$ terms such as $\phi\Box_X X$ and the mixed 
terms in $V_3$.
We can therefore use the reduced potential for backgrounds composed of decoupled  
solutions.
However, note that finding a minimum within the Higgs sector $V(\phi)$ in general implies
\eqref{eom-phi} \emph{only up to massive modes}. This 
is the reason why we will find exact solutions only in special cases within the maximal Higgs sector.

Furthermore, from the homogeneity of the full potential in either $X$ or $\phi$, we 
can infer that $4V_4 + 3 V_3 +2V_2= 0$ holds at a minimum within the Higgs 
sector, 
which implies as in \eqref{energy-homogeneous}
\begin{align}
V[\phi]\big|_{\text{sol. eom}} = \frac{1}{4} 
V_3[\phi]+\frac{1}{2} V_2[\phi] \ .
 \label{energy-homogeneous-higgs}
\end{align}
This is useful to compare the energy of different solutions.
%
%%%%%%%%%%%%%%%%%%%%%%%%%%%%%%%%%%%%%%%%%%%%%%%%%%%%%%%%%%%%%%%%
%%%%%%%%%%%%%%%%%%%%%%%%%%%%%%%%%%%%%%%%%%%%%%%%%%%%%%%%%%%%%%%%
%

\subsection{Massless versus massive solutions, Higgs condensate, and stability 
at \texorpdfstring{$M^*$}{M*}}
\label{sec:Higgs-stab-crit-M}
So far, we have restricted the attention mostly to the massless case, i.e.\ 
$M_i=0$. 
This has some immediate implications: 
\begin{compactenum}[(i)]
 \item The eom \eqref{eq:eom_new} can be 
solved immediately by the \emph{first integral} relations 
$\widetilde{B}_{ij}=0=D$, which are automatically satisfied on any $SU(3)$ 
representation $X_j^\pm = \pi(T_j^\pm)$.
\item The operator $\cO_V^X$ governing 
the dynamics of the bosonic fluctuations is positive semi-definite by virtue of 
representation theory \cite{Steinacker:2014lma}. This means that
there are no instabilities, and moreover there exists a 
classification of all regular zero modes.
\item The potential energy of the background can be 
computed by noting that $\widetilde{B_{ij}}=0$ implies 
$B_{ij}=\frac{1}{3}\varepsilon_{ijk} X_k^-$, so that 
\begin{align}
 V[X]= -\frac{2}{3}\sum_i \tr(X_i^- X_i^+) <0 \; .
\end{align}
\end{compactenum}
Now we would like to perturb the background $X$ with regular zero modes 
$\phi$ to obtain a solution to the full eom for $X+\phi$, keeping $M_i=0$ for 
the moment. Unfortunately, this is less accessible in general, because
\begin{compactenum}[(i)]
 \item Since zero modes $\phi$ 
do not satisfy the $\msu(3)$ Lie algebra,  it is not easy to find exact solutions. 
A notable exception are the maximal regular zero modes on $\cC[(n,0)]$.
\item The spectrum of the operator $\cO_V^{X+\phi}$ is not analytically 
understood, but numerical studies presented in sections 
\ref{sec:4D-branes}--\ref{sec:6D-branes}, and appendix 
\ref{app:solutions_spectrum} show the existence of instabilities. Moreover, we 
do not have a classification of appearing zero modes.
\item Assuming that both 
$X$ and $\phi$ satisfy the first integral relations 
$\widetilde{B}_{ij}=0=D$, the combined potential energy can be evaluate to read
\begin{align}
 V[X+\phi] = -\frac{2}{3}\sum_i \tr(X_i^- X_i^+) -\frac{2}{3}\sum_i 
\tr(\phi_i^- \phi_i^+) < V[X] \;.
\end{align}
Hence starting from a $\cC_N[\mu]$ background, the extension to a combined 
background definitely reduces the potential energy. This strongly suggests the existence of a 
Higgs condensate.

\end{compactenum}
Let us try to circumvent the appearance of instabilities by including uniform 
mass parameters $M_i \equiv M >0$ and adjusting the radii $R_i \equiv R$ in 
\eqref{basic-branes} accordingly. 
Starting with $X_i^\pm$ satisfying $\widetilde{B}_{ij}=0=D$,
the eom \eqref{eq:eom_new} for the ansatz $Y_i^\pm = R X_i^\pm$ reduce to
\begin{align}
0= R \big(R(R-1)+M^2\big) X_i^+ \quad \forall i
\end{align}
Thus there are three solutions 
\begin{align}
 R=0\,, \qquad R = \frac{1}{2} (1\pm \sqrt{1-4M^2}) 
\end{align}
for $|M|\leq \frac{1}{2}$. For these choices of $R$, the eom are satisfied and 
we can compute the potential from \eqref{eq:potential_rewritten}, which is
\begin{align}
 V[Y]=2R^2 \left( \left(R-\frac{2}{3}\right)^2 +2 \left(M^2 -\frac{2}{9}\right) 
\right) \sum_i \tr(X_i^- X_i^+) \; .
\end{align}
As discussed in appendix \ref{app:explicit_sol}, for $0<M\leq 
\frac{\sqrt{2}}{3}$ the solution $R(M) = \frac{1}{2} (1+ \sqrt{1-4M^2})$ 
describes 
the minimal energy configuration. This establishes 
the following statement (as stated in the introduction): 
any solution to $\widetilde{B}_{ij}=0=D$ gives rise to a 
solution to the eom with uniform mass parameter $M$, provided one rescales with 
$R(M)$.

Now we make the following very important observation: for the special mass value
\begin{align}
M^* \coloneqq \frac{\sqrt{2}}{3}
\label{M-star-def}
\end{align}
%%H
the full potential \eqref{eq:potential_rewritten} is 
positive semi-definite, and the potential vanishes if and only if
\begin{align}
 B_{ij}=0 = D \ .
\end{align}
Upon rescaling, this is equivalent to 
$\widetilde{B}_{ij}=0 = D$ with a radius $R=\frac{2}{3}$.
In particular, the spectrum of 
$\cO_V^{Y}$ is guaranteed to be free of instabilities for any such solution.
We will indeed find a number of nontrivial \emph{brane plus Higgs} solution of 
this type, which are thereby local minima up to a compact moduli space.
%%H
This statement is clearly reminiscent of the situation in supersymmetry, 
even though SUSY is explicitly broken by the potential. 

%
%%%%%%%%%%%%%%%%%%%%%%%%%%%%%%%%%%%%%%%%%%%%%%%%%%%%%%%%%%%%%%%%
%%%%%%%%%%%%%%%%%%%%%%%%%%%%%%%%%%%%%%%%%%%%%%%%%%%%%%%%%%%%%%%%
%
\subsection{Gauge fields and mode decomposition}
Now consider the gauge fields $A_\mu(y) \in \End(\cH)$ on $\R^4$, 
which together with the gauginos constitute the $\cN=1$ vector superfield in 
$\cN=4$ SYM.
The gauge fields take values in $\msu(N)$, and accordingly decompose 
\begin{align}
 A_\mu = A_{\mu;\L m}(y) Y_{\L m}
 \label{gauge-mode}
\end{align}
into eigenmodes $ Y_{\L m}$ of the matrix Laplacian $\Box_X$ on squashed 
$\cC_N[\mu]$ background.
These modes  acquire a mass due to the Higgs effect, given by \cite{Steinacker:2014lma}
\begin{align}
- \tr[X^a,Y_{\L m}][X_a,Y_{\L m}] &= \tr (Y_{\L m} ^\dagger \Box_X Y_{\L m})   
=  m^2_{\L,m} \nn\\[1ex]
   m^2_{\L,m}  &=  2 r^2\big((\L,\L+2\rho) - (m,m)\big) 
 \label{gaugemodes-mass}
\end{align}
assuming the normalization
\begin{align}
  \tr (Y_{\L' m'} ^\dagger Y_{\L m}) = \d_{\L m}^{\L' m'} .
  \label{harmonics-normaliz} 
\end{align}
In contrast to the scalar fields, there are no zero modes, and the 
gauge symmetry is broken completely on an irreducible brane.
The lowest of these modes are given by $ Y_{\L m} = \frac{1}{c_N} X_\a$,
and the corresponding KK mass scale is of order
\begin{align}
 m^2_{KK}  \ \sim  \ \frac{r^2}{g^2} m^2 
 \label{KK-mass-gauge}
\end{align}
%%H
re-inserting the YM coupling constant $g^2$ and the cubic coupling $m^2$  in the potential \eqref{V-soft}.

Now consider the coupling of
the above gauge modes  to the Higgs modes or to the fermions.
We recall from section \ref{sec:regular-zm} that the Higgs modes arise in different types,
in particular  (i) maximal (stringy) modes $\phi_\a^{\rm string}$,
and (ii) semi-classical modes $\phi_\a^{\rm low}$. 
A similar classification applies to the gauge modes and fermions.
The effective coupling strength depends strongly on the type of modes. 
For the semi-classical gauge modes 
(these will include $W$-like bosons discussed in section \ref{sec:standardmodel}), the coupling to
a stringy Higgs mode $\phi^{\rm string} = r_s |x\rangle\langle x'|$ 
\eqref{string-modes}  has the structure 
\begin{align}
 [Y_{\L m},\phi^{\rm string}]  \approx (Y_{\L m}(x) - Y_{\L m}(x')) \phi^{\rm string}
\end{align}
leading to the mass term 
\begin{align}
 - \tr[\phi^{\rm string},Y_{\L m}][\phi^{\rm string},Y_{\L m}] &\approx
 |Y_{\L m}(x)- Y_{\L m}(x')|^2 \tr(\phi^{\rm string}\phi^{\rm string})
  = |Y_{\L m}(x)- Y_{\L m}(x')|^2
\end{align}
Hence such stringy Higgs give a  contribution to the mass of nontrivial gauge boson
modes, which is proportional to the difference of the wave function $Y_{\L m}(x)$ 
at the two ends of the string.
However this contribution is suppressed\footnote{In contrast, the coupling 
of the $\phi^{\rm string}$ to the stringy fermions   is not suppressed.} 
by the localization (due to \eqref{harmonics-normaliz}),
and the main contribution to the mass of the gauge bosons arises from 
the background  \eqref{gaugemodes-mass}. 

These observations will apply in particular for the \emph{chiral} $A_\mu \sim 
\chi$ gauge mode \eqref{chiral-gauge-mode} on 6-dimensional branes.

%%H  work out carefully the KK scale, dependence on coupling!

\paragraph{Nonabelian case.}

As usual, unbroken nonabelian gauge fields arise on stacks of coincident branes.
E.g. on a stack of two coinciding branes, the gauge modes have the structure 
\begin{align}
 A_\mu = A_{\mu;\L m}(x) Y_{\L m} \otimes\sigma_i
\end{align}
where $\sigma_i \in \msu(2)$ act on the two branes.
%Then this $\msu(2)$ provides the dominant contribution to their commutators.
Using  $Y_{\L m} = \frac{1}{\sqrt{\dim\cH}}\one$ for 
the massless modes, the 
corresponding $\msu(2)$ coupling constant is found to be smaller than $g$ by a factor
\begin{align}
 \tilde g\sim\frac{g}{\sqrt{\dim\cH}} \ .
\end{align}
This reduction is somewhat related to \eqref{almost-comm}.
%%H

% 
% \paragraph{W-KK-hierarchy}
% 
% 
% We should check whether there is a big hierarchy between the $W$ mass and the KK 
% masses. 
% The $W$ mass $m^2_W \sim g^2 \phi^2 \sim g^2 N$
% arises from the various Higgses, and the trace doesn't contribute because the 
% Higgs 
% is a coherent state linking the corners. 
% This should be much smaller than the KK mass $m_l^2 \sim g^2 l^2 m^2$, which
% arise from the eigenvalues of the $SU(N)$ Casimir $\Box_X$, \emph{weighted by 
% the trace}.
% The latter contributes a huge factor, so that the KK masses should be much 
% larger. Work it out!.
% 
% ***
% Here $g \gg g \sim g/\sqrt{\dim\cH}$ \cite{} 
% is the bare coupling of the $SU(N)$ gauge theory. So this seems to work, good!
% 
% 
% However: the bare coupling $g$ is also the strong $SU(3)$ coupling constant 
% which arises from 3 baryonic point branes ... so it can't be all that strong??
% Problem? 
% 
% 
% Also, the $\phi_{L}^-$ might induce a smaller SSB than the $H$, since the latter 
% couples to corner states where $X^\pm$ is large...
% no, doesn't matter for the electroweak generators, they're $\sim\one$. 

%
%%%%%%%%%%%%%%%%%%%%%%%%%%%%%%%%%%%%%%%%%%%%%%%%%%%%%%%%%%%%%%%%%%%%%%%%%
%%%%%%%%%%%%%%%%%%%%%%%%%%%%%%%%%%%%%%%%%%%%%%%%%%%%%%%%%%%%%%%%%%%%%%%%%
%
\section{4-dimensional branes}
\label{sec:4D-branes}
After the formal discussion of the algebraic properties of the regular zero 
modes, it is now time to show the existence of exact solutions of the form 
brane background $\cC[\mu]$ plus some non-trivial Higgs modes. We begin with 
the 4-dimensional fuzzy branes $\cC[(N,0)]$ in combination with point branes.
\subsection{Single squashed \texorpdfstring{$\C P^2$}{CP2} brane \& Higgs}
\label{sec:single-brane-Higgs}
Now we discuss some of these branes in more detail.
We begin with a squashed brane $\cC[\mu]$ with $\mu = (N,0)$ and add 
some (regular) zero mode(s)
\begin{align}
  Y_\a =  X_\a +  \phi_\a ,
\end{align}
where $X_\a = R \pi_{\mu}(T_\a)$.
First we treat the massless case $M_i^2 = 0$.
\subsubsection{Minimal brane and instability towards \texorpdfstring{$S^2$}{S2}}
For the minimal branes $\cC[\mu]$ with $\mu = (1,0)$, the only non-trivial 
(regular) zero modes are given by
\begin{align}
 \phi_\a = r_\a\pi_{\mu}(T_{-\a}) \ \propto \  X_{-\a}
\end{align}
Since the potential \eqref{V-phi-eff} for $\phi_\a$ is the same as for $X_\a$, 
there is a non-trivial minimum at $r_i = \pm 1$, see \eqref{R-solution-equal}. 
Then the full matrix configuration is
\begin{align}
Y_\a = X_\a \pm  X_{-\a} = \pm Y_{-\a} \ .
\label{minimal-solutions-S2}
\end{align}
This must be an exact solution according to the discussion in section \ref{sec:interactions}, and indeed 
\begin{align}
 [Y_1,Y_2] &=  [X_{\a_1} + X_{-\a_1},X_{\a_2} +  X_{-\a_2}]
  = X_{\a_3} + X_{-\a_3} \
  = Y_3
\end{align}
is nothing but a fuzzy sphere $S^2_n$ with $n=3$, explaining related observations in \cite{Steinacker:2014lma}. 
Its energy is 
\begin{align}
V[X+\phi] = V[X] + V[\phi] = -16
\end{align}
using \eqref{V-minimalCP2} and \eqref{V-additivity}. 
This  appears to be the global  minimum of the classical potential for $N=3$.
We will see below that although squashed $\C P^2$ is not the global minimum, 
it can be (locally) stabilized by  adding a small positive mass term $M_i^2 > 0$. 

\subsubsection{Non-minimal brane with maximal Higgs}
\label{non-mini-brane-Higgs}
The situation is similar, but more interesting for $\cC[\mu]$ with $\mu=(N,0)$. 
Then the (regular) zero modes are given by $\phi_{\a}^{(l)} \propto  
(X_{-\a})^l$ such that the full matrix configuration looks like 
\begin{align}
  Y_\a &=  X_\a + \phi_{\a}^{(l)}.
\end{align}
As before, the $l=1$ mode leads to an exact solution given by an 
irreducible fuzzy 2-sphere with (negative) energy equal twice that of $\cC[\mu]$.
This explains how the known fuzzy $S^2$ solutions are related to squashed $\C P^2$.

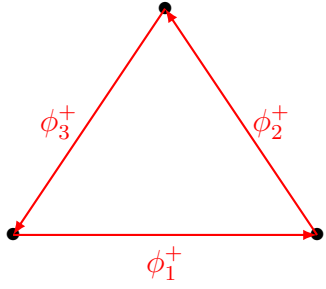
\begin{figure}
\centering
\begin{tikzpicture}
  \coordinate (Origin)   at (0,0);
  \coordinate (P1) at (-2,-1);
  \coordinate (P2) at (0,2);
  \coordinate (P3) at (2,-1);
  \node at (P1) {$\bullet$};
  \node at (P2) {$\bullet$};
  \node at (P3) {$\bullet$};
  \draw [thick,red,-latex] (P1) -- (P3) node[midway,below] {$\phi_1^{+}$};
  \draw [thick,red,-latex] (P3) -- (P2) node[midway, right]  {$\phi_2^{+}$};
  \draw [thick,red,-latex] (P2) -- (P1) node[midway, left]  {$\phi_3^{+}$};
\end{tikzpicture}%
\caption{$\cC[\mu]$ for $\mu=(N,0)$ with maximal Higgs $\phi_i^+$. The 
indicated dots $\bullet$ are the extremal weights of $(N,0)$, which are 
connected by the maximal Higgs $\in \End(\cH_\mu)$.}%
\label{fig:brane-Higgs-edge}%
\end{figure}

However, there are other, more interesting solutions corresponding to small perturbations of the $\C P^2$ branes
localized at their intersections.
For $l \gg 1$, the $\phi_\a^{(l)}$ are string-like modes connecting different sheets of  $\cC[\mu]$, 
confined to the vicinity of the origin.
The most interesting ones are the \emph{maximal} zero modes with $l=N$, 
denoted by 
\begin{align}
  \phi_{\a} &\coloneqq  -r_\a  \frac{1}{N!}{(X_{-\a})}^N \ = \  r_\a \tilde \pi(T_{\a}) , \nn\\
  K(\phi_{\a}) &= (N+1)\a
\end{align}
dropping the superscript $N$. 
Here $\tilde \pi(T_\a)$ is the fundamental representation\footnote{This fixes the normalization $\frac{1}{N!}$} of $\msu(3)$ acting on the 3 
extremal weight states of $(N,0)$. In other words, the $\phi_{\a}$ are connecting the 3 
coherent states on $\C P^2$ located at the origin on each of the 3 sheets of 
squashed $\C P^2$, cf.\ \cite{Steinacker:2014lma}.
More explicitly, denoting these states
as $\{|1,\mu\rangle,|2,\mu\rangle,|3,\mu\rangle\}$, the zero modes have the form
\begin{align} 
  \phi_i^+ &= r_i |i-1, \mu\rangle \langle i,\mu|, \qquad i \sim i+3 \; ,
\end{align}
which are depicted in figure \ref{fig:brane-Higgs-edge}.
Since the potential $V[\phi]$ has the same form as that for the full brane 
we obtain a new exact solution for $r_i  = \pm 1$, given by
\begin{align}
 Y^\a = X^\a + \phi^a
\end{align}
with energy $V[X+\phi] = V[X] -8$ (assuming $M_i=0)$.
This follows again from the full equations of motion \eqref{eom-X+Y} and 
\eqref{V-minimalCP2},
noting that both $X^\a$ and $\phi^\a$ satisfy the eom and the decoupling condition\footnote{
Note that the $X$ can be viewed as regular zero modes w.r.t.\ the $\phi$,
linking the extremal weight states of $\cH_\mu$  to the remaining weight states of $\cH_\mu$ 
(which are point branes w.r.t.\ $\phi_\a$), cf.\ section 
\ref{sec:brane-point-Higgs}.}.
The signs of the 
$r_i$ are of course chosen such that the potential is a minimum.
This solution describes 
a Higgs condensate in the $\C P^2$ background
localized at the intersections, as sketched in figure 
\ref{fig:brane-Higgs-edge} 
and \ref{fig:squashedCP2-Higgs}. The general solutions for the combined system 
of $\cC[\mu]$ brane background with maximal Higgs are discussed in appendix 
\ref{subsec:brane_eom}.
Again, even though that this solution is not the global minimum, we 
will see in section \ref{sec:masses} that it can be locally stabilized by adding a small mass term $M>0$.

We note that for $1 < l< N$, the zero modes $\phi_\a^{(l)}$ 
do not satisfy the $\msu(3)$ algebra, and there is no obvious solutions of the 
form $X + \phi_\a^{(l)}$ at the non-linear level.
\paragraph{Backreaction and exact solution.}
In general, one should worry about the backreaction of the Higgs $\phi$  on the 
background brane $X_\a$.
First, it is important to note that the (almost-) maximal Higgs $\phi$ with $r_\a=\pm 1$ are   
a small perturbation\footnote{This is in contrast to the minimal Higgs with $l=1$.} located at the origin of the $\C P^2$ background,
so that the backreaction is very small for branes $\cC[\mu]$ with large $\mu=(N,0)$.
To see this, 
we compare the matrix elements of $X_\a$ and $\phi_\a$ connecting the states $|i, \mu\rangle$.
For $r_\a = R_\a = \pm 1$,
these are easily see to be
\begin{align}
X_\a |_{|i, \mu\rangle} &=
 \pi_\mu(T_\a)|_{|i, \mu\rangle} = O(\sqrt{N})  \quad \gg \quad
 \phi_\a |_{|i, \mu\rangle} =  \tilde\pi_\mu( T_\a)|_{|i, \mu\rangle} = O(1) 
\,.
\end{align}
Hence, the background generators are larger by a factor of $\sqrt{N}$ and
the backreaction of the background becomes negligible for large $N$.
Moreover, the backreaction vanishes \emph{exactly} for the maximal Higgs modes, as discussed above.
\paragraph{Remark.}
The combined solution $Y^\a = X^\a + \phi^a$ is not only a solution of the potential with cubic term, but also a solution 
of the basic $\cN=4$ potential with a negative mass term, since
\begin{align}
 \Box_Y Y = 4 Y
\end{align}
cf. \eqref{eom-X+Y} and \eqref{Box-Sum-zeromodes}.
The point is that $(\Box_X + 2\Di_{ad}^X) \phi_\a = 0$ follows from the 
decoupling condition for regular zero modes due to \eqref{fluct-equation},
 and similarly $(\Box_\phi + 2\Di_{ad}^\phi) X_\a = 0$ because $X$ is a regular 
zero mode w.r.t.\ $\phi$.  This is quite remarkable, and it means that these 
solutions might arise even without the cubic terms in the action, if a 
negative mass term arises for these modes by quantum fluctuations.
\subsubsection{Mass terms, stabilization and mass-induced Higgs}
\label{sec:masses}
Now consider a single $\C P^2$ brane $(N,0)$ in the presence of a
(sufficiently small) positive mass terms $M_i^2 > 0$. 
Then the radii $R_i$ change according to \eqref{eom-general}. 
\paragraph{Equal masses $M_i^2 = M^2 < \frac{1}{4}$.}
As a first observation, we note that the squashed brane background is 
stabilized 
for sufficiently small equal masses $M_i^2 = M^2$, with 
\begin{align}
 R_M^2 = 1 - M^2 + O(M^4)
\end{align}
due to \eqref{soln-M-equal}. The quadratic potential for such a background is 
\begin{align}
V_{2}[\phi] &= \frac 12R_M^2\, \tr\, \phi^\a \Big(\Box_X + \frac 4{R_M^2} M^2 
+ 2\slashed{D}_{\rm diag} + 2\slashed{D}_{\rm mix} 
 - \frac 4{R_M} \slashed{D}_{\rm mix}\Big) \phi_\a \nn\\
 &= \frac 12R_M^2 \, \tr\, \phi^\a \Big(\Box_X  + \frac 4{R_M^2}M^2 + 2\slashed{D}_{\rm diag} - 2\slashed{D}_{\rm mix} 
 +4 \underbrace{\big(1- \frac 1{R_M}\big)}_{O(M^2)} \slashed{D}_{\rm mix}\Big) \phi_\a 
 \label{V2-M}
\end{align}
using \eqref{Dmix-explicit},
where $\Box_X$ and $\slashed{D}_{\ldots}$ are the \emph{same} operators as for 
$M^2=0$.
For sufficiently small $M^2$, 
the massive modes remain massive, because $\slashed{D}_{\rm mix} $ is clearly 
bounded.
For the regular zero modes, we observe that the condition $\Di_{\rm mix} 
\phi_\a^{(0)} = 0$ of \eqref{D-mix-vanish} is independent of the $R_i$. 
It follows that regular zero modes acquire a positive mass from the 
explicit $M^2$ contribution. 
For the exceptional zero modes, the above argument does not apply;
for instance, on $\C P^2$ the exceptional zero modes are the $SU(3)/U(1)\times 
U(1)$  Goldstone bosons, which remain massless even in the presence of 
explicit (equal) mass terms $M_i^2=M^2$. 
As a consequence, a single $\C P^2$ brane is stable in the presence of 
small equal masses $M_i^2 = M^2$, up to the flat directions due to Goldstone 
bosons.

For further verification, we performed a numerical analysis of the spectrum 
of the vector Laplacian 
$\cO_V^X$ on a $\cC[(N,0)]$ background for $N=1,\ldots,10$. For details, we 
refer to appendix \ref{subsec:brane}, and only summarize the results here.
\begin{itemize}
 \item In the massless case $M_i=0$, we observe $6(N+2)$ zero modes for the 
gauge-fixed $\cO_V^X$. These correpond to 6 Goldstone bosons plus $6(N+1)$ 
regular zero modes, due to \eqref{eq:decomposition_End-CP2}.
%%H
\item In the massive case $M_i \equiv M>0$, we observed $6$ zero modes and 
precisely $6(N+1)$ massive modes with eigenvalue $4 M^2$. See in particular 
figure \ref{fig:CP2-brane_0-modes}.
\end{itemize}
This confirms that small masses are sufficient to stabilize $\C P^2$ branes up 
to Goldstone bosons.

\paragraph{Different (positive) masses $M_i^2 \neq M_j^2$.}
If the $M_i^2$ are positive but distinct, the situation is more interesting.
The massive modes will stay massive for sufficiently small $M_i^2$, such that 
we may focus on the Higgs sector.
However, some of the (would-be) zero modes now acquire a \emph{negative} 
mass.
Continuing the computation \eqref{bosonic-zeromodes-ri}, the  adjusted 
radii lead to an induced mass originating from the vector Laplacian without 
explicit mass terms, i.e.
\begin{align}
 \cO_{V,M_i\equiv 0} \phi_{\a,\L'} 
  &= [H_R,\phi_{\a,\L'}] 
  = - \l(H_R)\,  \phi_{\a,\L'} \ \eqqcolon m_\a^2  \phi_{\a,\L'} 
 \label{bosonic-zeromodes-m}
\end{align}
with $\l$  as in \eqref{zeromodes-weights-Weyl}.
This has one or two negative eigenvalues $m_i$ if the $R_i$ are different, depending on the sign of $ \l(H_R)$.
In fact since $ \l_1+\l_2+\l_3 = 0$ (from the $\Z_3$ symmetry), we obtain the 
following sum rule:
\begin{align}
 m_1^2 + m_2^2 + m_3^2 = 0 \ . 
 \label{eq:sum_rule_ind-masses}
\end{align}
Note that $m_i$ depends on $R_i$ resp. $M_i$ through \eqref{eom-general}.
Taking into account the bare masses \eqref{V-full-expand}, the quadratic part of 
the potential for $\phi$ is
\begin{align}
 V_2[\phi] &= \frac 12 \sum_{i=1}^3 \tr \Big( (m_i^2 + 4M_i^2) 
\phi_{-\a_3,-\L'} \phi_{\a_3,\L'} + \text{h.c.} \Big) \,.
\end{align}
If the brane and $\L$ are sufficiently large, the $m_i$ will dominate $M_i$ 
such that one or two pairs of zero modes acquire a negative mass whenever the 
$M_i$ are different.
This is independent of the cubic term in the potential, and it works even positive bare masses $M_i^2>0$. 
Therefore at least one pair will definitely get a VEV $\langle 
\phi_{\pm\a_i}\rangle \neq 0$, which should in turn lead to a partial stabilization of the zero 
modes. Even though this scenario is interesting since it breaks the $\Z_3$ 
\emph{generation} symmetry, we will mostly 
focus on the case of equal masses in this paper.
%%H

For a $\C P^2_N$ brane, the weights of the zero modes are $\l_i = k \a_i$ with 
$k \leq N$, see \eqref{zeromodes-examples}. 
Then switching on one $M_3^2 > 0$ leads to two negative induced mass terms 
$m_1^2 = m_2^2 = - O(k M_3^2) < 0$, see for instance
\eqref{mi-induced-oneM}. 
\subsubsection{Stability of the brane-Higgs system}
\label{sec:Higgs-stability}

Now consider the non-trivial solutions $X + \phi^{(l)}$ 
found in section \ref{non-mini-brane-Higgs} involving
maximal zero modes $\phi_{\a}^{(l)}$.
An important question is whether these new solutions 
are stable, or if there are further zero modes or instabilities. 
Consider the brane $\cC[\mu]$ for $\mu = (N,0)$ with 
maximal Higgs solution $\phi_{\a} = c (X_{-a})^N$ as above, and add first  
some additional zero modes $\phi_{\a}^{(l)}$ which we assume \emph{not} to be 
maximal. Thus 
\begin{align}
 X_\a + \phi_\a + \phi_\a, \qquad \phi_\a = \sum c_{l,\a} \phi_{\a}^{(l)} , \qquad  l \leq N .
\end{align}
According to \eqref{V-phi-eff}, the effective potential for the combined 
perturbation is given by $V(\phi + \phi)$. 
Since $\phi$  has the structure of a squashed $\C P^2$ brane,
we know that the masses of $\phi$ arising from 
this background are non-negative  \cite{Steinacker:2014lma}.
However, we have to admit the most general fluctuations $\phi_\a$ here, not only 
zero modes. Then the linear term in $\phi$ still vanishes identically, because 
$X+\phi$ is an exact solution of \eqref{eom-X+Y}.
The quadratic part of the potential for $\phi$ is given by\footnote{We can 
assume that $f^2$ is canceled by the gauge fixing term.}
\begin{align}
 V_2^{X+\phi}[\phi] 
 = \frac{1}{2}\tr \Big(\phi\big(\cO_V^X + \cO_V^\phi + 2[X_\b,[\phi^\b,\cdot]] 
+ 2\slashed{D}_{\mathrm{ad}}^{X\phi}\big) \phi \Big)
\end{align}
using $[\phi^\b,X_\b] = 0$. 
%%H
In general, it is not clear and, according to numerical studies, also not true 
that this is a positive semi-definite bilinear form. 
However in the presence of suitable masses $M_i=M$, and notably for $M=M^*$, 
it follows using the results in section \ref{sec:Higgs-stab-crit-M} that the 
(brane +Higgs) solution is indeed a local minimum up to a compact moduli space.
\paragraph{Numerical results.}
We investigated the stability of $\cC[(N,0)]$ branes with maximal Higgs 
numerically by analyzing the spectrum of the 
vector Laplacian, see appendix \ref{subsec:brane} for the details.
We considered the massless case $M_i=0$ as well as the massive case $M_i=M$.
To summarize, we find the following for the fluctuations around $(X+\phi)$:
\begin{itemize}
 \item For the $\cC[(1,0)]$ brane with maximal regular zero modes, there are no 
negative modes, for both $M=0$ and $M>0$. This is clear because it is a fuzzy 
sphere. Moreover, for $M=0$ we find $11$ zero modes, while for $M>0$ there are 
$5$ zero modes. This is the expected number of Goldstone bosons for the fuzzy 
2-sphere as an $SU(2)$ symmetry is manifest, leaving $8-3=5$ broken generators 
for $SU(3)\slash SU(2)$.
 \item For the $\cC[(2,0)]$ brane with maximal regular zero modes, there are no 
negative modes. This is very remarkable. In addition, there are $20$ zero modes 
for $M=0$ and $8$ zero-modes for $M>0$.
  \item For the $\cC[(N,0)]$ brane with $N\geq 3$ with maximal regular zero 
modes, there exist $3(N-2)$ negative modes and $20$ zero modes for $M=0$. 
Although the number of negative modes increases with $N$, the amplitude of 
the negative eigenvalues decreases, 
%%H
and it appears that the mass $M$ required to lift them approaches zero for large $N$. 
This is exemplified in figure 
\ref{fig:neg_ev_n-0_brane} for $N =3,\ldots, 10$. Adding a mass term with 
$M\gtrsim 0.2$ lifts all negative modes for $N\geq 3$.
\end{itemize}
Therefore, uniform mass terms $M_i\equiv M$ are sufficient to stabilize the $\C 
P^2$ brane plus maximal Higgs system up to $8=6+2$ zero modes.
Remarkably, the properties such as number of zero modes, seem to be 
independent of the brane size in the massive case $M>0$.
Hence the (brane+Higgs) system has  a well-behaved scaling limit $N\to \infty$, 
and is stable up to a compact moduli space (which could be lifted by lifting the degeneracy of the $M_i$).
%%H
%with only a finite number of low-energy degrees of freedom. 
This is clearly a very interesting result, which will be seen throughout this paper.

%
%%%%%%%%%%%%%%%%%%%%%%%%%%%%%%%%%%%%%%%%%%%%%%%%%%%%%%%%%%%%%%%%%
%%%%%%%%%%%%%%%%%%%%%%%%%%%%%%%%%%%%%%%%%%%%%%%%%%%%%%%%%%%%%%%%%
%
\subsection{Single squashed \texorpdfstring{$\C P^2$}{CP2} brane with a point 
brane} 
\label{sec:brane-point-Higgs}
\begin{figure}[t!]
\centering
 \begin{subfigure}{0.32\textwidth}
\centering
\begin{tikzpicture}
  \coordinate (Origin)   at (0,0);
  \coordinate (P1) at (-1,-0.5);
  \coordinate (P2) at (0,1);
  \coordinate (P3) at (1,-0.5);
  \node at (Origin) {$\bigstar$};
  \node at (P1) {$\bullet$};
  \node at (P2) {$\bullet$};
  \node at (P3) {$\bullet$};
\draw[thick,red] (P1)--(P2)--(P3)--cycle;
\end{tikzpicture}%
\caption{}%
\label{fig:sol_CP2-point_Higgs}%
\end{subfigure}
 \begin{subfigure}{0.32\textwidth}
\centering
\begin{tikzpicture}
  \coordinate (Origin)   at (0,0);
  \coordinate (P1) at (-1,-0.5);
  \coordinate (P2) at (0,1);
  \coordinate (P3) at (1,-0.5);
  \node at (Origin) {$\bigstar$};
  \node at (P1) {$\bullet$};
  \node at (P2) {$\bullet$};
  \node at (P3) {$\bullet$};
\draw[thick,red] (P1)--(Origin)--(P2) -- (P3)-- (P1) --(P2) ;
\end{tikzpicture}%
\caption{}%
\label{fig:sol_CP2-point_interpolate}%
\end{subfigure}
 \begin{subfigure}{0.32\textwidth}
\centering
\begin{tikzpicture}
  \coordinate (Origin)   at (0,0);
  \coordinate (P1) at (-1,-0.5);
  \coordinate (P2) at (0,1);
  \coordinate (P3) at (1,-0.5);
  \node at (Origin) {$\bigstar$};
  \node at (P1) {$\bullet$};
  \node at (P2) {$\bullet$};
  \node at (P3) {$\bullet$};
\draw[dotted] (P1)--(P3)--(P2);
\draw[thick,red] (P1)--(Origin)--(P2)--cycle;
\end{tikzpicture}%
\caption{}%
 \label{fig:sol_CP2-point_triangle}%
\end{subfigure}
\caption{One-parameter family of solutions interpolating between the maximal 
Higgs solution (\subref{fig:sol_CP2-point_Higgs}) and the connected 
configuration (\subref{fig:sol_CP2-point_triangle}) between $\cD$ and 
$\cC[\mu$]. The intermediate states (\subref{fig:sol_CP2-point_interpolate}) 
have different magnitudes for the  fields involved.}
\label{fig:sol_CP2-point}
\end{figure}
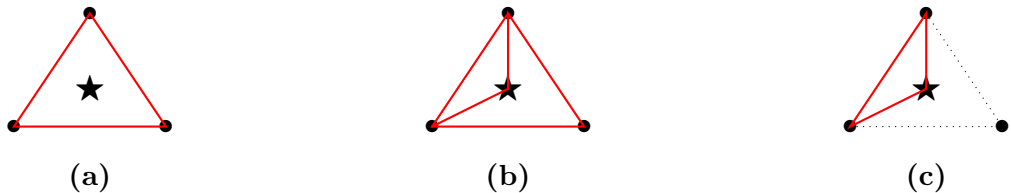

Now consider a $\cC[\mu]$ brane and add a point brane $\cD\equiv \cC[0]$.
We will show that there are non-trivial new vacua which involve a non-vanishing 
Higgs zero modes linking  $\cC[\mu]$ and  $\cD$.
We choose $\mu=(N,0)$ to be specific; the discussion for $\mu=(0,N)$ would be 
analogous.
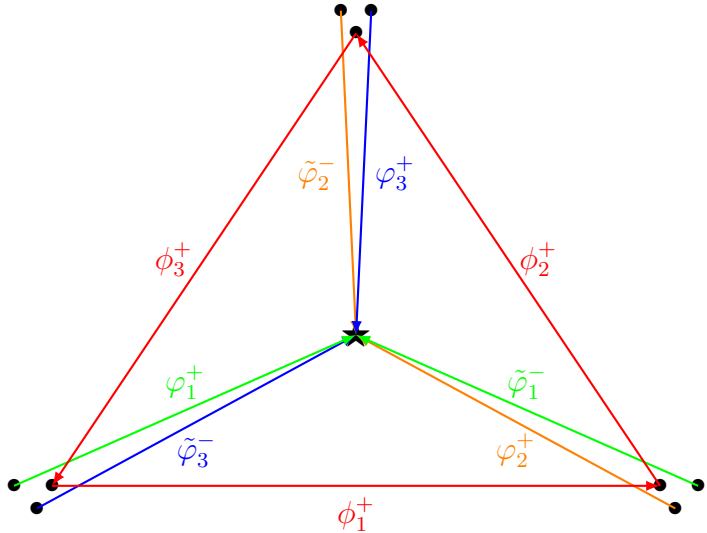
\begin{figure}[t!]
\centering
\begin{tikzpicture}
  \coordinate (Origin)   at (0,0);
  \coordinate (P1) at (-4,-2);
  \coordinate (P2) at (0,4);
  \coordinate (P3) at (4,-2);
  \coordinate (P1+a3) at (-4-0.2,-2-0.3);
  \coordinate (P1-a1) at (-4-0.5,-2);
  \coordinate (P2+a2) at (0-0.2,4+0.3);
  \coordinate (P2-a3) at (0+0.2,4+0.3);
  \coordinate (P3+a1) at (4+0.5,-2);
  \coordinate (P3-a2) at (4+0.2,-2-0.3);
  \node at (Origin) {$\bigstar$};
  \node at (P1) {$\bullet$};
  \node at (P2) {$\bullet$};
  \node at (P3) {$\bullet$};
  \node at (P1+a3) {$\bullet$};
  \node at (P1-a1) {$\bullet$};
  \node at (P2+a2) {$\bullet$};
  \node at (P2-a3) {$\bullet$};
  \node at (P3+a1) {$\bullet$};
  \node at (P3-a2) {$\bullet$};
  \draw [thick,blue,-latex] (P1+a3) -- (Origin) node[midway,below] 
{$\tilde{\varphi}_3^{-}$};%
  \draw [thick,orange,-latex] (P3-a2) -- (Origin) node[midway,below] 
{$\varphi_2^{+}$};%
  \draw [thick,green,-latex] (P1-a1) -- (Origin)node[midway,above] 
{$\varphi_1^{+}$};%
  \draw [thick,orange,-latex] (P2+a2) -- (Origin) node[midway,left] 
{$\tilde{\varphi}_2^{-}$};%
  \draw [thick,green,-latex] (P3+a1) -- (Origin) node[midway,above] 
{$\tilde{\varphi}_1^{-}$};%
  \draw [thick,blue,-latex] (P2-a3) -- (Origin) node[midway,right] 
{$\varphi_3^{+}$};%
  \draw [thick,red,-latex] (P1) -- (P3) node[midway,below] {$\phi_1^{+}$};
  \draw [thick,red,-latex] (P3) -- (P2) node[midway, right]  {$\phi_2^{+}$};
  \draw [thick,red,-latex] (P2) -- (P1) node[midway, left]  {$\phi_3^{+}$};
\end{tikzpicture}%
\caption{Various regular zero modes for 
$\cC[(N,0)]$ brane + point brane 
$\cD$. The $\bigstar$ represents $\cD$, while $\bullet$ denote the extremal 
weights of $(N,0)$. The two shifted two points in the vicinity 
of each extremal weight state indicate the polarization of the zero 
mode. The $\phi_i^+$ are maximal regular zero modes on $\cC[(N,0)]$, while 
$\varphi_i^+$ and $\tilde{\varphi}_i^-$ are  regular zero modes 
connecting $\cC[(N,0)]$ and $\cD$. Note that the conjugate modes, 
corresponding to arrows pointing the opposite directions, are not depicted 
here.}%
\label{fig:CP2-point}%
\end{figure}
The Higgs modes given by the regular 
zero modes discussed above are illustrated in figure \ref{fig:CP2-point}.
The inter-brane regular zero modes, originating from $\Hom(\cH_\mu,\C)$, 
separate into 3+3 independent Higgs $\varphi_i^+$
and \emph{mirror} Higgs $\tilde{\varphi}_i^-$, given by
\begin{align}
 \varphi_i^{+} &= |i\rangle\langle 0| \qquad   \in  \  (N,0)\ , \ \t=+1   
\nn\\
 \tilde{\varphi}_i^{-} &= |i-1\rangle\langle 0|  \ \in  \  (N,0)\ , \ 
\t=-1 \ .
\end{align}
distinguished by the $\tau$-parity \eqref{tau-parity}.
Note that these determine their conjugate modes $(\varphi_i^{+})^\dagger \sim |0\rangle\langle i|$ etc. living in 
$\Hom(\C,\cH_\mu)$.
Taking into account the maximal intra-brane Higgs $\phi_i^\pm$ on $\cC[\mu]$ as 
discussed above,
we find non-trivial solutions where such links $\varphi_i^+$ 
are switched on with different strength, i.e.\ the 
point brane is connected to the $\cC[\mu]$ brane, but the energy is the same. 
For example, we can switch on one triangle consisting of two $\varphi_i$, 
$\tilde{\varphi}_i$
linking $\cD$ with two corners of $\cC[\mu]$, connected by a maximal Higgs 
$\phi_i$ of $\cC[\mu]$ as in figure \ref{fig:sol_CP2-point}.
The existence of such  solutions is explained by an $SU(2)$ symmetry 
within the zero mode sector, rotating the point brane and one 
corner. This leads to a moduli space parametrized by some linking angles 
$\theta$.

The binding energy for all these configuration is $V[\phi] =-8$.
Numerical investigations indicate that these are indeed the global minima of 
this sub sector of the full zero mode sector.
In particular there seems to be no solution which respects the $\Z_3$ symmetry,
except for $\varphi = 0$.
%%H
In other words, the \emph{generation symmetry} $\Z_3$ is spontaneously broken.

This means that although the $\cD$ brane is connected to $\cC[\mu]$ by some 
Higgs as in figure \ref{fig:sol_CP2-point_triangle}, the energy is nevertheless 
degenerate 
to the case where only the maximal intra-brane Higgs on $\cC[\mu]$ are switched 
on, see figure 
\ref{fig:sol_CP2-point_Higgs}. This degeneracy is manifest in the existence of the
continuous family of solutions of figure 
\ref{fig:sol_CP2-point_interpolate} 
that interpolates between these configurations. Hence there is no binding 
energy, rather there is a flat 
direction in moduli space where the point brane may or may not be attached. The 
details of the exact solutions to the equations of motions are presented in 
appendix \ref{subsec:brane+point_eom}.

\paragraph{Numerical results.}
We studied numerically the stability of the combined solutions 
consisting of $\cC[(N,0)]$ brane plus point brane $\cD$ together with some 
Higgs modes, with respect to arbitrary fluctuations.
The details are provided in appendix \ref{subsec:brane+point_spectrum}. 
It turns out that for $M=0$ there are typically a number of negative modes, i.e.\ they are unstable 
towards some of the (originally massive) deformation modes. This is similar to 
the situation in section \ref{sec:Higgs-stability}.
Again, these instabilities can be stabilized by adding a (small) mass term $M$ to 
the background, see in particular figure \ref{fig:brane+pt_spectrum}. 
Then the above solutions, comprising the $\cC[\mu]$ brane plus 
point brane and several Higgs, are still exact solutions with adjusted radii 
according to \eqref{eom-general}. 
Here, we 
observe different qualitative behavior 
between the configurations of figure \ref{fig:sol_CP2-point_Higgs} and 
\ref{fig:sol_CP2-point_interpolate}, \ref{fig:sol_CP2-point_triangle}.
\begin{itemize}
 \item The triangle of maximal intra-brane Higgs of figure 
\ref{fig:sol_CP2-point_Higgs} can be stabilized with small masses, consistent with 
the results in section \ref{sec:Higgs-stability}. 
The number of zero modes becomes $14$ in the massive case, independent of the brane 
size.
  \item The interpolating state of figure \ref{fig:sol_CP2-point_interpolate} 
can only be stabilized with masses almost saturating the allowed values (and 
only for large enough branes $N\geq5$). In addition, the number of zero-modes 
in the massive case becomes $9$, independent of the brane size.
  \item The connected configuration of figure \ref{fig:sol_CP2-point_triangle} 
can also be stabilized only with relatively large mass values, but negativity 
of the eigenvalues is qualitatively different to the interpolating case. Again, 
the number of zero modes stabilizes at $9$ in the massive case.
\end{itemize}

Consistent with the general results in section \ref{sec:Higgs-stab-crit-M}, 
we observe that there are no instabilities for the critical mass $M^*$.
The number of zero modes exceeds the expected 6
zero modes corresponding to the 6 Goldstone bosons of the broken $SU(3)_R$. 
One may hope that these remaining zero modes are 
lifted by introducing different masses $M_i$, but we did not verify this 
explicitly.
% 
%
%%%%%%%%%%%%%%%%%%%%%%%%%%%%%%%%%%%%%%%%%%%%%%%%%%%%%%%%%%%%%%%%%%%
%%%%%%%%%%%%%%%%%%%%%%%%%%%%%%%%%%%%%%%%%%%%%%%%%%%%%%%%%%%%%%%%%%%
%
\subsection{Two squashed \texorpdfstring{$\C P^2$}{CP2} branes \& Higgs}
\label{sec:two-branes-Higgs}
Now consider a system of two non-identical $\C P^2$ branes, such as  $\cC[\mu_L]$ and  
$\cC[\mu_R]$. 
Then  
\begin{equation}
\label{eq:algebra_two_branes}
\begin{aligned}
 \End(\cH_L\oplus\cH_R) &= \cA_{LL} \oplus \cA_{RR} \oplus \cA_{LR} 
\oplus\cA_{RL} \, ,\\
\cA_{LL}&= \End(\cH_{\mu_L}) \, , \quad \cA_{LR}= \Hom(\cH_{\mu_L},\cH_{\mu_R}) 
\, , \quad \text{etc.} 
\end{aligned} 
\end{equation}
contains various types of zero modes.
Besides the intra-brane zero modes discussed above, there are additional modes 
$\phi_{LR} \in \cA_{LR}$ etc. connecting $\cC[\mu_L]$ with  $\cC[\mu_R]$; 
those are the most interesting ones as we will see. We will explicitly find such 
solutions.

\begin{figure}[t!]
\centering
\begin{tikzpicture}
  \coordinate (Q1) at (-1,-0.5);
  \coordinate (Q2) at (0,1);
  \coordinate (Q3) at (1,-0.5);
  \coordinate (P1) at (-4,-2);
  \coordinate (P2) at (0,4);
  \coordinate (P3) at (4,-2);
  \coordinate (P1+a3) at (-4-0.2,-2-0.3);
  \coordinate (P1-a1) at (-4-0.5,-2);
  \coordinate (P2+a2) at (0-0.2,4+0.3);
  \coordinate (P2-a3) at (0+0.2,4+0.3);
  \coordinate (P3+a1) at (4+0.5,-2);
  \coordinate (P3-a2) at (4+0.2,-2-0.3);
  \node at (Q1) {$\bullet$};
  \node at (Q2) {$\bullet$};
  \node at (Q3) {$\bullet$};
  \node at (P1) {$\bullet$};
  \node at (P2) {$\bullet$};
  \node at (P3) {$\bullet$};
  \node at (P1+a3) {$\bullet$};
  \node at (P1-a1) {$\bullet$};
  \node at (P2+a2) {$\bullet$};
  \node at (P2-a3) {$\bullet$};
  \node at (P3+a1) {$\bullet$};
  \node at (P3-a2) {$\bullet$};
  \draw [thick,green,-latex] (P1+a3) -- (Q3) node[midway,below] 
{$\varphi_1^{+}$};%
  \draw [thick,green,-latex] (P3-a2) -- (Q1) node[midway,below] 
{$\tilde{\varphi}_1^{-}$};%
  \draw [thick,blue,-latex] (P1-a1) -- (Q2) node[midway,left] 
{$\tilde{\varphi}_3^{-}$};%
  \draw [thick,blue,-latex] (P2+a2) -- (Q1) node[midway,left] 
{$\varphi_{3+}$};%
  \draw [thick,orange,-latex] (P3+a1) -- (Q2) node[midway,right] 
{$\varphi_2^{+}$};%
  \draw [thick,orange,-latex] (P2-a3) -- (Q3) node[midway,right] 
{$\tilde{\varphi}_2^{-}$};%
  \draw [thick,red,-latex] (P1) -- (P3) node[midway,below] {$\phi_1^{+}$};
  \draw [thick,red,-latex] (P3) -- (P2) node[midway, right]  {$\phi_2^{+}$};
  \draw [thick,red,-latex] (P2) -- (P1) node[midway, left]  {$\phi_3^{+}$};
  \draw [thick,-latex] (Q1) -- (Q3) node[midway,below] {$\phi_1^{+}$};
  \draw [thick,-latex] (Q3) -- (Q2) node[midway, left]  {$\phi_2^{+}$};
  \draw [thick,-latex] (Q2) -- (Q1) node[midway, below]  {$\phi_3^{+}$};
\end{tikzpicture}%
\caption{Parallel branes $\cC[\mu_R]$ and $\cC[\mu_L]$, with 
$\cC[\mu_L]$ being the outer brane. We only indicate the polarization for the 
outermost brane by two additional $\bullet$ next to each extremal weight state. 
The maximal intra-brane Higgs are denoted as 
$\phi_i^+$ and $\phi_i^+$ for $\cC[\mu_L]$ and $\cC[\mu_R]$, respectively. The 
maximal regular inter-brane zero modes are labeled by $\varphi_i^+$ and 
$\tilde{\varphi}_i^-$, and we only draw the $\cA_{LR}$ sector here.}%
\label{fig:Brane+brane}%
\end{figure}
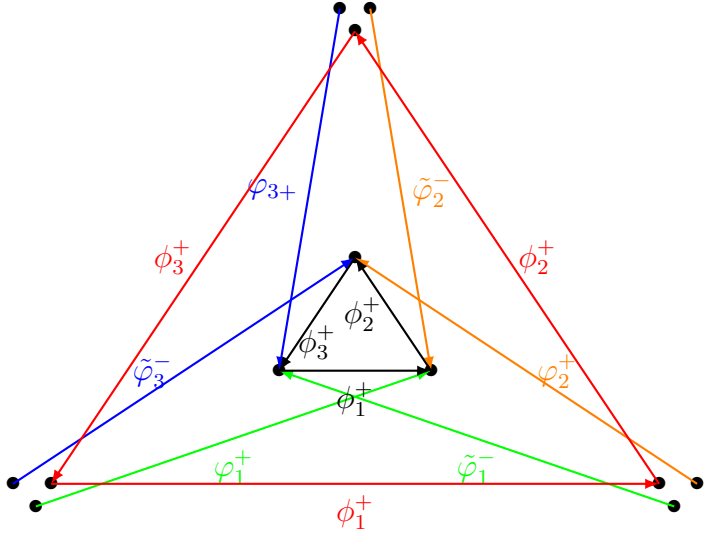

Consider the case of parallel branes $\cC[\mu_L]$ and $\cC[\mu_R]$ with  
$\mu_L=(N,0)$ and $\mu_R = (l,0)$,
possibly connected by some Higgs, as in figure \ref{fig:Brane+brane}.
This set-up is interesting because we will find maximal Higgs connecting 
different branes, which are exact solutions at the non-linear level. These Higgs
clearly breaks the $U(1)\times U(1)$ gauge symmetry on the two branes down to the diagonal $U(1)$,
and lead to various Yukawa couplings of the fermionic zero modes.
%%H
This is the mechanism we are interested in, even though  the present background 
may not yet be very interesting physically.
Note that $n$ identical branes would lead to $\mmu(n)$-valued fields.

Besides the algebras of functions  $\cA_{LL}$ and $\cA_{RR}$  on $\cC[\mu_L]$ 
and $\cC[\mu_R]$, respectively, the full 
algebra of functions contains the following intertwining part: 
\begin{align}
 \cA_{LR} =  (N,l) \oplus (N-1,l-1)\oplus  \ldots  \oplus (N-l,0) 
\end{align}
and similarly $\cA_{RL}$.
There is again a special class of $3+3$ \emph{maximal} Higgs  modes  arising 
from the extremal states 
of $(N,l)$, denoted by
\begin{align}
 \tilde{\varphi}_{i}^- &\in \ (N,l) \subset \cA_{LR} \qquad \mbox{mirror 
Higgs} \nn\\
 \varphi_{i}^+  &\in \ (N,l) \subset \cA_{LR} \qquad \mbox{chiral Higgs}
\end{align}
which connect the corners of $\mu_L$ and $\mu_R$. This is displayed in 
figure \ref{fig:Brane+brane}. Their conjugate modes are given by 
\begin{align}
 (\tilde{\varphi}_{i}^-)^\dagger = \tilde{\varphi}_{i}^+ \in    \cA_{RL} 
 \quad \text{and} \quad 
 (\varphi_{i}^+)^\dagger = \varphi_{i}^- \in    \cA_{RL} \; .
\end{align}
We will see below that each of these Higgs $\tilde{\varphi}_{i}^-$, 
$\varphi_{i}^+$  give rise to precisely one Yukawa coupling 
between fermionic zero modes linking a point brane with either of the branes.

Additionally, there exist the maximal intra-brane Higgs $\phi_i^\pm \in 
\cA_{LL}$ and  
$\phi_{i}^\pm \in \cA_{RR}$ along some edge of $\cC[\mu_L]$ and $\cC[\mu_R]$, 
respectively.
As summarized in figure \ref{fig:Brane+brane}, altogether we have the 
following 
maximal Higgs modes 
\begin{align}
 \tilde{\varphi}_{i}^- = (\tilde{\varphi}_{i}^+)^\dagger, \quad  
 \varphi_{i}^+ = (\varphi_{i}^-)^\dagger , \quad  
 \phi_i^+ = (\phi_i^-)^\dagger , \quad
 \phi_i^+ = (\phi_i^-)^\dagger
\end{align}
which connect the extremal weights and therefore form a closed 
algebra. 
Our aim is to find stable non-trivial solutions on top of $\cC[\mu_L]$ and 
$\cC[\mu_R]$, where some of these are switched on.
\paragraph{Solutions.}
It is clear that there are non-trivial solutions within the maximal Higgs 
sector involving one closed triangle and we have exemplified such cases in 
figure \ref{fig:2branes_1triangle}.
There exist several continuously parametrized solutions that 
interpolate between a maximal intra-brane Higgs on one brane and a closed 
triangle between the two branes. We refer to figures 
\ref{fig:2branes_1triangle_smallMaxHiggs}--\ref{%
fig:2branes_1triangle_smallBound} and 
figures \ref{fig:2branes_1triangle_smallMaxHiggs}--\ref{%
fig:2branes_1triangle_smallBound} for two representative cases.

In addition, as shown in figure \ref{fig:2branes_2triangle_maxHiggs}, there 
exists an exact solution of the 
form $X + \phi +\phi$ involving the full brane background plus maximal 
intra-brane Higgs on $\cC[\mu_L]$ and $\cC[\mu_R]$ simultaneously.
Moreover, one can non-trivially combine the triangular configurations of 
\ref{fig:2branes_1triangle_smallBound} and \ref{fig:2branes_1triangle_bigBound} 
and obtains another continuous family, see 
figure \ref{fig:2branes_2triangle}, which always corresponds to a configuration 
of 
two closed triangles. 

The details of the numerical combinations of the 
several Higgs fields that give rise to exact solutions to the equations of 
motions are presented in appendix \ref{subsec:brane+brane}.
As far as the potential energy is concerned, 
the (degenerate) configurations of figure \ref{fig:2branes_1triangle} count only as one closed 
triangle, whereas 
the (degenerate) configurations of figure \ref{fig:2branes_2triangle} have two 
independent triangles, and consequently have lower potential energy.
That being said and recalling the $\Z_3$ symmetry, it is clear that the states of lowest 
energy, i.e.\ those equivalent to two closed triangles in figure \ref{fig:2branes_2triangle}, are 
highly degenerate. These solutions should be interpreted as 2 branes linked by some Higgs, 
but the binding energy is again zero due to the degeneracy.

\begin{figure}[t!]
\centering
\begin{subfigure}{0.32\textwidth}
\centering
\begin{tikzpicture}
  \coordinate (Q1) at (-0.5,-0.25);
  \coordinate (Q2) at (0,0.5);
  \coordinate (Q3) at (0.5,-0.25);
  \coordinate (P1) at (-2,-1);
  \coordinate (P2) at (0,2);
  \coordinate (P3) at (2,-1);
  \node at (Q1) {$\bullet$};
  \node at (Q2) {$\bullet$};
  \node at (Q3) {$\bullet$};
  \node at (P1) {$\bullet$};
  \node at (P2) {$\bullet$};
  \node at (P3) {$\bullet$};
\draw[red,thick] (Q1)--(Q2)--(Q3)--cycle;
\draw[dotted] (P1)--(P2)--(P3)--cycle;
\end{tikzpicture}%
\caption{}%
\label{fig:2branes_1triangle_smallMaxHiggs}
\end{subfigure}
\begin{subfigure}{0.32\textwidth}
\centering
\begin{tikzpicture}
  \coordinate (Q1) at (-0.5,-0.25);
  \coordinate (Q2) at (0,0.5);
  \coordinate (Q3) at (0.5,-0.25);
  \coordinate (P1) at (-2,-1);
  \coordinate (P2) at (0,2);
  \coordinate (P3) at (2,-1);
  \node at (Q1) {$\bullet$};
  \node at (Q2) {$\bullet$};
  \node at (Q3) {$\bullet$};
  \node at (P1) {$\bullet$};
  \node at (P2) {$\bullet$};
  \node at (P3) {$\bullet$};
\draw[blue,thick] (Q1)--(P2)--(Q3);
\draw[red,thick] (Q1)--(Q2)--(Q3);
\draw[magenta,thick] (Q1)--(Q3);
\draw[dotted] (P2)--(P1)--(P3) --cycle;
\end{tikzpicture}%
\caption{}%
\label{fig:2branes_1triangle_smallInt}
\end{subfigure}
\begin{subfigure}{0.32\textwidth}
\centering
\begin{tikzpicture}
  \coordinate (Q1) at (-0.5,-0.25);
  \coordinate (Q2) at (0,0.5);
  \coordinate (Q3) at (0.5,-0.25);
  \coordinate (P1) at (-2,-1);
  \coordinate (P2) at (0,2);
  \coordinate (P3) at (2,-1);
  \node at (Q1) {$\bullet$};
  \node at (Q2) {$\bullet$};
  \node at (Q3) {$\bullet$};
  \node at (P1) {$\bullet$};
  \node at (P2) {$\bullet$};
  \node at (P3) {$\bullet$};
\draw[blue,thick] (Q1)--(P2)--(Q3)--cycle;
\draw[dotted] (Q1)--(Q2)--(Q3);
\draw[dotted] (P2)--(P1)--(P3) --cycle;
\end{tikzpicture}%
\caption{}%
\label{fig:2branes_1triangle_smallBound}
\end{subfigure}
\begin{subfigure}{0.32\textwidth}
\centering
\begin{tikzpicture}
  \coordinate (Q1) at (-0.5,-0.25);
  \coordinate (Q2) at (0,0.5);
  \coordinate (Q3) at (0.5,-0.25);
  \coordinate (P1) at (-2,-1);
  \coordinate (P2) at (0,2);
  \coordinate (P3) at (2,-1);
  \node at (Q1) {$\bullet$};
  \node at (Q2) {$\bullet$};
  \node at (Q3) {$\bullet$};
  \node at (P1) {$\bullet$};
  \node at (P2) {$\bullet$};
  \node at (P3) {$\bullet$};
\draw[red,thick] (P1)--(P2)--(P3)--cycle;
\draw[dotted] (Q2)--(Q1)--(Q3) --cycle;
\end{tikzpicture}%
\caption{}%
\label{fig:2branes_1triangle_bigMaxHiggs}
\end{subfigure}
\begin{subfigure}{0.32\textwidth}
\centering
\begin{tikzpicture}
  \coordinate (Q1) at (-0.5,-0.25);
  \coordinate (Q2) at (0,0.5);
  \coordinate (Q3) at (0.5,-0.25);
  \coordinate (P1) at (-2,-1);
  \coordinate (P2) at (0,2);
  \coordinate (P3) at (2,-1);
  \node at (Q1) {$\bullet$};
  \node at (Q2) {$\bullet$};
  \node at (Q3) {$\bullet$};
  \node at (P1) {$\bullet$};
  \node at (P2) {$\bullet$};
  \node at (P3) {$\bullet$};
\draw[red,thick] (P1)--(P2)--(P3);
\draw[blue,thick] (P1)--(Q2)--(P3);
\draw[magenta,thick] (P1)--(P3);
\draw[dotted] (Q2)--(Q1)--(Q3) --cycle;
\end{tikzpicture}%
\caption{}%
\label{fig:2branes_1triangle_bigInt}
\end{subfigure}
\begin{subfigure}{0.32\textwidth}
\centering
\begin{tikzpicture}
  \coordinate (Q1) at (-0.5,-0.25);
  \coordinate (Q2) at (0,0.5);
  \coordinate (Q3) at (0.5,-0.25);
  \coordinate (P1) at (-2,-1);
  \coordinate (P2) at (0,2);
  \coordinate (P3) at (2,-1);
  \node at (Q1) {$\bullet$};
  \node at (Q2) {$\bullet$};
  \node at (Q3) {$\bullet$};
  \node at (P1) {$\bullet$};
  \node at (P2) {$\bullet$};
  \node at (P3) {$\bullet$};
\draw[blue,thick] (P1)--(Q2)--(P3)--cycle;
\draw[dotted] (Q2)--(Q1)--(Q3) --cycle;
\draw[dotted] (P1)--(P2)--(P3);
\end{tikzpicture}%
\caption{}%
\label{fig:2branes_1triangle_bigBound}
\end{subfigure}
\caption{$\cC[\mu_R]$ and $\cC[\mu_L]$ brane plus intra and inter-brane 
Higgs. Here, we display example solutions involving only one triangular 
configuration. There exists a continuously parametrized solution interpolating 
between (\subref{fig:2branes_1triangle_smallMaxHiggs}) and 
(\subref{fig:2branes_1triangle_smallBound}), while all intermediate states can 
be 
thought of as in (\subref{fig:2branes_1triangle_smallInt}). In addition, there 
exists 
another continuously parametrized solution interpolating between 
(\subref{fig:2branes_1triangle_bigMaxHiggs}) and 
(\subref{fig:2branes_1triangle_bigBound}), while all intermediate states look 
like 
(\subref{fig:2branes_1triangle_bigInt}). Legs depicted with the same color 
have 
the same amplitude. In particular, the intermediate states have five legs with 
three different, but related amplitudes.}
\label{fig:2branes_1triangle}
\end{figure}
\begin{figure}[t!]
\centering
\begin{subfigure}{0.32\textwidth}
\centering
\begin{tikzpicture}
  \coordinate (Q1) at (-0.5,-0.25);
  \coordinate (Q2) at (0,0.5);
  \coordinate (Q3) at (0.5,-0.25);
  \coordinate (P1) at (-2,-1);
  \coordinate (P2) at (0,2);
  \coordinate (P3) at (2,-1);
  \node at (Q1) {$\bullet$};
  \node at (Q2) {$\bullet$};
  \node at (Q3) {$\bullet$};
  \node at (P1) {$\bullet$};
  \node at (P2) {$\bullet$};
  \node at (P3) {$\bullet$};
\draw[red,thick] (P1)--(P2)--(P3)--cycle;
\draw[red,thick] (Q2)--(Q1)--(Q3) --cycle;
\end{tikzpicture}%
\caption{}%
\label{fig:2branes_2triangle_maxHiggs}
\end{subfigure}
\begin{subfigure}{0.32\textwidth}
\centering
\begin{tikzpicture}
  \coordinate (Q1) at (-0.5,-0.25);
  \coordinate (Q2) at (0,0.5);
  \coordinate (Q3) at (0.5,-0.25);
  \coordinate (P1) at (-2,-1);
  \coordinate (P2) at (0,2);
  \coordinate (P3) at (2,-1);
  \node at (Q1) {$\bullet$};
  \node at (Q2) {$\bullet$};
  \node at (Q3) {$\bullet$};
  \node at (P1) {$\bullet$};
  \node at (P2) {$\bullet$};
  \node at (P3) {$\bullet$};
\draw[red,thick] (P1)--(P2)--(P3);
\draw[red,thick] (Q1)--(Q2)--(Q3);
\draw[blue,thick] (P1)--(Q2)--(P3);
\draw[blue,thick] (Q1)--(P2)--(Q3);
\draw[magenta,thick] (P1)--(P3);
\draw[magenta,thick] (Q1)--(Q3);
\end{tikzpicture}%
\caption{}%
\label{fig:2branes_2triangle_intermed}
\end{subfigure}
\begin{subfigure}{0.32\textwidth}
\centering
\begin{tikzpicture}
  \coordinate (Q1) at (-0.5,-0.25);
  \coordinate (Q2) at (0,0.5);
  \coordinate (Q3) at (0.5,-0.25);
  \coordinate (P1) at (-2,-1);
  \coordinate (P2) at (0,2);
  \coordinate (P3) at (2,-1);
  \node at (Q1) {$\bullet$};
  \node at (Q2) {$\bullet$};
  \node at (Q3) {$\bullet$};
  \node at (P1) {$\bullet$};
  \node at (P2) {$\bullet$};
  \node at (P3) {$\bullet$};
\draw[blue,thick] (P1)--(Q2)--(P3)--cycle;
\draw[blue,thick] (Q1)--(P2)--(Q3)--cycle;
\draw[dotted] (Q1)--(Q2)--(Q3) ;
\draw[dotted] (P1)--(P2)--(P3);
\end{tikzpicture}%
\caption{}%
\label{fig:2branes_2triangle_bound}
\end{subfigure}
\caption{$\cC[\mu_R]$ and $\cC[\mu_L]$ brane plus intra and inter-brane 
Higgs. Here, we display example solutions involving only two triangular 
configuration. There exists a continuously parametrized solution interpolating 
between (\subref{fig:2branes_2triangle_maxHiggs}) and 
(\subref{fig:2branes_2triangle_bound}), while all intermediate states can be 
thought of as in (\subref{fig:2branes_2triangle_intermed}).As before legs 
depicted with the same color have the same amplitude.}
\label{fig:2branes_2triangle}
\end{figure}
\paragraph{Spectrum.}
We have addressed questions about the spectrum of the vector Laplacian around 
such new background in appendix \ref{subsec:brane+brane}. Although the spectrum 
of $\cO_V^X$ in the background $\cC[\mu_L]+\cC[\mu_L]$ is known to be free of 
negative modes, in the novel backgrounds $\cO_V^{X+\phi}$ has a 
number of negative modes, see for instance figure \ref{fig:2branes_spectrum}. 
Turning on uniform masses $M_i \equiv M$  has two effects on the spectrum:
First, the number of negative modes decreases and all of these modes can be 
lifted consistently for mass $0.47 \lesssim M \leq \frac{\sqrt{2}}{3}$, and in particular
for $M^*$. This 
exemplifies the general results of section \ref{sec:massive_potential}.
Second, the number of zero modes stabilizes again to levels that are independent of 
the size of the system.
%  
%%%%%%%%%%%%%%%%%%%%%%%%%%%%%%%%%%%%%%%%%%%%%%%%%%%%%%%%%%%%%%%%%%%%%%%%%%%%%
%
\subsection{Two squashed \texorpdfstring{$\C P^2$}{CP2} branes with a point 
brane \& Higgs}
\label{sec:two-branes-point-Higgs}
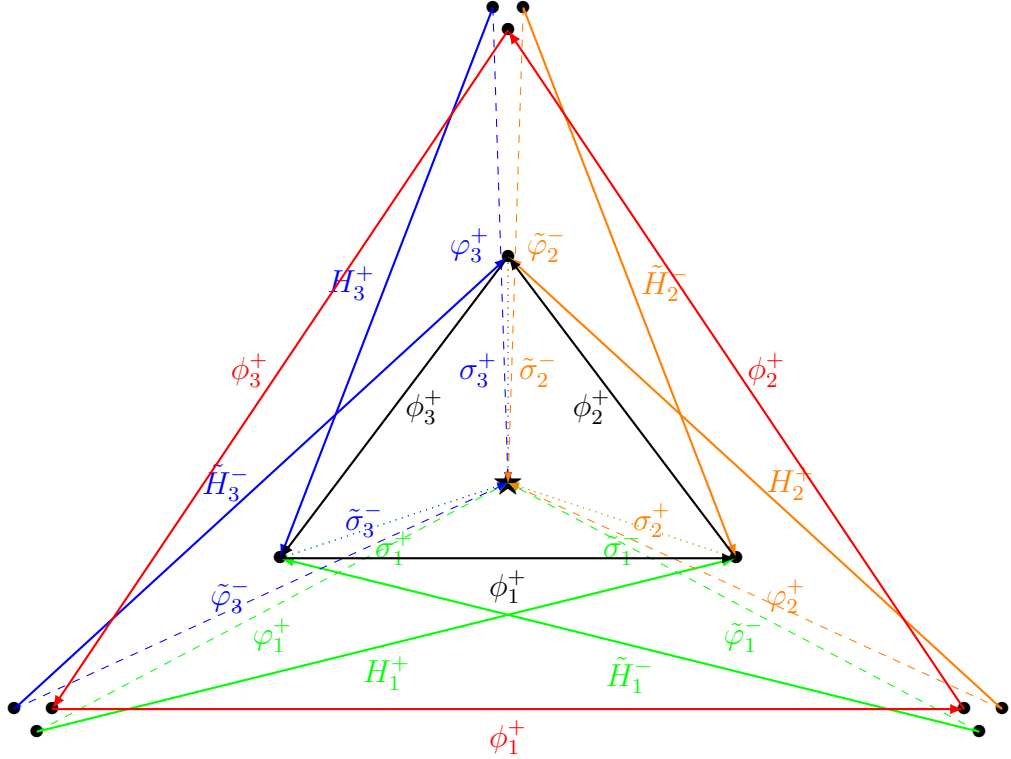
\begin{figure}[t!]
\centering
\begin{tikzpicture}
  \coordinate (Origin) at (0,0);
  \coordinate (Q1) at (-3,-1);
  \coordinate (Q2) at (0,3);
  \coordinate (Q3) at (3,-1);
  \coordinate (P1) at (-6,-3);
  \coordinate (P2) at (0,6);
  \coordinate (P3) at (6,-3);
  \coordinate (P1+a3) at (-6-0.2,-3-0.3);
  \coordinate (P1-a1) at (-6-0.5,-3);
  \coordinate (P2+a2) at (0-0.2,6+0.3);
  \coordinate (P2-a3) at (0+0.2,6+0.3);
  \coordinate (P3+a1) at (6+0.5,-3);
  \coordinate (P3-a2) at (6+0.2,-3-0.3);
  \node at (Origin) {$\bigstar$};
  \node at (Q1) {$\bullet$};
  \node at (Q2) {$\bullet$};
  \node at (Q3) {$\bullet$};
  \node at (P1) {$\bullet$};
  \node at (P2) {$\bullet$};
  \node at (P3) {$\bullet$};
  \node at (P1+a3) {$\bullet$};
  \node at (P1-a1) {$\bullet$};
  \node at (P2+a2) {$\bullet$};
  \node at (P2-a3) {$\bullet$};
  \node at (P3+a1) {$\bullet$};
  \node at (P3-a2) {$\bullet$};
  \draw [thick,green,-latex] (P1+a3) -- (Q3) node[midway,below] 
{$H_1^+$};%
  \draw [thick,green,-latex] (P3-a2) -- (Q1) node[midway,below] 
{$\tilde{H}_1^- $};%
  \draw [thick,blue,-latex] (P1-a1) -- (Q2) node[midway,left] 
{$\tilde{H}_3^-$};%
  \draw [thick,blue,-latex] (P2+a2) -- (Q1) node[midway,left] 
{$H_3^+$};%
  \draw [thick,orange,-latex] (P3+a1) -- (Q2) node[midway,right] 
{$H_2^+$};%
  \draw [thick,orange,-latex] (P2-a3) -- (Q3) node[midway,right] 
{$\tilde{H}_2^-$};%
  \draw [dashed,green,-latex] (P1+a3) -- (Origin) node[midway,below] 
{$\varphi_1^{+}$};%
  \draw [dashed,green,-latex] (P3-a2) -- (Origin) node[midway,below] 
{$\tilde{\varphi}_1^{-}$};%
  \draw [dashed,blue,-latex] (P1-a1) -- (Origin) node[midway,left] 
{$\tilde{\varphi}_3^{-}$};%
  \draw [dashed,blue,-latex] (P2+a2) -- (Origin) node[midway,left] 
{$\varphi_3^{+}$};%
  \draw [dashed,orange,-latex] (P3+a1) -- (Origin) node[midway,right] 
{$\varphi_2^{+}$};%
  \draw [dashed,orange,-latex] (P2-a3) -- (Origin) node[midway,right] 
{$\tilde{\varphi}_2^{-}$};%
  \draw [dotted,green,-latex] (Q1) -- (Origin) node[midway,below] 
{$\sigma_1^{+}$};%
  \draw [dotted,green,-latex] (Q3) -- (Origin) node[midway,below] 
{$\tilde{\sigma}_1^{-}$};%
  \draw [dotted,blue,-latex] (Q1) -- (Origin) node[midway,left] 
{$\tilde{\sigma}_3^{-}$};%
  \draw [dotted,blue,-latex] (Q2) -- (Origin) node[midway,left] 
{$\sigma_3^{+}$};%
  \draw [dotted,orange,-latex] (Q3) -- (Origin) node[midway,right] 
{$\sigma_2^{+}$};%
  \draw [dotted,orange,-latex] (Q2) -- (Origin) node[midway,right] 
{$\tilde{\sigma}_2^{-}$};%
  \draw [thick,red,-latex] (P1) -- (P3) node[midway,below] {$\phi_1^{+}$};
  \draw [thick,red,-latex] (P3) -- (P2) node[midway, right]  {$\phi_2^{+}$};
  \draw [thick,red,-latex] (P2) -- (P1) node[midway, left]  {$\phi_3^{+}$};
  \draw [thick,-latex] (Q1) -- (Q3) node[midway,below] {$\phi_1^{+}$};
  \draw [thick,-latex] (Q3) -- (Q2) node[midway, left]  {$\phi_2^{+}$};
  \draw [thick,-latex] (Q2) -- (Q1) node[midway, right]  {$\phi_3^{+}$};
\end{tikzpicture}%
\caption{The various Higgs fields for the background of $\cC[\mu_L]$ and 
$\cC[\mu_R]$ brane together with a point brane $\cD$. Here, we picture 
$\cC[\mu_L]$ as the outer most brane and place the point brane in the center.}%
\label{fig:Brane+brane+point}%
\end{figure}

Now consider again two parallel branes  $\cC[\mu_L]$ and $\cC[\mu_R]$ with  
$\mu_L=(N_1,0)$ and $\mu_R = (N_2,0)$, and a point brane $\cD \cong \cC[0]$.
The full algebra of functions $\End(\cH_{\mu_L}\oplus \cH_{\mu_R} \oplus \C)$ 
contains \ref{eq:algebra_two_branes}, but also exhibits additional intra-brane 
zero modes originating from $\cA_{L0}=\Hom(\cH_{\mu_L},\C)$, 
$\cA_{R0}=\Hom(\cH_{\mu_R},\C)$ and their conjugates. 
The corresponding regular zero modes are labeled  $\varphi_i^+, 
\tilde{\varphi}_i^- \in \cA_{L0}$ and $\sigma_i^+,\tilde{\sigma}_i^- \in 
\cA_{R0}$, respectively.
In addition, there exist Higgs modes $H_i^+$, $\tilde{H}_i^-$ linking 
$\cC[\mu_L]$ and $\cC[\mu_R]$. The collection of maximal regular zero modes 
(Higgs fields) 
is summarized in figure \ref{fig:Brane+brane+point}. As usual, the conjugate 
fields are 
\begin{equation*}
 \varphi_i^- = (\varphi_i^+)^\dagger , \quad
 \tilde{\varphi}_i^+ = (\tilde{\varphi}_i^-)^\dagger , \quad
 \sigma_i^- = (\sigma_i^+)^\dagger , \quad
 \tilde{\sigma}_i^+ = (\tilde{\sigma}_i^-)^\dagger, \quad
 H_i^- = (H_i^+)^\dagger , \quad
 \tilde{H}_i^+ = (\tilde{H}_i^-)^\dagger
\end{equation*}
and correspond to arrows in the opposite directions.
\paragraph{Solutions.}

\begin{figure}[t!]
\centering
\begin{subfigure}{0.48\textwidth}
\centering
\begin{tikzpicture}
  \coordinate (Origin) at (0,0);
  \coordinate (Q1) at (-1.5,-0.5);
  \coordinate (Q2) at (0,1.5);
  \coordinate (Q3) at (1.5,-0.5);
  \coordinate (P1) at (-3,-1.5);
  \coordinate (P2) at (0,3);
  \coordinate (P3) at (3,-1.5);
  \node at (Origin) {$\bigstar$};
  \node at (Q1) {$\bullet$};
  \node at (Q2) {$\bullet$};
  \node at (Q3) {$\bullet$};
  \node at (P1) {$\bullet$};
  \node at (P2) {$\bullet$};
  \node at (P3) {$\bullet$};
\draw[red,thick] (P3)--(Q1)--(Origin)--cycle;
\draw[dotted] (Q1)--(Q2)--(Q3)--cycle;
\draw[dotted] (P1)--(P2)--(P3)--cycle;
\end{tikzpicture}%
\caption{$\tilde{H}_1^+ \ \varphi_2^+ \ \tilde{\sigma}_3^+$}
\label{fig:2branes+point_chiral-Higgs_1}
\end{subfigure}%
\begin{subfigure}{0.48\textwidth}
\centering
\begin{tikzpicture}
  \coordinate (Origin) at (0,0);
  \coordinate (Q1) at (-1.5,-0.5);
  \coordinate (Q2) at (0,1.5);
  \coordinate (Q3) at (1.5,-0.5);
  \coordinate (P1) at (-3,-1.5);
  \coordinate (P2) at (0,3);
  \coordinate (P3) at (3,-1.5);
  \node at (Origin) {$\bigstar$};
  \node at (Q1) {$\bullet$};
  \node at (Q2) {$\bullet$};
  \node at (Q3) {$\bullet$};
  \node at (P1) {$\bullet$};
  \node at (P2) {$\bullet$};
  \node at (P3) {$\bullet$};
\draw[red,thick] (P1)--(Q3)--(Origin)--cycle;
\draw[dotted] (Q1)--(Q2)--(Q3)--cycle;
\draw[dotted] (P1)--(P2)--(P3)--cycle;
\end{tikzpicture}%
\caption{$H_1^+ \ \sigma_2^+ \ \tilde{\varphi}_3^+$}
\label{fig:2branes+point_chiral-Higgs_2}
\end{subfigure}%
\caption{Higgs configuration with non-vanishing cubic potential.}
\label{fig:two-branes-Higgs-point-chiral}
\end{figure}
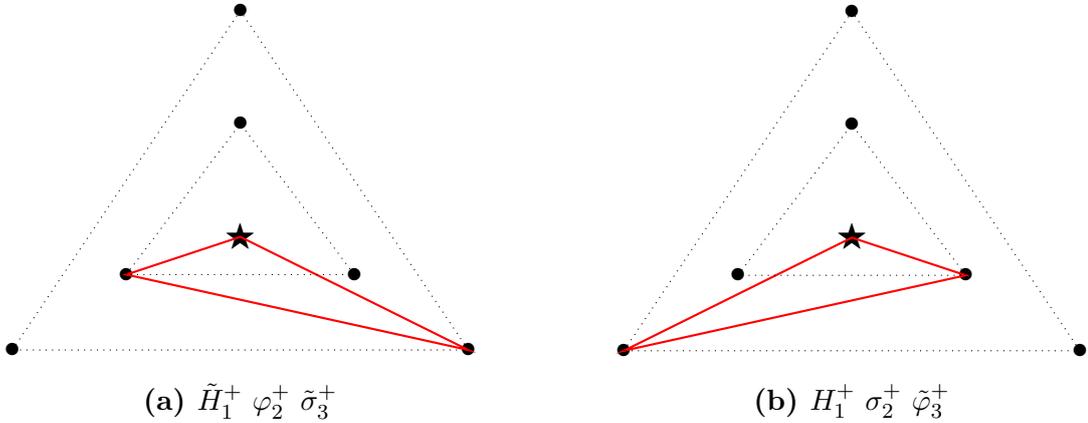

\begin{figure}[t!]
\centering
\begin{subfigure}{0.33\textwidth}
\centering
\begin{tikzpicture}
  \coordinate (Origin) at (0,0);
  \coordinate (Q1) at (-0.8,-0.4);
  \coordinate (Q2) at (0,1);
  \coordinate (Q3) at (0.8,-0.4);
  \coordinate (P1) at (-2,-1);
  \coordinate (P2) at (0,2);
  \coordinate (P3) at (2,-1);
  \node at (Origin) {$\bigstar$};
  \node at (Q1) {$\bullet$};
  \node at (Q2) {$\bullet$};
  \node at (Q3) {$\bullet$};
  \node at (P1) {$\bullet$};
  \node at (P2) {$\bullet$};
  \node at (P3) {$\bullet$};
\draw[red,thick] (P3)--(Q1)--(Origin)--cycle;
\draw[dotted] (Q1)--(Q2)--(Q3)--cycle;
\draw[dotted] (P1)--(P2)--(P3)--cycle;
\end{tikzpicture}%
\caption{}
\label{fig:2branes_1point_1triangle_1}
\end{subfigure}%
\begin{subfigure}{0.33\textwidth}
\centering
\begin{tikzpicture}
  \coordinate (Origin) at (0,0);
  \coordinate (Q1) at (-0.8,-0.4);
  \coordinate (Q2) at (0,1);
  \coordinate (Q3) at (0.8,-0.4);
  \coordinate (P1) at (-2,-1);
  \coordinate (P2) at (0,2);
  \coordinate (P3) at (2,-1);
  \node at (Origin) {$\bigstar$};
  \node at (Q1) {$\bullet$};
  \node at (Q2) {$\bullet$};
  \node at (Q3) {$\bullet$};
  \node at (P1) {$\bullet$};
  \node at (P2) {$\bullet$};
  \node at (P3) {$\bullet$};
\draw[red,thick] (P3)--(Q1)--(Origin);
\draw[blue,thick] (P3)--(P1)--(Origin);
\draw[magenta,thick] (P3)--(Origin);
\draw[dotted] (Q1)--(Q2)--(Q3)--cycle;
\draw[dotted] (P1)--(P2)--(P3)--cycle;
\end{tikzpicture}%
\caption{}
\label{fig:2branes_1point_1triangle_2}
\end{subfigure}%
\begin{subfigure}{0.33\textwidth}
\centering
\begin{tikzpicture}
  \coordinate (Origin) at (0,0);
  \coordinate (Q1) at (-0.8,-0.4);
  \coordinate (Q2) at (0,1);
  \coordinate (Q3) at (0.8,-0.4);
  \coordinate (P1) at (-2,-1);
  \coordinate (P2) at (0,2);
  \coordinate (P3) at (2,-1);
  \node at (Origin) {$\bigstar$};
  \node at (Q1) {$\bullet$};
  \node at (Q2) {$\bullet$};
  \node at (Q3) {$\bullet$};
  \node at (P1) {$\bullet$};
  \node at (P2) {$\bullet$};
  \node at (P3) {$\bullet$};
\draw[blue,thick] (P3)--(P1)--(Origin)--cycle;
\draw[dotted] (Q1)--(Q2)--(Q3)--cycle;
\draw[dotted] (P1)--(P2)--(P3)--cycle;
\end{tikzpicture}%
\caption{}
\label{fig:2branes_1point_1triangle_3}
\end{subfigure}%
% 
%%%%%%%%%%%%%%%%%%%%%%%%%%%%%%%%%%%%%%%%%%%%%%%%%%%%%
%%%%%%%%%%%%%%%%%%%%%%%%%%%%%%%%%%%%%%%%%%%%%%%%%%%%%
% 
\\
\begin{subfigure}{0.33\textwidth}
\centering
\begin{tikzpicture}
  \coordinate (Origin) at (0,0);
  \coordinate (Q1) at (-0.8,-0.4);
  \coordinate (Q2) at (0,1);
  \coordinate (Q3) at (0.8,-0.4);
  \coordinate (P1) at (-2,-1);
  \coordinate (P2) at (0,2);
  \coordinate (P3) at (2,-1);
  \node at (Origin) {$\bigstar$};
  \node at (Q1) {$\bullet$};
  \node at (Q2) {$\bullet$};
  \node at (Q3) {$\bullet$};
  \node at (P1) {$\bullet$};
  \node at (P2) {$\bullet$};
  \node at (P3) {$\bullet$};
\draw[red,thick] (P3)--(Q1)--(Origin)--cycle;
\draw[dotted] (Q1)--(Q2)--(Q3)--cycle;
\draw[dotted] (P1)--(P2)--(P3)--cycle;
\end{tikzpicture}%
\caption{}
\label{fig:2branes_1point_1triangle_4}
\end{subfigure}%
\begin{subfigure}{0.33\textwidth}
\centering
\begin{tikzpicture}
  \coordinate (Origin) at (0,0);
  \coordinate (Q1) at (-0.8,-0.4);
  \coordinate (Q2) at (0,1);
  \coordinate (Q3) at (0.8,-0.4);
  \coordinate (P1) at (-2,-1);
  \coordinate (P2) at (0,2);
  \coordinate (P3) at (2,-1);
  \node at (Origin) {$\bigstar$};
  \node at (Q1) {$\bullet$};
  \node at (Q2) {$\bullet$};
  \node at (Q3) {$\bullet$};
  \node at (P1) {$\bullet$};
  \node at (P2) {$\bullet$};
  \node at (P3) {$\bullet$};
\draw[red,thick] (Q1)--(P3)--(Origin);
\draw[blue,thick] (Q1)--(Q3)--(Origin);
\draw[magenta,thick] (Q1)--(Origin);
\draw[dotted] (Q1)--(Q2)--(Q3)--cycle;
\draw[dotted] (P1)--(P2)--(P3)--cycle;
\end{tikzpicture}%
\caption{}
\label{fig:2branes_1point_1triangle_5}
\end{subfigure}%
\begin{subfigure}{0.33\textwidth}
\centering
\begin{tikzpicture}
  \coordinate (Origin) at (0,0);
  \coordinate (Q1) at (-0.8,-0.4);
  \coordinate (Q2) at (0,1);
  \coordinate (Q3) at (0.8,-0.4);
  \coordinate (P1) at (-2,-1);
  \coordinate (P2) at (0,2);
  \coordinate (P3) at (2,-1);
  \node at (Origin) {$\bigstar$};
  \node at (Q1) {$\bullet$};
  \node at (Q2) {$\bullet$};
  \node at (Q3) {$\bullet$};
  \node at (P1) {$\bullet$};
  \node at (P2) {$\bullet$};
  \node at (P3) {$\bullet$};
\draw[blue,thick] (Q3)--(Q1)--(Origin)--cycle;
\draw[dotted] (Q1)--(Q2)--(Q3)--cycle;
\draw[dotted] (P1)--(P2)--(P3)--cycle;
\end{tikzpicture}%
\caption{}
\label{fig:2branes_1point_1triangle_6}
\end{subfigure}%
% 
%%%%%%%%%%%%%%%%%%%%%%%%%%%%%%%%%%%%%%%%%%%%%%%%%%%%%
%%%%%%%%%%%%%%%%%%%%%%%%%%%%%%%%%%%%%%%%%%%%%%%%%%%%%
% 
\\
\begin{subfigure}{0.33\textwidth}
\centering
\begin{tikzpicture}
  \coordinate (Origin) at (0,0);
  \coordinate (Q1) at (-0.8,-0.4);
  \coordinate (Q2) at (0,1);
  \coordinate (Q3) at (0.8,-0.4);
  \coordinate (P1) at (-2,-1);
  \coordinate (P2) at (0,2);
  \coordinate (P3) at (2,-1);
  \node at (Origin) {$\bigstar$};
  \node at (Q1) {$\bullet$};
  \node at (Q2) {$\bullet$};
  \node at (Q3) {$\bullet$};
  \node at (P1) {$\bullet$};
  \node at (P2) {$\bullet$};
  \node at (P3) {$\bullet$};
\draw[red,thick] (P3)--(Q1)--(Origin)--cycle;
\draw[dotted] (Q1)--(Q2)--(Q3)--cycle;
\draw[dotted] (P1)--(P2)--(P3)--cycle;
\end{tikzpicture}%
\caption{}
\label{fig:2branes_1point_1triangle_7}
\end{subfigure}%
\begin{subfigure}{0.33\textwidth}
\centering
\begin{tikzpicture}
  \coordinate (Origin) at (0,0);
  \coordinate (Q1) at (-0.8,-0.4);
  \coordinate (Q2) at (0,1);
  \coordinate (Q3) at (0.8,-0.4);
  \coordinate (P1) at (-2,-1);
  \coordinate (P2) at (0,2);
  \coordinate (P3) at (2,-1);
  \node at (Origin) {$\bigstar$};
  \node at (Q1) {$\bullet$};
  \node at (Q2) {$\bullet$};
  \node at (Q3) {$\bullet$};
  \node at (P1) {$\bullet$};
  \node at (P2) {$\bullet$};
  \node at (P3) {$\bullet$};
\draw[red,thick] (Q1)--(Origin)--(P3);
\draw[blue,thick] (P3)--(Q2)--(Q1);
\draw[magenta,thick] (Q1)--(P3);
\draw[dotted] (Q1)--(Q2)--(Q3)--cycle;
\draw[dotted] (P1)--(P2)--(P3)--cycle;
\end{tikzpicture}%
\caption{}
\label{fig:2branes_1point_1triangle_8}
\end{subfigure}%
\begin{subfigure}{0.33\textwidth}
\centering
\begin{tikzpicture}
  \coordinate (Origin) at (0,0);
  \coordinate (Q1) at (-0.8,-0.4);
  \coordinate (Q2) at (0,1);
  \coordinate (Q3) at (0.8,-0.4);
  \coordinate (P1) at (-2,-1);
  \coordinate (P2) at (0,2);
  \coordinate (P3) at (2,-1);
  \node at (Origin) {$\bigstar$};
  \node at (Q1) {$\bullet$};
  \node at (Q2) {$\bullet$};
  \node at (Q3) {$\bullet$};
  \node at (P1) {$\bullet$};
  \node at (P2) {$\bullet$};
  \node at (P3) {$\bullet$};
\draw[blue,thick] (Q1)--(Q2)--(P3)--cycle;
\draw[dotted] (Q1)--(Q2)--(Q3)--cycle;
\draw[dotted] (P1)--(P2)--(P3)--cycle;
\end{tikzpicture}%
\caption{}
\label{fig:2branes_1point_1triangle_9}
\end{subfigure}%
\caption{Continuous solutions that relate the closed $\tilde{H}_1^+ \ 
\varphi_2^+ \ \tilde{\sigma}_3^+$ triangle with other known one triangle 
solutions. The three figures in each row are related by an explicit 
one-parameter solution.}
\label{fig:2branes_1point_1triangle}
\end{figure}
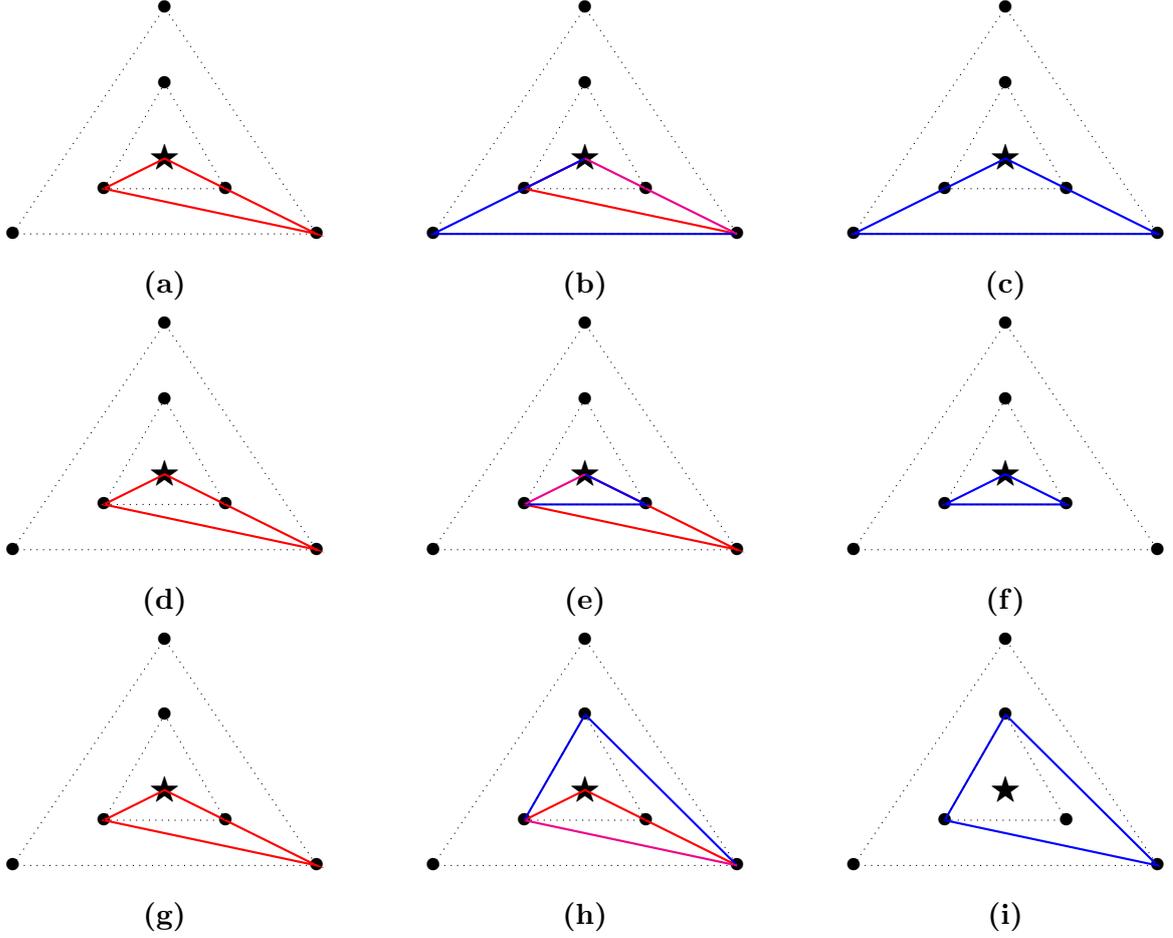

In this set-up, there is a novel type of Higgs solution which involve only links between different 
branes as indicated in figure \ref{fig:two-branes-Higgs-point-chiral}, due to 
cubic terms like $\sigma_i^+ H_j^+  \tilde{\varphi}_k^+$  and $\varphi_i^+ \tilde{H}_j^+ \tilde{\sigma}_k^+$.
This leads to a potentially interesting structure of Yukawa couplings and chiral fermions, as discussed in sections 
\ref{sec:fermions} and \ref{sec:standardmodel}.
The two different types of such triangles have opposite ``orientation``.
As further elaborated in appendix \ref{subsec:brane+brane+point}, one can find 
various one-parameter solutions that interpolate between the configurations in 
figure \ref{fig:two-branes-Higgs-point-chiral} and other solutions with one 
closed triangle. We summarize these in figure 
\ref{fig:2branes_1point_1triangle}. 
Starting from \ref{fig:2branes+point_chiral-Higgs_1} there exist at least three 
continuous families of exact solutions which are displayed in 
\ref{fig:2branes_1point_1triangle_1}--\ref{fig:2branes_1point_1triangle_3},
\ref{fig:2branes_1point_1triangle_4}--\ref{fig:2branes_1point_1triangle_6},
and 
\ref{fig:2branes_1point_1triangle_7}--\ref{fig:2branes_1point_1triangle_9}.
Due to our knowledge of the two parallel brane case, see figure 
\ref{fig:2branes_1triangle}, it follows that configuration 
\ref{fig:2branes_1point_1triangle_9}  %equivalent
%%H
can be deformed into to the maximal intra-brane Higgs configuration
$\phi_i^+$ on $\cC[\mu_R]$. Moreover, inspecting figure \ref{fig:sol_CP2-point} 
%from the $\C P^2$ brane plus point brane case 
reveals that configurations 
\ref{fig:2branes_1point_1triangle_3} and \ref{fig:2branes_1point_1triangle_6} 
can be deformed to closed maximal intra-brane Higgs triangles on $\cC[\mu_L]$ and 
$\cC[\mu_R]$, respectively. They all have the same energy.

Repeating the analogous analysis for \ref{fig:2branes+point_chiral-Higgs_2}, 
one concludes that all closed one triangle solutions are connected by 
continuous deformations, which all have the same  energy.

\begin{figure}[t!]
\begin{subfigure}{0.33\textwidth}
\centering
\begin{tikzpicture}
  \coordinate (Origin) at (0,0);
  \coordinate (Q1) at (-0.8,-0.4);
  \coordinate (Q2) at (0,1);
  \coordinate (Q3) at (0.8,-0.4);
  \coordinate (P1) at (-2,-1);
  \coordinate (P2) at (0,2);
  \coordinate (P3) at (2,-1);
  \node at (Origin) {$\bigstar$};
  \node at (Q1) {$\bullet$};
  \node at (Q2) {$\bullet$};
  \node at (Q3) {$\bullet$};
  \node at (P1) {$\bullet$};
  \node at (P2) {$\bullet$};
  \node at (P3) {$\bullet$};
\draw[red,thick] (P1)--(P2)--(Q3)--cycle;
\draw[blue,thick] (P3)--(Q1)--(Origin)--cycle;
\draw[dotted] (Q1)--(Q2)--(Q3)--cycle;
\draw[dotted] (P1)--(P3)--(P2);
\end{tikzpicture}%
\caption{}
\label{fig:2branes_1point_2triangle_1}
\end{subfigure}%
\begin{subfigure}{0.33\textwidth}
\centering
\begin{tikzpicture}
  \coordinate (Origin) at (0,0);
  \coordinate (Q1) at (-0.8,-0.4);
  \coordinate (Q2) at (0,1);
  \coordinate (Q3) at (0.8,-0.4);
  \coordinate (P1) at (-2,-1);
  \coordinate (P2) at (0,2);
  \coordinate (P3) at (2,-1);
  \node at (Origin) {$\bigstar$};
  \node at (Q1) {$\bullet$};
  \node at (Q2) {$\bullet$};
  \node at (Q3) {$\bullet$};
  \node at (P1) {$\bullet$};
  \node at (P2) {$\bullet$};
  \node at (P3) {$\bullet$};
\draw[red,thick] (P1)--(P2)--(P3)--cycle;
\draw[blue,thick] (Q3)--(Q1)--(Origin)--cycle;
\draw[dotted] (Q1)--(Q2)--(Q3);
\end{tikzpicture}%
\caption{}
\label{fig:2branes_1point_2triangle_2}
\end{subfigure}%
\begin{subfigure}{0.33\textwidth}
\centering
\begin{tikzpicture}
  \coordinate (Origin) at (0,0);
  \coordinate (Q1) at (-0.8,-0.4);
  \coordinate (Q2) at (0,1);
  \coordinate (Q3) at (0.8,-0.4);
  \coordinate (P1) at (-2,-1);
  \coordinate (P2) at (0,2);
  \coordinate (P3) at (2,-1);
  \node at (Origin) {$\bigstar$};
  \node at (Q1) {$\bullet$};
  \node at (Q2) {$\bullet$};
  \node at (Q3) {$\bullet$};
  \node at (P1) {$\bullet$};
  \node at (P2) {$\bullet$};
  \node at (P3) {$\bullet$};
\draw[red,thick] (P1)--(P2)--(P3)--cycle;
\draw[blue,thick] (Q3)--(Q1)--(Q2)--cycle;
\end{tikzpicture}%
\caption{}
\label{fig:2branes_1point_2triangle_3}
\end{subfigure}%
\caption{Three two triangle configurations which are equivalent by continuous 
deformations among each other.}
\label{fig:2branes_1point_2triangle}
\end{figure}
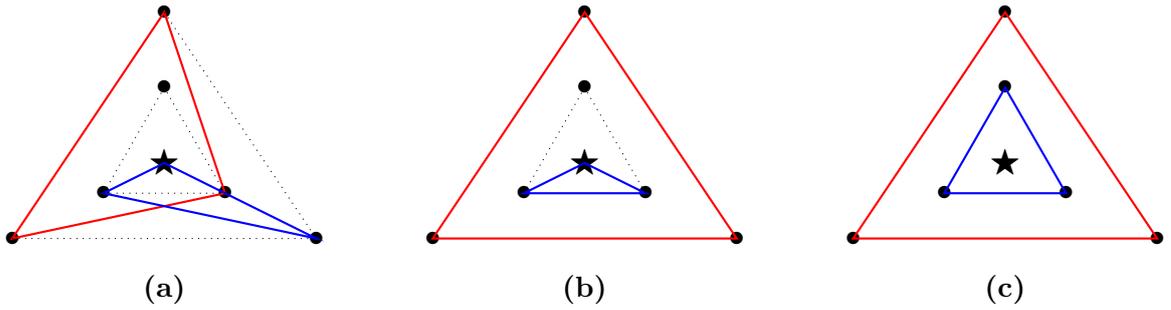

As in previous cases, solutions with more than one closed triangle have a lower 
potential energy, and we immediately recognize the 
solution with maximal intra-brane Higgs $\phi_i^+$ and $\phi_i^+$. 
Again, one can verify explicitly that 
solutions like the ones depicted in figure \ref{fig:2branes_1point_2triangle} 
exist, and we find continuous deformations that transform figure 
\ref{fig:2branes_1point_2triangle_1} into figure 
\ref{fig:2branes_1point_2triangle_2} as well as  
figure 
\ref{fig:2branes_1point_2triangle_2} into 
figure 
\ref{fig:2branes_1point_2triangle_3}. By analogous considerations, one infers that all 
 other configurations involving two closed triangles are 
degenerate in their potential energy and can be deformed into each other.
%%H
However, they lead to  distinct patterns of symmetry breaking and Yukawa couplings.

\paragraph{Spectrum.}
Having established the existence of numerous novel solutions consisting of brane 
backgrounds and maximal Higgs, we need to address the spectrum of the vector 
Laplacian. We illustrate the typical behavior for the one-parameter family 
connecting the configurations of figure \ref{fig:2branes_1point_2triangle_2} 
and \ref{fig:2branes_1point_2triangle_1}, and  summarize  the results in 
appendix \ref{subsec:brane+brane+point}, in particular figure 
\ref{fig:2branes+pt_spectrum}.

As in the previous cases, the combined backgrounds $X+\phi$ suffer from the 
presence of negative modes in the spectrum of $\cO_V^{X+\phi}$. Fortunately, 
one can lift all of these consistently by inclusion of uniform masses $M_i 
\equiv M$ with the choice $0.47 \lesssim M \leq \frac{\sqrt{2}}{3}$, including $M^*$.
As before, these non-trivial masses eliminate a large fraction of the zero 
modes, and stabilize their number to a level which is (roughly) independent of 
the system size.
\subsection{Three squashed \texorpdfstring{$\C P^2$}{CP2} branes \& Higgs}
\label{sec:three-branes}
\begin{figure}[t!]
\centering
\begin{subfigure}{0.32\textwidth}
\centering
\begin{tikzpicture}
  \coordinate (R1) at (-0.5,-0.25);
  \coordinate (R2) at (0,0.5);
  \coordinate (R3) at (0.5,-0.25);
  \coordinate (Q1) at (-1.5,-0.75);
  \coordinate (Q2) at (0,1.5);
  \coordinate (Q3) at (1.5,-0.75);
  \coordinate (P1) at (-2.5,-1.25);
  \coordinate (P2) at (0,2.5);
  \coordinate (P3) at (2.5,-1.25);
  \node at (R1) {$\bullet$};
  \node at (R2) {$\bullet$};
  \node at (R3) {$\bullet$};
  \node at (Q1) {$\bullet$};
  \node at (Q2) {$\bullet$};
  \node at (Q3) {$\bullet$};
  \node at (P1) {$\bullet$};
  \node at (P2) {$\bullet$};
  \node at (P3) {$\bullet$};
\draw[dotted] (P1)--(P2)--(P3)--cycle;
\draw[dotted] (Q1)--(Q2)--(Q3)--cycle;
\draw[dotted] (R1)--(R2)--(R3)--cycle;
\draw[thick,red] (P1)--(Q2)--(R3)--cycle;
\draw[thick,green] (P2)--(Q3)--(R1)--cycle;
\draw[thick,blue] (P3)--(Q1)--(R2)--cycle;
\end{tikzpicture}%
\caption{}%
\label{fig:3Branes_1}%
\end{subfigure}
\begin{subfigure}{0.32\textwidth}
\centering
\begin{tikzpicture}
  \coordinate (R1) at (-0.5,-0.25);
  \coordinate (R2) at (0,0.5);
  \coordinate (R3) at (0.5,-0.25);
  \coordinate (Q1) at (-1.5,-0.75);
  \coordinate (Q2) at (0,1.5);
  \coordinate (Q3) at (1.5,-0.75);
  \coordinate (P1) at (-2.5,-1.25);
  \coordinate (P2) at (0,2.5);
  \coordinate (P3) at (2.5,-1.25);
  \node at (R1) {$\bullet$};
  \node at (R2) {$\bullet$};
  \node at (R3) {$\bullet$};
  \node at (Q1) {$\bullet$};
  \node at (Q2) {$\bullet$};
  \node at (Q3) {$\bullet$};
  \node at (P1) {$\bullet$};
  \node at (P2) {$\bullet$};
  \node at (P3) {$\bullet$};
\draw[dotted] (P1)--(P2)--(P3)--cycle;
\draw[dotted] (Q1)--(Q2)--(Q3);
\draw[dotted] (R1)--(R2)--(R3)--cycle;
\draw[thick,red] (P1)--(Q2)--(Q3)--cycle;
\draw[thick,green] (P2)--(R3)--(R1)--cycle;
\draw[thick,blue] (P3)--(Q1)--(R2)--cycle;
\end{tikzpicture}%
\caption{}%
\label{fig:3Branes_2}%
\end{subfigure}
\begin{subfigure}{0.32\textwidth}
\centering
\begin{tikzpicture}
  \coordinate (R1) at (-0.5,-0.25);
  \coordinate (R2) at (0,0.5);
  \coordinate (R3) at (0.5,-0.25);
  \coordinate (Q1) at (-1.5,-0.75);
  \coordinate (Q2) at (0,1.5);
  \coordinate (Q3) at (1.5,-0.75);
  \coordinate (P1) at (-2.5,-1.25);
  \coordinate (P2) at (0,2.5);
  \coordinate (P3) at (2.5,-1.25);
  \node at (R1) {$\bullet$};
  \node at (R2) {$\bullet$};
  \node at (R3) {$\bullet$};
  \node at (Q1) {$\bullet$};
  \node at (Q2) {$\bullet$};
  \node at (Q3) {$\bullet$};
  \node at (P1) {$\bullet$};
  \node at (P2) {$\bullet$};
  \node at (P3) {$\bullet$};
\draw[dotted] (P1)--(P2)--(P3)--cycle;
\draw[dotted] (Q1)--(Q2)--(Q3);
\draw[dotted] (R1)--(R2)--(R3)--cycle;
\draw[thick,red] (P1)--(P2)--(Q3)--cycle;
\draw[thick,green] (Q2)--(R3)--(R1)--cycle;
\draw[thick,blue] (P3)--(Q1)--(R2)--cycle;
\end{tikzpicture}%
\caption{}%
\label{fig:3Branes_3}%
\end{subfigure}
\begin{subfigure}{0.32\textwidth}
\centering
\begin{tikzpicture}
  \coordinate (R1) at (-0.5,-0.25);
  \coordinate (R2) at (0,0.5);
  \coordinate (R3) at (0.5,-0.25);
  \coordinate (Q1) at (-1.5,-0.75);
  \coordinate (Q2) at (0,1.5);
  \coordinate (Q3) at (1.5,-0.75);
  \coordinate (P1) at (-2.5,-1.25);
  \coordinate (P2) at (0,2.5);
  \coordinate (P3) at (2.5,-1.25);
  \node at (R1) {$\bullet$};
  \node at (R2) {$\bullet$};
  \node at (R3) {$\bullet$};
  \node at (Q1) {$\bullet$};
  \node at (Q2) {$\bullet$};
  \node at (Q3) {$\bullet$};
  \node at (P1) {$\bullet$};
  \node at (P2) {$\bullet$};
  \node at (P3) {$\bullet$};
\draw[dotted] (Q1)--(Q2)--(Q3);
\draw[dotted] (R1)--(R2)--(R3);
\draw[thick,red] (P1)--(P2)--(P3)--cycle;
\draw[thick,green] (Q2)--(R3)--(R1)--cycle;
\draw[thick,blue] (Q3)--(Q1)--(R2)--cycle;
\end{tikzpicture}%
\caption{}%
\label{fig:3Branes_4}%
\end{subfigure}
\begin{subfigure}{0.32\textwidth}
\centering
\begin{tikzpicture}
  \coordinate (R1) at (-0.5,-0.25);
  \coordinate (R2) at (0,0.5);
  \coordinate (R3) at (0.5,-0.25);
  \coordinate (Q1) at (-1.5,-0.75);
  \coordinate (Q2) at (0,1.5);
  \coordinate (Q3) at (1.5,-0.75);
  \coordinate (P1) at (-2.5,-1.25);
  \coordinate (P2) at (0,2.5);
  \coordinate (P3) at (2.5,-1.25);
  \node at (R1) {$\bullet$};
  \node at (R2) {$\bullet$};
  \node at (R3) {$\bullet$};
  \node at (Q1) {$\bullet$};
  \node at (Q2) {$\bullet$};
  \node at (Q3) {$\bullet$};
  \node at (P1) {$\bullet$};
  \node at (P2) {$\bullet$};
  \node at (P3) {$\bullet$};
\draw[thick,red] (P1)--(P2)--(P3)--cycle;
\draw[thick,green] (R2)--(R3)--(R1)--cycle;
\draw[thick,blue] (Q3)--(Q1)--(Q2)--cycle;
\end{tikzpicture}%
\caption{}%
\label{fig:3Branes_5}%
\end{subfigure}
\caption{Configurations of three closed triangles for three parallel $\C P^2$ 
branes. All of the shown configurations can be deformed into each other by a 
continuous family of solutions. Hence, all configurations have identical 
potential energy, but the fluctuations around each background can differ.}
 \label{fig:three-CP2-branes-Higgs}
\end{figure}
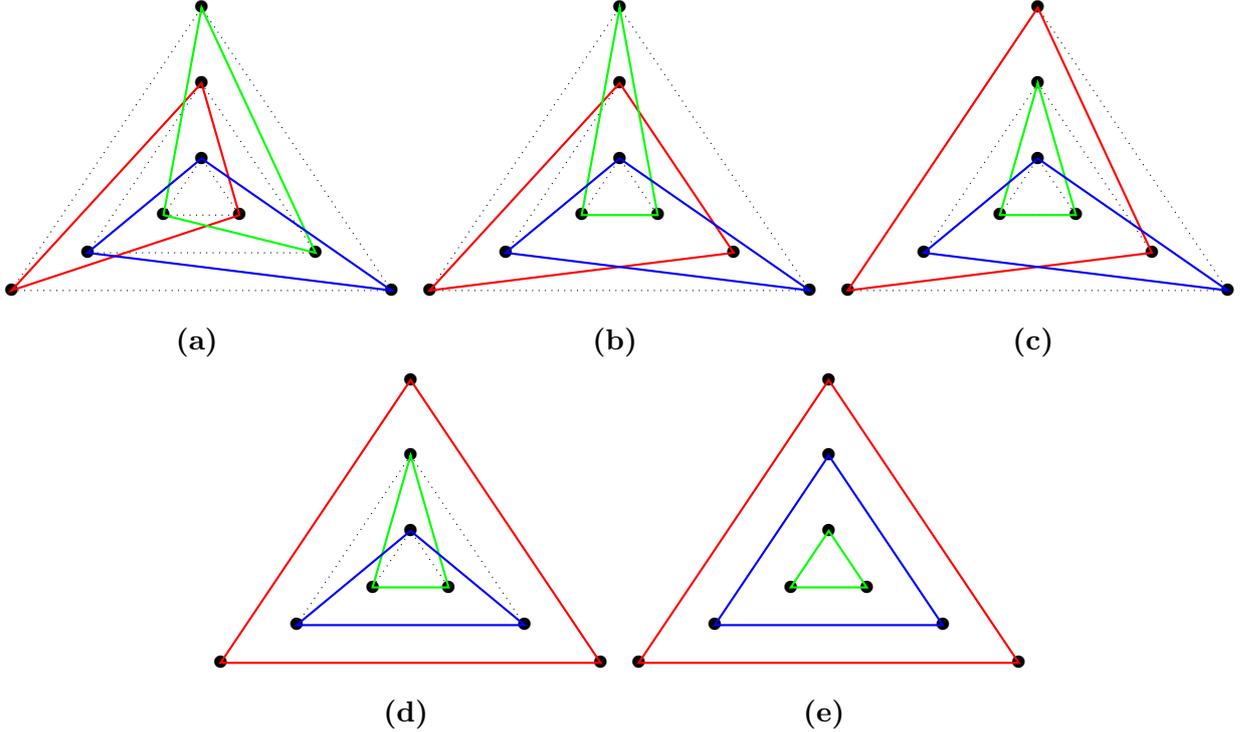

Finally, consider the case where also $\cD$ is not a point brane, but a $\C 
P^2$ brane. Since the sector of maximal regular zero modes is a straight 
forward generalization of the previous cases, we refrain from depicting them in 
full detail.
\paragraph{Solutions.}
In this set-up, one can again find configurations with three closed Higgs 
triangles connecting the different branes as in figures 
\ref{fig:3Branes_1}--\ref{fig:3Branes_3}. 
The energy of these configurations is conjectured to be minimal, and equal to the case 
where only the maximal intra-brane Higgs on the branes are switched on.  
Continuous deformations interpolating between these 
configurations are indicated in figure \ref{fig:three-CP2-branes-Higgs}. As we have 
already seen in the previous cases, the all solutions with fixed number of 
closed triangles (here $1$, $2$, or $3$) can be deformed into one another. 
Moreover, we know that solutions with the maximal number of closed triangles 
have the minimal potential energy. Hence these configurations are highly 
degenerate, but
%%H
they lead to  distinct patterns of symmetry breaking and Yukawa couplings.
\paragraph{Spectrum.}
Again, we studied the spectrum of the vector Laplacian in these combined 
background $X+\phi$, and provide the details in appendix 
\ref{subsec:brane+brane+brane}, see in particular figure 
\ref{fig:3branes_spectrum}.
We illustrate for the configurations of figure \ref{fig:3Branes_1} and 
\ref{fig:3Branes_5} the existence of negative modes in the massless case, and 
their uplifting via masses $M_i \equiv M$ with $0.47 \lesssim M 
\leq \frac{\sqrt{2}}{3}$ including $M^*$. Then the increasingly large number of zero 
modes in the massless case is reduced and stabilized to a level that appears to 
be independent of the systems size.

\subsection{Coinciding branes with Higgs}
%%H new

Now consider a stack of $n$ identical branes as above. 
Then the massless i.e.\ trivial gauge modes constitute an unbroken $U(n)$
(or $SU(n)$) gauge group. 
As in the case of distinct branes, there are numerous exact Higgs solutions, which may or 
may not link the different branes in various patterns. It is clear that this  leads
to various patterns of (partial or complete) symmetry breaking, and it is straightforward
in principle to work out the masses of the broken gauge bosons from the 
Higgs effect, cf.\ \eqref{gaugeboson-mass-phi}.

%M  todo: shorten that sentence
Taking into account the fermionic zero modes,
the question arises  how these  fermions 
couple to the broken and unbroken gauge fields, which Yukawa couplings arise, and whether some kind of chiral 
gauge theory  emerges.
%, where fermions with different chiralities couple in 
%inequivalent ways to these gauge bosons. 
This is quite natural as we will see, and
will be discussed in section \ref{sec:fermions}.

\subsection{Flipped minimal branes plus point brane as 
\texorpdfstring{$G_2$}{G2} brane.}
\label{sec:flipped-branes}
Now consider the case of two branes $\cC[\mu_{L,R}]$ of conjugate type, i.e.\ 
with  $\mu_L = (N,0), \ \mu_R=(0,N_R)$, together with a point brane $\cD$. 
Then the full set of inter-brane zero modes between the two squashed $\C P^2$ 
branes can  be obtained from the mode decomposition 
\begin{align}
 \Hom(\cH_{(N,0)},\cH_{(0,N_R)})  = \cH_{(N+N_R,0)} \oplus 
\cH_{(N+N_R-2,1)} \oplus \ldots \oplus \cH_{(\ldots,N_R)}
\end{align}
assuming $N \geq N_R$. The maximal regular inter-brane modes transform as 
$(N+N_R,0)$.
However, for these modes linking the extremal weights by the longest 
possible arrow, there exists no configuration with a non-trivial cubic term. 
Therefore we have no reason to assume that they acquire a VEV.
\begin{figure}[t!]
\centering
\begin{subfigure}{0.24\textwidth}
\centering
\begin{tikzpicture}
  \coordinate (P1) at (-1.5,-1);
  \coordinate (P2) at (0,2);
  \coordinate (P3) at (1.5,-1);
  \coordinate (P4) at (-1.5,1);
  \coordinate (P5) at (0,-2);
  \coordinate (P6) at (1.5,1);
  \coordinate (P7) at (0,0);
  \node at (P1) {$\bullet$};
  \node at (P2) {$\bullet$};
  \node at (P3) {$\bullet$};
  \node at (P4) {$\blacktriangle$};
  \node at (P5) {$\blacktriangle$};
  \node at (P6) {$\blacktriangle$};
  \node at (P7) {$\bigstar$};
% % 
\draw [thick,red,-latex] (P1) -- (P3) node[at end,below] {$\phi_1^+$};%
\draw [thick,red,-latex] (P3) -- (P2) node[at end,right] {$\phi_2^+$};%
\draw [thick,red,-latex] (P2) -- (P1) node[at end,left] {$\phi_3^+$};%
\draw [thick,blue,-latex] (P4) -- (P6) node[at end,above] {$\tilde{\phi}_1^+$};%
\draw [thick,blue,-latex] (P5) -- (P4) node[at end,left] {$\tilde{\phi}_2^+$};%
\draw [thick,blue,-latex] (P6) -- (P5) node[at end,right] {$\tilde{\phi}_3^+$};%
\end{tikzpicture}%
\caption{}%
\label{fig:flipped_brane+pt_intra_Higgs}%
\end{subfigure}%
\begin{subfigure}{0.24\textwidth}
\centering
\begin{tikzpicture}
  \coordinate (P1) at (-1.5,-1);
  \coordinate (P2) at (0,2);
  \coordinate (P3) at (1.5,-1);
  \coordinate (P4) at (-1.5,1);
  \coordinate (P5) at (0,-2);
  \coordinate (P6) at (1.5,1);
  \coordinate (P7) at (0,0);
  \node at (P1) {$\bullet$};
  \node at (P2) {$\bullet$};
  \node at (P3) {$\bullet$};
  \node at (P4) {$\blacktriangle$};
  \node at (P5) {$\blacktriangle$};
  \node at (P6) {$\blacktriangle$};
  \node at (P7) {$\bigstar$};
% % 
\draw [thick,red,-latex] (P1) -- (P6) node[at end,below] {$\varphi_1^+$};%
\draw [thick,blue,-latex] (P3) -- (P4) node[at end,below] 
{$\tilde{\varphi}_1^-$};%
\draw [dotted] (P1)--(P2)--(P3)--cycle;
\draw [dotted] (P4)--(P5)--(P6)--cycle;
\end{tikzpicture}%
\caption{}%
\label{fig:flipped_brane+pt_inter_Higgs_max1}%
\end{subfigure}%
\begin{subfigure}{0.24\textwidth}
\centering
\begin{tikzpicture}
  \coordinate (P1) at (-1.5,-1);
  \coordinate (P2) at (0,2);
  \coordinate (P3) at (1.5,-1);
  \coordinate (P4) at (-1.5,1);
  \coordinate (P5) at (0,-2);
  \coordinate (P6) at (1.5,1);
  \coordinate (P7) at (0,0);
  \node at (P1) {$\bullet$};
  \node at (P2) {$\bullet$};
  \node at (P3) {$\bullet$};
  \node at (P4) {$\blacktriangle$};
  \node at (P5) {$\blacktriangle$};
  \node at (P6) {$\blacktriangle$};
  \node at (P7) {$\bigstar$};
% % 
\draw [thick,red,-latex] (P3) -- (P4) node[at end,below] {$\varphi_2^+$};%
\draw [thick,blue,-latex] (P2) -- (P5) node[at end,right] 
{$\tilde{\varphi}_2^-$};%
\draw [dotted] (P1)--(P2)--(P3)--cycle;
\draw [dotted] (P4)--(P5)--(P6)--cycle;
\end{tikzpicture}%
\caption{}%
\label{fig:flipped_brane+pt_inter_Higgs_max2}
\end{subfigure}%
\begin{subfigure}{0.24\textwidth}
\centering
\begin{tikzpicture}
  \coordinate (P1) at (-1.5,-1);
  \coordinate (P2) at (0,2);
  \coordinate (P3) at (1.5,-1);
  \coordinate (P4) at (-1.5,1);
  \coordinate (P5) at (0,-2);
  \coordinate (P6) at (1.5,1);
  \coordinate (P7) at (0,0);
  \node at (P1) {$\bullet$};
  \node at (P2) {$\bullet$};
  \node at (P3) {$\bullet$};
  \node at (P4) {$\blacktriangle$};
  \node at (P5) {$\blacktriangle$};
  \node at (P6) {$\blacktriangle$};
  \node at (P7) {$\bigstar$};
% % 
\draw [thick,red,-latex] (P2) -- (P5) node[at end,right] {$\varphi_3^+$};%
\draw [thick,blue,-latex] (P1) -- (P6) node[at end,below] 
{$\tilde{\varphi}_3^-$};%
\draw [dotted] (P1)--(P2)--(P3)--cycle;
\draw [dotted] (P4)--(P5)--(P6)--cycle;
\end{tikzpicture}%
\caption{}%
\label{fig:flipped_brane+pt_inter_Higgs_max3}
\end{subfigure}%
%%%%%%%%%%%%%%%%%%%%%%%%%%%%%%%%%%%%%%%%%%%%%%%%%%%%%%%%
\\
%%%%%%%%%%%%%%%%%%%%%%%%%%%%%%%%%%%%%%%%%%%%%%%%%%%%%%%%
\begin{subfigure}{0.24\textwidth}
\centering
\begin{tikzpicture}
  \coordinate (P1) at (-1.5,-1);
  \coordinate (P2) at (0,2);
  \coordinate (P3) at (1.5,-1);
  \coordinate (P4) at (-1.5,1);
  \coordinate (P5) at (0,-2);
  \coordinate (P6) at (1.5,1);
  \coordinate (P7) at (0,0);
  \node at (P1) {$\bullet$};
  \node at (P2) {$\bullet$};
  \node at (P3) {$\bullet$};
  \node at (P4) {$\blacktriangle$};
  \node at (P5) {$\blacktriangle$};
  \node at (P6) {$\blacktriangle$};
  \node at (P7) {$\bigstar$};
% % 
\draw [thick,red,-latex] (P1) -- (P5) node[midway,below] {$\zeta_1^+$};%
\draw [thick,red,-latex] (P2) -- (P6);
\draw [thick,blue,-latex] (P3) -- (P5) node[midway,below] 
{$\tilde{\zeta}_1^-$};%
\draw [thick,blue,-latex] (P2) -- (P4);
\draw [dotted] (P1)--(P2)--(P3)--cycle;
\draw [dotted] (P4)--(P5)--(P6)--cycle;
\end{tikzpicture}%
\caption{}%
\label{fig:flipped_brane+pt_inter_Higgs_next-to1}
\end{subfigure}%
\begin{subfigure}{0.24\textwidth}
\centering
\begin{tikzpicture}
  \coordinate (P1) at (-1.5,-1);
  \coordinate (P2) at (0,2);
  \coordinate (P3) at (1.5,-1);
  \coordinate (P4) at (-1.5,1);
  \coordinate (P5) at (0,-2);
  \coordinate (P6) at (1.5,1);
  \coordinate (P7) at (0,0);
  \node at (P1) {$\bullet$};
  \node at (P2) {$\bullet$};
  \node at (P3) {$\bullet$};
  \node at (P4) {$\blacktriangle$};
  \node at (P5) {$\blacktriangle$};
  \node at (P6) {$\blacktriangle$};
  \node at (P7) {$\bigstar$};
% % 
\draw [thick,red,-latex] (P1) -- (P4) node[midway,left] {$\zeta_2^+$};%
\draw [thick,red,-latex] (P3) -- (P6);
\draw [thick,blue,-latex] (P1) -- (P5) node[midway,below] 
{$\tilde{\zeta}_2^-$};%
\draw [thick,blue,-latex] (P2) -- (P6);
\draw [dotted] (P1)--(P2)--(P3)--cycle;
\draw [dotted] (P4)--(P5)--(P6)--cycle;
\end{tikzpicture}%
\caption{}%
\label{fig:flipped_brane+pt_inter_Higgs_next-to2}
\end{subfigure}%
\begin{subfigure}{0.24\textwidth}
\centering
\begin{tikzpicture}
  \coordinate (P1) at (-1.5,-1);
  \coordinate (P2) at (0,2);
  \coordinate (P3) at (1.5,-1);
  \coordinate (P4) at (-1.5,1);
  \coordinate (P5) at (0,-2);
  \coordinate (P6) at (1.5,1);
  \coordinate (P7) at (0,0);
  \node at (P1) {$\bullet$};
  \node at (P2) {$\bullet$};
  \node at (P3) {$\bullet$};
  \node at (P4) {$\blacktriangle$};
  \node at (P5) {$\blacktriangle$};
  \node at (P6) {$\blacktriangle$};
  \node at (P7) {$\bigstar$};
% % 
\draw [thick,red,-latex] (P2) -- (P4) node[midway,above] {$\zeta_3^+$};%
\draw [thick,red,-latex] (P3) -- (P5);
\draw [thick,blue,-latex] (P1) -- (P4) node[midway,left] 
{$\tilde{\zeta}_3^-$};%
\draw [thick,blue,-latex] (P3) -- (P6);
\draw [dotted] (P1)--(P2)--(P3)--cycle;
\draw [dotted] (P4)--(P5)--(P6)--cycle;
\end{tikzpicture}%
\caption{}%
\label{fig:flipped_brane+pt_inter_Higgs_next-to3}
\end{subfigure}%
% 
%%%%%%%%%%%%%%%%%%%%%%%%%%%%%%%%%%%%%%%%%%%%%%%%%%%%%%%
\caption{Set-up $\cC[(1,0)]+\cC[(0,1)]+\cD$. Extremal weights for $(1,0)$ are 
denoted by $\bullet$, while $(0,1)$ extremal weights are indicated by  
$\blacktriangle$, and $\cD$ is represented by $\bigstar$.
The regular intra-brane Higgs 
modes $\phi$ on $\cC[(1,0)]$ and $\tilde{\phi}$ on $\cC[(0,1)]$ are shown in 
(\subref{fig:flipped_brane+pt_intra_Higgs}). In 
(\subref{fig:flipped_brane+pt_inter_Higgs_max1})--(\subref{%
fig:flipped_brane+pt_inter_Higgs_max3}) we display the maximal regular 
inter-brane Higgs $\varphi$, 
$\tilde{\varphi}$ between $\cC[(1,0)]$  and $\cC[(0,1)]$, while the 
next-to-maximal regular inter-brane Higgs $\zeta$, $\tilde{\zeta}$ between 
$\cC[(1,0)]$ and $\cC[(0,1)]$ are shown in 
(\subref{fig:flipped_brane+pt_inter_Higgs_next-to1})--(\subref{%
fig:flipped_brane+pt_inter_Higgs_next-to3})}
\label{fig:flipped_brane+point_Higgs_partI}
\end{figure}

\begin{figure}[t!]
\centering
 \begin{subfigure}{0.485\textwidth}
\centering
\begin{tikzpicture}
  \coordinate (P1) at (-1.5,-1);
  \coordinate (P2) at (0,2);
  \coordinate (P3) at (1.5,-1);
  \coordinate (P4) at (-1.5,1);
  \coordinate (P5) at (0,-2);
  \coordinate (P6) at (1.5,1);
  \coordinate (P7) at (0,0);
  \node at (P1) {$\bullet$};
  \node at (P2) {$\bullet$};
  \node at (P3) {$\bullet$};
  \node at (P4) {$\blacktriangle$};
  \node at (P5) {$\blacktriangle$};
  \node at (P6) {$\blacktriangle$};
  \node at (P7) {$\bigstar$};
  \draw [dotted] (P1)--(P2)--(P3)--cycle;
  \draw [dotted] (P4)--(P5)--(P6)--cycle;
% % 
\draw [thick,red,-latex] (P5) -- (P3);
\draw [thick,red,-latex] (P4) -- (P2) node[midway,above] {$\tilde{\zeta}_1^+$};
\draw [thick,blue,-latex] (P5) -- (P1);
\draw [thick,blue,-latex] (P6) -- (P2) node[midway,above] {$\tilde{\zeta}_2^+$};
\draw [thick,green,-latex] (P4) -- (P1);
\draw [thick,green,-latex] (P6) -- (P3) node[midway,right] 
{$\tilde{\zeta}_3^+$};
\draw [thick,dashed,red,-latex] (P1) -- (P7) node[midway,above] {$\varrho_1^+$};
\draw [thick,dashed,red,-latex] (P7) -- (P6) node[midway,below] 
{$\tilde{\sigma}_1^+$};
\draw [thick,dashed,blue,-latex] (P3) -- (P7) node[midway,below] 
{$\varrho_2^+$};
\draw [thick,dashed,blue,-latex] (P7) -- (P4) node[midway,above] 
{$\tilde{\sigma}_2^+$};
\draw [thick,dashed,green,-latex] (P2) -- (P7) node[midway,right] 
{$\varrho_3^+$};
\draw [thick,dashed,green,-latex] (P7) -- (P5)node[midway,left] 
{$\tilde{\sigma}_3^+$};
\end{tikzpicture}%
\caption{}%
\label{fig:flipped_brane+pt_G2-sol1}
\end{subfigure}%
\begin{subfigure}{0.485\textwidth}
\centering
\begin{tikzpicture}
  \coordinate (P1) at (-1.5,-1);
  \coordinate (P2) at (0,2);
  \coordinate (P3) at (1.5,-1);
  \coordinate (P4) at (-1.5,1);
  \coordinate (P5) at (0,-2);
  \coordinate (P6) at (1.5,1);
  \coordinate (P7) at (0,0);
  \node at (P1) {$\bullet$};
  \node at (P2) {$\bullet$};
  \node at (P3) {$\bullet$};
  \node at (P4) {$\blacktriangle$};
  \node at (P5) {$\blacktriangle$};
  \node at (P6) {$\blacktriangle$};
  \node at (P7) {$\bigstar$};
  \draw [dotted] (P1)--(P2)--(P3)--cycle;
  \draw [dotted] (P4)--(P5)--(P6)--cycle;
% % 
\draw [thick,red,-latex] (P1) -- (P5);
\draw [thick,red,-latex] (P2) -- (P6) node[midway,above] {$\zeta_1^+$};
\draw [thick,blue,-latex] (P1) -- (P4);
\draw [thick,blue,-latex] (P3) -- (P6) node[midway,right] {$\zeta_2^+$};
\draw [thick,green,-latex] (P3) -- (P5);
\draw [thick,green,-latex] (P2) -- (P4) node[midway,above] 
{$\zeta_3^+$};
\draw [thick,dashed,red,-latex] (P4) -- (P7) node[midway,above] {$\sigma_1^+$};
\draw [thick,dashed,red,-latex] (P7) -- (P3) node[midway,below] 
{$\tilde{\varrho}_1^+$};
\draw [thick,dashed,blue,-latex] (P5) -- (P7) node[midway,left] 
{$\sigma_2^+$};
\draw [thick,dashed,blue,-latex] (P7) -- (P2) node[midway,right] 
{$\tilde{\varrho}_2^+$};
\draw [thick,dashed,green,-latex] (P6) -- (P7) node[midway,below] 
{$\sigma_3^+$};
\draw [thick,dashed,green,-latex] (P7) -- (P1) node[midway,above] 
{$\tilde{\varrho}_3^+$};
\end{tikzpicture}%
\caption{}%
\label{fig:flipped_brane+pt_G2-sol2}
\end{subfigure}%
 \caption{Set-up $\cC[(1,0)]+\cC[(0,1)]+\cD$. By turning on various 
intra-brane zero modes, we can construct an exact solution which realizes 
the 7 dimensional irrep of $G_2$. } 
 \label{fig:G2-branes}
\end{figure}
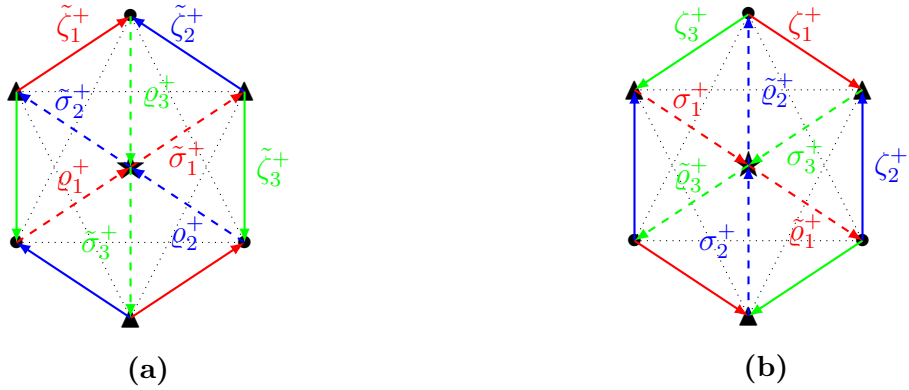

Specializing to the case of $N=N_R$, the next most interesting regular 
zero modes transform in $(0,N)$. These modes connect the parallel edges of 
the irreps $(N,0)$ and $(0,N)$. 
For minimal branes, they form a closed algebra and lead to an exact 
solution. Let us discuss this in more detail:
Let $\cC[(1,0)] +\cC[(1,0)] + \cC[(0,0)]$ be the background solution. There are 
 various regular zero modes to be taken into account. To begin with, the 
regular intra-brane zero modes on $\cC[(1,0)]$ and $\cC[(0,1)]$ are depicted 
in figure \ref{fig:flipped_brane+pt_intra_Higgs}. From earlier arguments, it is 
clear that the background plus the intra-brane triangular configuration lead to 
exact solutions of the equations of motion.
In addition, the maximal inter-brane regular zero modes between 
$\cC[(1,0)]$ and $\cC[(0,1)]$ are shown in figure 
\ref{fig:flipped_brane+pt_inter_Higgs_max1}--\ref{%
fig:flipped_brane+pt_inter_Higgs_max3}, while the next-to-maximal intra-brane 
Higgs are displayed in figures 
\ref{fig:flipped_brane+pt_inter_Higgs_next-to1}--\ref{%
fig:flipped_brane+pt_inter_Higgs_next-to3}.
Lastly, there exists the class of inter-brane regular zero modes between each 
squashed $\C P^2$ brane and the point brane. The regular zero modes between 
$\cC[(1,0)]$ and $\cD$ are displayed in figure 
\ref{fig:flipped_brane+pt_inter_Higgs_1-0_to_Pt1}--\ref{%
fig:flipped_brane+pt_inter_Higgs_1-0_to_Pt3}; whereas the inter-brane Higgs 
between $\cC[(0,1)]$ and $\cD$ are shown in figures 
\ref{fig:flipped_brane+pt_inter_Higgs_0-1_to_Pt1}--\ref{%
fig:flipped_brane+pt_inter_Higgs_0-1_to_Pt3}.
\begin{figure}[t!]
\centering
%%%%%%%%%%%%%%%%%%%%%%%%%%%%%%%%%%%%%%%%%%%%%%%%%%%%%%%
\begin{subfigure}{0.24\textwidth}
\centering
\begin{tikzpicture}
  \coordinate (P1) at (-1.5,-1);
  \coordinate (P2) at (0,2);
  \coordinate (P3) at (1.5,-1);
  \coordinate (P4) at (-1.5,1);
  \coordinate (P5) at (0,-2);
  \coordinate (P6) at (1.5,1);
  \coordinate (P7) at (0,0);
  \node at (P1) {$\bullet$};
  \node at (P2) {$\bullet$};
  \node at (P3) {$\bullet$};
  \node at (P4) {$\blacktriangle$};
  \node at (P5) {$\blacktriangle$};
  \node at (P6) {$\blacktriangle$};
  \node at (P7) {$\bigstar$};
% % 
\draw [thick,red,-latex] (P1) -- (P7) node[midway,below] {$\varrho_1^+$};%
\draw [thick,blue,-latex] (P3) -- (P7) node[midway,below] 
{$\tilde{\varrho}_1^-$};%
\draw [dotted] (P1)--(P2)--(P3)--cycle;
\draw [dotted] (P4)--(P5)--(P6)--cycle;
\end{tikzpicture}%
\caption{}%
\label{fig:flipped_brane+pt_inter_Higgs_1-0_to_Pt1}%
\end{subfigure}%
\begin{subfigure}{0.24\textwidth}
\centering
\begin{tikzpicture}
  \coordinate (P1) at (-1.5,-1);
  \coordinate (P2) at (0,2);
  \coordinate (P3) at (1.5,-1);
  \coordinate (P4) at (-1.5,1);
  \coordinate (P5) at (0,-2);
  \coordinate (P6) at (1.5,1);
  \coordinate (P7) at (0,0);
  \node at (P1) {$\bullet$};
  \node at (P2) {$\bullet$};
  \node at (P3) {$\bullet$};
  \node at (P4) {$\blacktriangle$};
  \node at (P5) {$\blacktriangle$};
  \node at (P6) {$\blacktriangle$};
  \node at (P7) {$\bigstar$};
% % 
\draw [thick,red,-latex] (P3) -- (P7) node[midway,below] {$\varrho_2^+$};%
\draw [thick,blue,-latex] (P2) -- (P7) node[midway,left] 
{$\tilde{\varrho}_2^-$};%
\draw [dotted] (P1)--(P2)--(P3)--cycle;
\draw [dotted] (P4)--(P5)--(P6)--cycle;
\end{tikzpicture}%
\caption{}%
\label{fig:flipped_brane+pt_inter_Higgs_1-0_to_Pt2}
\end{subfigure}%
\begin{subfigure}{0.24\textwidth}
\centering
\begin{tikzpicture}
  \coordinate (P1) at (-1.5,-1);
  \coordinate (P2) at (0,2);
  \coordinate (P3) at (1.5,-1);
  \coordinate (P4) at (-1.5,1);
  \coordinate (P5) at (0,-2);
  \coordinate (P6) at (1.5,1);
  \coordinate (P7) at (0,0);
  \node at (P1) {$\bullet$};
  \node at (P2) {$\bullet$};
  \node at (P3) {$\bullet$};
  \node at (P4) {$\blacktriangle$};
  \node at (P5) {$\blacktriangle$};
  \node at (P6) {$\blacktriangle$};
  \node at (P7) {$\bigstar$};
% % 
\draw [thick,red,-latex] (P2) -- (P7) node[midway,left] {$\varrho_3^+$};%
\draw [thick,blue,-latex] (P1) -- (P7) node[midway,above] 
{$\tilde{\varrho}_3^-$};%
\draw [dotted] (P1)--(P2)--(P3)--cycle;
\draw [dotted] (P4)--(P5)--(P6)--cycle;
\end{tikzpicture}%
\caption{}%
\label{fig:flipped_brane+pt_inter_Higgs_1-0_to_Pt3}
\end{subfigure}%
%%%%%%%%%%%%%%%%%%%%%%%%%%%%%%%%%%%%%%%%%%%%%%%%%%%%%%%
\\
%%%%%%%%%%%%%%%%%%%%%%%%%%%%%%%%%%%%%%%%%%%%%%%%%%%%%%%
\begin{subfigure}{0.24\textwidth}
\centering
\begin{tikzpicture}
  \coordinate (P1) at (-1.5,-1);
  \coordinate (P2) at (0,2);
  \coordinate (P3) at (1.5,-1);
  \coordinate (P4) at (-1.5,1);
  \coordinate (P5) at (0,-2);
  \coordinate (P6) at (1.5,1);
  \coordinate (P7) at (0,0);
  \node at (P1) {$\bullet$};
  \node at (P2) {$\bullet$};
  \node at (P3) {$\bullet$};
  \node at (P4) {$\blacktriangle$};
  \node at (P5) {$\blacktriangle$};
  \node at (P6) {$\blacktriangle$};
  \node at (P7) {$\bigstar$};
% % 
\draw [thick,red,-latex] (P4) -- (P7) node[midway,below] {$\sigma_1^+$};%
\draw [thick,blue,-latex] (P6) -- (P7) node[midway,below] 
{$\tilde{\sigma}_1^-$};%
\draw [dotted] (P1)--(P2)--(P3)--cycle;
\draw [dotted] (P4)--(P5)--(P6)--cycle;
\end{tikzpicture}%
\caption{}%
\label{fig:flipped_brane+pt_inter_Higgs_0-1_to_Pt1}
\end{subfigure}%
\begin{subfigure}{0.24\textwidth}
\centering
\begin{tikzpicture}
  \coordinate (P1) at (-1.5,-1);
  \coordinate (P2) at (0,2);
  \coordinate (P3) at (1.5,-1);
  \coordinate (P4) at (-1.5,1);
  \coordinate (P5) at (0,-2);
  \coordinate (P6) at (1.5,1);
  \coordinate (P7) at (0,0);
  \node at (P1) {$\bullet$};
  \node at (P2) {$\bullet$};
  \node at (P3) {$\bullet$};
  \node at (P4) {$\blacktriangle$};
  \node at (P5) {$\blacktriangle$};
  \node at (P6) {$\blacktriangle$};
  \node at (P7) {$\bigstar$};
% % 
\draw [thick,red,-latex] (P5) -- (P7) node[midway,left] {$\sigma_2^+$};%
\draw [thick,blue,-latex] (P4) -- (P7) node[midway,above] 
{$\tilde{\sigma}_2^-$};%
\draw [dotted] (P1)--(P2)--(P3)--cycle;
\draw [dotted] (P4)--(P5)--(P6)--cycle;
\end{tikzpicture}%
\caption{}%
\label{fig:flipped_brane+pt_inter_Higgs_0-1_to_Pt2}
\end{subfigure}%
\begin{subfigure}{0.24\textwidth}
\centering
\begin{tikzpicture}
  \coordinate (P1) at (-1.5,-1);
  \coordinate (P2) at (0,2);
  \coordinate (P3) at (1.5,-1);
  \coordinate (P4) at (-1.5,1);
  \coordinate (P5) at (0,-2);
  \coordinate (P6) at (1.5,1);
  \coordinate (P7) at (0,0);
  \node at (P1) {$\bullet$};
  \node at (P2) {$\bullet$};
  \node at (P3) {$\bullet$};
  \node at (P4) {$\blacktriangle$};
  \node at (P5) {$\blacktriangle$};
  \node at (P6) {$\blacktriangle$};
  \node at (P7) {$\bigstar$};
% % 
\draw [thick,red,-latex] (P6) -- (P7) node[midway,above] {$\sigma_3^+$};%
\draw [thick,blue,-latex] (P5) -- (P7) node[midway,left] 
{$\tilde{\sigma}_3^-$};%
\draw [dotted] (P1)--(P2)--(P3)--cycle;
\draw [dotted] (P4)--(P5)--(P6)--cycle;
\end{tikzpicture}%
\caption{}%
\label{fig:flipped_brane+pt_inter_Higgs_0-1_to_Pt3}
\end{subfigure}%
\caption{Set-up $\cC[(1,0)]+\cC[(0,1)]+\cD$. The inter-brane Higgs 
$\varrho$, $\tilde{\varrho}$ between $\cC[(1,0)]$ and point brane $\cD$ are 
displayed in 
(\subref{fig:flipped_brane+pt_inter_Higgs_1-0_to_Pt1})--(\subref{%
fig:flipped_brane+pt_inter_Higgs_1-0_to_Pt3}), while the inter-brane Higgs 
$\sigma$, $\tilde{\sigma}$ 
between $\cC[(0,1)]$ and point brane $\cD$ are displayed in 
(\subref{fig:flipped_brane+pt_inter_Higgs_0-1_to_Pt1})--(\subref{%
fig:flipped_brane+pt_inter_Higgs_0-1_to_Pt3}).}
\label{fig:flipped_brane+point_Higgs_partII}
\end{figure}
\paragraph{Solutions.}
As a nice illustration, we can construct  an exact 
solution which has the structure of the 7-dimensional irrep of $G_2$. We 
accomplish this by combining the rank 2 inter-brane zero modes between 
the minimal brane and its conjugate with the inter-brane modes between the 
minimal branes and the point brane. Actually there exist two realizations of  
such a solution, as shown in figure 
\ref{fig:G2-branes}. Following appendix \ref{subsec:flipped_min_brane+point}, 
one  realizes that the coefficients of the involved zero modes are 
such that they realize the  short roots of $G_2$, while the long roots are realized 
by the background $X_\a$.
Since $G_2$ is a Lie algebra, this solution
suggests a vast generalization based on  higher representations of $G_2$. 
Indeed, the long and the short roots of $G_2$ satisfy the decoupling 
condition\footnote{HS, unpublished; useful discussions with G.\ Zoupanos are 
acknowledged.}.
This should be elaborated in more detail elsewhere.

For non-minimal branes $\cC[(N,0)]$, it is difficult to find 
analogous exact solutions, because the regular zero modes  no longer satisfy 
a closed algebra.
Nevertheless, one would expect that similar  solutions might  exist for 
flipped non-minimal branes with a point brane. This construction is in a sense dual to the 
discussion of 6-dimensional branes in the next section.

\paragraph{Stability.}
The new combined solutions of $G_2$-type have been obtained in the massless 
case. However, we can transfer them to massive solutions as before. This turns 
out to be necessary to obtain a instability-free spectrum of the vector 
Laplacian around this combined backgrounds. As in previous cases, a uniform 
mass parameter of order $0.45 \lesssim M \leq \frac{\sqrt{2}}{3}$ is sufficient 
to achieve this, and details are provided in appendix 
\ref{subsec:flipped_min_brane+point}.
% 
% 
% 
%
%%%%%%%%%%%%%%%%%%%%%%%%%%%%%%%%%%%%%%%%%%%%%%%%%%%%%%%%%%%%%%%%%%%%%%
%%%%%%%%%%%%%%%%%%%%%%%%%%%%%%%%%%%%%%%%%%%%%%%%%%%%%%%%%%%%%%%%%%%%%%
%
\section{6-dimensional branes}
\label{sec:6D-branes}
Now consider the case of general $\cC[(N,M)]$ branes. These are 6-dimensional quantized coadjoint orbits
embedded in $\R^6$, which decompose into $3_L+3_R$ chiral sheets with opposite flux measured 
by the gauge mode $\chi$ \eqref{chi-def}. 
Because these branes have maximal dimension (in contrast to the above $\C P^2$ 
branes), this leads to $3_L+3_R$ zero modes between a point brane and $\cC[(N,M)]$, 
whose chirality is measured by $\chi$. This is the basis for a chiral gauge 
theory.
\paragraph{Chirality generator and  chiral sheets.}
For 6-dimensional (fuzzy) branes, the operator
\begin{align}
 \chi \coloneqq  & \frac{\im}{8} \varepsilon^{\a\b \ldots \g\d}[X_\a,X_\b] 
\ldots
[X_\g,X_\d] \ \sim  \mathrm{Pf}(\theta^{\mu\nu}) \nn\\
  & \quad \in \ (3,0)  + (0,3) \subset    (1,1)^{\otimes_{\mathrm{sym}}^3} 
  \label{chi-def}
 \end{align}
reduces to the Pfaffian of the Poisson tensor in the semi-classical limit, and
therefore it should be a good observable to define the chiral $L$ and $R$ 
sheets, cf.\ \cite{Steinacker:2014lma,Steinacker:2014eua,Steinacker:2015mia}.  
It is easy to see \cite{Steinacker:2014lma} that on the six extremal weight states $w|\mu\rangle$
of $\cC[\mu]$ located at the origin,  $\chi$ takes the values
\begin{align}
 \chi =  (-1)^{|w|}(\a_1,\mu) (\a_2,\mu) (\a_3,\mu)  \,,
\end{align}
where $|w|$ is the signature of the appropriate Weyl group element, and $\a_i$ are the 
simple roots.
In the semi-classical limit,
$\chi$ can be identified with a function on the brane (more precisely a polynomial of degree 3), which 
takes positive or negative values 
corresponding to the orientation of the 3+3 sheets of $\cC[\mu]$. This function is odd under $\cW$.
Thus on the extremal weight states, $\chi$ has the simple structure
\begin{align}
 \chi \sim \one_L - \one_R \,
 \label{chi-structure}
\end{align}
because it has weight zero, it is invariant under $\Z_3$ and odd under Weyl reflections.
More precisely,
since $\chi$ is a cubic totally symmetric function on $\cC[\mu]$, it lives in 
$(1,1)^{\otimes_{\mathrm{sym}}^3}$. On the other hand,
$\chi = \chi^{38}$ is the weight 0 state in $(1,1)^{\wedge^2} = (3,0) + (0,3) + 
(1,1)$
of the $\msu(3)$ intertwiner
\begin{align}
 \chi^{gh} &=  \varepsilon^{ab \ldots gh}[T_a,T_b] \ldots [T,T] \in 
(1,1)^{\otimes_{\mathrm{sym}}^3} \cap (1,1)^{\wedge^2}\ = (3,0) + (0,3) + 
(1,1)\qquad     ,
\nn\\
 \chi &= \chi^{38} = - \chi^{83} \ .
\end{align}
The $(1,1)$ contribution can be excluded\footnote{It would have to be $c^{38}_e T^e$ where $c^{ab}_c$ are the structure constants, which vanishes.}, 
so that $\chi$ is the hermitian weight 0 combination in
\begin{align} 
 \chi \ \in \ (0,3) + (3,0) \ .
\end{align}
\paragraph{Chiral Higgs.}
We have seen that the extremal states of 6-dimensional branes decompose into 
$3_L + 3_R$ sets of states with definite chirality.
Now we consider the regular Higgs mode on such a brane, and single out those 
Higgs
$\tilde\phi_\a$  which respect the chirality, i.e.\
\begin{align}
 [ \tilde\phi^\a,\chi] \sim 0 . 
 \label{chiral-Higgs}
\end{align}
These are easy to identify for the $(1,1)$ brane in figure 
\ref{fig:brane_1-1_Higgs}. While the maximal regular zero modes depicted 
in figure \ref{fig:brane_1-1_rank1} are not chiral,  the 
next-to-maximal regular zero modes  $\tilde\phi_\a\in \cH_{(3,0)} + \cH_{(0,3)} \subset \End(\cH)$  depicted in figures 
\ref{fig:brane_1-1_rank2_1}--\ref{fig:brane_1-1_rank2_3} are in fact chiral, and have a 
string-like structure 
\begin{align}
 \tilde\phi^\a  \sim |x_L^i\rangle\langle x_L^j| \pm |x_R^i\rangle\langle x_R^j|
 \label{chiral-Higgs-decomp}
\end{align}
 relating extremal states with the same chirality; then \eqref{chiral-Higgs} 
follows from \eqref{chi-structure}.

For the simplest case $\cC[(1,1)]$, these chiral Higgs modes form a closed algebra 
and will lead to an exact solution, as  
discussed in more detail in section \ref{sec:1-1_brane}. 
This justifies the hypothesis that these $\tilde\phi^\a$ acquire a VEV.
The implications  for the fermions 
and their Yukawa couplings will be discussed in 
section \ref{sec:fermions}.

\begin{figure}[t!]
\centering
\begin{subfigure}{0.33\textwidth}
\centering
\begin{tikzpicture}
  \coordinate (P1) at (-2,0);
  \coordinate (P2) at (-1,1);
  \coordinate (P3) at (-1,-1);
  \coordinate (P5) at (1,1);
  \coordinate (P7) at (1,-1);
  \coordinate (P8) at (2,0);
  \node at (P1) {$\bullet$};
  \node at (P2) {$\bullet$};
  \node at (P3) {$\bullet$};
  \node at (P5) {$\bullet$};
  \node at (P7) {$\bullet$};
  \node at (P8) {$\bullet$};
% % 
\draw [thick,red,-latex] (P1) -- (P8) node[near end,below] {$\phi_1^+$};%
\draw [thick,red,-latex] (P7) -- (P2) node[near end,right] {$\phi_2^+$};%
\draw [thick,red,-latex] (P5) -- (P3) node[near end,left] {$\phi_3^+$};%
\draw[dotted] (P1)--(P3)--(P7)--(P8)--(P5)--(P2)--cycle;
\draw [black] (P1) node[left] {$R$};
\draw [black] (P2) node[above] {$L$};
\draw [black] (P3) node[below] {$L$};
\draw [black] (P5) node[above] {$R$};
\draw [black] (P7) node[below] {$R$};
\draw [black] (P8) node[right] {$L$};
\end{tikzpicture}%
\caption{}%
\label{fig:brane_1-1_rank1}%
\end{subfigure}%
\\
\begin{subfigure}{0.33\textwidth}
\centering
\begin{tikzpicture}
  \coordinate (P1) at (-2,0);
  \coordinate (P2) at (-1,1);
  \coordinate (P3) at (-1,-1);
  \coordinate (P5) at (1,1);
  \coordinate (P7) at (1,-1);
  \coordinate (P8) at (2,0);
  \node at (P1) {$\bullet$};
  \node at (P2) {$\bullet$};
  \node at (P3) {$\bullet$};
  \node at (P5) {$\bullet$};
  \node at (P7) {$\bullet$};
  \node at (P8) {$\bullet$};
% % 
\draw [thick,red,-latex] (P1) -- (P7) node[midway,above] {$\varphi_1^{+}$};%
\draw [thick,red,-latex] (P2) -- (P8) ;%
\draw [thick,blue,-latex] (P1) -- (P5);%
\draw [thick,blue,-latex] (P3) -- (P8)  node[midway,above] {$\sigma_1^{+}$};%
\draw[dotted] (P1)--(P3)--(P7)--(P8)--(P5)--(P2)--cycle;
\end{tikzpicture}%
\caption{}%
\label{fig:brane_1-1_rank2_1}%
\end{subfigure}%
% \\
\begin{subfigure}{0.33\textwidth}
\centering
\begin{tikzpicture}
  \coordinate (P1) at (-2,0);
  \coordinate (P2) at (-1,1);
  \coordinate (P3) at (-1,-1);
  \coordinate (P5) at (1,1);
  \coordinate (P7) at (1,-1);
  \coordinate (P8) at (2,0);
  \node at (P1) {$\bullet$};
  \node at (P2) {$\bullet$};
  \node at (P3) {$\bullet$};
  \node at (P5) {$\bullet$};
  \node at (P7) {$\bullet$};
  \node at (P8) {$\bullet$};
% % 
\draw [thick,red,-latex] (P3) -- (P2) ;%
\draw [thick,red,-latex] (P7) -- (P5) node[midway,left] {$\varphi_2^{+}$};%
\draw [thick,blue,-latex] (P7) -- (P1) node[midway,above] {$\sigma_2^{+}$};%
\draw [thick,blue,-latex] (P8) -- (P2) ;%
\draw[dotted] (P1)--(P3)--(P7)--(P8)--(P5)--(P2)--cycle;
\end{tikzpicture}%
\caption{}%
\label{fig:brane_1-1_rank2_2}%
\end{subfigure}%
\begin{subfigure}{0.33\textwidth}
\centering
\begin{tikzpicture}
  \coordinate (P1) at (-2,0);
  \coordinate (P2) at (-1,1);
  \coordinate (P3) at (-1,-1);
  \coordinate (P5) at (1,1);
  \coordinate (P7) at (1,-1);
  \coordinate (P8) at (2,0);
  \node at (P1) {$\bullet$};
  \node at (P2) {$\bullet$};
  \node at (P3) {$\bullet$};
  \node at (P5) {$\bullet$};
  \node at (P7) {$\bullet$};
  \node at (P8) {$\bullet$};
% % 
\draw [thick,red,-latex] (P5) -- (P1) node[midway,below] {$\varphi_3^{+}$};%
\draw [thick,red,-latex] (P8) -- (P3) ;%
\draw [thick,blue,-latex] (P2) -- (P3) ;%
\draw [thick,blue,-latex] (P5) -- (P7) node[midway,left] {$\sigma_3^{+}$};%
\draw[dotted] (P1)--(P3)--(P7)--(P8)--(P5)--(P2)--cycle;
\end{tikzpicture}%
\caption{}%
\label{fig:brane_1-1_rank2_3}%
\end{subfigure}%
\caption{Regular zero modes of rank $1$ or $2$ for $\cC[(1,1)]$ brane. 
The maximal Higgs $\phi_i^+$ of (\subref{fig:brane_1-1_rank1}) are not chiral, 
but the rank $2$ Higgs $\varphi_i^+$, $\sigma_i^+$ of 
(\subref{fig:brane_1-1_rank2_1})--(\subref{fig:brane_1-1_rank2_3}) are chiral.}
\label{fig:brane_1-1_Higgs}
\end{figure}
% 
%%H add some chirality labels  
% 

% 
\paragraph{Chiral gauge field $A_\mu$ and its mass.}
Among all the gauge fields on the $\cC[(N,M)]$ brane,
consider the $\chi$-valued gauge field
\begin{align}
A_\mu = A_\mu(x) \chi \ .
\label{chiral-gauge-mode}
\end{align}
We call it \emph{chiral}, because it
measures the chirality of the  $L$ and $R$  sheets according to 
\eqref{chi-structure}, and 
therefore couples accordingly to chiral fermions.
We are going to argue that $A_\mu$ may be the lightest non-trivial gauge mode
in the presence of a chiral Higgs VEV as above,  and describes a  gauge field
in a spontaneously broken $U(1)_L \times U(1)_R$ chiral gauge theory.

With this in mind, 
the mass of the gauge boson $A_\mu$ in the presence of some Higgs $\phi_\a$ arises as usual from
\begin{align}
 D_\mu (X^a + \phi^a) D^\mu (X_a + \phi_a)
  &= \del_\mu \phi_a \del^\mu \phi^a + A_\mu(x)A^\mu(x) [\chi,X^a + 
\phi^a][\chi,X_a + \phi_a] + \ldots  \nn\\
  &=  \del_\mu \phi_a \del^\mu \phi^a + A_\mu A^\mu \Big( m^2_{(3,0),0}  \chi \chi
   + [\chi,\phi^a][\chi,\phi_a]\Big) 
\label{gaugeboson-mass-phi}
\end{align}
assuming $[X^a,\phi_a] =0$  (see \cite{Aschieri:2006uw}). 
Since $\chi$ is a weight zero mode in $\cH_{(3,0)}$, the contribution form the 
 brane $X$ is obtained from \eqref{gaugemodes-mass}  
\begin{align}
 m^2_{(3,0),0} = 2(\L,\L+2\rho) 
 = 2 (5,2)^T G (3,0) = 2\cdot 27 \ .
\end{align}
Assuming that a chiral  Higgs $\tilde\phi$ takes a VEV (as justified below), 
it is natural to expect that $A_\mu \sim \chi$
will be the lightest gauge boson, 
because  the mass contribution $[\chi,\tilde\phi^a][\chi, \tilde\phi_a]$ vanishes
due to its defining property \eqref{chiral-Higgs}.
All other non-chiral gauge modes $A'_\mu \sim \r$  acquire an extra  mass 
 $\tr [\tilde\phi_\a,\r]^2$.

Explicitly, the masses of the lightest gauge modes (with zero weight) 
due to the background are as follows:
\begin{align}
 m^2_{(3,0),0} &= 2 (5,2)^T G (3,0) = 2\cdot 36 ,  \nn\\  
 m^2_{(1,1),0} &= 2(3,3)^T G (1,1) = 2\cdot 18 , \nn\\  
 m^2_{(2,2),0} &= 2(4,4)^T G (2,2) = 2\cdot 48,   
 \label{masses-gauge-modes}
\end{align}
using $\r=(1,1)$. Here $G={\scriptsize \begin{pmatrix}
                           2 & 1 \\1 & 2
                          \end{pmatrix}}$ is the metric on $\msu(3)$ weight space.
We restrict ourselves to gauge fields with weight zero
here\footnote{The modes with nonzero weight acquire additional  
 mass terms from $\tilde\phi$ beyond the ones discussed here,
 which are difficult to evaluate. Moreover for large branes,
 we expect that the currents of fermionic links to a point brane 
 have either large or zero weight, and therefore do not couple to light 
gauge modes with  nonzero weight.};
note that e.g.\ $\L=(2,0)$ contains no weight zero modes.

For the minimal  $\cC[(1,1)]$ brane, we can also compute  the mass contribution from the chiral Higgs 
$\tilde\phi_\a$ solution, which lives on the chiral triangles of the 
$(1,1)$ branes as in figure \ref{fig:brane_1-1_plus_Higgs}. 
To this end, we decompose the $(1,1)$ brane and its Hilbert space
into the two triangles consisting of the $L$ and $R$ states, which can be viewed as $(1,0)$ and $(0,1)$ 
branes defined by the chiral Higgs $\tilde\phi_\a$.
Then all gauge modes with weight zero live in  $End(1,0) = (1,1) + (0,0)$ and $End(0,1) = 
(1,1) + (0,0)$
w.r.t.\ these triangles.  
Hence the mass contribution for the chiral gauge field $A_\mu\sim\chi$ with $\L=(3,0)$ vanishes  (because $[\chi,\tilde\phi] = 0$), 
while the mass contribution for the weight zero $\L=(1,1)$ mode is
\begin{align}
 m^2_{(1,1),0} &= 2(3,3)^T G (1,1) = 2\cdot 18  .
\end{align} 
Adding this to \eqref{masses-gauge-modes}, we see that indeed the chiral boson $A_\mu \sim\chi$ 
as well as the two $\L=(1,1)$ gauge bosons are the 
lightest, degenerate gauge bosons  (within weight zero), with mass $m^2 = 2\cdot36$. 
Note that  the contribution from the chiral Higgs  to the 
non-chiral gauge bosons such as $m^2_{(1,1),0} $ should be larger on larger branes, while 
 \eqref{masses-gauge-modes} is universal.
It is then plausible that  the chiral gauge mode $A_\mu$ \eqref{chiral-gauge-mode}  becomes the lightest mode, 
which would entail a chiral gauge theory
with a spontaneously broken $U(1)_L \times U(1)_R$ gauge field and 3 generations.

On larger branes, the details are more complicated, because there are many 
chiral Higgs which may contribute to the mass. 
However, the underlying geometrical mechanism is very clear: the $3_L+3_R$ sheets at the origin lead a priori to a 
$U(3)_L \times U(3)_R$ gauge theory, which
is broken not only by the global connectedness of the brane leading to 
\eqref{masses-gauge-modes}, but also by the chiral 
Higgs modes which link the L and R sheets among themselves. These in turn
break the symmetry to $U(1)_L\times U(1)_R$ with 3 generations, and hopefully leave
the chiral gauge field $A_\mu$ as lightest non-trivial gauge boson.
Note also that in general, non-trivial Higgs configurations may lead to some back-reaction on the brane 
(cf. the discussion in section \ref{sec:rigged-CP2}),  which may lead  to a relative shift between the $L$ and $R$ 
branes in target space, thus
amplifying the effects on the symmetry breaking. This should be kept in mind in 
the discussion about approaching the Standard Model in section 
\ref{sec:standardmodel}.
\subsection{\texorpdfstring{$\cC[(1,1)]$}{(1,1)} brane with chiral Higgs 
solution}
\label{sec:1-1_brane}
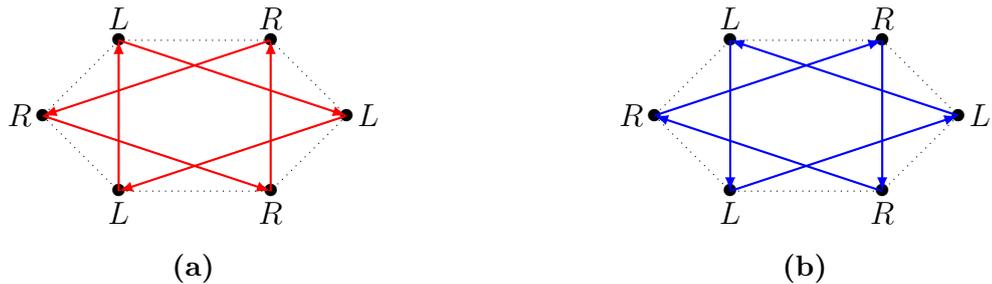
\begin{figure}[t!]
\centering
\begin{subfigure}{0.48\textwidth}
\centering
\begin{tikzpicture}
  \coordinate (P1) at (-2,0);
  \coordinate (P2) at (-1,1);
  \coordinate (P3) at (-1,-1);
  \coordinate (P5) at (1,1);
  \coordinate (P7) at (1,-1);
  \coordinate (P8) at (2,0);
  \node at (P1) {$\bullet$};
  \node at (P2) {$\bullet$};
  \node at (P3) {$\bullet$};
  \node at (P5) {$\bullet$};
  \node at (P7) {$\bullet$};
  \node at (P8) {$\bullet$};
% % 
\draw [thick,red,-latex] (P5) -- (P1);%
\draw [thick,red,-latex] (P8) -- (P3);%
\draw [thick,red,-latex] (P3) -- (P2);%
\draw [thick,red,-latex] (P7) -- (P5);%
\draw [thick,red,-latex] (P1) -- (P7);%
\draw [thick,red,-latex] (P2) -- (P8);%
\draw[dotted] (P1)--(P3)--(P7)--(P8)--(P5)--(P2)--cycle;
\draw [black] (P1) node[left] {$R$};
\draw [black] (P2) node[above] {$L$};
\draw [black] (P3) node[below] {$L$};
\draw [black] (P5) node[above] {$R$};
\draw [black] (P7) node[below] {$R$};
\draw [black] (P8) node[right] {$L$};
\end{tikzpicture}%
\caption{}%
\label{fig:brane_1-1_config_1}%
\end{subfigure}%
\begin{subfigure}{0.48\textwidth}
\centering
\begin{tikzpicture}
  \coordinate (P1) at (-2,0);
  \coordinate (P2) at (-1,1);
  \coordinate (P3) at (-1,-1);
  \coordinate (P5) at (1,1);
  \coordinate (P7) at (1,-1);
  \coordinate (P8) at (2,0);
  \node at (P1) {$\bullet$};
  \node at (P2) {$\bullet$};
  \node at (P3) {$\bullet$};
  \node at (P5) {$\bullet$};
  \node at (P7) {$\bullet$};
  \node at (P8) {$\bullet$};
% %
\draw [thick,blue,-latex] (P2) -- (P3);%
\draw [thick,blue,-latex] (P5) -- (P7);%
\draw [thick,blue,-latex] (P7) -- (P1);%
\draw [thick,blue,-latex] (P8) -- (P2);%
\draw [thick,blue,-latex] (P1) -- (P5);%
\draw [thick,blue,-latex] (P3) -- (P8);%
\draw[dotted] (P1)--(P3)--(P7)--(P8)--(P5)--(P2)--cycle;
\draw [black] (P1) node[left] {$R$};
\draw [black] (P2) node[above] {$L$};
\draw [black] (P3) node[below] {$L$};
\draw [black] (P5) node[above] {$R$};
\draw [black] (P7) node[below] {$R$};
\draw [black] (P8) node[right] {$L$};
\end{tikzpicture}%
\caption{}%
\label{fig:brane_1-1_config_2}%
\end{subfigure}%
\caption{$\cC[(1,1)]$ brane with non-maximal regular zero modes. (a) Brane 
background together with a configuration involving $\varphi_i^{\pm}$. (b) 
Brane background together with a configuration consisting of $\sigma_i^{\pm}$.}
\label{fig:brane_1-1_plus_Higgs}
\end{figure}

On the  $\cC[(1,1)]$ brane, we have indeed an exact brane plus chiral Higgs
solution.
The underlying rank two regular zero modes have already been presented in 
figures 
\ref{fig:brane_1-1_rank2_1}--\ref{fig:brane_1-1_rank2_3}.
Because they form a closed algebra, we can combine these to form new exact 
solutions of the form $X +\varphi$ and $X + \sigma$, which are depicted in 
figure \ref{fig:brane_1-1_plus_Higgs}. The details of how to arrange this to 
get an exact solution to the equations of motion are delegated to appendix 
\ref{subsec:1-1_brane}.

Having found two such equivalent solutions, we analyzed the spectrum of the 
vector Laplacian around these exact solutions. As it turns out, there are a 
number of negative modes, indicating potential instabilities. However
by including a mass terms $M_i\equiv M$, one can again eliminate all instabilities 
for  $0.47\lesssim M \leq \frac{\sqrt{2}}{3}$, see for instance figure 
\ref{fig:spectrum_1-1_brane}. In this massive case there remain 8 zero modes of 
$\cO_V^{X+\varphi}$ or $\cO_V^{X+\sigma}$, which are again understood in terms of a compact moduli space, 
which could presumably  be lifted by introducing different masses.
%%H
% 
% 
\subsection{Rigged \texorpdfstring{$\C P^2$}{CP2} branes}
\label{sec:rigged-CP2}

\begin{figure}[t!]
\centering
\begin{tikzpicture}
  \coordinate (P1) at (1,-1);
  \coordinate (P3) at (-1,-1);
  \coordinate (P4) at (1.5,0);
  \coordinate (P7) at (-1.5,0);
  \coordinate (P14) at (0.5,2);
  \coordinate (P15) at (-0.5,2);
  \node at (P1) {$\bullet$};
  \node at (P3) {$\bullet$};
  \node at (P4) {$\bullet$};
  \node at (P7) {$\bullet$};
  \node at (P14) {$\bullet$};
  \node at (P15) {$\bullet$};
  \draw[dotted] (P1)--(P3)--(P7)--(P15)--(P14)--(P4)--cycle;
% % % 
\draw [thick,blue,-latex] (P1) -- (P7) -- (P14) -- cycle;%
\draw [thick,red,-latex]  (P3) -- (P4) -- (P15) -- cycle;%
\draw [black] (P3) node[left] {$L$};
\draw [black] (P4) node[right] {$L$};
\draw [black] (P15) node[left] {$L$};
\draw [black] (P1) node[right] {$R$};
\draw [black] (P7) node[left] {$R$};
\draw [black] (P14) node[right] {$R$};
\end{tikzpicture}%
 \caption{Rigged $(N,1)$ brane as two chiral branes, stabilized by chiral Higgs 
(red \& blue).}
 \label{fig:inst-diagram}
\end{figure}
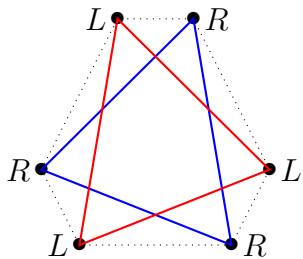

Now consider $(N,1)$ branes. These can be viewed as a stack of 
two $\C P^2$ branes linked by a minimal fuzzy sphere $S^2_2$, see for instance 
\cite{Grosse:2004wm}. We illustrate this set-up in figure 
\ref{fig:inst-diagram}.
The minimal fuzzy 2-sphere plays the role of a Higgs linking the two $\C P^2$ 
branes. 
This can be understood noting that the extra $S^2$ fiber is 
embedded transversal to the $\C P^2$, leading to a 6-dimensional geometry.
This is also similar to the flipped branes connected by Higgs in section \ref{sec:flipped-branes}, 
which is now realized as an exact solution with non-trivial intrinsic topology.
A decomposition into $(N,0)$ branes can be obtained explicitly in a suitable basis, see \cite{Grosse:2004wm}. 
This also leads to Yukawa couplings  which remove certain 
fermions from the massless sector, resulting in the typical structure found on 
6-dimensional branes.

\begin{figure}[t!]
\centering
\begin{subfigure}{0.25\textwidth}
\centering
\begin{tikzpicture}
  \coordinate (P1) at (1,-1);
  \coordinate (P2) at (0,-1);
  \coordinate (P3) at (-1,-1);
  \coordinate (P4) at (1.5,0);
  \coordinate (P7) at (-1.5,0);
  \coordinate (P10) at (1,1);
  \coordinate (P12) at (-1,1);
  \coordinate (P14) at (0.5,2);
  \coordinate (P15) at (-0.5,2);
  \node at (P1) {$\bullet$};
  \node at (P2) {$\circ$};
  \node at (P3) {$\bullet$};
  \node at (P4) {$\bullet$};
  \node at (P7) {$\bullet$};
  \node at (P10) {$\circ$};
  \node at (P12) {$\circ$};
  \node at (P14) {$\bullet$};
  \node at (P15) {$\bullet$};
  \draw[dotted] (P1)--(P2)--(P3)--(P7)--(P12)--(P15)--(P14)--(P10)--(P4)--cycle;
% % % 
\draw [thick,red,-latex] (P1) -- (P15) node[at end,right] {$\phi_1^+$};%
\draw [thick,red,-latex] (P14) -- (P3) node[at end,right] {$\phi_2^+$};%
\draw [thick,red,-latex] (P7) -- (P4) node[at end,above] {$\phi_3^+$};%
\end{tikzpicture}%
\caption{}%
\label{fig:brane_2-1_rank1}%
\end{subfigure}%
\begin{subfigure}{0.25\textwidth}
\centering
\begin{tikzpicture}
  \coordinate (P1) at (1,-1);
  \coordinate (P2) at (0,-1);
  \coordinate (P3) at (-1,-1);
  \coordinate (P4) at (1.5,0);
  \coordinate (P7) at (-1.5,0);
  \coordinate (P10) at (1,1);
  \coordinate (P12) at (-1,1);
  \coordinate (P14) at (0.5,2);
  \coordinate (P15) at (-0.5,2);
  \node at (P1) {$\bullet$};
  \node at (P2) {$\circ$};
  \node at (P3) {$\bullet$};
  \node at (P4) {$\bullet$};
  \node at (P7) {$\bullet$};
  \node at (P10) {$\circ$};
  \node at (P12) {$\circ$};
  \node at (P14) {$\bullet$};
  \node at (P15) {$\bullet$};
  \draw[dotted] (P1)--(P2)--(P3)--(P7)--(P12)--(P15)--(P14)--(P10)--(P4)--cycle;
% % 
\draw [thick,red,-latex] (P4) -- (P15) node[midway,left] {$\varphi_1^{+}$};%
\draw [thick,red,-latex] (P1) -- (P12) ;%
\draw [thick,blue,-latex] (P1) -- (P14);%
\draw [thick,blue,-latex] (P2) -- (P15)  node[midway,right] {$\sigma_1^{+}$};%
\end{tikzpicture}%
\caption{}%
\label{fig:brane_2-1_rank2_1}%
\end{subfigure}%
% \\
\begin{subfigure}{0.25\textwidth}
\centering
\begin{tikzpicture}
  \coordinate (P1) at (1,-1);
  \coordinate (P2) at (0,-1);
  \coordinate (P3) at (-1,-1);
  \coordinate (P4) at (1.5,0);
  \coordinate (P7) at (-1.5,0);
  \coordinate (P10) at (1,1);
  \coordinate (P12) at (-1,1);
  \coordinate (P14) at (0.5,2);
  \coordinate (P15) at (-0.5,2);
  \node at (P1) {$\bullet$};
  \node at (P2) {$\circ$};
  \node at (P3) {$\bullet$};
  \node at (P4) {$\bullet$};
  \node at (P7) {$\bullet$};
  \node at (P10) {$\circ$};
  \node at (P12) {$\circ$};
  \node at (P14) {$\bullet$};
  \node at (P15) {$\bullet$};
  \draw[dotted] (P1)--(P2)--(P3)--(P7)--(P12)--(P15)--(P14)--(P10)--(P4)--cycle;
% % 
\draw [thick,red,-latex] (P14) -- (P2) ;%
\draw [thick,red,-latex] (P15) -- (P3) node[midway,right] {$\varphi_2^{+}$};%
\draw [thick,blue,-latex] (P10) -- (P3) node[midway,left] {$\sigma_2^{+}$};%
\draw [thick,blue,-latex] (P14) -- (P7) ;%
\end{tikzpicture}%
\caption{}%
\label{fig:brane_2-1_rank2_2}%
\end{subfigure}%
\begin{subfigure}{0.25\textwidth}
\centering
\begin{tikzpicture}
  \coordinate (P1) at (1,-1);
  \coordinate (P2) at (0,-1);
  \coordinate (P3) at (-1,-1);
  \coordinate (P4) at (1.5,0);
  \coordinate (P7) at (-1.5,0);
  \coordinate (P10) at (1,1);
  \coordinate (P12) at (-1,1);
  \coordinate (P14) at (0.5,2);
  \coordinate (P15) at (-0.5,2);
  \node at (P1) {$\bullet$};
  \node at (P2) {$\circ$};
  \node at (P3) {$\bullet$};
  \node at (P4) {$\bullet$};
  \node at (P7) {$\bullet$};
  \node at (P10) {$\circ$};
  \node at (P12) {$\circ$};
  \node at (P14) {$\bullet$};
  \node at (P15) {$\bullet$};
  \draw[dotted] (P1)--(P2)--(P3)--(P7)--(P12)--(P15)--(P14)--(P10)--(P4)--cycle;
% % 
\draw [thick,red,-latex] (P12) -- (P4) node[midway,below] {$\varphi_3^{+}$};%
\draw [thick,red,-latex] (P7) -- (P1) ;%
\draw [thick,blue,-latex] (P7) -- (P10) node[midway,above right] 
{$\sigma_3^{+}$};%
\draw [thick,blue,-latex] (P3) -- (P4);%
\end{tikzpicture}%
\caption{}%
\label{fig:brane_2-1_rank2_3}%
\end{subfigure}%
\caption{The rank 1 and 2 regular zero modes for the $\cC[(2,1)]$ brane, 
exemplifying the $(N,1)$ case. The maximal Higgs $\phi_i^+$ of 
(\subref{fig:brane_2-1_rank1}) connect extremal weight states, while the 
next-to-maximal Higgs $\varphi_i^+$, $\sigma_i^+$ of 
(\subref{fig:brane_2-1_rank2_1})--(\subref{fig:brane_2-1_rank2_3}) also relate 
extremal and non-extremal weight states. Only the latter are (approximately) chiral.}
%%H these Higgs is exactly chiral.
\label{fig:brane_2-1_Higgs}
\end{figure}
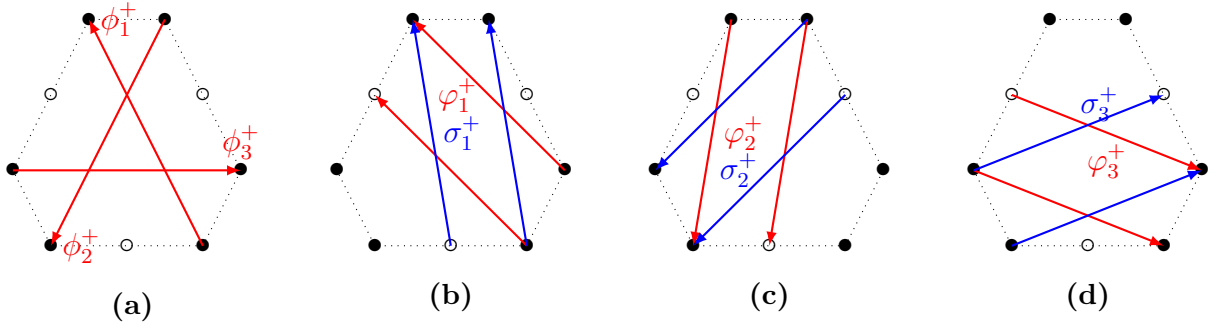

\paragraph{Equations of motion.}
For these $(N,1)$ branes, the \emph{next-to maximal} Higgs modes play again 
the role of 
chiral stringy Higgs modes. In view of figure 
(\ref{fig:brane_2-1_rank2_1})--(\ref{fig:brane_2-1_rank2_3}), 
these modes have the structure 
\begin{align}
 \varphi_i^+ &= \tilde{\varphi}_{i}^+ + H_{i}^+ 
 \label{chiral-Higgs-rigged}
 \end{align}
 where 
 \begin{align}
 \label{eq:Higgs_rigged_CP2}
  \tilde{\varphi}_{i}^+ &= \a_N|x_L^i\rangle\langle x_L^j|,   \qquad 
   H_{i}^+ = h_N |x_R^i\rangle\langle {x'}^j|,  \qquad
    h_N \sim \frac{1}{\sqrt{N}}\a_N
%  (X_i^-)^{N+1} X_j^- + \a_N X_j^- (X_i^-)^{N+1}, \qquad [X_j^+,\tilde \phi_i^+] = 0
\end{align}
because 
\begin{align}
0 =  [X_i^+ ,\varphi_j^-] = X_i^+ \tilde\varphi_j^- - \tilde\varphi_j^-X_i^+ 
\sim (\a_N - h_N \sqrt{N}) \phi_j^- \ .
\end{align}
Note that $\varphi_{i}$ relates again extremal states with the same chirality, 
but they no longer form a closed algebra
because of the sub-leading $H$ contribution.
Hence the $\varphi_i$ do not yield an exact solution within the Higgs 
sector. Nevertheless they lower the energy of the brane, and 
we expect that there exist slightly \emph{deformed} solutions with similar 
properties. Presumably, such solutions would involve small admixtures of 
other zero modes (and possibly massive modes). For large $N$, 
this deformation should be negligible.
The argument for the $\sigma_j$ modes is completely analogous.

\subsection{\texorpdfstring{$\cC[(N,N)]$}{C[(N,N)]} branes}
\begin{figure}[t!]
\centering
\begin{subfigure}{0.25\textwidth}
\centering
\begin{tikzpicture}
  \coordinate (P1) at (1,-2);
  \coordinate (P2) at (0,-2);
  \coordinate (P3) at (-1,-2);
  \coordinate (P4) at (1.5,-1);
  \coordinate (P7) at (-1.5,-1);
  \coordinate (P8) at (2,0);
  \coordinate (P12) at (-2,0);
  \coordinate (P18) at (1.5,1);
  \coordinate (P21) at (-1.5,1);
  \coordinate (P25) at (1,2);
  \coordinate (P26) at (0,2);
  \coordinate (P27) at (-1,2);
  \node at (P1) {$\bullet$};
  \node at (P2) {$\circ$};
  \node at (P3) {$\bullet$};
  \node at (P4) {$\circ$};
  \node at (P7) {$\circ$};
  \node at (P8) {$\bullet$};
  \node at (P12) {$\bullet$};
  \node at (P18) {$\circ$};
  \node at (P21) {$\circ$};
  \node at (P25) {$\bullet$};
  \node at (P26) {$\circ$};
  \node at (P27) {$\bullet$}; 
  \draw[dotted] (P1)--(P3)--(P12)--(P27)--(P25)--(P8)--cycle;
% % % 
\draw [thick,red,-latex] (P1) -- (P27) node[at end,right] {$\phi_1^+$};%
\draw [thick,red,-latex] (P25) -- (P3) node[at end,right] {$\phi_2^+$};%
\draw [thick,red,-latex] (P12) -- (P8) node[at end,above] {$\phi_3^+$};%
\end{tikzpicture}%
\caption{}%
\label{fig:brane_2-2_rank1}%
\end{subfigure}%
\begin{subfigure}{0.25\textwidth}
\centering
\begin{tikzpicture}
  \coordinate (P1) at (1,-2);
  \coordinate (P2) at (0,-2);
  \coordinate (P3) at (-1,-2);
  \coordinate (P4) at (1.5,-1);
  \coordinate (P7) at (-1.5,-1);
  \coordinate (P8) at (2,0);
  \coordinate (P12) at (-2,0);
  \coordinate (P18) at (1.5,1);
  \coordinate (P21) at (-1.5,1);
  \coordinate (P25) at (1,2);
  \coordinate (P26) at (0,2);
  \coordinate (P27) at (-1,2);
  \node at (P1) {$\bullet$};
  \node at (P2) {$\circ$};
  \node at (P3) {$\bullet$};
  \node at (P4) {$\circ$};
  \node at (P7) {$\circ$};
  \node at (P8) {$\bullet$};
  \node at (P12) {$\bullet$};
  \node at (P18) {$\circ$};
  \node at (P21) {$\circ$};
  \node at (P25) {$\bullet$};
  \node at (P26) {$\circ$};
  \node at (P27) {$\bullet$}; 
  \draw[dotted] (P1)--(P3)--(P12)--(P27)--(P25)--(P8)--cycle;
% % 
\draw [thick,red,-latex] (P1) -- (P21) node[midway,left] {$\varphi_1^{+}$};%
\draw [thick,red,-latex] (P4) -- (P27) ;%
\draw [thick,blue,-latex] (P1) -- (P26);%
\draw [thick,blue,-latex] (P2) -- (P27)  node[midway,right] {$\sigma_1^{+}$};%
\end{tikzpicture}%
\caption{}%
\label{fig:brane_2-2_rank2_1}%
\end{subfigure}%
% \\
\begin{subfigure}{0.25\textwidth}
\centering
\begin{tikzpicture}
    \coordinate (P1) at (1,-2);
  \coordinate (P2) at (0,-2);
  \coordinate (P3) at (-1,-2);
  \coordinate (P4) at (1.5,-1);
  \coordinate (P7) at (-1.5,-1);
  \coordinate (P8) at (2,0);
  \coordinate (P12) at (-2,0);
  \coordinate (P18) at (1.5,1);
  \coordinate (P21) at (-1.5,1);
  \coordinate (P25) at (1,2);
  \coordinate (P26) at (0,2);
  \coordinate (P27) at (-1,2);
  \node at (P1) {$\bullet$};
  \node at (P2) {$\circ$};
  \node at (P3) {$\bullet$};
  \node at (P4) {$\circ$};
  \node at (P7) {$\circ$};
  \node at (P8) {$\bullet$};
  \node at (P12) {$\bullet$};
  \node at (P18) {$\circ$};
  \node at (P21) {$\circ$};
  \node at (P25) {$\bullet$};
  \node at (P26) {$\circ$};
  \node at (P27) {$\bullet$}; 
  \draw[dotted] (P1)--(P3)--(P12)--(P27)--(P25)--(P8)--cycle;% % 
\draw [thick,red,-latex] (P25) -- (P2) ;%
\draw [thick,red,-latex] (P26) -- (P3) node[midway,right] {$\varphi_2^{+}$};%
\draw [thick,blue,-latex] (P25) -- (P7) node[midway,left] {$\sigma_2^{+}$};%
\draw [thick,blue,-latex] (P18) -- (P3) ;%
\end{tikzpicture}%
\caption{}%
\label{fig:brane_2-2_rank2_2}%
\end{subfigure}%
\begin{subfigure}{0.25\textwidth}
\centering
\begin{tikzpicture}
  \coordinate (P1) at (1,-2);
  \coordinate (P2) at (0,-2);
  \coordinate (P3) at (-1,-2);
  \coordinate (P4) at (1.5,-1);
  \coordinate (P7) at (-1.5,-1);
  \coordinate (P8) at (2,0);
  \coordinate (P12) at (-2,0);
  \coordinate (P18) at (1.5,1);
  \coordinate (P21) at (-1.5,1);
  \coordinate (P25) at (1,2);
  \coordinate (P26) at (0,2);
  \coordinate (P27) at (-1,2);
  \node at (P1) {$\bullet$};
  \node at (P2) {$\circ$};
  \node at (P3) {$\bullet$};
  \node at (P4) {$\circ$};
  \node at (P7) {$\circ$};
  \node at (P8) {$\bullet$};
  \node at (P12) {$\bullet$};
  \node at (P18) {$\circ$};
  \node at (P21) {$\circ$};
  \node at (P25) {$\bullet$};
  \node at (P26) {$\circ$};
  \node at (P27) {$\bullet$}; 
  \draw[dotted] (P1)--(P3)--(P12)--(P27)--(P25)--(P8)--cycle;% % 
\draw [thick,red,-latex] (P21) -- (P8) node[midway,below] {$\varphi_3^{+}$};%
\draw [thick,red,-latex] (P12) -- (P4) ;%
\draw [thick,blue,-latex] (P12) -- (P18) node[midway,above right] 
{$\sigma_3^{+}$};%
\draw [thick,blue,-latex] (P7) -- (P8);%
\end{tikzpicture}%
\caption{}%
\label{fig:brane_2-2_rank2_3}%
\end{subfigure}%
\\
\begin{subfigure}{0.25\textwidth}
\centering
\begin{tikzpicture}
  \coordinate (P1) at (1,-2);
  \coordinate (P2) at (0,-2);
  \coordinate (P3) at (-1,-2);
  \coordinate (P4) at (1.5,-1);
  \coordinate (P7) at (-1.5,-1);
  \coordinate (P8) at (2,0);
  \coordinate (P12) at (-2,0);
  \coordinate (P18) at (1.5,1);
  \coordinate (P21) at (-1.5,1);
  \coordinate (P25) at (1,2);
  \coordinate (P26) at (0,2);
  \coordinate (P27) at (-1,2);
  \node at (P1) {$\bullet$};
  \node at (P2) {$\circ$};
  \node at (P3) {$\bullet$};
  \node at (P4) {$\circ$};
  \node at (P7) {$\circ$};
  \node at (P8) {$\bullet$};
  \node at (P12) {$\bullet$};
  \node at (P18) {$\circ$};
  \node at (P21) {$\circ$};
  \node at (P25) {$\bullet$};
  \node at (P26) {$\circ$};
  \node at (P27) {$\bullet$}; 
  \draw[dotted] (P1)--(P3)--(P12)--(P27)--(P25)--(P8)--cycle;
% % 
\draw [thick,red,-latex] (P8) -- (P27) node[midway,above] 
{$\tilde{\varphi}_1^{+}$};%
\draw [thick,red,-latex] (P4) -- (P21) ;%
\draw [thick,red,-latex] (P1) -- (P12) ;%
\draw [thick,blue,-latex] (P1) -- (P25);%
\draw [thick,blue,-latex] (P2) -- (P26);%
\draw [thick,blue,-latex] (P3) -- (P27)  node[midway,right] 
{$\tilde{\sigma}_1^{+}$};%
\end{tikzpicture}%
\caption{}%
\label{fig:brane_2-2_rank3_1}%
\end{subfigure}%
% \\
\begin{subfigure}{0.25\textwidth}
\centering
\begin{tikzpicture}
    \coordinate (P1) at (1,-2);
  \coordinate (P2) at (0,-2);
  \coordinate (P3) at (-1,-2);
  \coordinate (P4) at (1.5,-1);
  \coordinate (P7) at (-1.5,-1);
  \coordinate (P8) at (2,0);
  \coordinate (P12) at (-2,0);
  \coordinate (P18) at (1.5,1);
  \coordinate (P21) at (-1.5,1);
  \coordinate (P25) at (1,2);
  \coordinate (P26) at (0,2);
  \coordinate (P27) at (-1,2);
  \node at (P1) {$\bullet$};
  \node at (P2) {$\circ$};
  \node at (P3) {$\bullet$};
  \node at (P4) {$\circ$};
  \node at (P7) {$\circ$};
  \node at (P8) {$\bullet$};
  \node at (P12) {$\bullet$};
  \node at (P18) {$\circ$};
  \node at (P21) {$\circ$};
  \node at (P25) {$\bullet$};
  \node at (P26) {$\circ$};
  \node at (P27) {$\bullet$}; 
  \draw[dotted] (P1)--(P3)--(P12)--(P27)--(P25)--(P8)--cycle;% % 
\draw [thick,red,-latex] (P27) -- (P3) node[midway,left] 
{$\tilde{\varphi}_2^{+}$};%
\draw [thick,red,-latex] (P26) -- (P2) ;%
\draw [thick,red,-latex] (P25) -- (P1) ;
\draw [thick,blue,-latex] (P25) -- (P12) ;%
\draw [thick,blue,-latex] (P18) -- (P7) ;%
\draw [thick,blue,-latex] (P8) -- (P3) node[midway,below] 
{$\tilde{\sigma}_2^{+}$};%
\end{tikzpicture}%
\caption{}%
\label{fig:brane_2-2_rank3_2}%
\end{subfigure}%
\begin{subfigure}{0.25\textwidth}
\centering
\begin{tikzpicture}
  \coordinate (P1) at (1,-2);
  \coordinate (P2) at (0,-2);
  \coordinate (P3) at (-1,-2);
  \coordinate (P4) at (1.5,-1);
  \coordinate (P7) at (-1.5,-1);
  \coordinate (P8) at (2,0);
  \coordinate (P12) at (-2,0);
  \coordinate (P18) at (1.5,1);
  \coordinate (P21) at (-1.5,1);
  \coordinate (P25) at (1,2);
  \coordinate (P26) at (0,2);
  \coordinate (P27) at (-1,2);
  \node at (P1) {$\bullet$};
  \node at (P2) {$\circ$};
  \node at (P3) {$\bullet$};
  \node at (P4) {$\circ$};
  \node at (P7) {$\circ$};
  \node at (P8) {$\bullet$};
  \node at (P12) {$\bullet$};
  \node at (P18) {$\circ$};
  \node at (P21) {$\circ$};
  \node at (P25) {$\bullet$};
  \node at (P26) {$\circ$};
  \node at (P27) {$\bullet$}; 
  \draw[dotted] (P1)--(P3)--(P12)--(P27)--(P25)--(P8)--cycle;% % 
\draw [thick,red,-latex] (P27) -- (P8) node[midway,above] 
{$\tilde{\varphi}_3^{+}$};%
\draw [thick,red,-latex] (P21) -- (P4) ;%
\draw [thick,red,-latex] (P12) -- (P1) ;%
\draw [thick,blue,-latex] (P12) -- (P25) node[midway,below] 
{$\tilde{\sigma}_3^{+}$};%
\draw [thick,blue,-latex] (P7) -- (P18);%
\draw [thick,blue,-latex] (P3) -- (P8);%
\end{tikzpicture}%
\caption{}%
\label{fig:brane_2-2_rank3_3}%
\end{subfigure}%
\caption{The rank one, two, and three regular zero modes for the $\cC[(2,2)]$ 
brane, 
exemplifying the $(N,N)$ case. The maximal Higgs $\phi_i^+$ of 
(\subref{fig:brane_2-2_rank1}) connect extremal weight states. The 
next-to-maximal Higgs $\varphi_i^+$, $\sigma_i^+$ of 
(\subref{fig:brane_2-2_rank2_1})--(\subref{fig:brane_2-2_rank2_3}) also relate 
extremal 
and non-extremal weight states. Neither the rank one nor the rank two zero 
modes are chiral. Only the rank three zero modes of 
(\subref{fig:brane_2-2_rank3_1})--(\subref{fig:brane_2-2_rank3_3}) are chiral 
modes.}
\label{fig:brane_2-2_Higgs}
\end{figure}
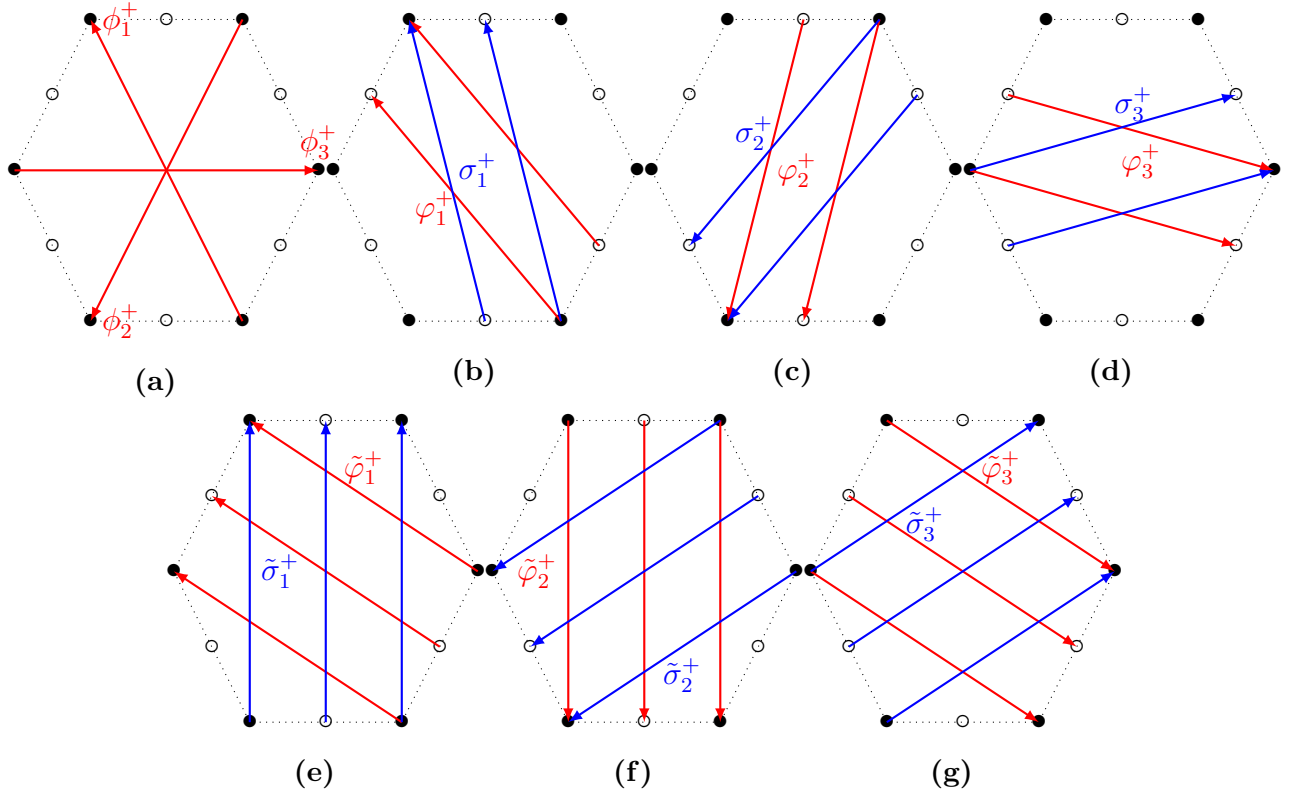

Finally, we briefly consider $\cC[(N,N)]$ branes 
as sketched in figure \ref{fig:brane_2-2_Higgs}. 
Among the Higgs modes we mention the maximal zero modes 
$\phi_i^+$ in $(2N,2N)$, and the $(3N,0)$ modes 
$\tilde{\varphi}_i^+$, $\tilde{\sigma}_i^+$ (and similarly $(0,3N)$ modes)
connecting the opposite edges of $(N,N)$. These
are again approximately chiral Higgs modes
(similar to the $(1,1)$ case), which do not quite form a closed algebra, 
but clearly lower the energy of the brane.
Hence we expect that there are
nearby deformed solutions.
\subsection{Nonabelian case: Stacks of 6-dimensional branes and point-branes}
\label{sec:nonabel-Higgs-6D}
Now consider a stack of two identical 6-dimensional branes as above, each with 
chiral Higgs $\tilde \phi$ switched on, and add also
an extra point brane $\cD$ to make it more interesting.
This clearly leads to an unbroken $U(2)$ gauge group. However, the results of 
the last sections lead to a more refined statement:
the two branes lead to a $U(2)_L\times U(2)_R$ gauge theory which is 
spontaneously broken to 
$U(2)_{\diag}$, and massive chiral gauge bosons $A_\mu$ taking values in 
$\mmu(2)_L - \mmu(2)_R$. 
We will see in section \ref{sec:fermions} that there are also fermionic zero 
modes linking $\cD$ with the $3_L + 3_R$ sheets on each brane,
leading to 3 generations of chiral fermions transforming in the fundamental of 
$U(2)_L$ and $U(2)_R$, respectively,
which have opposite charges under $A_\mu$ according to their chirality.
This is a chiral gauge theory in a broken phase, reminiscent of the $SU(2)_L 
\times SU(2)_R$ Pati-Salam-type electroweak model.
Further suitable (maximal, non-chiral) Higgs  
between the two 6-dimensional branes may break the symmetry to $U(1)$ and lead 
to patterns quite close to the Standard Model, which
will be discussed  in section \ref{sec:standardmodel}.
%
%%%%%%%%%%%%%%%%%%%%%%%%%%%%%%%%%%%%%%%%%%%%%%%%%%%%%%%
%%%%%%%%%%%%%%%%%%%%%%%%%%%%%%%%%%%%%%%%%%%%%%%%%%%%%%%
% 
\section{Fermions on branes with Higgs}
\label{sec:fermions}
The Dirac operator on a squashed background $\cC[\mu]$ is given by 
\begin{equation}
 \slashed{D}^X = \sqrt{2} \sum_{j=1}^3 \left(\Delta_j^- [X_j^+,\cdot] + 
\Delta_j^+ [X_j^-,\cdot]\right)
\end{equation}
acting on $\End(\cH) \otimes \cS$, where $\cS\cong \C^8$
accommodates the spinors. 
The $\Delta_j^\pm$ are fermionic ladder operators which satisfy 
$\{\Delta_j^-,\Delta_k^+\}=\delta_{j,k}$. For the 
squashed background $\cC[\mu]$, the $X_j^\pm$ act as ladder operators for the 
preserved $U(1)_{K_i}$ charges. With this input, it was shown in
\cite{Steinacker:2014lma,Steinacker:2014eua,Steinacker:2015mia}
that the fermionic zero modes $\Psi_{\a,\L'}$ on $\cC[\mu]$ 
correspond to the extremal weight states of each 
irrep $\cH_\L$ appearing in $\End(\cH)=\oplus_\L \cH_\L$, and are
in one-to-one correspondence to the 
regular zero modes $\phi_{\a,\L'}$ of the scalar fields discussed in section 
\ref{sec:regular-zm}. 
The proof in \cite{Steinacker:2014lma} is based on the extremal weight 
properties, while a proof in the spirit of index theory was given in 
\cite{Steinacker:2015mia}.
Consequently, the zero modes are again labeled by their $U(1)_{K_i}$ quantum 
numbers $\L'$, and 
their chirality is determined by the 
parity $\t=\pm 1$ of (the Weyl chamber of) $\L'$.
In addition, there exist two trivial gaugino zero modes on each brane.

The results can be summarized by stating that a quiver gauge theory arises on
stacks of squashed branes $\oplus n_i\cC[\mu_i]$, with gauge group $U(n_i)$ on 
each node $\mu_i$
and arrows corresponding to chiral superfields $\phi_{\a,\L'}$
labeled by the extremal weights $\L'$ corresponding to
the multiplets $\cH_\L \subset \Hom(\cH_{\mu_i},\cH_{\mu_j})$.
The trivial modes $\L=0$ on each node lead to $\cN=4$ supermultiplets.

More specifically, on a given stack of squashed $\cC[\mu_i]$ branes,
fermionic zero modes arise in two ways: first, as \emph{intra-brane fermions} 
$\Psi\in \End(\cH_{\mu_i}) \otimes \cS$ on some given brane $\cC[\mu_i]$. 
The intra-brane fermions are uncharged under the gauge groups arising on the 
(stacks of) different branes, but they are chiral and charged under 
$U(1)_{K_i}$. The latter two features protect them 
from acquiring any mass terms on the $\cC[\mu_i]$ background, because opposite 
charges have opposite chirality.
Nevertheless, these modes may acquire masses in the presence of 
non-vanishing Higgs modes due to Yukawa couplings discussed below.
In contrast, the two trivial gaugino modes with $\L=0$ are unprotected, and
are therefore expected to acquire a large mass due to the soft SUSY breaking, 
either at tree level or through loop corrections.

Second and more interestingly, fermionic zero modes also arise as \emph{inter-brane fermions} 
$\Psi\in \Hom(\cH_{\mu_i},\cH_{\mu_j}) \otimes \cS$ linking two different 
(stacks of) branes.
These are  charged under the gauge groups arising on the 
(stacks of) different branes, chiral, and protected by their $U(1)_{K_i}$ charges. 
Hence they can acquire a mass only through Yukawa couplings in the presence of 
 Higgs modes linking different branes.
Note that due to the 9+1-dimensional Majorana-Weyl condition, the 
$\Psi\in \Hom(\cH_{\mu_i},\cH_{\mu_j}) \otimes \cS$ are related via charge conjugation 
to the $\Psi\in \Hom(\cH_{\mu_j},\cH_{\mu_i}) \otimes \cS$. This is important to 
avoid over-counting 
e.g. in \eqref{fermion-matrix}, and to obtain a chiral gauge theory.

%%H 
% 

\paragraph{Yukawa couplings.}
Assume that $\cC[\mu_L]$ and $\cC[\mu_R]$ are connected with some Higgs 
$\phi_{\a}$. Then Yukawa couplings of two such 
fermionic zero modes arise from the $\cN=4$ action \eqref{N=4SYM}, with 
structure
\begin{align}
  \tr \left( \overline{\Psi} \gamma_5 \Delta^\a[\phi_\a,\Psi] \right)\ .
 \label{Yukawa}
\end{align}
This respects the $U(3)_R$ symmetry and the $U(1)_{K_i}$ symmetry. 
Consequently, the non-vanishing Yukawa couplings in the zero-mode sector have 
the same structure as the cubic term $V_{\mathrm{soft}}$ in the potential,
\begin{align}
 \tr \left( \overline{\Psi}_{-\a_i} \gamma_5 
\Delta^{\a_j}[\phi_{\a_j},\Psi_{\a_k}] 
\right) \ \sim\ \varepsilon_{ijk} 
\label{tau-cons-yuk}
\end{align}
and its conjugate. These Yukawa couplings are (non-)vanishing if and only if the 
corresponding cubic term $\tr ( \phi_{\a_i} [\phi_{\a_j},\phi_{\a_k}])$
(with the same $U(1)_{K_i}$ quantum numbers) is (non-)vanishing.
This requires in particular that the $U(1)^K_i$ charges $\L'$ of $\phi_{\a_j}$ 
and $\Psi_{\a_k}$ add up to that of $\Psi_{\a_i}$.
In particular, \emph{the $\tau$-parities of $\a_i,\a_j,\a_k$ must be equal}.
\paragraph{Fermionic zero modes on branes with Higgs.}
For a combined brane plus Higgs background $Y=X+\phi$, the above 
classification of fermionic (and bosonic) zero modes  does not apply 
any more. The reason is that the $Y_j^\pm$ are typically
no longer ladder operators and do not satisfy any Lie algebra relations.
As illustrated in various settings in appendix 
\ref{app:solutions_spectrum}, the spectrum of $\cO_V^{Y}$ and $\Di^Y$ may 
behave  utterly different compared to their spectrum on the squashed backgrounds. 

Nevertheless, one can understand in many cases the fate of the fermionic zero modes, and 
obtain a qualitative understanding of the remaining low-energy sector.
We will argue that some of the chiral fermionic zero modes 
of $\Di^X$ are coupled by the Yukawa couplings induced by $\phi$ and acquire a 
mass. Hence, they disappear from the low-energy spectrum on the combined 
background $X+\phi$. 
On the other hand, some other fermionic zero modes are protected and remain massless.
Adding a point brane $\cD$ to the combined background 
solution, the inter-brane fermionic zero modes may lead to a very interesting low-energy physics,
reproducing  ingredients of the Standard Model.

Due to the complicated setting, most of these arguments are only qualitative at this point, and 
we do not have a complete understanding in all cases. 
The detailed numerical results are given in appendix 
\ref{app:solutions_spectrum}.
% 
%%%%%%%%%%%%%%%%%%%%%%%%%%%%%%%%%%%%%%%%%%%%%%%%%%%%%%%%%%%%%%%
%
\subsection{Fermions on \texorpdfstring{$\cC[(N,0)]$}{C[N,0]} branes with 
maximal Higgs}
For a single $\cC[(N,0)]$ background, there are $6(N+1)+2$ fermionic zero 
modes. These consist of $6(N+1)$ modes from the decomposition 
$\End(\cH_{(N,0)})=\oplus_{l=0}^N \cH_{(l,l)}$, plus two trivial gaugino
modes. Turning on the maximal regular bosonic zero modes $\phi$, the spectrum 
of $\Di^{X+\phi}$ contains only 14 zero modes independent of $N$, see figure 
\ref{fig:CP2-brane+Higgs_Dirac-modes}. This reduction clearly arises from the Yukawa couplings 
due to the maximal Higgs $\phi$, which leaves the $6+2$ zero modes from $\cH_{(0,0)}$, plus 6 extra zero
modes whose origin is obscure.

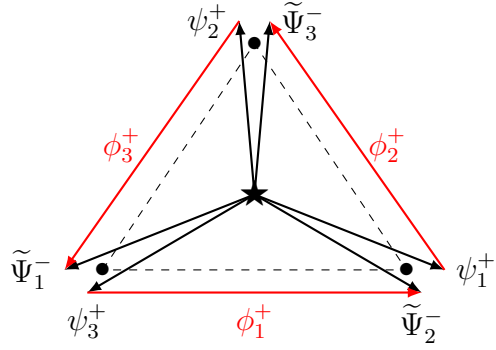
\begin{figure}[t!]
\centering
\begin{tikzpicture}
  \coordinate (Origin)   at (0,0);
  \coordinate (P1) at (-2,-1);
  \coordinate (P2) at (0,2);
  \coordinate (P3) at (2,-1);
  \coordinate (P1+a3) at (-2-0.2,-1-0.3);
  \coordinate (P1-a1) at (-2-0.5,-1);
  \coordinate (P2+a2) at (0-0.2,2+0.3);
  \coordinate (P2-a3) at (0+0.2,2+0.3);
  \coordinate (P3+a1) at (2+0.5,-1);
  \coordinate (P3-a2) at (2+0.2,-1-0.3);
  \node at (Origin) {$\bigstar$};
  \node at (P1) {$\bullet$};
  \node at (P2) {$\bullet$};
  \node at (P3) {$\bullet$};
  \draw [thick,-latex] (Origin)--(P1+a3) node[at end,below] {$\psi_3^+$};
  \draw [thick,-latex] (Origin)--(P1-a1) node[at end,left] 
{$\widetilde{\Psi}_1^-$};
  \draw [thick,-latex] (Origin)--(P2+a2) node[at end,left] {$\psi_2^+$};
  \draw [thick,-latex] (Origin)--(P2-a3) node[at end,right] 
{$\widetilde{\Psi}_3^-$};
  \draw [thick,-latex] (Origin)--(P3+a1) node[at end,right] {$\psi_1^+$};
  \draw [thick,-latex] (Origin)--(P3-a2) node[at end,below] 
{$\widetilde{\Psi}_2^-$};
\draw [dashed] (P1)--(P2)--(P3)--cycle;
 \draw [red,thick,-latex] (P3+a1) -- (P2-a3) node[midway,right] {$\phi_2^+$};
 \draw [red,thick,-latex] (P2+a2) -- (P1-a1) node[midway,left] {$\phi_3^+$};
 \draw [red,thick,-latex] (P1+a3) -- (P3-a2) node[midway,below] {$\phi_1^+$};
\end{tikzpicture}%
\caption{Chiral inter-brane fermions $\Psi_{\a,\L'}$ and 
$\widetilde{\Psi}_{\a,\L'}$ linking $\cC[(N,0)]$  with a point-brane $\cD$, 
which is represented by $\bigstar$. 
Their chirality is indicated by the 
sign $\pm$, which is inherited from the Weyl chamber. Red arrows correspond to 
maximal Higgs $\phi_j^+$.}
\label{fig:CP2-brane-center-fermions}
\end{figure}

Adding a point brane $\cD$ to this $X+\phi$, the 
number of fermionic zero modes is 22 according to figure 
\ref{fig:Dirac-modes_brane+point_x=0+1_2}, 
which differs by 8 from the 14 modes on $\cC[(N,0)]$ with maximal Higgs.
We can understand this as follows:
A priori, there are $2\cdot6$ fermionic inter-brane zero modes between 
$\cC[(N,0)]$ and $\cD$,
which arise from $\Hom(\cH_{(N,0)},\C) \cong \cH_{(N,0)}$. 
However, upon switching on the maximal Higgs, 
Yukawa couplings $\tr(\widetilde{\Psi}_3^+ [\phi_2^+,\Psi_1^+])$ arise as 
depicted in figure \ref{fig:CP2-brane-center-fermions}, which couple these zero modes and give them a mass.
This leaves only the $6+2$ intra-brane fermions on $\cD$, which remain massless.
This  explains the numerical findings.
% 
%%%%%%%%%%%%%%%%%%%%%%%%%%%%%%%%%%%%%%%%%%%%%%%%%%%%%%%%%%%%%%
%%%%%%%%%%%%%%%%%%%%%%%%%%%%%%%%%%%%%%%%%%%%%%%%%%%%%%%%%%%%%%
% 
% 
\subsection{Fermions on \texorpdfstring{$\cC[(1,1)]$}{C[1,1]} brane with 
chiral Higgs, and point brane}
\label{sec:ferm-11-chiral}
Now we come to our most interesting solution: start with 
one of the two exact solutions $X + \varphi$ or $X+\sigma$
of figure \ref{fig:brane_1-1_plus_Higgs}, corresponding to a $\cC[(1,1)]$ brane with chiral Higgs.
As shown in figure \ref{fig:spectrum_1-1_brane}, the number of 
fermionic zero modes is reduced to 20, in comparison to the 38 zero modes of the 
$\cC[(1,1)]$ brane without Higgs.

Now, add a  point brane $\cD$ to the above $\cC[(1,1)]$ with chiral Higgs. 
Then on top of the above 20 fermionic zero modes from $\cC[(1,1)]$ with chiral Higgs, 
the numerical analysis reveals 20 additional fermionic zero modes. These are understood as follows:
$2 \cdot 6$ zero-modes arise from inter-brane zero modes $\Psi_{\cC,\cD}$ 
corresponding to $\Hom(\cH_{(1,1)},\C)$ and its conjugate. 
In addition, there are 8 trivial intra-brane fermions on $\cD$, 
which consist of 6 modes in $\End(\cH_{(0,0)})\cong \C$ and 2 gaugino modes.

The  interesting point is that these $2\cdot6$ inter-brane zero modes do 
\emph{not} acquire a mass, even in the presence of the 
chiral Higgs on $\cC[(1,1)]$. The key word here is \emph{chiral} Higgs, which 
by definition link the extremal weight states 
of $\cC[(1,1)]$ with the same chirality, see  figure 
\ref{fig:brane_1-1_plus_Higgs} and \eqref{chiral-Higgs-decomp}. 
Since the $\Psi_{\cC,\cD}$ linking $\cD$ with these states have the same 
chirality,  
they cannot form a mass term, and remain massless.
We view them as (toy-versions of) left-and right-handed leptons, 
since they couple with opposite charges to the 
chiral gauge field $A_\mu$ of \eqref{chiral-gauge-mode},
\begin{align}
 A_\mu \sim \chi \sim \one_L - \one_R \, \sim \g_5
 \label{chi-structure-g5}
\end{align}
and come in 3 generations.
If some extra Higgs mode is switched on 
which links the states with opposite chirality, e.g.\ the maximal Higgs mode, 
then these left-and right-handed leptons 
would acquire a mass,  as in the Standard Model. Such scenarios will be 
discussed 
further in section \ref{sec:standardmodel}.
\subsection{Fermions on rigged \texorpdfstring{$\C P^2$}{CP2} brane}
\label{sec:ferm-1N-chiral}
The story for the $\cC[(N,1)]$ branes from 
section \ref{sec:rigged-CP2} is very similar to the $(1,1)$ case, but should be 
even more interesting as far as physics and the scales are concerned.
As before, we are mostly interested  in 
a point brane $\cD$ added to a $\cC[(N,1)]$ brane with Higgs. 
The drawback for this rigged $\C P^2$ scenario is that we lack an exact solution 
which would reflect the configuration of figure \ref{fig:inst-diagram}. 
Nonetheless, some qualitative statements can be made. 
There are again the $3+3$ fundamental chiral zero modes linking 
$\cD$ to $\cC[(N,1)]$, which are attached to the $3_L+3_R$ extremal weight  
(coherent) states, and are viewed as 3 generations of \emph{leptons}. 
These leptons will survive in the presence of the chiral Higgs as before, and 
couple to
the chiral gauge field $A_\mu \sim \chi \sim \g_5$ of \eqref{chi-structure-g5}.
Due to the large $N$, one may hope that this chiral $A_\mu$ is now indeed the 
lightest non-trivial gauge boson on $\cC[(N,1)]$, as discussed in section 
\ref{sec:rigged-CP2}.
The resulting physics is that of 3 generations of leptons coupled to $A_\mu \sim\g_5$. 
This demonstrates how a chiral gauge theory can arise from softly broken 
$\cN=4$ SYM in a suitable vacuum corresponding to space-filling branes with 
fluxes.

\subsection{Fermions on flipped minimal branes plus point brane --- the 
\texorpdfstring{$G_2$}{G2} brane}
\label{sec:ferm-G2-chiral}
Let us consider the fermionic zero modes around the combined $G_2$-type 
backgrounds of figure \ref{fig:G2-branes}. 
The analysis presented in appendix \ref{subsec:flipped_min_brane+point} shows 
that from the original 84 fermionic zero modes on $\cC[(1,0)]+\cC[(0,1)]+\cD$ 
(without any Higgs), only 14 remain on the combined background.

Adding an extra point brane $\cD$ to the $G_2$-type solution,  
one might  expect  (i) inter-brane fermions between $\cD$ 
and the $G_2$-type solution, and (ii) intra-brane fermions on $\cD$. The numerical
analysis of the Dirac spectrum, however, shows that the combined system has 
only 22 fermion zero modes, which is only 8 more than on the $G_2$ solution.
This means that among the $2\cdot6$ inter-brane fermions 
and the $6+2$ trivial fermionic modes from $\cD$, only 8 remain 
massless, while the 
remaining ones pair up and form massive states. 
The reason is that the $G_2$ orbits have higher dimension, and 
the chirality properties of the 6-dimensional solutions no longer apply.
Therefore the present $G_2$ solution is less interesting for the application 
in section \ref{sec:standardmodel}, but it may give hints how to find non-trivial Higgs 
solutions on several $SU(3)$ branes.
% 
%%%%%%%%%%%%%%%%%%%%%%%%%%%%%%%%%%%%%%%%%%%%%%%%%%%%%%%%%%%%%%
%%%%%%%%%%%%%%%%%%%%%%%%%%%%%%%%%%%%%%%%%%%%%%%%%%%%%%%%%%%%%%
% 
\subsection{Fermions on two squashed \texorpdfstring{$\C P^2$}{CP2} branes with 
a point brane \& Higgs}
\label{sec:two-CP2-point-Higgs}
Finally consider two parallel branes  $\cC[\mu_L]$ and $\cC[\mu_R]$ with  
$\mu_L=(N_L,0)$ and $\mu_R = (N_R,0)$ and a point brane $\cD$. In 
section \ref{sec:two-branes-point-Higgs}, we found various Higgs solutions, linking the 
extremal weight states of the various branes, which break the $U(1)\times U(1)$ gauge fields 
$A^{L,R}_\mu$ arising from the $\cC[\mu_{L,R}]$.
There are again fermionic zero modes $\Psi_{\cD \cC_L}$ and  $\Psi_{\cD \cC_R}$ linking $\cD$ to the two branes,
and we want to see if the Yukawa couplings to  the Higgs may lead to an interesting chiral low-energy 
gauge theory, where fermions with different chiralities have different couplings to the gauge fields $A^{L,R}_\mu$
as in sections \ref{sec:ferm-11-chiral} and  \ref{sec:ferm-1N-chiral}.

The task is trickier here, because $\Psi_{\cD \cC_L}$, for instance, provides 
two fermions connecting to the same 
corner of $\cH_{((N_L,0)}$, whose chirality is given by their $\tau$-parity. They will be paired up by the intra-brane Higgs 
connecting these corners. However, by inspection, all solutions found in section 
\ref{sec:two-branes-point-Higgs}  
break the $\Z^3$ symmetry, and they also typically involve intra-brane Higgs 
(except for the solutions in section \ref{sec:three-branes}). 
Hence, even though the surviving massless fermions are chiral and have different 
couplings to $A^L_\mu$ and $A^R_\mu$,
the generation symmetry $\Z_3$ is not respected. This leads to a somewhat strange low-energy theory far from the 
Standard Model. Nevertheless, it is conceivable --- and even reasonable --- 
that the inter-brane Higgs acquire by some other mechanism a VEV 
which does respect $\Z_3$, as indicated in figure \ref{fig:two-CP2-brabes-Yukawas}.
\begin{figure}[t!]
\centering
\begin{tikzpicture}
  \coordinate (Origin) at (0,0);
  \coordinate (Q1) at (-1,-0.5);
  \coordinate (Q2) at (0,1);
  \coordinate (Q3) at (1,-0.5);
  \coordinate (P1) at (-3,-3/2);
  \coordinate (P2) at (0,3);
  \coordinate (P3) at (3,-3/2);
  \coordinate (Q1+a3) at (-1-0.2,-0.5-0.3);
  \coordinate (Q1-a1) at (-1-0.5,-0.5);
  \coordinate (Q2+a2) at (0-0.2,1+0.3);
  \coordinate (Q2-a3) at (0+0.2,1+0.3);
  \coordinate (Q3+a1) at (1+0.5,-0.5);
  \coordinate (Q3-a2) at (1+0.2,-0.5-0.3);
  \coordinate (P1+a3) at (-3-0.2,-3/2-0.3);
  \coordinate (P1-a1) at (-3-0.5,-3/2);
  \coordinate (P2+a2) at (0-0.2,3+0.3);
  \coordinate (P2-a3) at (0+0.2,3+0.3);
  \coordinate (P3+a1) at (3+0.5,-3/2);
  \coordinate (P3-a2) at (3+0.2,-3/2-0.3);
  \node at (Q1) {$\bullet$};
  \node at (Q2) {$\bullet$};
  \node at (Q3) {$\bullet$};
  \node at (P1) {$\bullet$};
  \node at (P2) {$\bullet$};
  \node at (P3) {$\bullet$};
  \draw[-latex,orange] (Origin)--(Q1+a3) node[at end,below] 
{$\psi_3^{+}$};
  \draw[-latex] (Origin)--(Q1-a1) node[at end,left] 
{$\widetilde{\psi}_1^{-}$};
  \draw[-latex,orange] (Origin)--(Q2+a2) node[at end,left] 
{$\psi_2^{+}$};
  \draw[-latex] (Origin)--(Q2-a3) node[at end,right] 
{$\widetilde{\psi}_3^{-}$};
  \draw[-latex,orange] (Origin)--(Q3+a1) node[at end,right] 
{$\psi_1^{+}$};
  \draw[-latex] (Origin)--(Q3-a2) node[at end,below] 
{$\widetilde{\psi}_2^{-}$};
  \draw[-latex] (Origin)--(P1+a3) node[at end,below] 
{$\Psi_3^{+}$};
  \draw[-latex,orange] (Origin)--(P1-a1) node[at end,left] 
{$\widetilde{\Psi}_1^{-}$};
  \draw[-latex] (Origin)--(P2+a2) node[at end,left] 
{$\Psi_2^{+}$};
  \draw[-latex,orange] (Origin)--(P2-a3) node[at end,right] 
{$\widetilde{\Psi}_3^{-}$};
  \draw[-latex] (Origin)--(P3+a1) node[at end,right] 
{$\Psi_1^{+}$};
  \draw[-latex,orange] (Origin)--(P3-a2) node[at end,below] 
{$\widetilde{\Psi}_2^{-}$};
\draw[dotted] (P1)--(P2)--(P3)--cycle;
\draw[dotted] (Q1)--(Q2)--(Q3)--cycle;
\node at (Origin) {$\bigstar$};
  \draw [thick,red,-latex] (P3+a1) -- (Q2-a3) node[midway,above] 
{$H_2^{+}$};
  \draw [thick,red,-latex] (P2+a2) -- (Q1-a1) node[midway,left]  
{$H_3^{+}$};
  \draw [thick,red,-latex] (P1+a3) -- (Q3-a2) node[midway,below]  
{$H_1^{+}$};
\end{tikzpicture}%
 \caption{}
 \label{fig:two-CP2-brabes-Yukawas}
\caption{Set-up $\cC[\mu_L]+\cC[\mu_R] + \cD$ with a $\Z_3$-invariant Higgs configuration, where 
the point brane $\cD$ is depicted in the center and $\cC[\mu_L]$ corresponds to 
the outermost brane. We sketch the chiral intra-brane fermions $\Psi$, 
$\widetilde{\Psi}$ between $\cC[\mu_L]$ and $\cD$ as well as $\psi$, 
$\widetilde{\psi}$ between 
$\cC[\mu_R]$ and $\cD$. The fermions $\Psi$ and $\widetilde{\psi}$ can be 
linked by maximal inter-brane Higgs $H$, $\tilde{H}$ between $\cC[\mu_{L,R}]$. 
The chirality of the fermionic zero modes is indicated by the sign $\pm$, which 
is inherited from the Weyl chamber.}
\end{figure}
Then it is possible to have a situation similar as in sections  \ref{sec:ferm-11-chiral} and  \ref{sec:ferm-1N-chiral},
where e.g. all $\Psi_{\cD \cC_L}$ are left-handed and all $\Psi_{\cD \cC_R}$ are right-handed. 

Since we do not have a dynamical justification for such a $\Z_3$-invariant 
Higgs configuration,
we will focus on the configurations in sections \ref{sec:ferm-11-chiral} and  \ref{sec:ferm-1N-chiral}
in the following discussion towards the Standard Model.
%
%%%%%%%%%%%%%%%%%%%%%%%%%%%%%%%%%%%%%%%%%%%%%%%%%%%%%%%%%%%%%%%%
%%%%%%%%%%%%%%%%%%%%%%%%%%%%%%%%%%%%%%%%%%%%%%%%%%%%%%%%%%%%%%%%
%
\section{Approaching the Standard Model}
\label{sec:standardmodel}
At first sight, it may seem impossible to get anything resembling the Standard 
Model from deformed $\cN=4$ SYM. After all the Standard Model is chiral, while 
$\cN=4$ SYM is not.
In fact, any low-energy gauge theory arising in some vacuum of a 
deformation of $\cN=4$, as considered here, will have index zero, cf.\ 
\cite{Steinacker:2013eya}.
However, the Standard Model extended by right-handed neutrinos $\nu_R$ does 
have index zero, and this is what we aim to approach with  \emph{sterile} 
$\nu_R$, which are uncharged under the gauge group of the SM.
The scenario to be discussed will be reminiscent of 
(a supersymmetric extension of) the Pati-Salam model \cite{Pati:1974yy} in the 
broken phase.
This is a refinement of the brane configuration proposed in \cite{Steinacker:2015mia}, using the 
results of the previous sections.

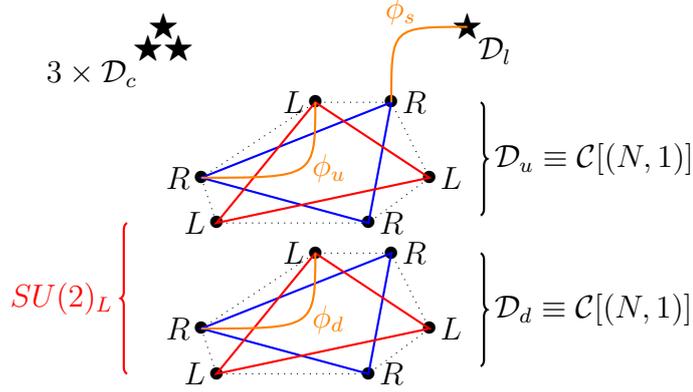
\begin{figure}[t!]
\centering
\begin{tikzpicture}
%Brane 1 
  \coordinate (P1) at (0.7,-0.6);
  \coordinate (P3) at (-1.3,-0.6);
  \coordinate (P4) at (1.5,0);
  \coordinate (P7) at (-1.5,0);
  \coordinate (P14) at (1,1);
  \coordinate (P15) at (0,1);
  \node at (P1) {$\bullet$};
  \node at (P3) {$\bullet$};
  \node at (P4) {$\bullet$};
  \node at (P7) {$\bullet$};
  \node at (P14) {$\bullet$};
  \node at (P15) {$\bullet$};
  \draw[dotted] (P1)--(P3)--(P7)--(P15)--(P14)--(P4)--cycle;
% % % 
\draw [thick,blue,-latex] (P1) -- (P7) -- (P14) -- cycle;%
\draw [thick,red,-latex]  (P3) -- (P4) -- (P15) -- cycle;%
\draw [black] (P3) node[left] {$L$};
\draw [black] (P4) node[right] {$L$};
\draw [black] (P15) node[left] {$L$};
\draw [black] (P1) node[right] {$R$};
\draw [black] (P7) node[left] {$R$};
\draw [black] (P14) node[right] {$R$};
% 
%define label
\draw[decoration={brace,mirror,raise=5pt},decorate,thick]
  (2,-0.5) -- node[right=6pt] {$\cD_u \equiv \cC[(N,1)]$} (2,1.0);
%   
%draw phi_u
\coordinate (phiU) at (0,0);
\draw[orange,thick] (P7) .. controls (phiU) .. (P15) 
node[midway,right] {$\phi_u$};
% 
%Brane 2 
  \coordinate (R1) at (0.7,-0.6-2);
  \coordinate (R3) at (-1.3,-0.6-2);
  \coordinate (R4) at (1.5,0-2);
  \coordinate (R7) at (-1.5,0-2);
  \coordinate (R14) at (1,1-2);
  \coordinate (R15) at (0,1-2);
  \node at (R1) {$\bullet$};
  \node at (R3) {$\bullet$};
  \node at (R4) {$\bullet$};
  \node at (R7) {$\bullet$};
  \node at (R14) {$\bullet$};
  \node at (R15) {$\bullet$};
  \draw[dotted] (R1)--(R3)--(R7)--(R15)--(R14)--(R4)--cycle;
% % % 
\draw [thick,blue,-latex] (R1) -- (R7) -- (R14) -- cycle;%
\draw [thick,red,-latex]  (R3) -- (R4) -- (R15) -- cycle;%
\draw [black] (R3) node[left] {$L$};
\draw [black] (R4) node[right] {$L$};
\draw [black] (R15) node[left] {$L$};
\draw [black] (R1) node[right] {$R$};
\draw [black] (R7) node[left] {$R$};
\draw [black] (R14) node[right] {$R$};
% 
%
%define label
\draw[decoration={brace,mirror,raise=5pt},decorate,thick]
  (2,-2.5) -- node[right=6pt] {$\cD_d \equiv \cC[(N,1)]$} (2,-1.0);
% 
%draw phi_d
\coordinate (phiD) at (0,0-2);
\draw[orange,thick] (R7) .. controls (phiD) .. (R15) 
node[midway,right] {$\phi_d$};
% 
%leptonic point brane
\coordinate (L) at (2,2);
\node at (L) {$\bigstar$};
\draw [black] (L) node[below right] {$\cD_l$};
% 
%color point branes
\coordinate (C1) at (-2,2);
\coordinate (C2) at (-2+0.2,2-0.3);
\coordinate (C3) at (-2-0.2,2-0.3);
\node at (C1) {$\bigstar$};
\node at (C2) {$\bigstar$};
\node at (C3) {$\bigstar$};
\draw [black] (C3) node[below left] {$3\times\cD_c$};
% 
% 
%draw SU2L
\draw[decoration={brace,mirror,raise=5pt},decorate,red,thick]
  (-1.3-1,-0.6) -- node[left=6pt] {$SU(2)_L$} (-1.3-1,-0.6-2);
% 
%draw Phi_S
\coordinate (phi) at (1,2);
\draw[orange,thick] (L) .. controls (phi) .. (P14) node[midway,above] 
{$\phi_s$};
\end{tikzpicture}%
 \caption{Brane configuration towards the Standard Model. Both $\cD_u$ and 
$\cD_d$ can be realized e.g.\ by rigged $(N,1)$ brane solutions, which 
decompose into $L$ and $R$ sheets with 3 generations.}
 \label{fig:SM-branes}
\end{figure}
Consider the brane configuration of figure \ref{fig:SM-branes}, which
is constructed in terms of our squashed brane solutions as follows:
The $\cD_d$ brane is realized by a \emph{rigged brane} $\cC[(N,1)]$, 
which decomposes into two chiral $\cD_{Ld}+\cD_{Rd}$ branes, with chiral Higgs 
$\tilde\phi$ switched on as in section \ref{sec:rigged-CP2}.
The $\cD_u$ has the same structure\footnote{this is 
motivated e.g. by the structure of the traceless electric charge generator $Q$ \eqref{Q-def}.} as $\cD_d$,
which  decomposes into two chiral $\cD_{Lu}+\cD_{Ru}$ branes, again with chiral Higgs $\tilde\phi$. 
%%H
Finally add 4 point branes, denoted as $3 \times \cD_c + \cD_l$ for reasons which will become 
clear soon.
The overall brane configuration $\cD_l + 3 \times \cD_c + (\cD_u + \cD_d)$ in 
the $SU(N)$ SYM model leads to a $SU(4) \times SU(2) \times U(1)$ gauge group.

Next, we \emph{assume} that there exists a non-vanishing ''Pati-Salam``  Higgs\footnote{Note 
that $\phi_S$ corresponds to the $(4,1,2)$ Higgs in the Pati-Salam model 
\cite{Pati:1974yy}.  
While there are indeed suitable regular zero modes  which link $\cD_l$ with the 
3 extremal $R$ states of $\cD_u$, 
we did not find an exact solution of this type, because they do not form closed 
triangles.
It might be possible to find such solutions in conjunction with the Higgs 
$H_{u,d}$ \eqref{Higgs-doublets-2},
or along the lines of the $G_2$ solutions in section \ref{sec:flipped-branes}.} 
$\phi_S$ linking $\cD_{Ru}$ with $\cD_l$, which breaks the gauge group to
\begin{align}
 SU(3)_c \times U(1)_Q \times U(1)_{B'} \ .
 \label{symm-SM-branes}
\end{align}
This breaking might also be achieved or accompanied 
by a displacement of the two $R$ branes, while the $L$ branes 
remain coincident.
Here $Q$ is the electric charge,  $\cL$ is the lepton number,  $B$ the baryon number, and
\begin{align}
 Q &\coloneqq \frac 12\big(\one_{{u}} -\one_{d}  + \cL - B \big) \ ,  \\
    B &= \frac 13 \one_c, \qquad B' = B - c \one \nn\\
    \cL &= \one_l \ ,
\label{Q-def}
\end{align}
such that $B'$ is traceless.
Note that $Q$ is traceless provided $\dim\cH_{u} = \dim\cH_{d}$,
which strongly suggests that the $\cD_u$ and $\cD_d$ have the same structure.
We also introduce the weak hypercharge
\begin{align}
 Y \coloneqq \one_{Ru} -\one_{Rd} + \cL - B \ .  
 \label{Y-def} 
\end{align}
$Q$ and $Y$ will reproduce the correct charge assignments of the Standard Model, 
and 
we recover the  Gell-Mann-Nishjima formula
\begin{align}
 Q- \frac 12 Y = \frac 12(\one_{Lu} - \one_{Ld}) \eqqcolon  T^3_L \ .
\end{align}
\paragraph{Fermions.}
Now consider the off-diagonal fermions linking these branes,
which arise as  zero modes $\Psi_{ij} \in \End(\cH_i,\cH_j)$ linking the 
extremal weight states of the  branes $\cC[\mu_i]$ and $\cC[\mu_j]$. 
Consider first the fermions between the point branes 
$\cD_l, \cD_c$ and $\cD_u, \cD_d$.
Since the former are point branes and
$\cD_u$ as well as $\cD_d$ have the structure of $\cC[(N,1)]$ branes (by 
assumption), 
we can simply apply the results of sections \ref{sec:ferm-11-chiral} and \ref{sec:ferm-1N-chiral}. 
Recalling that the extremal weight states of $\cD_{u}$
separate into $3_L + 3_R$ chiral extremal weight 
states $|\mu^i_{Lu}\rangle, |\mu^i_{Ru}\rangle$ and similarly for $\cD_d$,
we obtain 3 generations of chiral leptons  linking $\cD_l$ with $\cD_u, \cD_d$, and
 3 generations of chiral quarks linking $\cD_l$ with $\cD_u, \cD_d$.
In the basis $(|\mu^i_{Lu}\rangle,|\mu^i_{Ld}\rangle,|\mu^i_{Ru}\rangle,|\mu^i_{Rd}\rangle,
|0\rangle_l,|0^j\rangle_c)$, we denote the inter-brane fermions as 
\begin{align}
 \Psi =
 \begin{pmatrix}
  *_{2} &  \tilde H_u  & \tilde H_d  &  l_L & Q_L \\
   &  *  & e' & \nu_R & u_R \\
   &    & * & e_R & d_R  \\
   & & ~~~ & * &  u' \\
&  & & & *_3
\end{pmatrix} .
\label{fermion-matrix}
\end{align}
Here the left-handed quarks and leptons
\begin{align}
Q_L = \begin{pmatrix} u_L \\ d_L   \end{pmatrix}, \qquad
l_L = \begin{pmatrix} \nu_L \\ e_L   \end{pmatrix}, 
\end{align} 
arise as $SU(2)$ doublet acting on the $\cD_u,\cD_d$ branes. This $SU(2)$ is 
broken by $\phi_S$, which will be discussed in more detail below.
Similarly, the right-handed quarks and leptons also arise as $SU(2)$ doublets
\begin{align}
Q_R = \begin{pmatrix} u_R \\ d_R   \end{pmatrix}, \qquad
l_R = \begin{pmatrix} \nu_R \\ e_R   \end{pmatrix} .
\end{align} 
Note that the entries below the diagonal are not independent but related to the upper entries by charge conjugation, 
see section \ref{sec:fermions}.
%%H
The charge generators \eqref{Q-def}, \eqref{Y-def} are given explicitly by
\begin{equation}
\begin{aligned}
 Q &= \frac{1}{2} \diag\left(1,-1,1,-1,1,-\frac{1}{3}\right)\, , \\
  Y &= \diag\left(0,0,1,-1,1, - \frac{1}{3}\right) 
\end{aligned} 
\end{equation}
which results in the following quantum numbers for the off-diagonal modes 
\begin{align}
 (Q,Y)|_\Psi =
 \begin{pmatrix}
  * \ \ &   \begin{pmatrix} (0,-1) \\ (-1,-1) \end{pmatrix} & \begin{pmatrix} 
(1,1) \\ (0,1) \end{pmatrix}  & \begin{pmatrix} (0,-1) \\ (-1,-1) \end{pmatrix}  
& \begin{pmatrix} (\frac{2}{3},\frac{1}{3}) \\ 
(-\frac{1}{3},\frac{1}{3})\end{pmatrix}  \\[2ex]
   &  *  & (-1,-2)  & (0,0)  & (\frac{2}{3},\frac{4}{3})  \\[1ex]
   &     & *        &  (-1,-2) & (-\frac{1}{3},-\frac{2}{3})  \\[1ex]
   & & ~~~ & * & (\frac 23,\frac 43) \\
&  & & & *
\end{pmatrix} 
\label{charges}
\end{align}
dropping the obvious $SU(3)_c$ assignment.
All quantum numbers of the Standard Model are correctly reproduced,
and three families arise automatically due to the $\Z_3$ symmetry\footnote{The $\Z_3$ symmetry 
may of course be broken e.g.\ by the Yukawa couplings 
induced by different Higgs VEVs, or by a deformation of the background.}.
There are also some extra modes, including Higgsinos $\tilde H_{u,d}$ 
(as in the MSSM), the $\nu_R$ which is uncharged under the SM gauge group, 
gauginos, Winos, and some sterile diagonal fermionic modes. 
Furthermore, there are also extra modes $e'$ with the same quantum numbers as 
$e_R$, and 
$u'$ with the same quantum numbers as $u_R$. Their fate depends on the 
detailed structure of e.g.\ $\phi_S$ and will not be discussed here.
This way of obtaining the correct SM charges is familiar from the context of matrix models 
\cite{Grosse:2010zq,Chatzistavrakidis:2011gs,Steinacker:2014fja,Steinacker:2015mia} 
and  from intersecting brane constructions in string theory 
\cite{Aldazabal:2000cn,Antoniadis:2002qm,Blumenhagen:2005mu}.

%%H fix refs !

\paragraph{Higgs sector.}
In the present background, all the above fermions are exactly massless, even though the 
unbroken symmetry is only \eqref{symm-SM-branes}. The reason is that the 
$U(2)_L \times U(2)_R$ symmetry of the two $L$ and $R$ sheets 
of the 6-dimensional branes $\cD_u$ and $\cD_d$ is broken because the sheets 
are connected (see section \ref{sec:6D-branes}). This
can be viewed as breaking via some background 
''Higgs``,
%%H do keep "
which however does not couple to the above fermions. On the 
other hand, recall that
all the above fermionic zero modes have scalar superpartners,
as discussed in the first part of the paper.
In particular, this includes the \emph{electroweak} Higgs doublets $H_u$, $H_d$ 
with  $Y(H_d) = 1$ (as in the SM) and
$Y(H_u) = -1$ (as in the MSSM), which fit into the above matrix structure as 
\begin{align}
 \phi^a = \begin{pmatrix}
         0_2 & H_u & H_d & 0 & 0 \\
         & 0 & 0 & 0 & 0\\
         && 0 & \varphi_S & 0\\
         &&& 0 & 0 \\
         &&&&0
        \end{pmatrix} \ .
\label{higgs-matrix}
\end{align}
It is reasonable to assume that these are intra-brane Higgs modes within 
$\cD_u$ and within $\cD_d$, realized by bosonic
zero modes $\varphi_u$  and $\varphi_d$ as discussed in section 
\ref{sec:rigged-CP2}.
This means that they take the following VEVs
\begin{align}
 H_d = \begin{pmatrix}
         \varphi_d \\ 0
        \end{pmatrix}, \qquad 
 H_u = \begin{pmatrix}
        0 \\ \varphi_u 
        \end{pmatrix}
  \label{Higgs-doublets-2}
\end{align} 
with $Q=0$. 
More explicitly, these Higgs modes have the structure
\begin{align}
 H_d &\sim \sum \big(0\cdot |\mu_{L}\rangle_u + \varphi_{d} \cdot 
|\mu_{L}\rangle_d\big)
                   \langle\mu_{Rd}|_d  \cong \begin{pmatrix}
                                         0 \\ \varphi_{d} 
                                        \end{pmatrix}  \langle\mu_{Rd}|_d  \nn\\
 H_u &\sim \sum \big(\varphi_{u} \cdot|\mu_{L}\rangle_u  +  
0 \cdot |\mu_{L}\rangle_d\big)
                \langle\mu_{Ru}|_u \cong \begin{pmatrix}
                                         \varphi_{u} \\ 0
                                        \end{pmatrix} \langle\mu_{Ru}|_u \ .
 \label{Higgs-doublets}
\end{align}
Since they are intra-brane modes, they do not induce any further symmetry 
breaking; nonetheless, they do induce the desired Yukawa couplings between the 
left-and right-handed 
leptons and quarks, as in sections \ref{sec:ferm-11-chiral}, \ref{sec:ferm-1N-chiral}. 
Then the low-energy phenomenology should be fairly close 
to that of the Standard Model, extended by the various extra fields as above.

These  $H_{u,d}$ should be viewed as part of the combined background, 
which gives mass in particular to the $W^\pm$ bosons discussed below, 
as explained in section \ref{sec:6D-branes}. 
The lowest \emph{fluctuations} of this vacuum will involve all the 
constituents and 
may behave similar to a SM Higgs, while the detailed composition 
of the background will  enter only via the various gauge and Yukawa couplings.
We note that this is somewhat similar to a Pati-Salam model in the broken 
phase  \cite{Pati:1974yy}.
Whether or not such a scenario may be realistic is another issue, and perhaps 
there is a better background. 
The main point here is to show that one may come surprisingly close to SM-like 
low-energy physics, with only few reasonable assumptions on the VEVs of the zero 
modes.

Now consider the $\nu_R$ in more detail. Its fate is clearly affected by the presence 
of the Higgs link $\phi_S$ between $\cD_l$ and the $R$ states of $\cD_u$,
which leads to extra Yukawa couplings of the $\nu$ and some 
intra-brane fermions.
This would entail a mixture of fermionic states, which 
is too complicated to be fully analyzed here.
\paragraph{Gauge bosons.}
It is well-known that 
a background consisting of stacks of branes leads to massless $U(n_i)$ gauge 
bosons within stacks of $n_i$ coinciding identical branes. This yields 
\eqref{symm-SM-branes} in the present background. 

The story becomes more interesting if we account for the lightest massive gauge 
bosons. Assume for the moment that there is no $\phi_S$. 
Then the $\cD_u + \cD_d$ would define a $U(2)$ gauge symmetry which enhances to
$U(2)_L \times U(2)_R$ by taking the chiral $A_\mu \sim \chi \sim \g_5$ of
\eqref{chi-structure-g5} on the $\cC[(N,1)]$ branes into account. 
Together with the $SU(4)$ from the $\cD_l + 3 \times\cD_c$,
this is reminiscent of a Pati-Salam model in the broken phase. 
Clearly the $SU(2)_L$ along with the $Y$ 
contributes the $W^\pm$ and $Z$ gauge bosons of the electroweak sector, and
the $U(2)_R$ is broken in the presence of $\phi_S$, leading to 
\eqref{symm-SM-branes}. 
This provides the basic structure of an extended Standard Model.
One may hope that the various extra fields acquire a sufficiently high mass to be 
negligible at low energies, but this is beyond the scope of this paper.
\paragraph{Discussion.}
Let us briefly address some of the numerous open questions.
One concern is that we had to assume that the $\phi_S$ and the $H_{u,d}$ 
acquire the 
appropriate VEVs, without having an exact solution. However, if the $H_{u,d}$ 
are realized as links along the edges of $\cC[(N,1)]$, then there are in fact 
non-vanishing cubic terms involving e.g.\ $\tr(H_u \phi_S \phi_S')$,
which lower the energy. 
It is, therefore, plausible that there is such a solution, but it would 
presumably have 
a non-vanishing back-reaction on the brane, which we cannot compute. This is a 
non-trivial problem, and quantum effects might play an important role.

Another question is whether a suitable hierarchy could arise between the 3 generations.
Even though the background under consideration has an exact $\Z_3$  symmetry,
this could easily be broken in the Higgs sector, or possibly by introducing different
mass parameters $M_i$.

A further issue is the extra massless $U(1)_{B'}$
gauge field, which amounts to baryon number. 
%%H
Even though this protects from proton decay, there should not be any such massless gauge field, and 
it is not  clear how to remove this in the present 
field-theoretic setting. However, it might 
disappear via a St\"uckelberg-type mechanism in an analogous 
matrix model setting with an axion, cf.\ 
\cite{Morelli:2009ev,Coriano:2006xh,Kors:2005uz,Anastasopoulos:2006cz}
and the discussion in \cite{Steinacker:2015mia}.
In fact, all results of the present paper carry over immediately to the 
IKKT matrix model, where noncommutative $U(N)$ $\cN=4$ SYM arises on a stack of $N$ 
$(3+1)$-dimensional noncommutative brane solutions, while the internal 
structure 
of the present paper are unchanged. Thus, the present paper 
can also be seen as a possible way to obtain interesting particle physics
from the IKKT model, cf. \cite{Aoki:2014cya,Steinacker:2014fja}.

Finally, we note that instead of realizing, for instance, the $\cD_u$ 
branes via $\cC[(N,1)]$, we could alternatively use separate 
$L$ and $R$ branes $\cC[(N_L,0)]$ and $\cC[(N_R,0)]$, linked by suitable Higgs 
as in section 
\ref{sec:two-CP2-point-Higgs}. These links would give mass to the \emph{mirror 
fermions}, which disappear from the 
low-energy theory. The remaining discussion follows the logic employed above.
We recall, that this has been the setup proposed in \cite{Steinacker:2015mia}. 
The links are essentially an integral part of the 
$\cC[(N,1)]$ branes linking their chiral sheets, realized in an exact solution.

% *** discuss gauge couplings and mass gap etc ***

%
%%%%%%%%%%%%%%%%%%%%%%%%%%%%%%%%%%%%%%%%%%%%%%%%%%%%%%%%%%%%%%%%%
%%%%%%%%%%%%%%%%%%%%%%%%%%%%%%%%%%%%%%%%%%%%%%%%%%%%%%%%%%%%%%%%%
%
\section{Discussion  and conclusion}
\label{sec:conclusion}
Building on previous work \cite{Steinacker:2014lma,Steinacker:2015mia}, we 
studied vacua corresponding to squashed fuzzy coadjoint $SU(3)$ branes in 
$\cN=4$ SYM, softly broken by a $SU(3)$-invariant cubic potential and masses.
We found a rich class of novel vacua that include non-trivial condensates of 
the zero modes. The condensates are interpreted as string-like Higgs modes 
linking the self-intersecting branes.  
\paragraph{Summary.}
On a formal level, we showed that the potential can be rewritten in terms of complete squares 
\eqref{eq:potential_rewritten}, so that the equations of motion follow from a set of 
first integral equations \eqref{eq:1st-integral}. These  observations allow to extend
solutions from the massless to the massive case, and to establish the absence of any instabilities for 
a preferred mass parameter $M^*$, at the classical level. 
Furthermore, we give a useful new characterization of the regular zero modes 
(the \emph{Higgs modes}) on the $SU(3)$ branes in terms 
of a \emph{decoupling condition} \eqref{extremal-3}. As a consequence, the
full potential and equations of motion decouple completely for a combination 
$X+\phi$ of brane plus Higgs mode. 
This is the basis for establishing a rich class of new exact solutions of this 
type,
which lead to a spontaneously broken gauge theory with non-trivial Higgs VEVs 
and corresponding Yukawa couplings,
realizing the ideas put forward in \cite{Steinacker:2015mia}.

The decoupling conditions and the related rewriting of the potential in terms of complete squares are in many ways 
reminiscent of supersymmetry. It provides non-trivial moduli spaces of vacua, and establishes their stability. 
However there is no underlying symmetry, so that at present we cannot extend these structures to the quantum level.

At a more technical level, the first integral equations are somewhat weaker than 
$\msu(3)$ Lie algebra relations \eqref{Cartan-Weyl}. Nonetheless, $\msu(3)$ 
representations provide a valuable starting point, and most --- but not all --- 
of our solutions are based on $\msu(3)$
in some way. Clearly the basic branes $\cC[\mu]$ arise directly from $\msu(3)$. Using the decoupling properties of 
the zero modes, we constructed numerous novel combined brane plus Higgs 
solutions including the following:
\begin{compactenum}[(i)]
\item The fuzzy 2-sphere $S^2_N$ arises from the minimal regular zero mode $\phi_i^+ = 
- X_i^-$ on any $\cC[\mu]$ brane background. 
However, these are large perturbations, which completely change the geometry of the brane.
\item The maximal regular zero modes $\phi_i^+ = -\frac{1}{N!}(X_i^-)^N$ on 
$\cC[(N,0)]$ furnish the fundamental representation
$(1,0)$ of $\msu(3)$ and correspond the string-like operators. We have 
discussed these solutions at length in section \ref{sec:4D-branes}.
\item The next-to-maximal regular zero modes on $\cC[(1,1)]$ furnish interesting
chiral Higgs solutions which stabilize the chiral sheets, as shown in section \ref{sec:1-1_brane}.
A generalization to  $\cC[(N,1)]$ is conjectured.
\item As an elaborate example, various zero modes on 
$\cC[(1,0)]+\cC[(0,1)]+\cD$ are used to form the 7-dimensional fundamental 
representation of $\mathfrak{g}_2 \supset \msu(3)$, as discussed in section 
\ref{sec:flipped-branes}.
\end{compactenum}
The discussion of the fermion sector in section \ref{sec:fermions} is mostly qualitative. 
In the presence of several Yukawa couplings 
the situation is complicated, and a full 
treatment is beyond our scope. However in many cases we are able to explain the 
fermionic zero modes of $\Di^{X+\phi}$, notably for the 
$\cC[(1,1)]$ brane with chiral Higgs and point brane, which is used in our 
approach to the Standard Model. This is based on the detailed computations of 
appendix \ref{app:solutions_spectrum}.

\paragraph{Discussion.}
On a more physical level, two of the most interesting features of the new 
brane plus Higgs vacua are the following: (i) they have a mass gap, and 
(ii) typically only a small number of zero modes exists. The number 
of zero modes is independent of the rank $N \gg 1$ of the underlying $SU(N)$ 
gauge theory and of the size $\dim \cH_\L$ of the brane $\cC[\L]$. 
This observation is  remarkable, since the starting point is a 
gauge theory with large $N$, which is typically organized in terms of a t'Hooft 
 $\frac 1N$ genus expansion with effective coupling  $\l = g^2 N$.
In contrast, the $\cC[(N_1,N_2)]$ vacua  with large $N_i$ behave as 
semi-classical, large  branes. The fluctuation spectrum on these vacua 
consists of a small number of zero modes with typical coupling strength $g$, and 
a large tower of typically weakly interacting KK modes with a finite mass gap independent of $N$.
Hence, the original large $N$ gauge theory reduces to an
effective low-energy theory with few modes and an interesting geometric structure.
This should provide sufficient motivation to study them in more detail.

The most interesting aspect of these vacua is that they can lead to a chiral 
gauge theory at low energies, with interesting properties not far from the 
Standard Model. 
The point is that the underlying branes are locally space-filling in the 6 extra dimensions and carry a flux, so that 
bi-fundamental fermions are charged and expected to have chiral zero modes. This is precisely what happens, although the 
overall index is  bound to be zero. Elaborating on a previous proposal by using 
our new solutions, 
we discussed in section \ref{sec:standardmodel} a brane configuration 
which comes fairly close to the Standard Model. The set-up is 
reminiscent of the Pati-Salam model in the broken phase and of 
intersecting brane constructions in string theory. Even though we do not claim 
that this is realistic, it certainly provides strong motivation for further 
work.

\paragraph{Future directions.}

%%H
There are many open issues which should be addressed in future work. 
One task is to understand better the 
chiral Higgs solution \eqref{chiral-Higgs-rigged} on the $(N,1)$ branes and the associated chiral gauge bosons, 
which seem to be particularly interesting for physics. 
Another is to justify a non-vanishing ''Pati-Salam`` $\phi_S$  in our Standard Model approach, 
and to see how close to real physics one may come in this way. 
More generally, it would be desirable to have a more systematic understanding and perhaps 
a classification of the solutions of the  first-order equations of motion 
\eqref{firstorder-eom}.

At some point of course, these
classical considerations will no longer suffice.
We have argued that a classical treatment should be justifiable to some extent on large branes, due 
to the existence of a gap with few remaining low-energy modes, and  the mild UV 
behavior of the softly broken $\cN=4$ model. 
Nevertheless, quantum effects need to be taken into account at some point.
Due to the global $SU(3)$ symmetry, the relevant terms in the effective potential
must preserve the form \eqref{V-soft}. Moreover, the critical mass $M^*$ of 
\eqref{M-star-def} marks the transition 
between trivial and non-trivial stable vacua; hence, it should play a special 
place in the full quantum theory,  
possibly as an RG fixed point.
%%H
A direct loop computation by summing over all higher KK modes seems far too 
complicated. 
For a deformed $\cN=4$ model, one strategy might be to invoke holography; 
however, this is questionable since (i) the model is not conformal and (ii) the 
vacuum is highly non-trivial. 
Fortunately, a suitable alternative technique was recently proposed in 
\cite{Steinacker:2016nsc}, which is based on string-like states on the fuzzy brane backgrounds. 
This geometric approach has been applied successfully in the 
\emph{purely fuzzy} context of \cite{Steinacker:2016nsc,Steinacker:2016vgf}, 
but not yet in the present field-theoretic setting with 
fuzzy extra dimensions. Hence, this needs to be developed elsewhere.

%\vspace{0.2cm}
%\paragraph{Conclusion.}
\paragraph{}
The take-home message it that a simple deformation of $\cN=4$ SYM with large $N$ 
can reduce at low energy to an effective gauge theory with few  string-like  
Higgs modes coupled to chiral fermions, interpreted in terms of a gauge theory on 
intersecting branes in 6 extra dimensions. 
Moreover, the possible quantum numbers  include those of the Standard Model.
This should provide sufficient motivation to study such scenarios in more 
detail and at the quantum level, and it will be interesting to see how far 
these solutions can reach.

\section*{Acknowledgements}
This work was funded by the Austrian Science 
Fund (FWF) grant P28590. The Action MP1405 QSPACE from the European
Cooperation in Science and Technology (COST) also provided  support. 
Related discussions with Pascal Anastasopoulos,  Amihay Hanany, 
Cumrun Vafa and  George Zoupanos  are gratefully acknowledged. 
% 
%
%%%%%%%%%%%%%%%%%%%%%%%%%%%%%%%%%%%%%%%%%%%%%%%%%%%%%%%%%%%%%%%%%
%%%%%%%%%%%%%%%%%%%%%%%%%%%%%%%%%%%%%%%%%%%%%%%%%%%%%%%%%%%%%%%%%
%
\appendix 
%
%%%%%%%%%%%%%%%%%%%%%%%%%%%%%%%%%%%%%%%%%%%%%%%%%%%%%%%%%%%%%%%%%
%%%%%%%%%%%%%%%%%%%%%%%%%%%%%%%%%%%%%%%%%%%%%%%%%%%%%%%%%%%%%%%%%
%
\section{Solutions to equations of motion}
\label{app:explicit_sol}
\subsection{Preliminaries}
\paragraph{Potential.}
The full potential for an ansatz like \eqref{basic-branes} reads
\begin{align}
 V(R_i)= c_N \left( 
 16 \sum_i M_i^2 R_i^2  + 4 \sum_i R_i^4  
 + 4 \left(R_1^2 R_2^2 + R_1^2 R_3^2  +R_2^2 R_3^2 \right)
 - 32 R_1 R_2 R_3
 \right) \,.
 \label{eq:potential}
\end{align}
Following \cite[section 9]{Steinacker:2014lma}, the constant $c_N = c_N[\mu]$ 
 for irreducible branes $\cC[\mu]$ reads as follows:
\begin{align}
 c_N[\mu]=\frac{\dim(\cH_\mu)}{12}\left( m_1^2 +m_2^2 +m_1 m_2 +3m_1 +3m_2 
\right) \,, \quad \mu =(m_1,m_2)\, .
\end{align}
\paragraph{Phase degeneracy.}
For any configuration $({R}_1,{R}_2,{R}_3)$ which solves 
\eqref{eom-general}, we can freely change two of the three phases $R_i \to R_i e^{i\vartheta_i}$, due to the 
$U(1)\times U(1)$ symmetry of the potential. The third phase is then fixed by the equations of motion, and 
it is easy to see that if two $R_i$ are real then so is the third.
We will not spell out this trivial degeneracy in the solutions below.
Note that an overall sign flip ${R}_i \mapsto - {R}_i$ does  
in general not map  solutions into solutions.
\subsection{Solutions \texorpdfstring{$M_i=0$}{Mi=0}}
The only  solutions  are 
\begin{alignat}{2}
 &\mathcal{S}_0 : \qquad & R_1&=R_2=R_3=0  \; ,\\
 &\mathcal{S}_1 : \qquad & R_1&=R_2 = R_3=1 \; 
\end{alignat}
up to phases.
Observe that $V(\mathcal{S}_1)=-8<V(\mathcal{S}_0)=0$.
% 
%%%%%%%%%%%%%%%%%%%%%%%%%%%%%%%%%%%%%%%%%%%%%%%%%%%%%%%%%5
%
\subsection{Solutions \texorpdfstring{$M_i=M$}{Mi=M}}
\begin{figure}[t!]
\centering
 \includegraphics[width=0.65\textwidth]{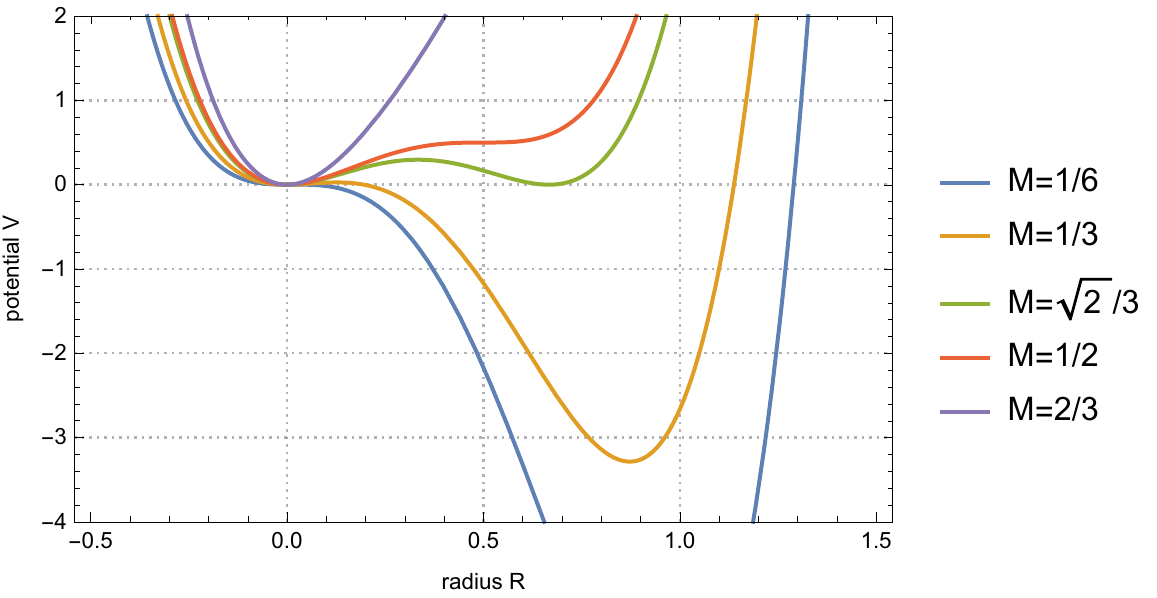}
 \caption{Potential for equal masses $M=M_i$ and equal radii $R=R_i$.}
 \label{fig:potential_equal_masses}
\end{figure}

For uniform masses $M_i \equiv M$, there are three cases to be considered, cf. 
figure \ref{fig:potential_equal_masses}.
\paragraph{\texorpdfstring{$\boldsymbol{M>\frac{1}{2}}$}{M>0.5}.}
The only solution is the trivial one $R_i=0$.
\paragraph{\texorpdfstring{$\boldsymbol{M=\frac{1}{2}}$}{M=0.5}.}
The only solutions  are
\begin{alignat}{2}
 &\mathcal{S}_0 : \qquad & R_1&=R_2=R_3=0  \; ,\\
 &\mathcal{S}_1 : \qquad & R_1&=R_2 = R_3=M \; 
\end{alignat}
up to phases.
In this case, we observe that 
$V(\mathcal{S}_1)=\frac{1}{2}>V(\mathcal{S}_0)=0$. Hence, $\mathcal{S}_0$
is the minimum.
\paragraph{\texorpdfstring{$\boldsymbol{M<\frac{1}{2}}$}{M<0.5}.}
There are nine three types of solutions, given by
\begin{alignat}{2}
&\mathcal{S}_0:  & \qquad R_1&=R_2=R_3=0 \\
&\mathcal{S}_1:  & \qquad
R_1 &=R_2=R_3= \frac{1}{2} \left( 1+ \sqrt{1-4M^2} \right) \, = 1-M^2 + O(M^4) , \label{soln-M-equal}\\
&\mathcal{S}_2:  & \qquad
R_1 &=R_2=R_3= \frac{1}{2} \left( 1- \sqrt{1-4M^2} \right) \, = M^2 + O(M^4) 
\end{alignat}
up to phases. We can compute the potential and find
\begin{align}
 V(\mathcal{S}_0)&=0\\
 V(\mathcal{S}_1) &= 
 4 \left(-6 M^4+\left(4 \sqrt{1-4 M^2}+6\right) M^2-\sqrt{1-4 M^2}-1\right)
\\ 
 V(\mathcal{S}_2) &=  
 -4 \left(6 M^4+\left(4 \sqrt{1-4 M^2}-6\right) M^2-\sqrt{1-4 M^2}+1\right) \ .
\end{align}
One can verify that $V(\mathcal{S}_1)< V(\mathcal{S}_2)$  
 for $0<M<\frac{1}{2}$, but $V(\mathcal{S}_1) < V(\mathcal{S}_0)=0$  
only for 
\begin{align}
 0<M< \frac{\sqrt{2}}{3} =: M^* \ .
\end{align}
Also, observe that $\mathcal{S}_1=\mathcal{S}_2$ for $M=\frac{1}{2}$. Hence
$\mathcal{S}_1$ is a (relative) minimum for $0<M<\frac{1}{2}$ and the absolute 
minimum for $0<M< M^*$, see also figure 
\ref{fig:potential_equal_masses}.
Note that within each solution $\mathcal{S}_i$ the radii are equal,  
$|R_1|=|R_2|=|R_3|\eqqcolon R(\mathcal{S}_i)$. We also observe
\begin{equation}
 R(\mathcal{S}_1) > R(\mathcal{S}_2) > R(\mathcal{S}_0)\;.
\end{equation}
Hence the radius with the lowest energy in this regime is given by
\begin{align}
 R(M) := \frac{1}{2} \left( 1+ \sqrt{1-4M^2} \right) \ .
 \label{R-M-def}
\end{align}
Note that the critical mass $M^*$ \eqref{M-star-def} marks the transition 
between trivial and non-trivial stable vacua.
\subsection{Solutions \texorpdfstring{$M_1=M_2=0$}{M1=M2=0} and 
\texorpdfstring{$M_3 
>0$}{M3>0}}
\label{subsec:M3>0}
For $0<M_3 < M^*$, we find the following  solutions:
\begin{alignat}{2}
&\mathcal{S}_0:  & \qquad R_1&=R_2=R_3=0 \,,\\
&\mathcal{S}_1:  & \qquad
R_1 &= R_2= \sqrt{-\frac{29}{6}+M_3^2 + \frac{7}{6} \sqrt{25-12 M_3^2}} \ = 
1-\frac{M_3^2}{5} + O(M_3^4) \; , \\
 & & R_3 &= \frac{1}{2} \left(  \sqrt{25 - 12 M_3^2}-3\right) \ = 
1-\frac{3M_3^2}{5} + O(M_3^4) \; ,
\end{alignat}
up to phases. Moreover, $|R_3|<|R_1|=|R_2|$ for  
$\mathcal{S}_1$.
Computing the potential, we find
\begin{align}
 V(\mathcal{S}_0)&=0 \; ,\\
 V(\mathcal{S}_1) &= 
 -\frac{4}{3} \left(18 M_3^4 -102M_3^2 +131
 +(12 M_3^2 -25) \sqrt{25-12 M_3^2}
 \right) \;,
\\
V(\mathcal{S}_1) &< V(\mathcal{S}_0)\qquad \text{for } 
0<M_3< M^* \; .
\end{align}
% 
%%%%%%%%%%%%%%%%%%%%%%%%%%%%%%%%%%%%%%%%%%%%%%%%%
%
\subsection{Induced Higgs mass terms}
\label{sec:Higgs-diff-masses}
Starting from section \ref{subsec:M3>0} we compute the mass of the regular zero 
modes
in $\mathcal{C}[(N,0)]$, which
are given by $\phi_\alpha^{(l)} \propto (X_{-\alpha})^l $ with
$SU(3)$ weights $\lambda(\phi_\alpha^{(l)}) = - l \alpha$.
These masses \eqref{bosonic-zeromodes-m} are given by 
\begin{align}
m^2(\phi_{\alpha,\lambda}) = (\lambda, \sum_i R_i^2 \alpha_i )
\end{align}
For the solution in section \ref{subsec:M3>0} we can simplify this to
\begin{align}
m^2(\phi_{\alpha,\lambda}^{(l)}) = -l  (R_1^2 -R_3^2) (\alpha, \alpha_1+ 
\alpha_2 ) 
\end{align}
such that 
\begin{align}
m^2(\phi_{\alpha_1,\lambda}^{(l)}) &\equiv m_1^2 = -l  (R_1^2 -R_3^2) <0  \\
m^2(\phi_{\alpha_2,\lambda}^{(l)}) &\equiv m_2^2 = -l  (R_1^2 -R_3^2) <0 \\
m^2(\phi_{\alpha_3,\lambda}^{(l)}) &\equiv m_3^2 = 2l  (R_1^2 -R_3^2) >0
\end{align}
which satisfy $\sum_i m_i^2 =0$. Note that
\begin{align}
 R_1^2 -R_3^2 \equiv \Delta R^2= \frac 43 \big(3M_3^2 + 2 \sqrt{25-12M_3^2} - 10\big) = \frac{4 M_3^2}{5} + O(M_3^4) \ >0
\end{align}
so that 
\begin{align}
 m_1^2 = m_2^2 &= -4l\,\frac{M_3^2}{5} + O(M_3^4) \ < 0 \,, \\
 m_3^2 &= 8l\,\frac{M_3^2}{5} + O(M_3^4) \ > 0 \, .
 \label{mi-induced-oneM}
\end{align}
In particular, the maximal Higgs $\phi_\alpha = r_\a 
\tilde\pi(T_{\a}) \propto (X_{-\alpha})^N$ in 
$\cC[(N,0)]$ provide a solution of the form
\begin{align}
 Y_i^+ = m\left( R_i \pi(T_i^+) + r_i \tilde\pi(T_i^{+}) \right)\;.  
 \label{eq:brane+maxHiggs}
\end{align}
Taking into account the cubic terms arising from the soft potential,
the $r_i$ have to satisfy the following eom
\begin{equation}
\begin{aligned}
 -N\Delta R^2  r_1+
 r_1 \left(2 r_1^2+r_2^2+r_3^2\right)
 -4 r_2 r_3&=0 \, ,\\
-N \Delta R^2  r_2
+r_2 \left(r_1^2+2 r_2^2+r_3^2\right)
-4 r_1 r_3&=0 \, ,\\
(4 M_3^2  +2N  \Delta R^2 ) r_3
+r_3 \left(r_1^2+r_2^2+2 r_3^2\right)
-4 r_1 r_2&=0 \,.
\end{aligned}
\label{eq:EOM_maxHiggs}
\end{equation}
One can check numerically that solutions to \eqref{eq:EOM_maxHiggs} do exist,
and the radii have different magnitude.
% 
%%%%%%%%%%%%%%%%%%%%%%%%%%%%%%%%%%%%%%%%%%%%%%%%%%%%%%%%%%%%%%%%%%%%
%%%%%%%%%%%%%%%%%%%%%%%%%%%%%%%%%%%%%%%%%%%%%%%%%%%%%%%%%%%%%%%%%%%%
%
\section{Combined backgrounds: solutions \& spectrum}
\label{app:solutions_spectrum}
In this section, we provide details on various solutions of the eom as discussed 
in sections \ref{sec:4D-branes} and \ref{sec:6D-branes},
and the spectra of 
the vector Laplacian and the Dirac operator on these solution. Let us 
briefly describe the set-up for the numerical work. Calculations are performed 
with \texttt{Mathematica}, and the $\msu(3)$ representation matrices $\lambda_a $ 
for an irrep $(N,K)$ are obtained from the package \texttt{BProbe}\footnote{
Lukas Schneiderbauer. (2016, January 20). BProbe: a Wolfram Mathematica 
package. Zenodo. http://doi.org/10.5281/zenodo.45045}.
For a fixed $(N,K)$ we obtain $8$ representation matrices 
$\lambda_a $ and define the generators via
\begin{align}
 T_1^{\pm} = \lambda_4 \pm \im \lambda_5  \, , \quad 
T_2^{\pm} =   \lambda_6 \mp \im \lambda_7  \, , \quad
T_3^{\pm} =   \lambda_1 \mp \im \lambda_2  \; ,
\end{align}
which satisfy the  Lie algebra 
relations \eqref{Cartan-Weyl}, with all required normalizations.

\paragraph{Single brane.}
For a single brane, we define the background as in \eqref{basic-branes}, where we only 
consider equal radii $R_i \equiv R(M) = \frac{1}{2} \left(1 + \sqrt{1 - 4M^2} 
\right)$ for $0< M<\frac{1}{2}$. The 6 hermitian matrices $X_a$ are then obtained by inverting 
\eqref{X-T-definition}.
The definition of the (gauge-fixed) vector Laplacian as
\begin{subequations}
\begin{align}
 \mathcal{O}_V^{X} &= \Box_X \delta_{ab} + 4M^2
 + 2 \left[ \left( [X_a,X_b] -2\im g_{abc}X_c \right) ,  \cdot  \right] -
[X_a,[X_b,\cdot]] \\
 \mathcal{O}_{V,\mathrm{fix}}^{X} &= \Box_X \delta_{ab} + 4M^2
 + 2 \left[ \left( [X_a,X_b] -2\im g_{abc}X_c \right) ,  \cdot  \right] 
\end{align}
is implemented, with structure constants \eqref{cubic-flux} determined by
\begin{align}
 g_{abc}= -\im \frac{\tr(T_a [T_b,T_c])}{\tr(T_1 T_1)} \; .
\end{align}
\end{subequations}
\paragraph{Multiple branes.}
Multiple branes $(N_l,K_l)$ for $l=1,\ldots,d$ are realized as a straightforward extension of the previous paragraph,
described by
\begin{align}
 X_a \equiv \oplus_{l=1}^d  X_{a}^{(N_l,K_l)} \cong  \diag 
\left( X_{a}^{(N_1,K_1)}, \ldots , X_{a}^{(N_d,K_D)} \right) \; .
\end{align}
\subsection{\texorpdfstring{$(N,0)$}{(N,0)} brane}
\label{subsec:brane}
\subsubsection{Solution to eom}
\label{subsec:brane_eom}
With the notation shown in figure \ref{fig:brane-Higgs-edge} we use the 
following ansatz:
\begin{align}
 Y_j^{+}=  X_{j}^{+} + f_j \ \phi_j^{+}  \; .
\label{eq:ansatz_brane}
\end{align}
Here and in the following, the radius 
$R=R(M)$ \eqref{R-M-def} is always fixed by solving the eom for equal masses 
$M_i = M$.
Then we find the following family of solutions to the eom:
\begin{align}
  f_1 &= e^{i\vartheta_1}, \quad f_2 = e^{i\vartheta_2}, \quad f_3  =\bar{f_2} \bar{f_1}  \,,
 \label{eq:solution_brane}
\end{align}
This exhibits  the above-mentioned flat direction corresponding to two $U(1)$ phases 
within the Higgs sector, due to \eqref{V-additivity}.
\subsubsection{Fluctuation spectrum}
 \label{subsec:brane_spectrum}
\begin{figure}[t!]
\centering
\begin{subfigure}{0.485\textwidth}
\centering
 \includegraphics[width=\textwidth]{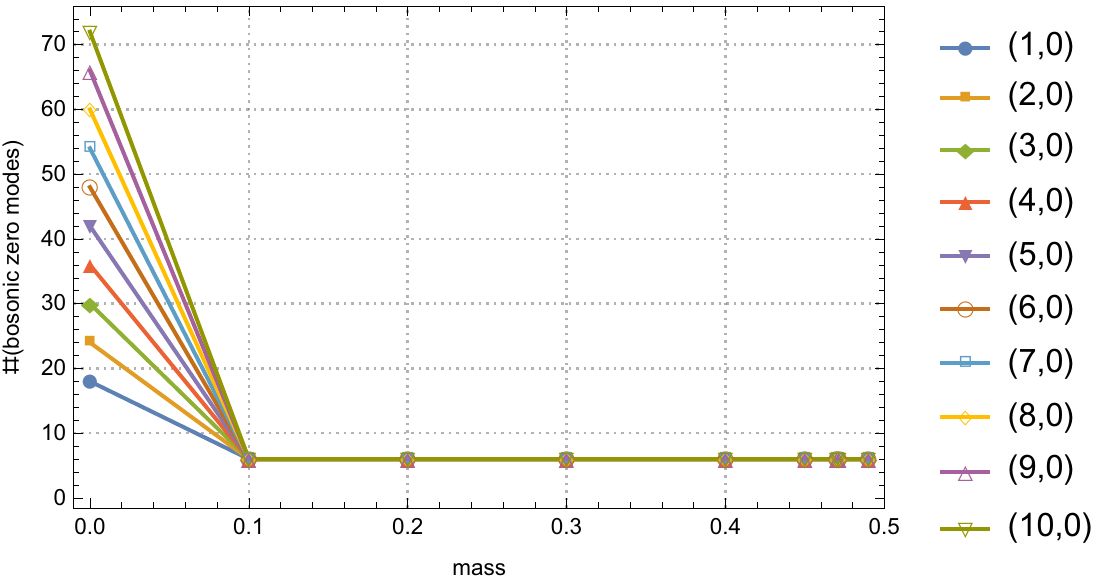}
 \caption{}
 \label{fig:CP2-brane_0-modes}
\end{subfigure}
\begin{subfigure}{0.485\textwidth}
\centering
 \includegraphics[width=\textwidth]{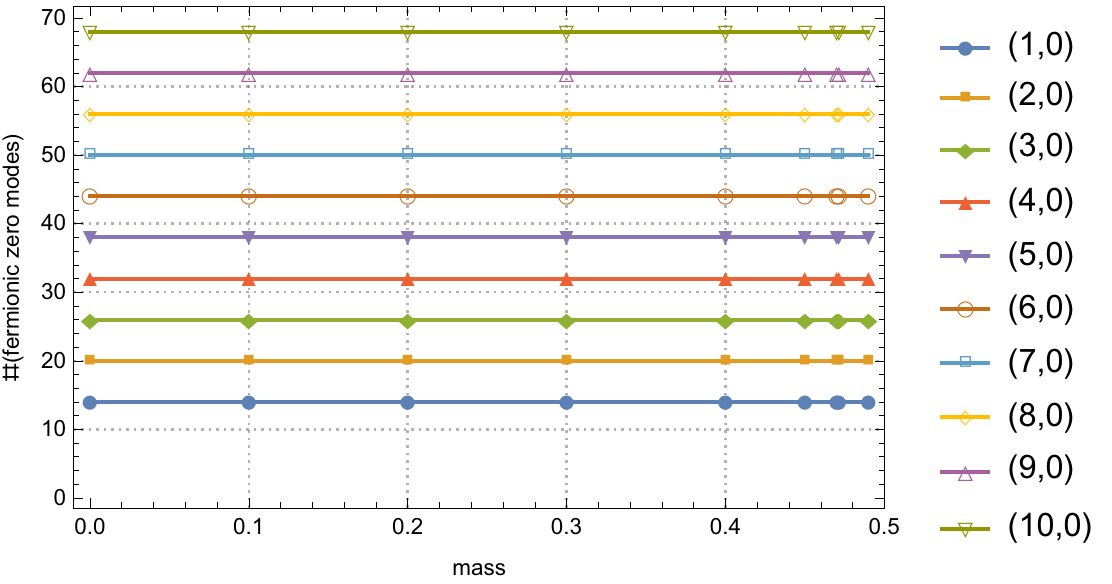}
 \caption{}
 \label{fig:CP2-brane_Dirac-modes}
\end{subfigure}
% 
%%%%%%%%%%%%%%%%%%%%%%%%%%%%%%%%%%%%%%%%%%%%%%%%%%%% 
% 
\begin{subfigure}{0.485\textwidth}
\centering
 \includegraphics[width=\textwidth]{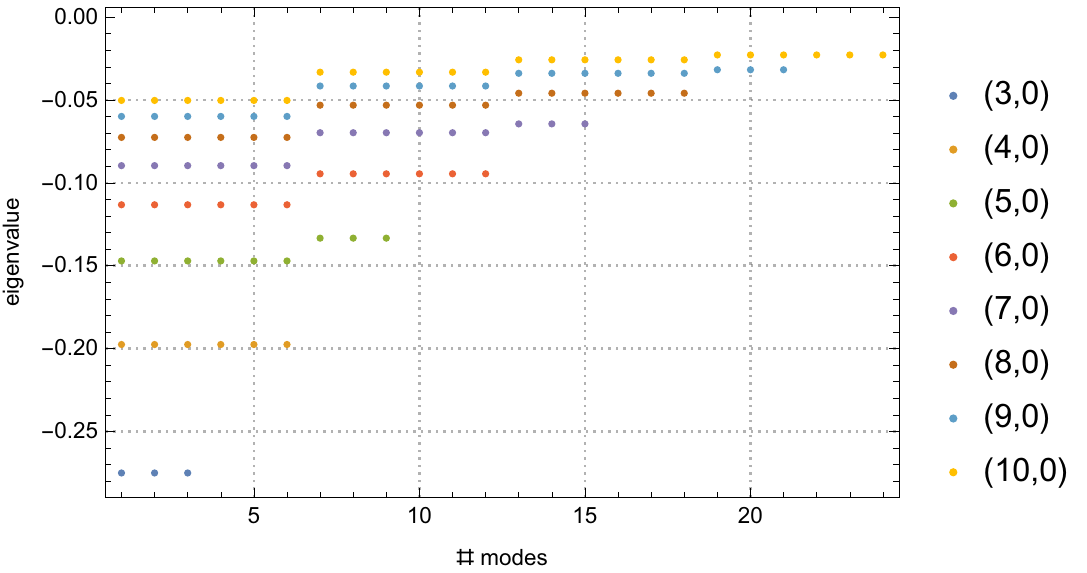}
 \caption{}
 \label{fig:neg_ev_n-0_brane}
 \end{subfigure}
 \begin{subfigure}{0.485\textwidth}
  \centering
  \includegraphics[width=\textwidth]{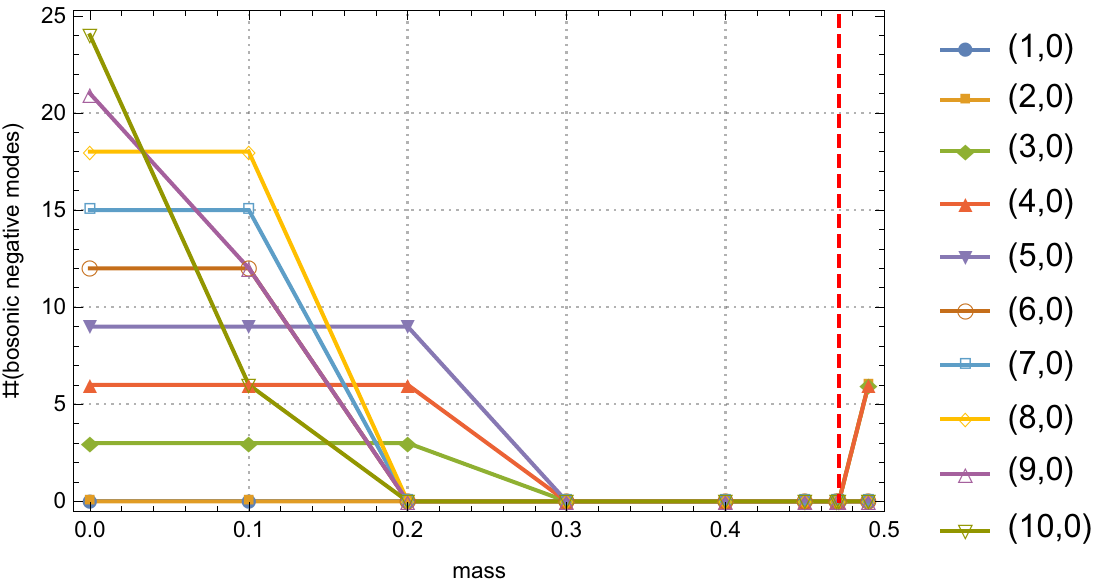}
  \caption{}
  \label{fig:CP2-brane+Higgs_neg_modes}
 \end{subfigure}
% 
%%%%%%%%%%%%%%%%%%%%%%%%%%%%%%%%%%%%%%%%%%%%%%%%%%%% 
%  
 \begin{subfigure}{0.485\textwidth}
  \centering
  \includegraphics[width=\textwidth]{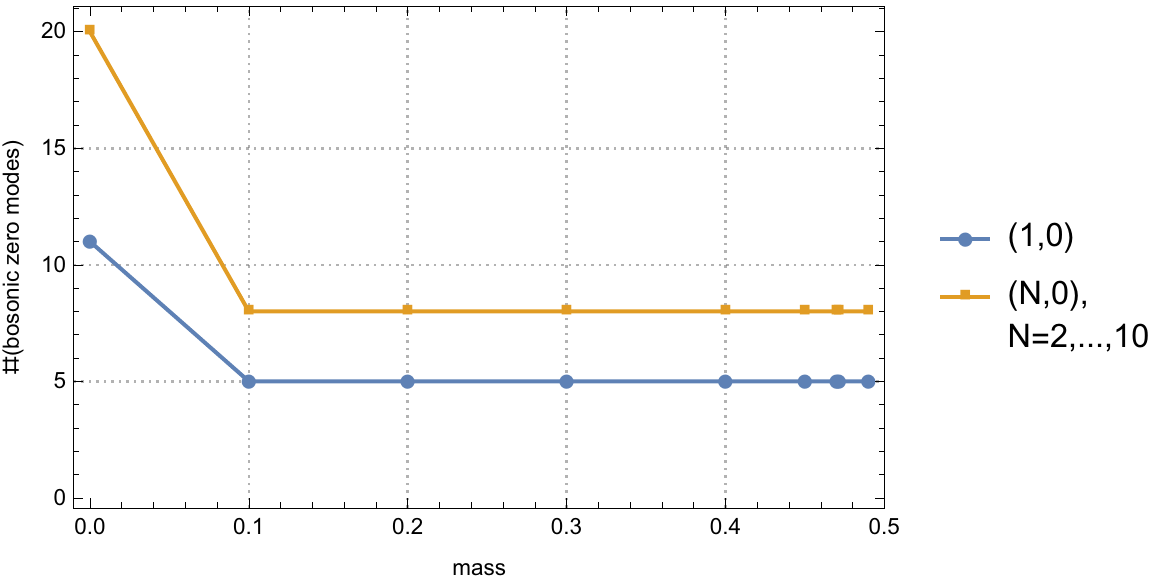}
  \caption{}
  \label{fig:CP2-brane+Higgs_0-modes}
 \end{subfigure}
 \begin{subfigure}{0.485\textwidth}
  \centering
  \includegraphics[width=\textwidth]{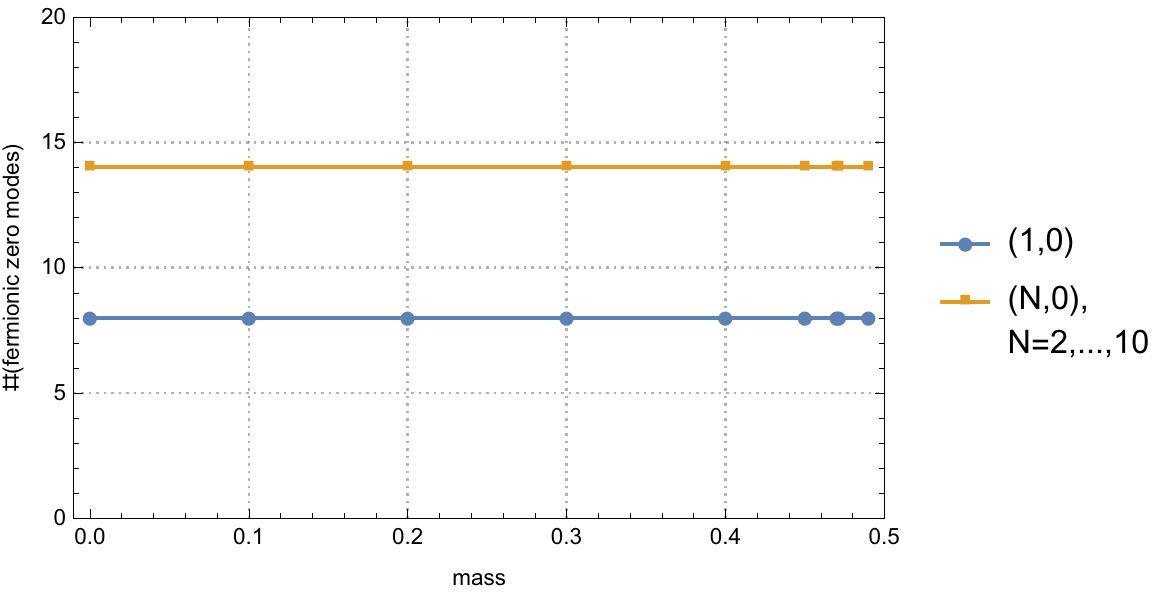}
  \caption{}
  \label{fig:CP2-brane+Higgs_Dirac-modes}
 \end{subfigure}
 \caption{$\cC[(N,0)]$ brane with or without maximal Higgs. 
For pure  $\cC[(N,0)]$ background, the number of zero modes of 
$\mathcal{O}_V^{X}$ are shown in (\subref{fig:CP2-brane_0-modes}), whereas the 
number of zero modes for $\slashed{D}^X$ is shown in 
(\subref{fig:CP2-brane_Dirac-modes}).
For combined background $\cC[(N,0)]$ plus maximal Higgs, the negative modes of 
$\mathcal{O}_V^{X}$ (identical to $\mathcal{O}_{V,\mathrm{fix}}^{X}$) with their 
corresponding eigenvalue are shown in (\subref{fig:neg_ev_n-0_brane}). For this 
set-up with uniform mass parameter $M$, we display the number of negative modes 
of $\cO^{X+\phi}$ 
in (\subref{fig:CP2-brane+Higgs_neg_modes}) and zero modes in 
(\subref{fig:CP2-brane+Higgs_0-modes}), whereas the zero modes of 
$\slashed{D}^{X+\phi}$ are show in 
(\subref{fig:CP2-brane+Higgs_Dirac-modes}). The red dashed vertical line in 
(\subref{fig:CP2-brane+Higgs_neg_modes}) indicates the critical mass value 
$M^*=\sqrt{2}\slash 3$ where the potential vanishes for all these solutions, see figure 
\ref{fig:potential_equal_masses}.}
\end{figure}
\paragraph{$\C P^2$ brane.}
For a single $\cC[(N,0)]$ brane, $\cO_V^{X}$ has a positive semi-definite 
spectrum, but the number of zero modes (in the gauge fixed case) increases with 
the brane size as $6(N+2)$. As shown in 
\cite{Steinacker:2014lma,Steinacker:2014eua,Steinacker:2015mia}, there are six 
regular zero modes for each irrep appearing in the endomorphism space. Here, 
$\End(\cH_{(N,0)})\cong \oplus_{l=0}^N \cH_{(l,l)}$ and we expect $6(N+1)$ 
regular zero modes. Additionally, there are six Goldstone bosons coming from 
$SU(3) \slash U(1)^2$, as the background only preserves the $U(1)_{K_i}$ 
symmetries. Numerically, we find indeed $6(N+2)$ zero modes in the massless 
case.
As depicted in figure \ref{fig:CP2-brane_0-modes}, 
turning on uniform masses $M_i \equiv M < \frac{1}{2}$ allows to lift all zero 
modes except the 6 modes associated to the Goldstone bosons.

Similarly, the zero modes of the Dirac operator $\Di^X$ are shown in 
\ref{fig:CP2-brane_Dirac-modes} and were classified in 
\cite{Steinacker:2014lma,Steinacker:2014eua,Steinacker:2015mia}. We observe indeed
$6(N+1)+2$ fermionic zero modes, among which $6(N+1)$ originate from the 
one-to-one correspondence with regular bosonic zero modes, and the remaining two are trivial gaugino modes. 
The spectrum of $\Di^X$ is  independent of the bosonic mass.
\paragraph{$\C P^2$ brane with maximal Higgs.}
Analyzing the spectrum of $\cO_V^{X+\phi}$ for a single $\cC[(N,0)]$ 
brane with maximal Higgs as combined background shows a 
number of negative modes in the massless case, 
as displayed in figure \ref{fig:neg_ev_n-0_brane}. We observe that the number 
of 
negative modes is $3(N-2)$ (valid for $N\geq 2$), but the magnitude of their 
eigenvalues decreases with $N$. 
Turning on equal masses $M_i \equiv M \leq M^*$ allows to lift 
all negative modes for $N>3$ by choosing $M$ appropriately, as shown in figure 
\ref{fig:CP2-brane+Higgs_neg_modes}. Note that $N=2$ is entirely free of negative modes.
Moreover, we observe from figure \ref{fig:CP2-brane+Higgs_0-modes} that the 
number of zero modes of $\cO_V^{X+\phi}$  is reduced to $8$ in the 
presence of $M>0$, independently of the brane size. We can understand these as 6 Goldstone bosons 
plus the two phases in \eqref{eq:solution_brane}.

Similarly, the spectrum of $\Di^{X+\phi}$ shows that the number 
of fermionic zero modes is $14$ for $N\geq 2$ and $8$ for the minimal brane. We emphasize again 
that the number of bosonic and fermionic zero 
modes for $\cC[(N,0)]$ plus maximal Higgs is independent of the size $N$.
\paragraph{Minimal brane.}
The combination of minimal brane $\cC[(1,0)]$ together with its maximal Higgs 
reduces to the fuzzy 2-sphere. We find that $\cO_V^{X}$ is positive 
semi-definite and $\cO_{V,\mathrm{fix}}^{X}$ has $11$ zero modes. 
Including masses partially lifts them and one finds $5$ remaining zero modes. 
These are the Goldstone bosons of $SU(3) \slash (U(1){\times} SU(2))$, 
because the fuzzy 2-sphere preserves a full $SU(2)$. In addition, the Dirac 
operator has $8$ zero modes.
% 
%
%%%%%%%%%%%%%%%%%%%%%%%%%%%%%%%%%%%%%%%%%%%%%%%%%%%%%%%%%%%
%%%%%%%%%%%%%%%%%%%%%%%%%%%%%%%%%%%%%%%%%%%%%%%%%%%%%%%%%%%
%
\subsection{\texorpdfstring{$(N,0)$}{(N,0)} brane + point brane}
\label{subsec:brane+point}
\subsubsection{Solution to eom}
\label{subsec:brane+point_eom}
We use the notation of figure \ref{fig:CP2-point} and the following ansatz:
\begin{align}
 Y_j^{+}=  X_{j}^{+} + f_j \ \phi_j^{+} + r_j \ \varphi_j^{+} 
+ s_j \ (\tilde{\varphi}_j^{-})^\dagger 
\label{eq:ansatz_brane+point}
\end{align}
for $f_j, r_j, s_j \in \C$ and $Y^{j,-} = (Y^{j,+})^\dagger$.
The setup necessarily contains the above solutions including the 
trivial solution, solutions with maximal intra-brane Higgs i.e. $f_j$  only,
and triangular solutions formed out of 
$(f_1,r_2,s_3)$, $(f_2,r_3,s_1)$, or $(f_3,r_1,s_2)$.
The question is whether there are more general solutions or whether two or more 
triangles can exist simultaneously.
\paragraph{Triangular solutions.}
Consider for instance the triangular system comprised of $f_1, r_2, 
s_3 \in \C$, where all remaining coefficients 
vanish. Then  we 
find the following  solutions:
\begin{align}
 f_1 = e^{i\vartheta_1}, \quad r_2 = e^{i\vartheta_2}, \quad  s_3 &= \bar{f_1} \bar{r_2} \,
\end{align}
\paragraph{Maximal Higgs + triangular system.}

in addition to the above triangular solutions, 
we also find new four-parameter solutions that involve combinations of maximal Higgs 
and inter-brane Higgs. 
In detail, we solved the system for $f_1, f_2 ,f_3, r_2, 
s_3 \in \C$ and all remaining coefficients 
vanish.
Then for $f_3\in \C$, $z,y \in \R$ with the constraints $1\geq |f_3|^2 
+ y^2 $ and $|f_3|^2  -z^2 \geq0 $, we find the following four solutions:
\begin{subequations}
\label{eq:sol_max_Higgs+triangle}
\begin{alignat}{4}
f_1 &= \frac{\bar{f_2}}{f_3}\,, \quad &
f_2 &= z - \im \sqrt{|f_3|^2  -z^2}\,, \quad &
r_2 &= y - \im \sqrt{1 -|f_3|^2 -y^2}\,, \quad &
s_3 &=  \frac{ \bar{r_2}  f_2 }{\bar{f_3}} \,,\\
f_1 &= \frac{\bar{f_2}}{f_3}\,, \quad &
f_2 &= z - \im \sqrt{ |f_3|^2 -z^2}\,, \quad &
r_2 &= y+\im \sqrt{1 -|f_3|^2 -y^2}\,, \quad &
s_3 &= \frac{ \bar{r_2}  f_2}{\bar{f_3}} \,,\\
f_1 &= \frac{\bar{f_2}}{f_3}\,, \quad &
f_2 &= z + \im \sqrt{ |f_3|^2 -z^2}\,, \quad &
r_2 &= y - \im \sqrt{ 1 -|f_3|^2 -y^2}\,, \quad &
s_3 &= -\frac{ \bar{r_2}  f_2}{\bar{f_3}} \,,\\
f_1 &= \frac{\bar{f_2}}{f_3 }\,, \quad &
f_2 &= z + \im \sqrt{ |f_3|^2 -z^2}\,, \quad &
r_2 &= y + \im \sqrt{1 -|f_3|^2 -y^2}\,, \quad &
s_3 &=  -\frac{ \bar{r_2}  f_2 }{\bar{f_3}} \,.
\end{alignat}
\end{subequations}
For these configurations we can verify that $|f_2|=|f_3|$, $|f_1|=1$ and 
$|r_2|=|s_3|=\sqrt{1-|f_3|^2}$. Analogous solutions exist for $f_1, f_2 ,f_3, 
r_1, s_2 \in \C$ and $f_1, f_2 ,f_3, r_3, s_1 \in \C$.

They can be understood as generalized triangles where one 
extremal weight state $|\mu_i\rangle$ of $(N,0)$ is replaced 
by a superposition  $\a|\mu_i\rangle + \b |0\rangle$
with the point brane. Such superpositions give again regular zero modes
because the latter form a vector space.

%%H
% 
% 
% 
\subsubsection{Spectrum}
\label{subsec:brane+point_spectrum}
\begin{figure}[t!]
 \centering
 \begin{subfigure}{0.485\textwidth}
  \centering
\includegraphics[width=\textwidth]{%
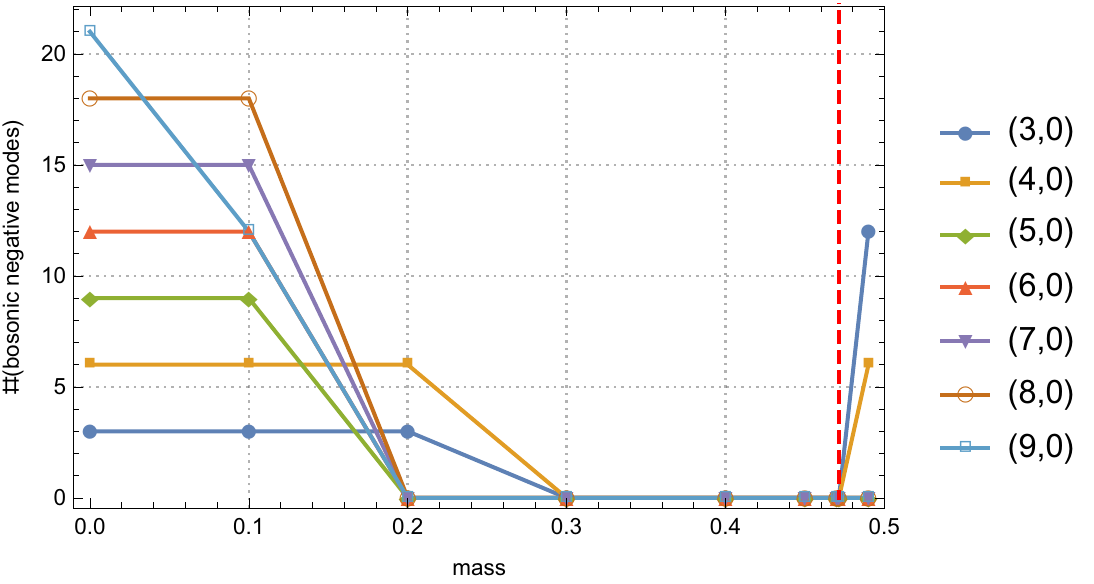}
  \caption{$x=0$}
  \label{fig:neg_modes_brane+point_x=0}
 \end{subfigure}
 \begin{subfigure}{0.485\textwidth}
  \centering
\includegraphics[width=\textwidth]{%
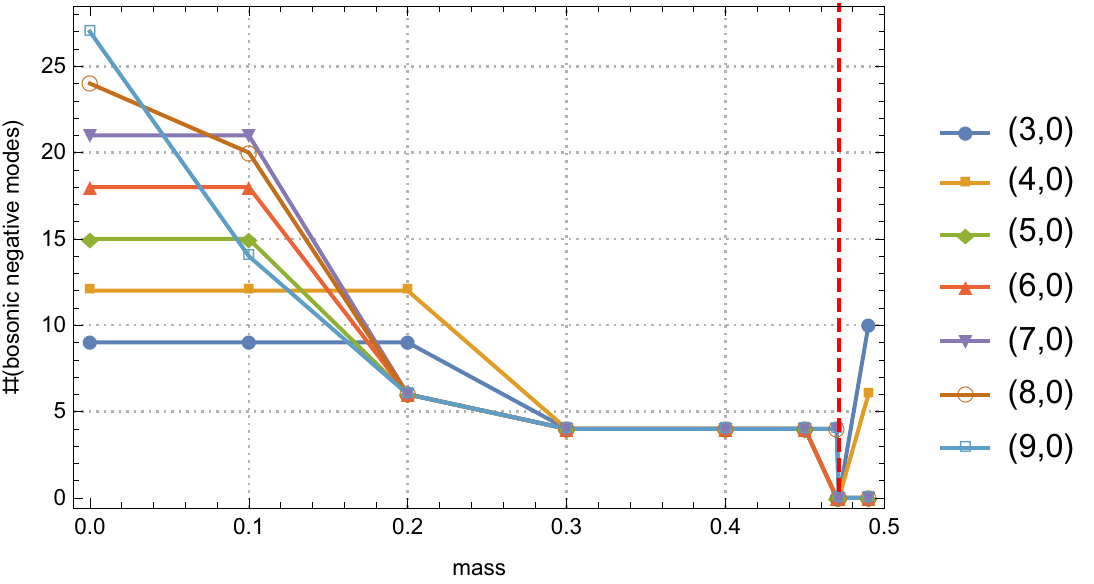}
  \caption{$x=\frac{1}{2}$}
  \label{fig:neg_modes_brane+point_x=1_2}
 \end{subfigure}
%  
%%%%%%%%%%%%%%%%%%%%%%%%%%%%%%%%%%%%%%%%%%%%%%%%%%%%%%% 
%
  \begin{subfigure}{0.485\textwidth}
  \centering
\includegraphics[width=\textwidth]{%
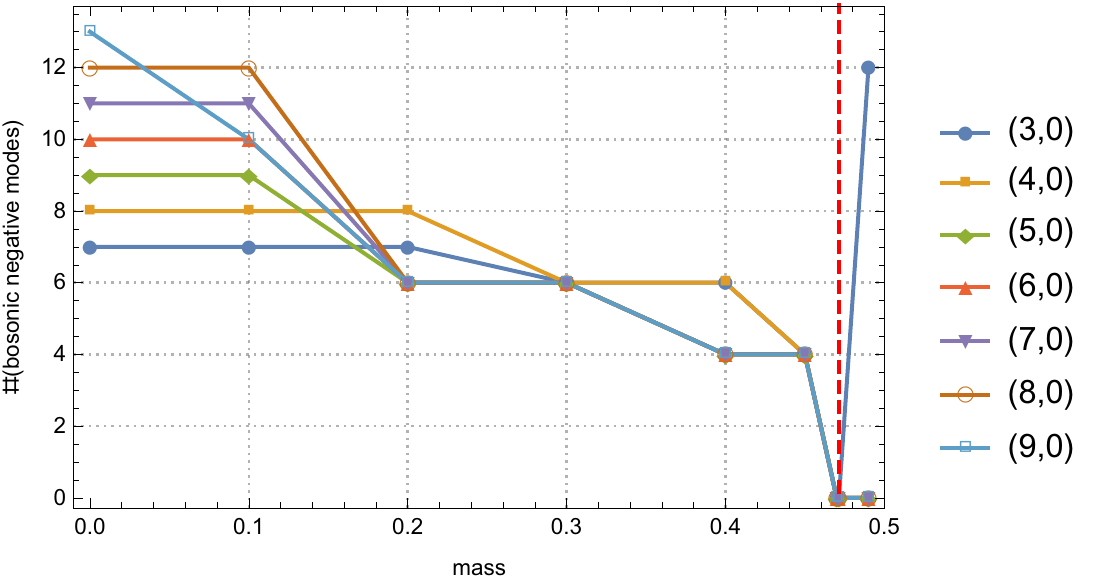}
  \caption{$x=1$}
  \label{fig:neg_modes_brane+point_x=1}
 \end{subfigure}
 \caption{$\cC[(N,0)]+\cD$ plus Higgs. 
For the configuration \eqref{eq:ex_family_brane+point}, the number of negative 
modes in the spectrum of $\cO_{V,\mathrm{fix}}^X$ for three representative 
$x$-values is shown in 
(\subref{fig:neg_modes_brane+point_x=0})--(\subref{%
fig:neg_modes_brane+point_x=1}) 
against a varying mass parameter $M$. The red dashed vertical line indicates the 
critical mass value $M^*$.}
\label{fig:brane+pt_spectrum}
\end{figure}

\begin{figure}[t!]
 \centering
  \begin{subfigure}{0.485\textwidth}
  \centering
\includegraphics[width=\textwidth]{%
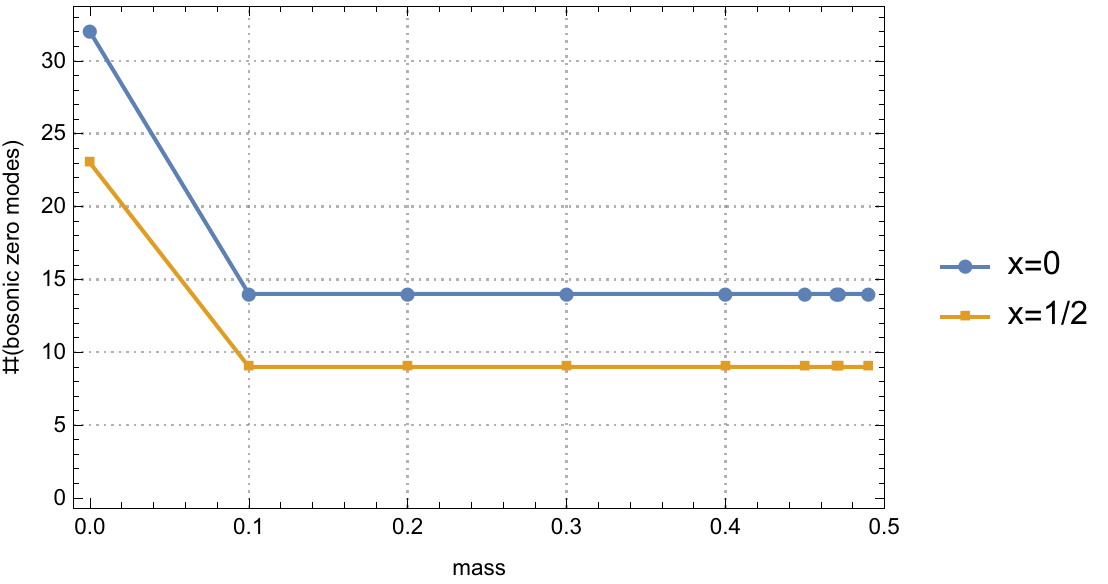}
  \caption{$x=0$}
  \label{fig:0-modes_brane+point_x=0+1_2}
 \end{subfigure}
  \begin{subfigure}{0.485\textwidth}
  \centering
\includegraphics[width=\textwidth]{%
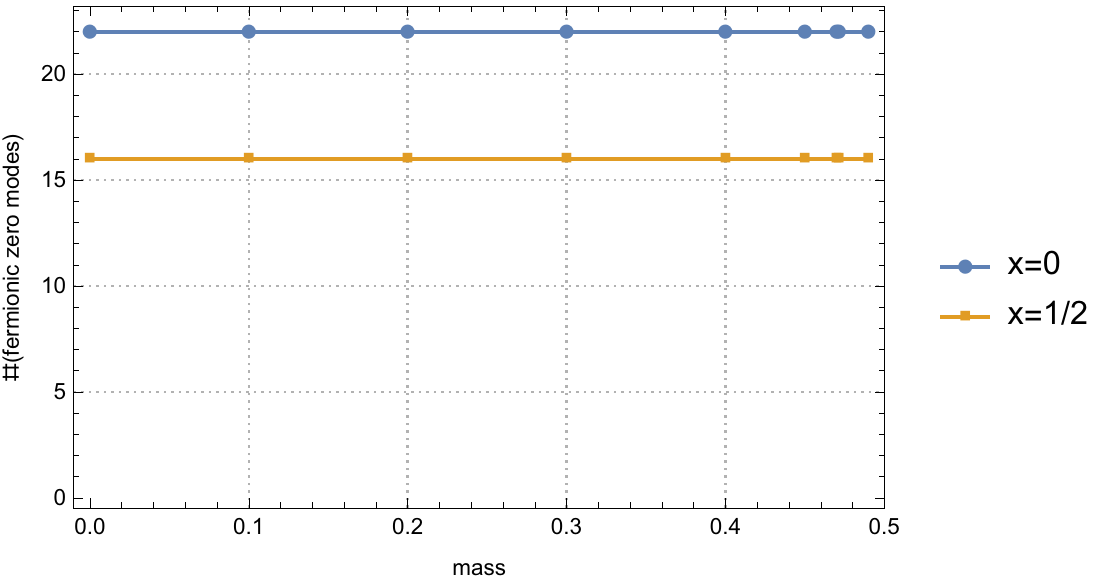}
  \caption{$x=0$}
  \label{fig:Dirac-modes_brane+point_x=0+1_2}
 \end{subfigure}
%
%%%%%%%%%%%%%%%%%%%%%%%%%%%%%%%%%%%%%%%%%%%%%%%%%%%%%
%  
 \begin{subfigure}{0.485\textwidth}
  \centering
\includegraphics[width=\textwidth]{%
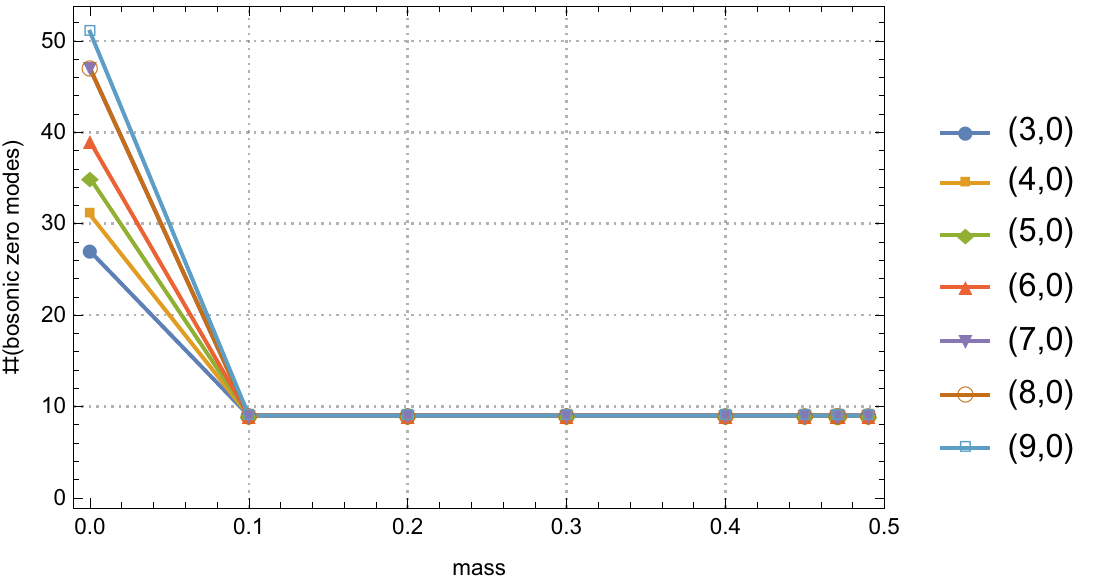}
  \caption{$x=1$}
  \label{fig:0-modes_brane+point_x=1}
 \end{subfigure}
 \begin{subfigure}{0.485\textwidth}
  \centering
\includegraphics[width=\textwidth]{%
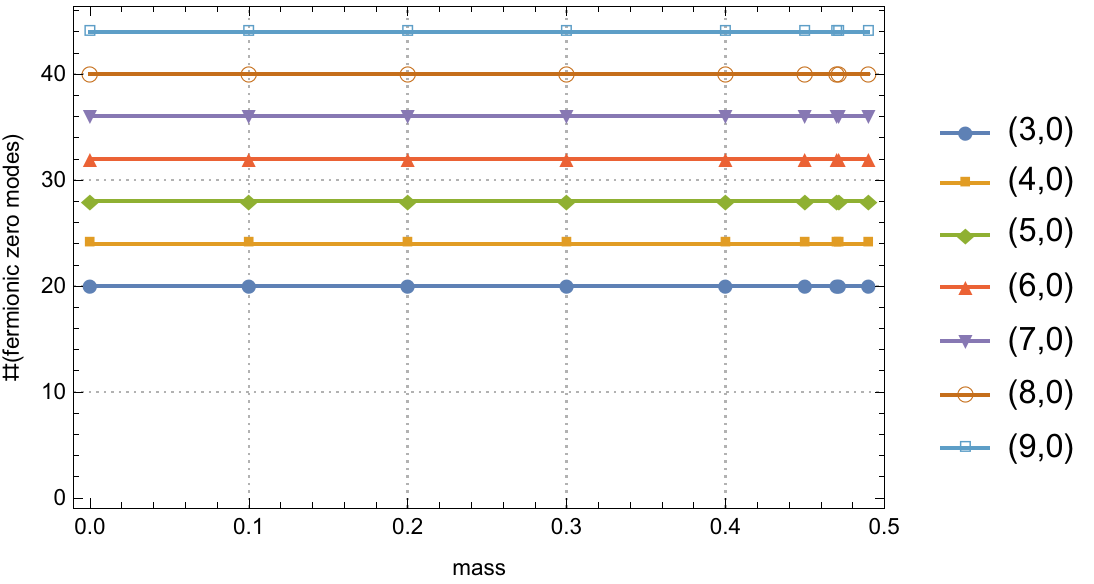}
  \caption{$x=1$}
  \label{fig:Dirac-modes_brane+point_x=1}
 \end{subfigure}
 \caption{$\cC[(N,0)]+\cD$ with Higgs configuration 
\eqref{eq:ex_family_brane+point}. 
We provide the number of zero modes of $\cO_V^{X+\phi}$ in  
(\subref{fig:0-modes_brane+point_x=0+1_2}) and 
(\subref{fig:0-modes_brane+point_x=1}), while the number of zero modes of 
$\Di^{X+\phi}$ are displayed in (\subref{fig:Dirac-modes_brane+point_x=0+1_2}) 
and (\subref{fig:Dirac-modes_brane+point_x=1}).
}
\label{fig:brane+pt_0-modes}
\end{figure}

As special case of \eqref{eq:sol_max_Higgs+triangle}, for $f_3=z=\sqrt{1-x^2}$ 
and $y=x$, consider the following 
1-parameter family of solutions
\begin{equation}
 f_1=1 
 \, , \quad 
 f_2=f_3 = \sqrt{1-x^2} 
 \, , \quad 
 r_2=s_3=x 
 \, , \quad
r_1=r_3=s_1=s_2=0 \; ,
 \label{eq:ex_family_brane+point}
\end{equation}
with $x\in \R$, $|x|\leq1$. 
We computed the spectrum of the vector Laplacian and found that a number 
of negative modes exist in the massless case, see figure 
\ref{fig:brane+pt_spectrum}. 
There are several observations:
\begin{itemize}
 \item In the massless case, the number of negative modes for $x=0$ increases 
like $3(N-2)$, for $x=1$ as $(N-2)+6$, and for $0<x<1$ as $3(N-2)+6$. Hence the 
number of negative modes is smallest for $x=1$ for large $N$.
\item However, the eigenvalues of the negative modes behave peculiar for $x>0$. 
We observe from figures \ref{fig:neg_modes_brane+point_x=1_2}, 
\ref{fig:neg_modes_brane+point_x=1} that a small number of negative modes 
acquire  relatively large negative eigenvalues that can only be lifted by mass 
parameters $M_i \equiv M$ approaching the limiting value $M^* = \sqrt{2}\slash 
3$. Nonetheless, for all these configurations one can lift all negative modes. 
\item As shown in figure \ref{fig:neg_modes_brane+point_x=0}, the negative 
modes for $x=0$ can be lifted by relatively small masses. This is to be 
expected as the $x=0$ configuration is the direct sum of $\cC[(N,0)]$ plus 
maximal intra-brane Higgs with a independent point brane $\cD$. In other words, 
we can compare to figure \ref{fig:CP2-brane+Higgs_neg_modes}.
\end{itemize}
Turning our attention to the number of zero modes of $\cO^{X+\phi}$ in this 
combined background, 
we summarize our numerical findings in figures 
\ref{fig:0-modes_brane+point_x=0+1_2} and \ref{fig:0-modes_brane+point_x=1}. As 
before for the single $\C P^2$ brane, 
with or without Higgs modes, the number of zero modes of $\mathcal{O}_V^X$ 
stabilizes upon inclusion of mass parameters $M_i \equiv M$. 
\begin{itemize}
 \item For the $\cC[(N,0)]$ plus maximal intra brane Higgs configuration 
$x=0$, there are 14 zero modes in the gauge fixed case. Comparing this to 
the 8 zero modes of figure \ref{fig:CP2-brane+Higgs_0-modes}, we explain the 
extra 6 modes by the Goldstone modes on $\cD$. Likewise, the number of zero 
modes in the massless case is 32, wherein 20 stem from $\cC[(N,0)]$ with 
maximal Higgs, as in figure \ref{fig:CP2-brane+Higgs_0-modes}, and 12 originate 
from $\cD$. These 12 are $6$ regular zero modes on $\cD$ and 6 Goldstone bosons.
 \item For the $\cC[(N,0)]$ plus maximal intra and inter brane Higgs 
configuration $x>0$, 
there are 9 zero modes in the gauge fixed case.
\end{itemize}
Again, these numbers are then independent of the size of the system, i.e.\ the 
value of $N$.
Considering the zero modes of $\Di^{X+\phi}$, we observe the following:
\begin{itemize}
 \item For $x=0$ as in figure \ref{fig:Dirac-modes_brane+point_x=0+1_2}, there 
exist 22 fermion zero modes, which we can explain by 14 fermion zero modes from 
$\cC[(N,0)]$ with maximal Higgs plus 8 zero modes from the point brane $\cD$. 
From these 8 modes, 6 correspond to the regular bosonic zero modes on $\cD$ and 
2 are trivial gaugino modes.
\item For $x\in (0,1)$, the situation remains qualitatively the same. The 
properties remain independent of the brane size $N$.
\item For $x=1$ as in figure \ref{fig:Dirac-modes_brane+point_x=1}, the 
behavior changes drastically on the fermionic side, as the number of zero modes 
equals $4(N+2)$.
\end{itemize}

% 
% % 
%
% 
%%%%%%%%%%%%%%%%%%%%%%%%%%%%%%%%%%%%%%%%%%%%%%%%%%%%%
%  
%
% 
%%%%%%%%%%%%%%%%%%%%%%%%%%%%%%%%%%%%%%%%%%%%%%%%%%%
%%%%%%%%%%%%%%%%%%%%%%%%%%%%%%%%%%%%%%%%%%%%%%%%%%%%

% 
\subsection{\texorpdfstring{$(N_1,0)$}{(N1,0)} brane + 
\texorpdfstring{$(N_2,0)$}{(N2,0)} brane}
\label{subsec:brane+brane}
Consider the set-up of figure \ref{fig:Brane+brane}, with the 
parametrization 
\begin{align}
 Y^{j,+}=  X_j^{+} + f_j \ \phi_j^{+} + h_j \ \phi_j^{+} + r_j 
\ \varphi^{j,+} + s_j \ (\tilde{\varphi}_j^{-})^\dagger 
\end{align}
for $f_j, h_j, r_j, s_j \in \C$ and $Y^{j,-} = (Y^{j,+})^\dagger$. Due to the 
considerable number of free complex parameters and their nonlinear appearance 
in the full eom, we can only probe a subset of solutions. Motivated by the 
solutions found in section \ref{subsec:brane} and \ref{subsec:brane+point}, 
we assume that we can restrict to solutions with real coefficients.
\subsubsection{Solution to eom}
There are the following types of solutions:
\paragraph{Maximal Higgs.}
Solve for $f_j, h_j \in \C$, all others coefficients vanish. Then $(f_j)$ and 
$(h_j)$ can be any solution of \eqref{eq:solution_brane} and any combination 
thereof is an exact solution for the two brane case.

\paragraph{Triangular system type I.}
Solving for $h_1,r_3,s_2 \in \R$ with all other coefficients vanishing, we find 
\begin{align}
(h_1,r_3,s_2) = (1,1,1)
 \label{eq:brane+brane_triangle_1}
\end{align}
up to phases, which is shown in figure \ref{fig:2branes_1triangle_smallBound}.
Similarly, there are solutions for $h_2,r_1,s_3 \in \R$ and $h_3,r_2,s_1 \in 
\R$.
\paragraph{Triangular system type I + maximal Higgs on $\cC[(N_2,0)]$.}
Solving for $h_j,r_3,s_2 \in \R$ and all other coefficients 
vanishing, we find for $|h_3|\leq1$
\begin{subequations}
\label{eq:sol_2brane_typeI-all}
\begin{alignat}{4}
h_1&=1, \quad &
h_2 &= h_3, \quad  &
r_3 &= \sqrt{1-h_3^2}, \quad &
s_2 &= \sqrt{1-h_3^2}  \,,
\label{eq:sol_2brane_typeI-1}
\\
h_1 &=1, \quad &
h_2 &=h_3, \quad  &
r_3 &=-\sqrt{1-h_3^2}, \quad &
s_2 &=-\sqrt{1-h_3^2} \,,
\label{eq:sol_2brane_typeI-2}
\\
h_1 &=-1, \quad  &
h_2 &= -h_3, \quad &  
r_3 &= \sqrt{1-h_3^2}, \quad &
s_2 &=-\sqrt{1-h_3^2} \,,
\label{eq:sol_2brane_typeI-3}
\\
h_1 &=-1,\quad &
h_2 &=-h_3, \quad  &
r_3 &= -\sqrt{1-h_3^2}, \quad&
s_2 &= \sqrt{1-h_3^2} \,.
\label{eq:sol_2brane_typeI-4}
\end{alignat}
\end{subequations}
This is depicted in figure 
\ref{fig:2branes_1triangle_smallMaxHiggs}--\ref{%
fig:2branes_1triangle_smallBound}.
\paragraph{Triangular system type II.}
Solving for $f_1, r_2, s_3 \in \R$ with all other coefficients vanishing, we 
find 
\begin{align}
(f_1,r_2,s_3) = (1,1,1)
 \label{eq:brane+brane_triangle_2}
\end{align}
up to phases,
which is shown in figure \ref{fig:2branes_1triangle_bigBound}.
Likewise, there exist solutions for $f_2, r_3, s_1 \in \R$ and $f_3, r_1, s_2 
\in \R$.
\paragraph{Triangular system type II + maximal Higgs on $\cC[(N_1,0)]$.}
Solving for $f_1, f_2, f_3 ,r_2,s_3 \in \R$ with all other coefficients 
vanishing, we find for $|f_3|\leq1$ 
\begin{subequations}
\label{eq:sol_2brane_typeII-all}
\begin{alignat}{4}
f_1 &=1, \quad  &
f_2 &= f_3, \quad    &
r_2 &= \sqrt{1-f_3^2}, \quad  &
s_3 &= \sqrt{1-f_3^2}  \,,
\label{eq:sol_2brane_typeII-1}
\\
f_1 &=1, \quad &
f_2 &=f_3, \quad & 
r_2 &=-\sqrt{1-f_3^2}, \quad &
s_3 &=-\sqrt{1-f_3^2} \,,
\label{eq:sol_2brane_typeII-2}
\\
f_1 &=-1, \quad & 
f_2 &= -f_3, \quad  &
r_2 &= \sqrt{1-f_3^2}, \quad  &
s_3 &=-\sqrt{1-f_3^2}\,,
\label{eq:sol_2brane_typeII-3}
\\
f_1 &=-1,\quad  &
f_2 &=-f_3, \quad &  
r_2 &= -\sqrt{1-f_3^2}, \quad  &
s_3 &= \sqrt{1-f_3^2} \,.
\label{eq:sol_2brane_typeII-4}
\end{alignat}
\end{subequations}
This is depicted in figure 
\ref{fig:2branes_1triangle_bigMaxHiggs}--\ref{fig:2branes_1triangle_bigBound}.
\paragraph{Triangular system of  type I and II combined.}
Solving for $f_1,h_1,r_2,r_3,s_2,s_3 \in \R $ with the remaining coefficients 
vanishing, we find that any combination of \eqref{eq:brane+brane_triangle_1} 
and \eqref{eq:brane+brane_triangle_2} is a non-trivial solution, which is 
depicted on figure \ref{fig:2branes_2triangle_bound}.
\paragraph{Triangular systems type I and II combined with maximal Higgs.}
Imposing only $r_1=s_1=0$, we can combine all solutions 
\eqref{eq:sol_2brane_typeI-all} and \eqref{eq:sol_2brane_typeII-all} as show in 
table \ref{tab:sol_2brane_typeI+II}. Note that these describe continuous 
solutions with constant potential energy. The entire configuration is sketched 
in figure \ref{fig:2branes_2triangle}.
\begin{table}
\centering
 \begin{tabular}{c|c|c||c|c|c}
  Sol. type I & Sol. type II & combined & Sol. type I & Sol. type II & combined 
\\ \hline 
\eqref{eq:sol_2brane_typeI-1} & \eqref{eq:sol_2brane_typeII-1} & $h_3=-f_3$ &
\eqref{eq:sol_2brane_typeI-3} & \eqref{eq:sol_2brane_typeII-1} & $h_3=f_3$ \\
\eqref{eq:sol_2brane_typeI-1} & \eqref{eq:sol_2brane_typeII-2} & $h_3=f_3$ & 
\eqref{eq:sol_2brane_typeI-3} & \eqref{eq:sol_2brane_typeII-2} & $h_3=-f_3$ \\
\eqref{eq:sol_2brane_typeI-1} & \eqref{eq:sol_2brane_typeII-3} & $h_3=-f_3$ &
\eqref{eq:sol_2brane_typeI-3} & \eqref{eq:sol_2brane_typeII-3} & $h_3=f_3$ \\
\eqref{eq:sol_2brane_typeI-1} & \eqref{eq:sol_2brane_typeII-4} & $h_3=f_3$ &
\eqref{eq:sol_2brane_typeI-3} & \eqref{eq:sol_2brane_typeII-4} & $h_3=-f_3$ \\ 
\hline
\eqref{eq:sol_2brane_typeI-2} & \eqref{eq:sol_2brane_typeII-1} & $h_3=f_3$ &
\eqref{eq:sol_2brane_typeI-4} & \eqref{eq:sol_2brane_typeII-1} & $h_3=-f_3$ \\
\eqref{eq:sol_2brane_typeI-2} & \eqref{eq:sol_2brane_typeII-2} & $h_3=-f_3$ & 
\eqref{eq:sol_2brane_typeI-4} & \eqref{eq:sol_2brane_typeII-2} & $h_3=f_3$ \\
\eqref{eq:sol_2brane_typeI-2} & \eqref{eq:sol_2brane_typeII-3} & $h_3=f_3$ &
\eqref{eq:sol_2brane_typeI-4} & \eqref{eq:sol_2brane_typeII-3} & $h_3=-f_3$ \\
\eqref{eq:sol_2brane_typeI-2} & \eqref{eq:sol_2brane_typeII-4} & $h_3=-f_3$ &
\eqref{eq:sol_2brane_typeI-4} & \eqref{eq:sol_2brane_typeII-4} & $h_3=f_3$
\end{tabular}
\caption{Combination of the one-parameter families 
\eqref{eq:sol_2brane_typeI-all} and \eqref{eq:sol_2brane_typeII-all} to a exact 
solution describing two non-trivial triangles. Note that the end points 
$f_3=\pm1$ correspond the maximal intra-brane Higgs solutions.}
\label{tab:sol_2brane_typeI+II}
\end{table}
\subsubsection{Spectrum}
\begin{figure}[t!]
\centering
 \begin{subfigure}{0.485\textwidth}
 \centering
\includegraphics[width=\textwidth]{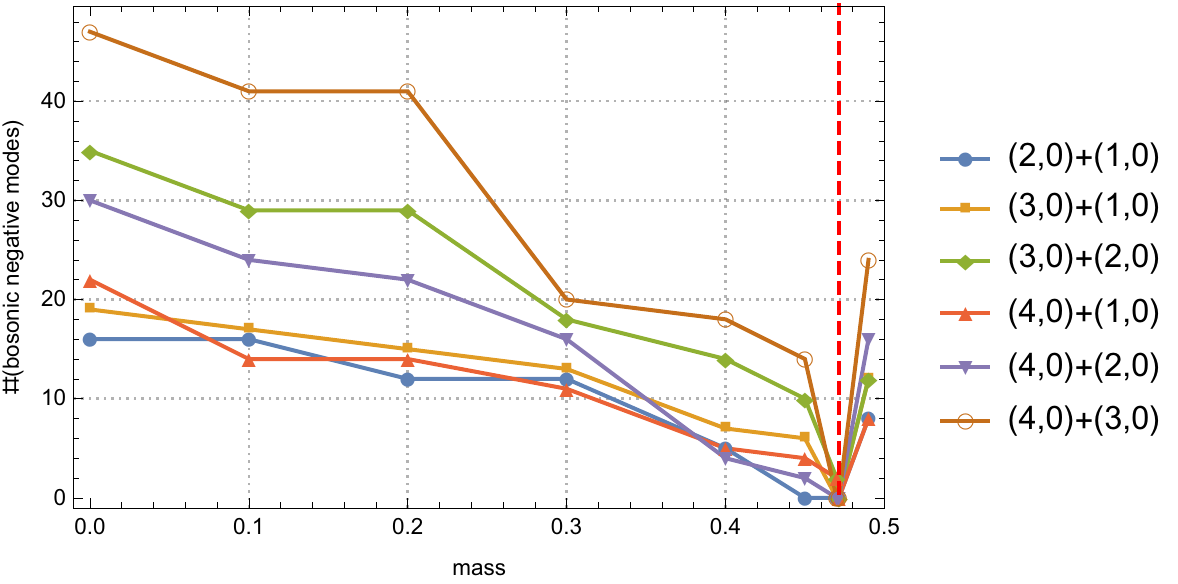}  
\caption{$x=0$}
\label{fig:2branes_neg_modes_x=0}
 \end{subfigure}
  \begin{subfigure}{0.485\textwidth}
  \centering
\includegraphics[width=\textwidth]{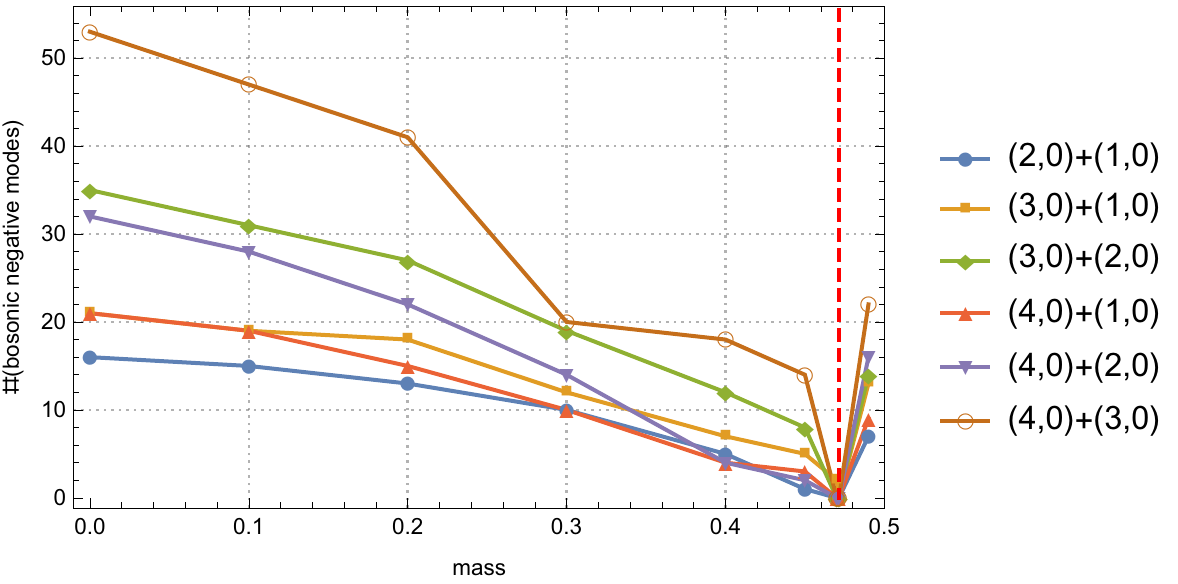}  
\caption{$x=\frac{1}{2}$}
\label{fig:2branes_neg_modes_x=1_2}
 \end{subfigure}
 \begin{subfigure}{0.485\textwidth}
 \centering
\includegraphics[width=\textwidth]{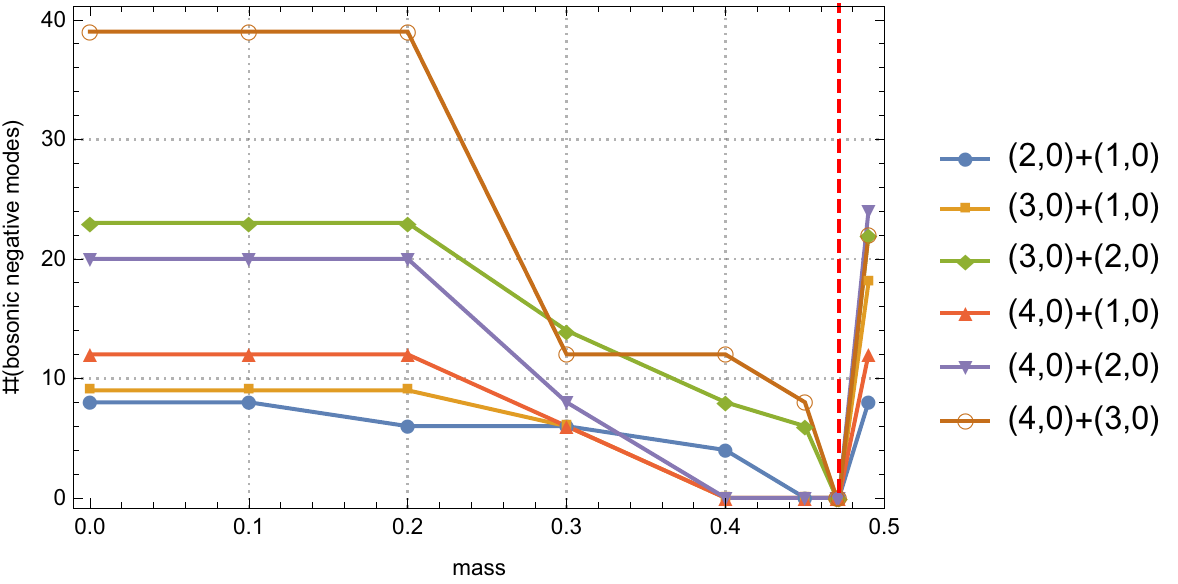}  
\caption{$x=1$}
\label{fig:2branes_neg_modes_x=1}
 \end{subfigure}
  \caption{$\cC[(N_1,0)]+\cC[(N_2,0)]$ with Higgs configuration
\eqref{eq:ex_family_2branes}. We display the number of negative modes in the 
spectrum of $\cO^{X+\phi}$  in 
(\subref{fig:2branes_neg_modes_x=0})--(\subref{fig:2branes_neg_modes_x=1}). The 
red dashed vertical lines indicates the mass value $M^*=\sqrt{2}\slash3$.}
\label{fig:2branes_spectrum}
\end{figure}

As before, we can analyze the spectrum of the vector Laplacian and of the 
Dirac operator in one of these 
new backgrounds. For concreteness, we consider the following one-parameter 
family of solution describing a configuration with two closed triangles:
\begin{equation}
\begin{aligned}
 f_1 &= h_1 = 1 \, , \quad &
 f_2 &= f_3 = -h_2 = -h_3= x \, , \\
 r_1&= s_1= 0 \, , \quad &
 r_2&=r_3 = s_2 = s_3= \sqrt{1-x^2} \,,
\end{aligned} 
\quad \text{with} \quad x\in \R, |x|\leq 1 \,.
\label{eq:ex_family_2branes}
\end{equation}
The configuration for $x=1$ describes figure 
\ref{fig:2branes_2triangle_maxHiggs}, whereas $x=0$ corresponds to figure 
\ref{fig:2branes_2triangle_bound}, and intermediate values can be thought of as 
in figure \ref{fig:2branes_2triangle_intermed}.
As in previous cases, the spectrum of $\cO_V^{X+\phi}$ has negative modes on 
the combined background $X +\phi$, but one can lift these by inclusion of mass 
terms $M_i \equiv M$ as shown in figure \ref{fig:2branes_spectrum}. It is 
quite apparent that only masses near the critical value $M^*=\sqrt{2}\slash 3$ 
allow to lift all negative modes.
% 
%%%%%%%%%%%%%%%%%%%%%%%%%%%%%%%%%%%%%%%%%%%%%%%%%%%%%%%%%%%%%%%  
%

\begin{figure}[t!]
\centering
\begin{subfigure}{0.485\textwidth}
 \centering
\includegraphics[width=\textwidth]{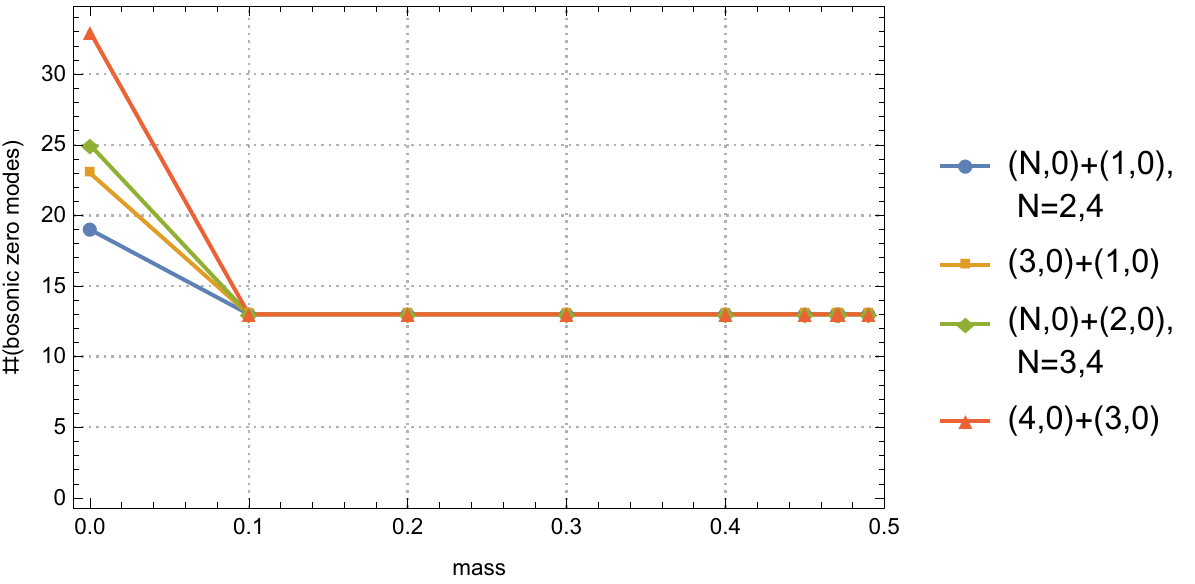}  
\caption{$x=0$}
\label{fig:2branes_0-modes_x=0}
 \end{subfigure}
 \begin{subfigure}{0.485\textwidth}
 \centering
\includegraphics[width=\textwidth]{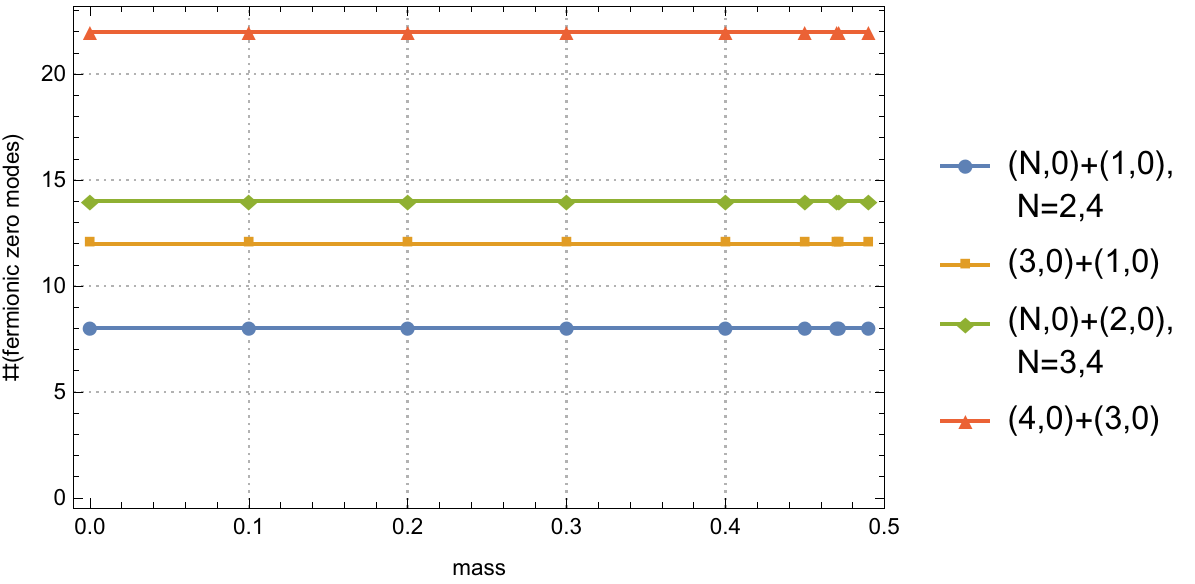}  
\caption{$x=0$}
\label{fig:2branes_Dirac-modes_x=0}
 \end{subfigure}
% 
%%%%%%%%%%%%%%%%%%%%%%%%%%%%%%%%%%%%%%%%%%%%%%%%%%%%%%%%%%%%%%%  
%
  \begin{subfigure}{0.485\textwidth}
  \centering
\includegraphics[width=\textwidth]{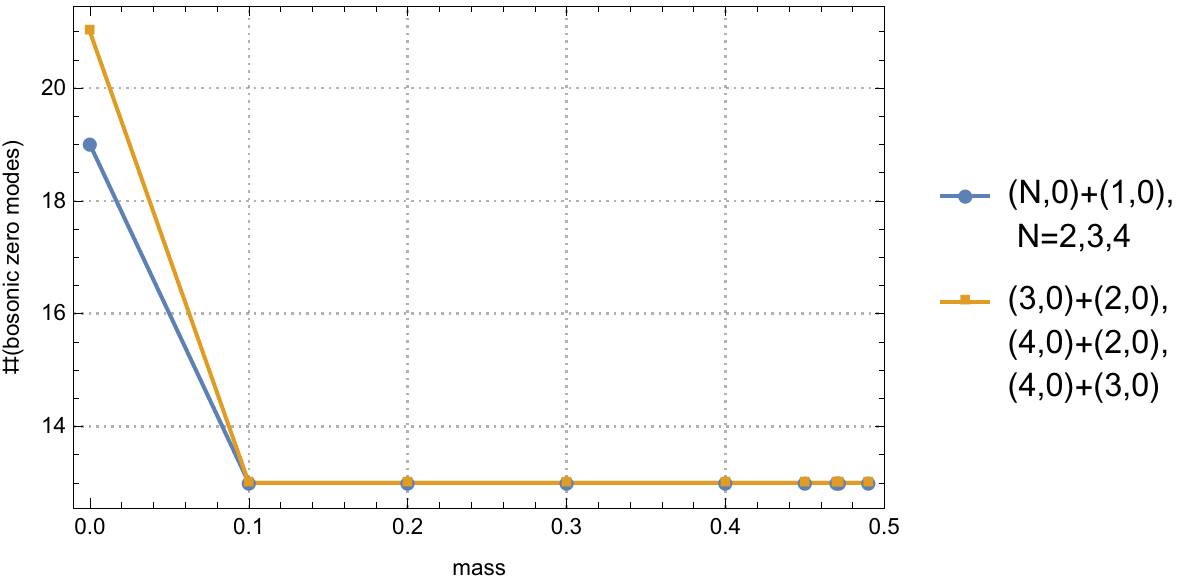}  
\caption{$x=\frac{1}{2}$}
\label{fig:2branes_0-modes_x=1_2}
 \end{subfigure}
 \begin{subfigure}{0.485\textwidth}
  \centering
\includegraphics[width=\textwidth]{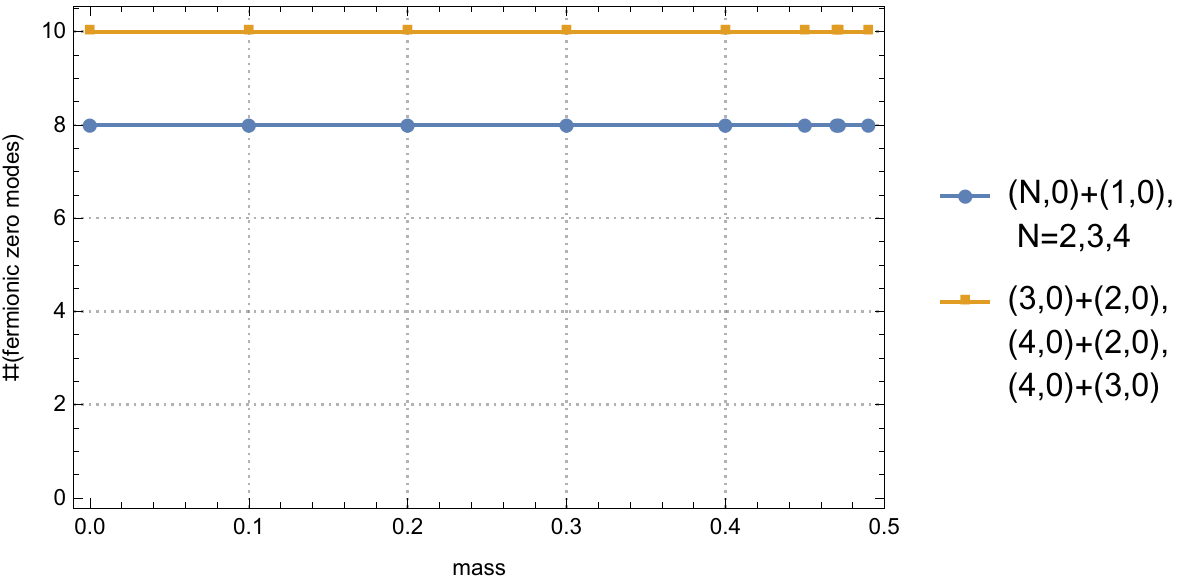}  
\caption{$x=\frac{1}{2}$}
\label{fig:2branes_Dirac-modes_x=1_2}
 \end{subfigure}
% 
%%%%%%%%%%%%%%%%%%%%%%%%%%%%%%%%%%%%%%%%%%%%%%%%%%%%%%%%%%%%%%%  
% 
 \begin{subfigure}{0.485\textwidth}
 \centering
\includegraphics[width=\textwidth]{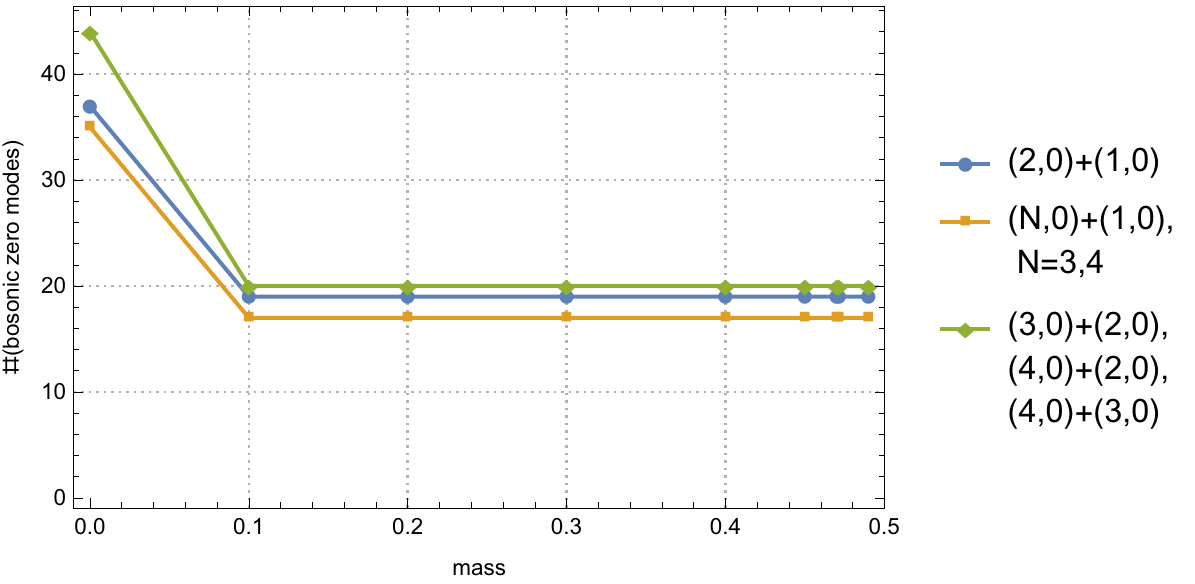}  
\caption{$x=1$}
\label{fig:2branes_0-modes_x=1}
 \end{subfigure}
 \begin{subfigure}{0.485\textwidth}
 \centering
\includegraphics[width=\textwidth]{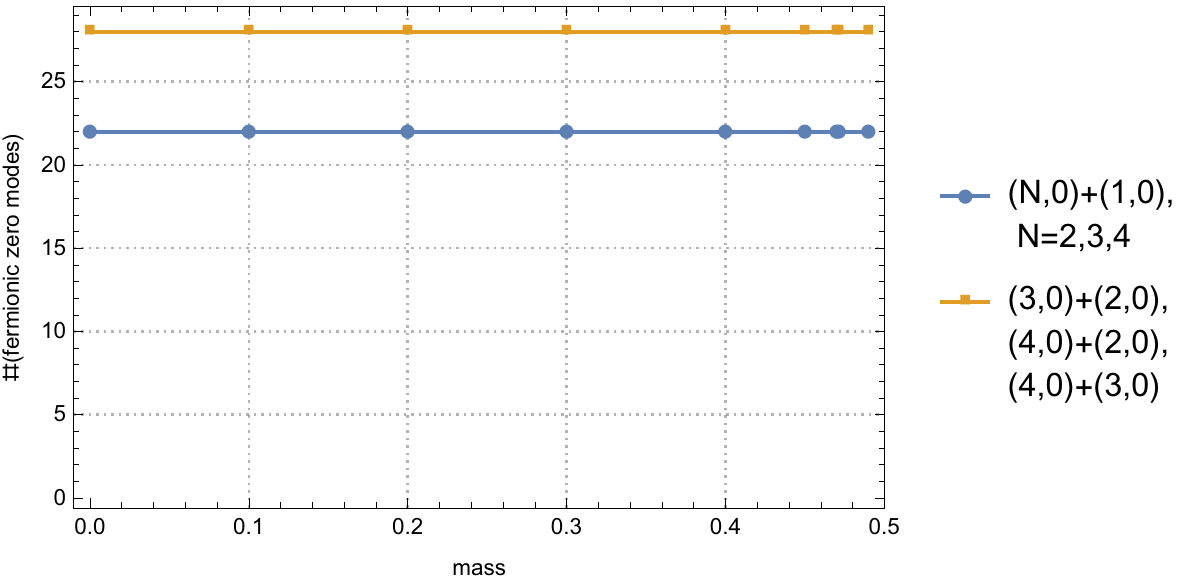}  
\caption{$x=1$}
\label{fig:2branes_Dirac-modes_x=1}
 \end{subfigure}
  \caption{$\cC[(N_1,0)]+\cC[(N_2,0)]$ with Higgs configuration
\eqref{eq:ex_family_2branes}. The number of zero modes in the spectrum of 
$\cO_V^{X+\phi}$ is shown in (\subref{fig:2branes_0-modes_x=0}), 
(\subref{fig:2branes_0-modes_x=1_2}), and (\subref{fig:2branes_0-modes_x=1}),
while 
the number of zero modes of $\Di^{X+\phi}$ are displayed in 
(\subref{fig:2branes_Dirac-modes_x=0}), 
(\subref{fig:2branes_Dirac-modes_x=1_2}), and 
(\subref{fig:2branes_Dirac-modes_x=1}).
}
\label{fig:2branes_0-modes}
\end{figure}

The number of bosonic zero modes in the combined background including masses 
stabilizes to a constant value of $13$ for $x<1$, independent of the size of 
the system, 
as shown in figures \ref{fig:2branes_0-modes_x=0}, 
\ref{fig:2branes_0-modes_x=1_2}. For $x=1$, however, which is the configuration 
of maximal intra brane Higgs, the number of zero modes for 
$\cC[(N,0)] + \cC[(1,0)]$ for $N\geq 3$ is found to be $17$, while for $\cC[(2,0)] + 
\cC[(1,0)]$ there are $19$ and for all other cases we observe $20$ zero modes. 
For details see figure 
\ref{fig:2branes_0-modes_x=1}. Since the maximal intra-brane configuration can 
be thought of as (roughly independent) tensor products of a $\cC[(N,0)]$ plus 
maximal Higgs configuration, it is tempting to interpret the smaller number of 
zero modes for branes involving $(1,0)$ in terms of a fuzzy $2$-sphere developing on 
the minimal brane with maximal Higgs.

The fermionic zero modes for the configuration $x=1$ can be understood from the 
previous case of $\cC[(N,0)]$ brane plus maximal intra-brane Higgs. As shown in 
figure \ref{fig:2branes_Dirac-modes_x=1}, combinations involving the minimal 
brane have 22 zero modes, which are 14 modes for $\cC[(N,0)]$ and 8 modes from 
the minimal brane, as in figure \ref{fig:CP2-brane+Higgs_Dirac-modes}. For two 
non-minimal $\C P^2$ brane, the number is 28, which comes from the 14 modes 
on each brane.

The other $x$ configurations behave similarly for $x>0$, in the sense that the 
properties are largely $N$ independent, cf.\ figure 
\ref{fig:2branes_Dirac-modes_x=1_2}. For the choice $x=0$, we observe a marked 
$N$ dependence in the fermionic zero mode spectrum, see figure 
\ref{fig:2branes_Dirac-modes_x=0}.
%
%%%%%%%%%%%%%%%%%%%%%%%%%%%%%%%%%%%%%%%%%%%%%%%%%%%%%%%%%%%%%%%%%%%%%%%%
%%%%%%%%%%%%%%%%%%%%%%%%%%%%%%%%%%%%%%%%%%%%%%%%%%%%%%%%%%%%%%%%%%%%%%%%%
%
\subsection{\texorpdfstring{$(N_1,0)$}{(N1,0)} brane + 
\texorpdfstring{$(N_2,0)$}{(N2,0)} brane + point brane}
\label{subsec:brane+brane+point}
\subsubsection{Solution to eom}
Consider the set-up of figure \ref{fig:Brane+brane+point} and let us choose the 
parametrization
\begin{equation}
\begin{aligned}
 Y_j^{+}=   X_j^{+} &+ f_j \ \phi_j^{+} + h_j \ \phi_j^{+} 
 + p_j \ H_j^{+} + q_j \ (\tilde{H}_j^{-})^\dagger \\
 &+ r_j \ \sigma_j^{+} + s_j \ (\tilde{\sigma}_j^{-})^\dagger
 + u_j \ \varphi_j^{+} + v_j \ (\tilde{\varphi}_j^{-})^\dagger 
\end{aligned} 
\end{equation}
for $f_j, h_j,p_j, q_j, r_j, s_j, u_j, v_j \in \C$ and $Y_j^{-} = 
(Y_j^{+})^\dagger$. 
Due to the considerable number of free complex parameters and their nonlinear 
appearance in the full eom, we can only probe a subset of solutions. As before, 
we assume that we can restrict to solutions with real coefficients.
\paragraph{Maximal Higgs.}
As in the previous cases, one finds solutions to the eom which contain only 
the maximal Higgs, i.e.\ non-trivial values for $f_j, h_j$.
\paragraph{Triangle of type I.}
Solving the eom for $q_1, u_2, s_3 \in \R$ and all other coefficients 
vanishing, one finds the solutions \eqref{eq:solution_brane}.
We exemplified this in figure \ref{fig:2branes+point_chiral-Higgs_1}.
One finds analogous solutions for $q_2, u_3, s_1$ or $q_3, u_1, s_2$.
\paragraph{Triangle of type II.}
Solving the eom for $p_1, r_2, v_3 \in \R$ and all other coefficients 
vanishing, one finds the usual solutions of \eqref{eq:solution_brane}.
We exemplified this in figure \ref{fig:2branes+point_chiral-Higgs_2}.
\paragraph{Two triangular subsystems combined.}
One can show that the following two one-parameter families
\begin{subequations}
\label{eq:2brane_1pt_ex1}
 \begin{alignat}{4}
  f_1 &= f_2 \, \quad &
  f_3 &= 1 \, \quad &
  p_1 &= \sqrt{1-f_2^2} \, \quad &
  q_2 &= \sqrt{1-f_2^2} \,,
  \label{eq:2brane_1pt_ex1_1}\\
 f_1 &= f_2 \, \quad &
 f_3 &= 1 \, \quad &
 p_1 &= -\sqrt{1-f_2^2} \, \quad &
 q_2 &= -\sqrt{1-f_2^2} \,,
 \label{eq:2brane_1pt_ex1_2}\\
 f_1 &= -f_2 \, \quad &
 f_3 &= -1 \, \quad &
 p_1 &= \sqrt{1-f_2^2} \, \quad &
 q_2 &= -\sqrt{1-f_2^2} \,,
 \label{eq:2brane_1pt_ex1_3}\\
f_1 &= -f_2 \, \quad &
f_3 &= -1 \, \quad &
p_1 &= -\sqrt{1-f_2^2} \, \quad &
q_2 &= \sqrt{1-f_2^2} \,,
\label{eq:2brane_1pt_ex1_4}
 \end{alignat}
\end{subequations}
with $f_2 \in \R$, $|f_2|\leq 1$ and, for $u_2 \in \R$, $|u_2| \leq 1$,
\begin{subequations}
\label{eq:2brane_1pt_ex2}
 \begin{alignat}{4}
  q_1 &= -u_2\, \quad &
  s_3 &= 1\, \quad &
  h_1 &= \sqrt{1-u_2^2}\, \quad &
  r_2 &= \sqrt{1-u_2^2} \,,
  \label{eq:2brane_1pt_ex2_1}\\
q_1 &= -u_2 \, \quad &
s_3 &= 1 \, \quad &
h_1 &= -\sqrt{1-u_2^2} \, \quad &
r_2 &= -\sqrt{1-u_2^2} \,,
\label{eq:2brane_1pt_ex2_2}\\
q_1 &= u_2 \, \quad &
s_3 &= -1 \, \quad &
h_1 &= \sqrt{1-u_2^2} \, \quad &
r_2 &= -\sqrt{1-u_2^2} \,,
\label{eq:2brane_1pt_ex2_3}\\
q_1 &=u_2 \, \quad &
s_3 &= -1 \, \quad &
h_1 &= -\sqrt{1-u_2^2}\, \quad &
r_2 &= \sqrt{1-u_2^2} \,,
\label{eq:2brane_1pt_ex2_4}
 \end{alignat}
\end{subequations}
can be combined non-trivially. The possible solutions are summarized in table 
\ref{tab:sol_2brane_1pt_typeI+II}.
\begin{table}
\centering
 \begin{tabular}{c|c|c|c||c|c|c|c}
  type I & type II & \multicolumn{2}{c||}{combined $u_2 {=} \kappa F[f_2]$} & 
type I & type II & 
 \multicolumn{2}{c}{combined $u_2 {=} \kappa F[f_2]$}\\
 & & $f_2\in[0,1]$ & $f_2\in [-1,0]$& & & $f_2\in[0,1]$ & $f_2\in [-1,0]$
\\ \hline 
\eqref{eq:2brane_1pt_ex1_1} & \eqref{eq:2brane_1pt_ex2_1} & 
$\kappa{=}+1$ & $\kappa{=}-1$ &
\eqref{eq:2brane_1pt_ex1_3} & \eqref{eq:2brane_1pt_ex2_1} & 
$\kappa{=}-1$ & $\kappa{=}+1$ \\
\eqref{eq:2brane_1pt_ex1_1} & \eqref{eq:2brane_1pt_ex2_2} & 
$\kappa{=}-1$ & $\kappa{=}+1$ & 
\eqref{eq:2brane_1pt_ex1_3} & \eqref{eq:2brane_1pt_ex2_2} & 
$\kappa{=}+1$ & $\kappa{=}-1$ \\
\eqref{eq:2brane_1pt_ex1_1} & \eqref{eq:2brane_1pt_ex2_3} & 
$\kappa{=}-1$ & $\kappa{=}+1$ &
\eqref{eq:2brane_1pt_ex1_3} & \eqref{eq:2brane_1pt_ex2_3} & 
$\kappa{=}+1$ & $\kappa{=}-1$\\
\eqref{eq:2brane_1pt_ex1_1} & \eqref{eq:2brane_1pt_ex2_4} & 
$\kappa{=}+1$ & $\kappa{=}-1$ &
\eqref{eq:2brane_1pt_ex1_3} & \eqref{eq:2brane_1pt_ex2_4} & 
$\kappa{=}-1$ & $\kappa{=}+1$\\ 
\hline
\eqref{eq:2brane_1pt_ex1_2} & \eqref{eq:2brane_1pt_ex2_1} & 
$\kappa{=}-1$ &$\kappa{=}+1$ &
\eqref{eq:2brane_1pt_ex1_4} & \eqref{eq:2brane_1pt_ex2_1} & 
$\kappa{=}+1$ & $\kappa{=}-1$\\
\eqref{eq:2brane_1pt_ex1_2} & \eqref{eq:2brane_1pt_ex2_2} & 
$\kappa{=}+1$ & $\kappa{=}-1$ & 
\eqref{eq:2brane_1pt_ex1_4} & \eqref{eq:2brane_1pt_ex2_2} & 
$\kappa{=}-1$  & $\kappa{=}+1$\\
\eqref{eq:2brane_1pt_ex1_2} & \eqref{eq:2brane_1pt_ex2_3} & 
$\kappa{=}+1$ & $\kappa{=}-1$ &
\eqref{eq:2brane_1pt_ex1_4} & \eqref{eq:2brane_1pt_ex2_3} & 
$\kappa{=}-1$  & $\kappa{=}+1$\\
\eqref{eq:2brane_1pt_ex1_2} & \eqref{eq:2brane_1pt_ex2_4} & 
$\kappa{=}-1$ & $\kappa{=}+1$ &
\eqref{eq:2brane_1pt_ex1_4} & \eqref{eq:2brane_1pt_ex2_4} & 
$\kappa{=}+1$ & $\kappa{=}-1$
\end{tabular}
\caption{The combination of the two one-parameter triangular solutions 
\eqref{eq:2brane_1pt_ex1} and  \eqref{eq:2brane_1pt_ex2} is consistent for a 
few choices, resulting in one-parameter families of exact solutions. Here
$F[f_2] = \sqrt{1-f_2^2}$.}
\label{tab:sol_2brane_1pt_typeI+II}
\end{table}

\subsubsection{Spectrum}
We exemplify the spectrum of the vector Laplacian and the Dirac operator around 
a background involving several Higgs fields by means of the following 
one-parameter family of exact 
solutions (see table \ref{tab:sol_2brane_1pt_typeI+II}):
\begin{equation}
\begin{aligned}
 f_1&=f_2= h_1=r_2=  x \,,  \quad
 f_3 = s_3= 1 \,,  \quad
 p_1 = -q_1 = q_2 = u_2 = \sqrt{1-x^2} \, ,\\
 h_2&= h_3= p_2=p_3=q_3=s_1=s_2=u_1=u_3=r_1=r_3=v_j= 0 \;,
\end{aligned} 
\label{eq:ex_family_2branes+pt}
\end{equation}
where $x\in [0,1]$. The configuration for $x=0$ is depicted in figure 
\ref{fig:2branes_1point_2triangle_1}, and figure 
\ref{fig:2branes_1point_2triangle_2} corresponds to $x=1$.
\begin{figure}[t!]
\centering
 \begin{subfigure}{0.485\textwidth}
 \centering
\includegraphics[width=\textwidth]{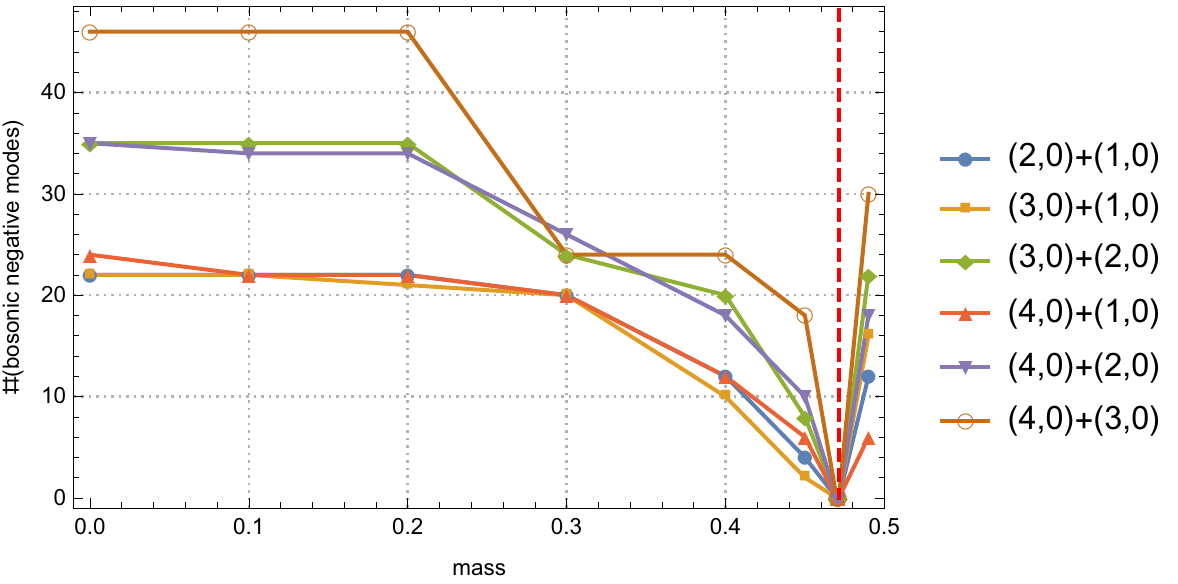}  
\caption{$x=0$}
\label{fig:2branes+pt_neg_modes_x=0}
 \end{subfigure}
  \begin{subfigure}{0.485\textwidth}
  \centering
\includegraphics[width=\textwidth]{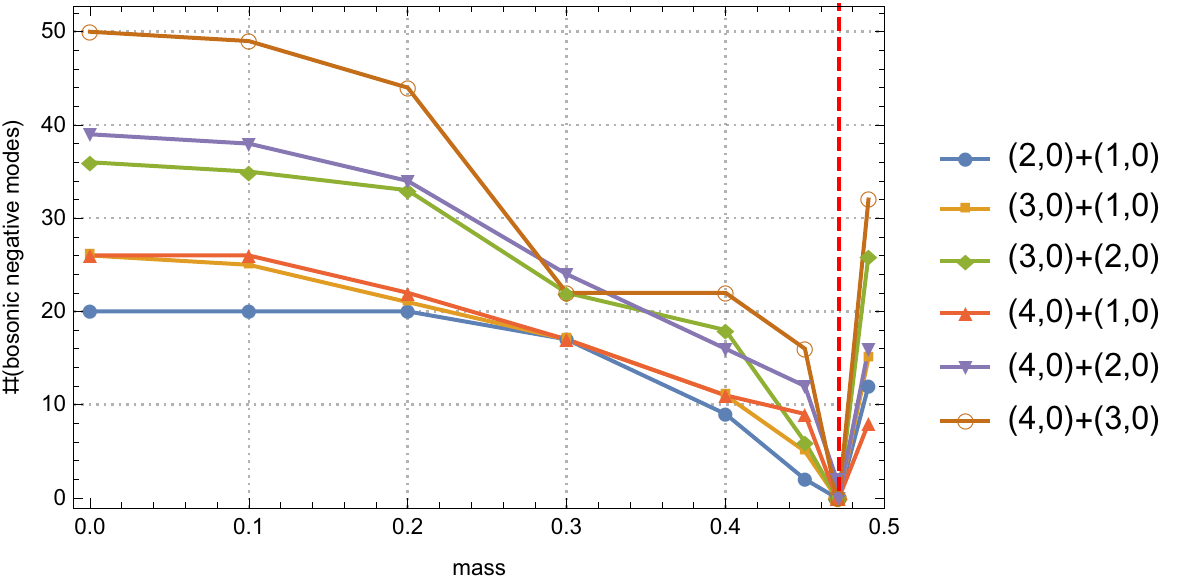}  
\caption{$x=\frac{1}{2}$}
\label{fig:2branes+pt_neg_modes_x=1_2}
 \end{subfigure}
 \begin{subfigure}{0.485\textwidth}
 \centering
\includegraphics[width=\textwidth]{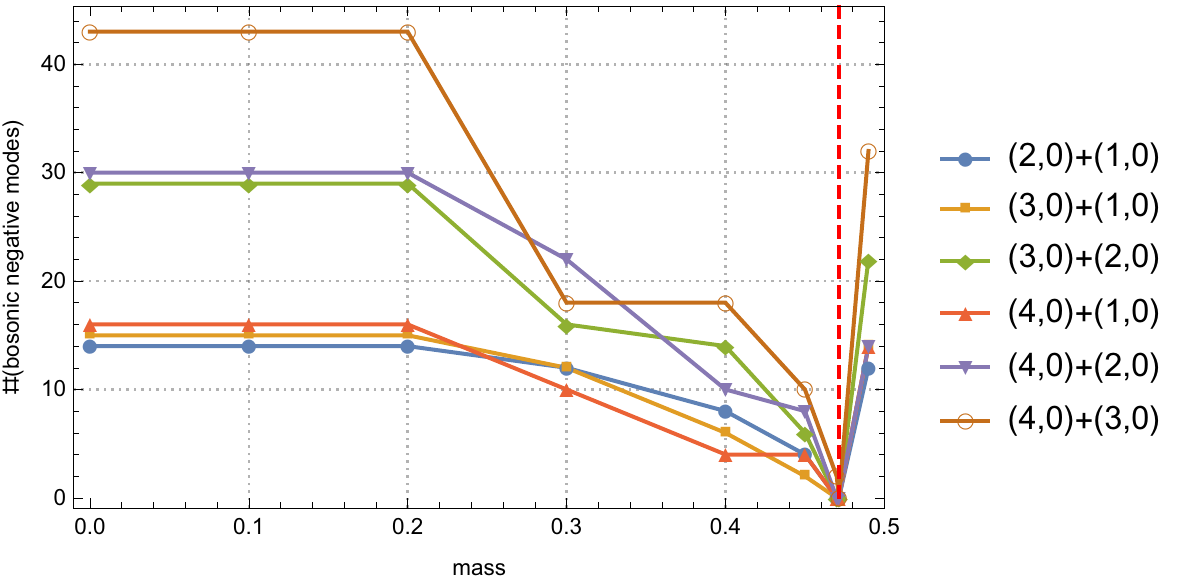}  
\caption{$x=1$}
\label{fig:2branes+pt_neg_modes_x=1}
 \end{subfigure} 
  \caption{$\cC[(N_1,0)] + \cC[(N_2,0)]+ \cD$ with Higgs configuration
\eqref{eq:ex_family_2branes+pt}. The number of negative modes for 
$\cO_V^{X+\phi}$, for each choice of mass value $M$, are depicted in
(\subref{fig:2branes+pt_neg_modes_x=0})--(\subref{%
fig:2branes+pt_neg_modes_x=1}). 
The red dashed vertical lines indicates the mass value $M=\sqrt{2}\slash3$.}
\label{fig:2branes+pt_spectrum}
\end{figure}
The bosonic spectrum then behaves similarly to the case of two 
parallel branes of section \ref{subsec:brane+brane}. The appearing negative 
modes of $\cO_V^{X+\phi}$ can be 
lifted by sufficiently large mass values $0.47 \lesssim M \leq 
\frac{\sqrt{2}}{3}$, which we exemplify in figure 
\ref{fig:2branes+pt_spectrum}.

% 
%%%%%%%%%%%%%%%%%%%%%%%%%%%%%%%%%%%%%%%%%%%%%%%%%%%%%%%%%%%%%%%  
%  
\begin{figure}[t!]
\centering
 \begin{subfigure}{0.485\textwidth}
 \centering
\includegraphics[width=\textwidth]{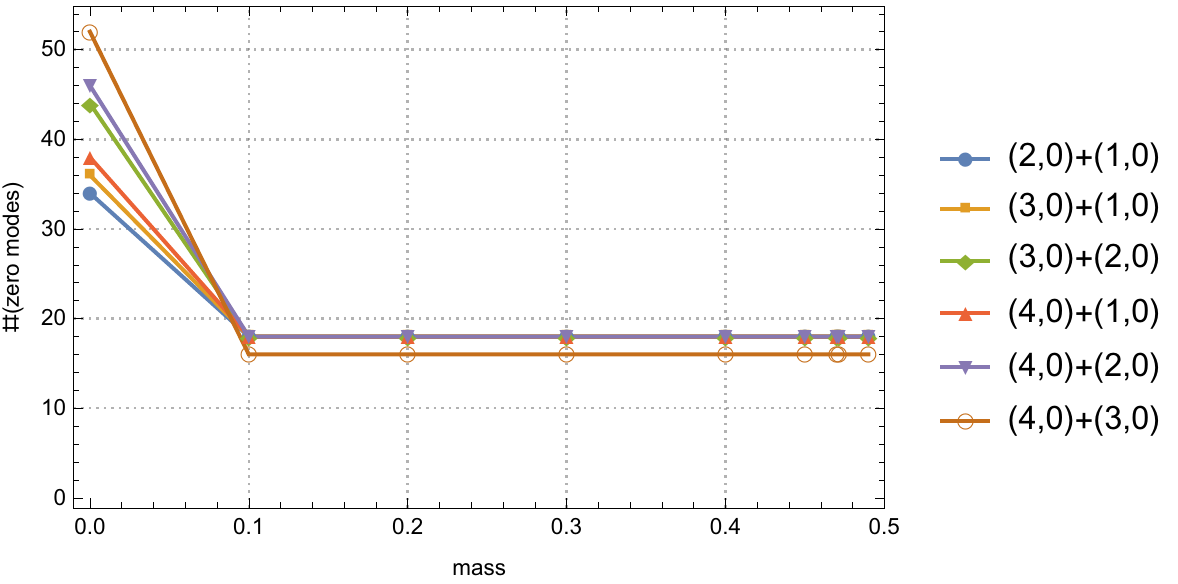}  
\caption{$x=0$}
\label{fig:2branes+pt_0-modes_x=0}
 \end{subfigure}
 \begin{subfigure}{0.485\textwidth}
 \centering
\includegraphics[width=\textwidth]{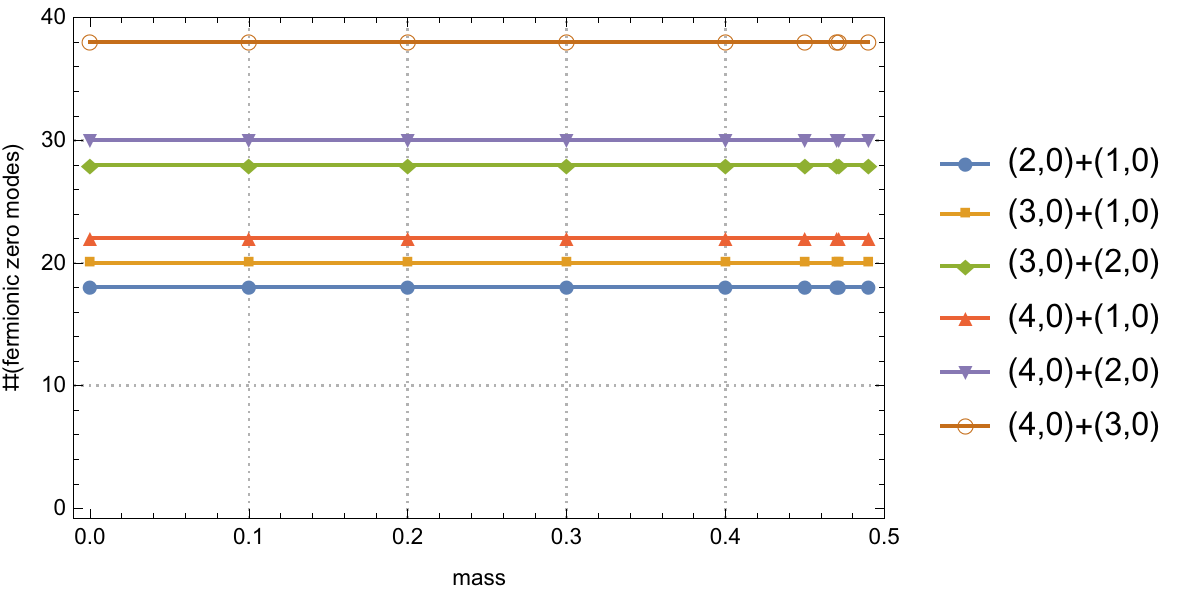}  
\caption{$x=0$}
\label{fig:2branes+pt_Dirac-modes_x=0}
 \end{subfigure}
% 
%%%%%%%%%%%%%%%%%%%%%%%%%%%%%%%%%%%%%%%%%%%%%%%%%%%%%%%%%%%%%%%  
% 
  \begin{subfigure}{0.485\textwidth}
  \centering
\includegraphics[width=\textwidth]{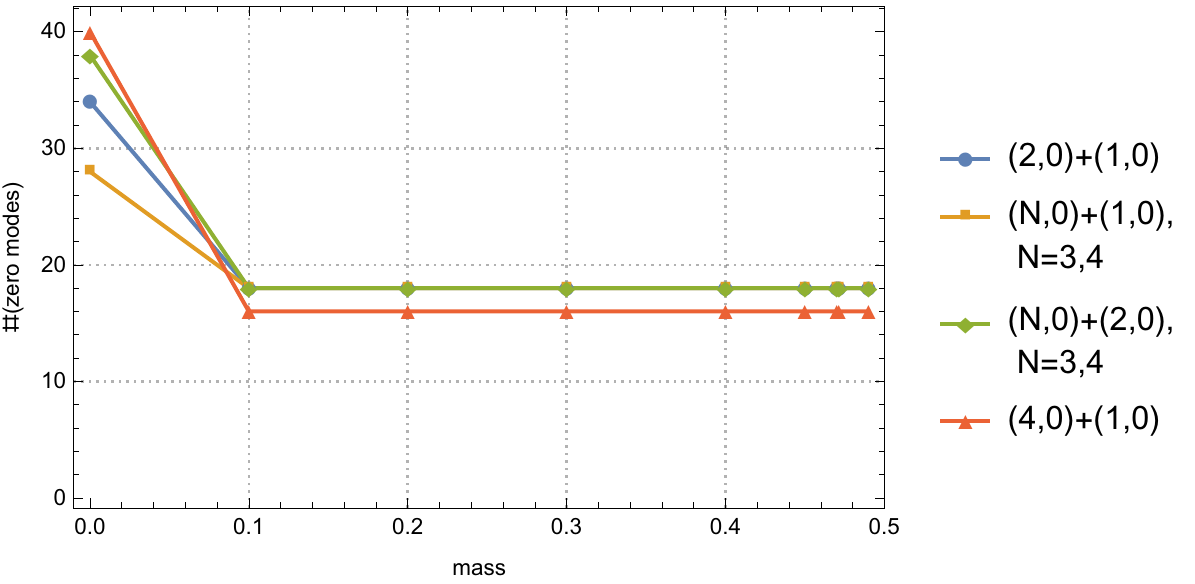}  
\caption{$x=\frac{1}{2}$}
\label{fig:2branes+pt_0-modes_x=1_2}
 \end{subfigure}
 \begin{subfigure}{0.485\textwidth}
  \centering
\includegraphics[width=\textwidth]{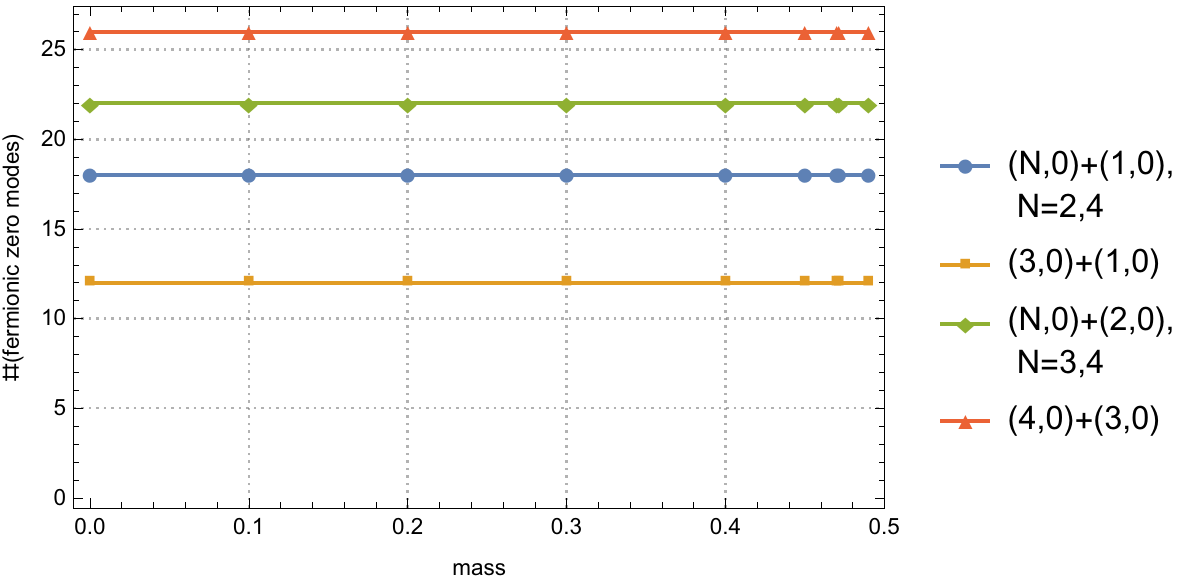}
\caption{$x=\frac{1}{2}$}
\label{fig:2branes+pt_Dirac-modes_x=1_2}
 \end{subfigure}
% 
%%%%%%%%%%%%%%%%%%%%%%%%%%%%%%%%%%%%%%%%%%%%%%%%%%%%%%%%%%%%%%%  
%  
 \begin{subfigure}{0.485\textwidth}
 \centering
\includegraphics[width=\textwidth]{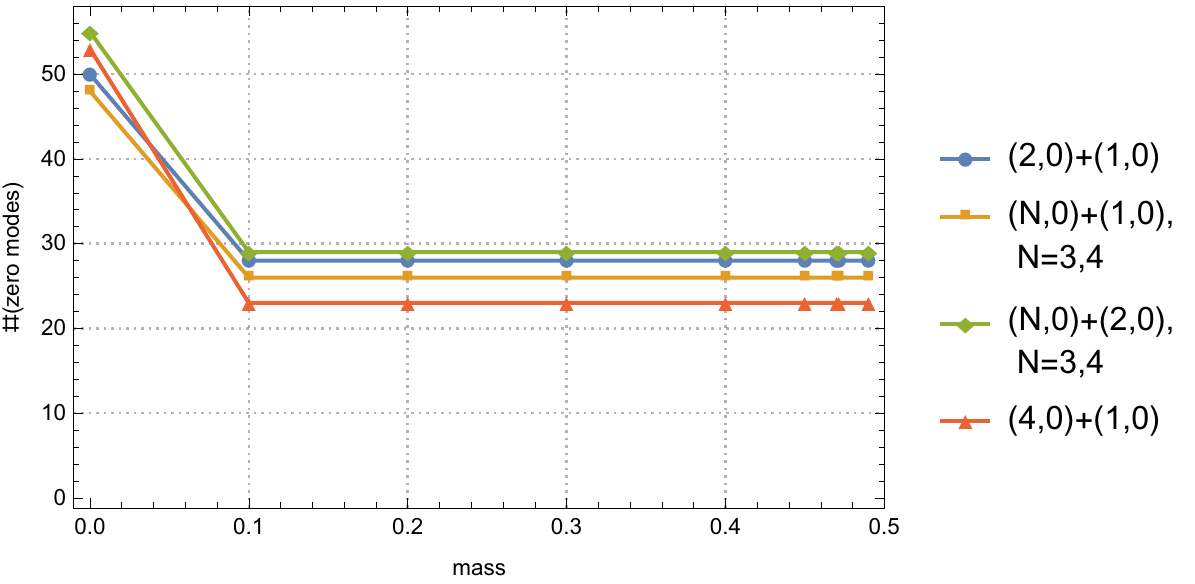}  
\caption{$x=1$}
\label{fig:2branes+pt_0-modes_x=1}
 \end{subfigure}
\begin{subfigure}{0.485\textwidth}
 \centering
\includegraphics[width=\textwidth]{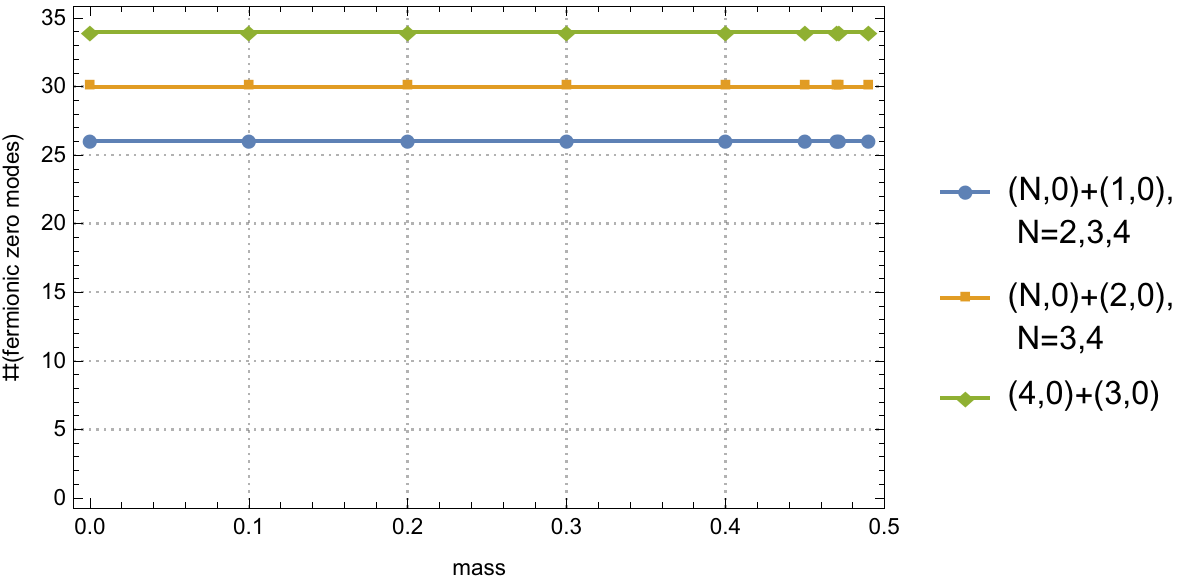}  
\caption{$x=1$}
\label{fig:2branes+pt_Dirac-modes_x=1}
 \end{subfigure}
  \caption{$\cC[(N_1,0)] + \cC[(N_2,0)]+ \cD$ with Higgs configuration
\eqref{eq:ex_family_2branes+pt}. The number of bosonic zero modes is shown in 
(\subref{fig:2branes+pt_0-modes_x=0}), 
(\subref{fig:2branes+pt_0-modes_x=1_2}), and 
(\subref{fig:2branes+pt_0-modes_x=1}),
while the number of zero modes of $\cD^{X+\phi}$ are provided in 
(\subref{fig:2branes+pt_Dirac-modes_x=0}), 
(\subref{fig:2branes+pt_Dirac-modes_x=1_2}), and 
(\subref{fig:2branes+pt_Dirac-modes_x=1}).}
\label{fig:2branes+pt_0-modes}
\end{figure}

The configuration for $x=1$ can be understood as direct sum of 
$\cC[(N_1,0)]$ with maximal intra-brane Higgs plus a bound version of 
$\cC[(N_2,0)]+\cD$ with some Higgs. As in appendix \ref{subsec:brane+point}, 
the \emph{bound state} of $\C P^2$ brane with a point brane leads to an $N$ 
dependent number of fermionic zero modes. Nonetheless, the fermionic zero 
modes, see figure \ref{fig:2branes+pt_Dirac-modes_x=1}, can be understood as 
linear combination of the two cases. The bosonic zero modes, cf.\ figure 
\ref{fig:2branes+pt_0-modes_x=1} do not follow this simple nature.

The more involved the connection of the two $\C P^2$ branes to the point brane 
becomes, i.e. $0\leq x < 1$, the more does the fermionic zero modes spectrum 
exhibit a pronounced dependence on the system size $N_i$. The bosonic 
zero modes  exhibit this only in the massless case, and the massive case 
seems less $N_i$ dependent.
% 
%%%%%%%%%%%%%%%%%%%%%%%%%%%%%%%%%%%%%%%%%%%%%%%%%%%%%%%%%%%%%%%%%%%%%%%%%%
%%%%%%%%%%%%%%%%%%%%%%%%%%%%%%%%%%%%%%%%%%%%%%%%%%%%%%%%%%%%%%%%%%%%%%%%%
%
\subsection{\texorpdfstring{$(N_1,0)$}{(N1,0)} brane + 
\texorpdfstring{$(N_2,0)$}{(N2,0)} brane + \texorpdfstring{$(N_3,0)$}{(N3,0)} 
brane}
\label{subsec:brane+brane+brane}
\subsubsection{Solution to eom}
Consider the following parametrization
\begin{equation}
\begin{aligned}
 Y_j^{+}=  X_j^{+} 
 &+ f_j \ \phi_{j,3}^{+} +  g_j \ \phi_{j,2}^{+}  + h_j \ \phi_{j,1}^{+} %\\
  + p_j \ \varphi^{+}_{j,3,2} + q_j \ (\varphi^{-}_{j,3,2})^\dagger \\
%  
%  &
 &+ r_j \ \varphi^{+}_{j,3,1} + s_j \ (\varphi^{-}_{j,3,1})^\dagger %\\
%  
%  & 
 + u_j \ \varphi^{+}_{j,2,1} + v_j \ (\varphi^{-}_{j,2,1})^\dagger 
\end{aligned} 
\end{equation}
with $f_j, g_j, h_j, p_j, q_j, r_j,s_j, u_j, v_j \in \C $. The subscript in
$\phi_{j,1}$ means maximal intra-brane Higgs on $\cC[(N_1,0)]$, while 
$\varphi^{+}_{j,3,2}$ means maximal inter-brane Higgs from $\cC[(N_3,0)]$ to 
$\cC[(N_2,0)]$ and so forth.
\paragraph{Maximal Higgs.}
Consider the configuration with only maximal Higgs, i.e.\ solve $f_j,g_j,h_j 
\in \C $ and all other coefficients vanish. Then we find that all possible 
combinations of $(f_j),(g_j),(h_j)$ independently taking the form of 
\eqref{eq:solution_brane} are in fact solutions in the three brane case.
Note that the potential energy for the case $f_j$, $g_j$, and $h_j$ 
simultaneously non-zero is the smallest.
\paragraph{Single triangles.}
We solve the eom for $p_1,u_2,s_3 \in \R$ and all other coefficients vanish. 
We find the following solutions:
\begin{align}
(p_1,u_2,s_3)  \in \left\{  
(1,1,1),(1,1,-1),(1,-1,1),(-1,1,1),(-1,-1,-1),(0,0,0) \right\} 
\label{eq:3branes_single_triangle}
\end{align}
Note that this is not the usual family obtained from 2 phases. 
Similarly, there are analogous solutions for $p_2,u_3,s_1 \in \R$ or 
$p_3,u_1,s_2 \in \R$. 
Additionally, one can consider the triangles going the other way around, i.e.\ 
$q_1,v_3,r_2 \in \R$, $q_2,v_1,r_3 \in \R$, or $q_3,v_2,r_1 \in \R$.
\paragraph{Multiple triangular subsystem of type I.}
Consider the configuration in figure \ref{fig:3Branes_1} such that we solve the 
eom for $p_j, u_j, s_j \in \R$ and all other coefficients vanish. 
As solutions we find all possible combinations of 
\eqref{eq:3branes_single_triangle}, i.e.\ we find $5^3 = 125$ real solutions.
Note that these configurations have the same potential energy as the 
configuration with all maximal Higgs non-vanishing.
\paragraph{Multiple triangular subsystem of type II.}
Consider the configuration in figure \ref{fig:3Branes_2}; i.e.\ we solve the 
eom for 
$g_2, h_3,p_1, p_3, q_3, r_2, s_1, s_2, u_1 \in \R$  and all other coefficients 
vanish.
Since the solutions is comprised of three independent triangles, we find 
$5^3=125$ cases, again.
Note that the configurations with all triangles non-trivial have the same 
potential energy as the configuration with three non-vanishing maximal Higgs 
configurations.
\subsubsection{Spectrum}
\begin{figure}[t!]
\centering
 \begin{subfigure}{0.495\textwidth}
 \centering
\includegraphics[width=\textwidth]{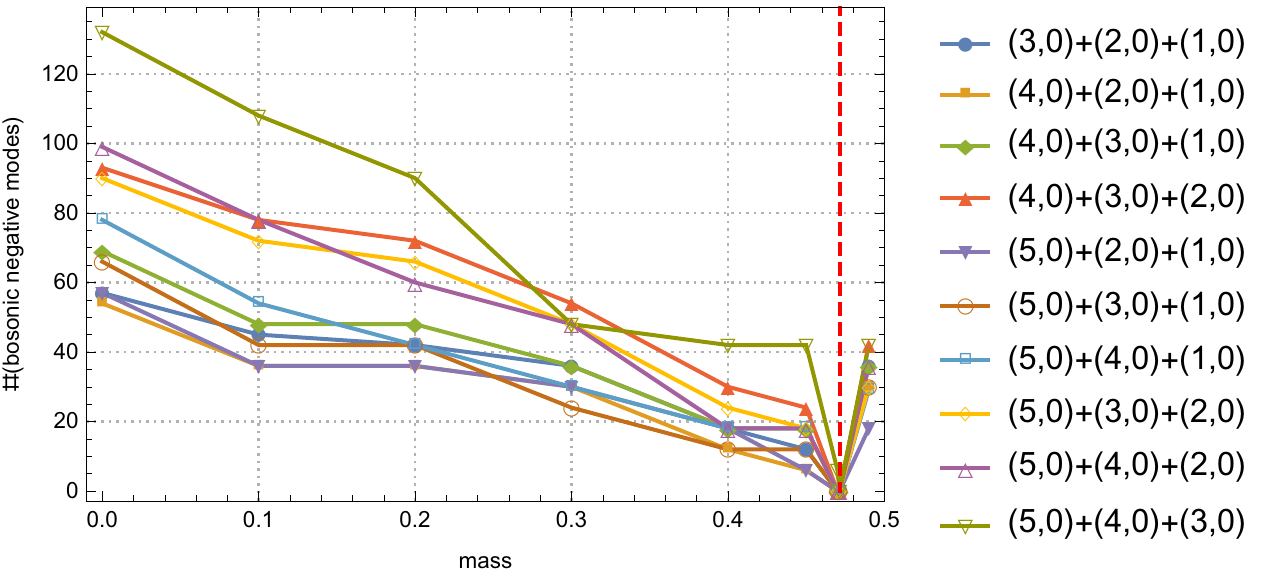}  
\caption{Configuration \ref{fig:3Branes_1}}
\label{fig:3branes_neg_modes_x=1}
 \end{subfigure}
\begin{subfigure}{0.495\textwidth}
 \centering
\includegraphics[width=\textwidth]{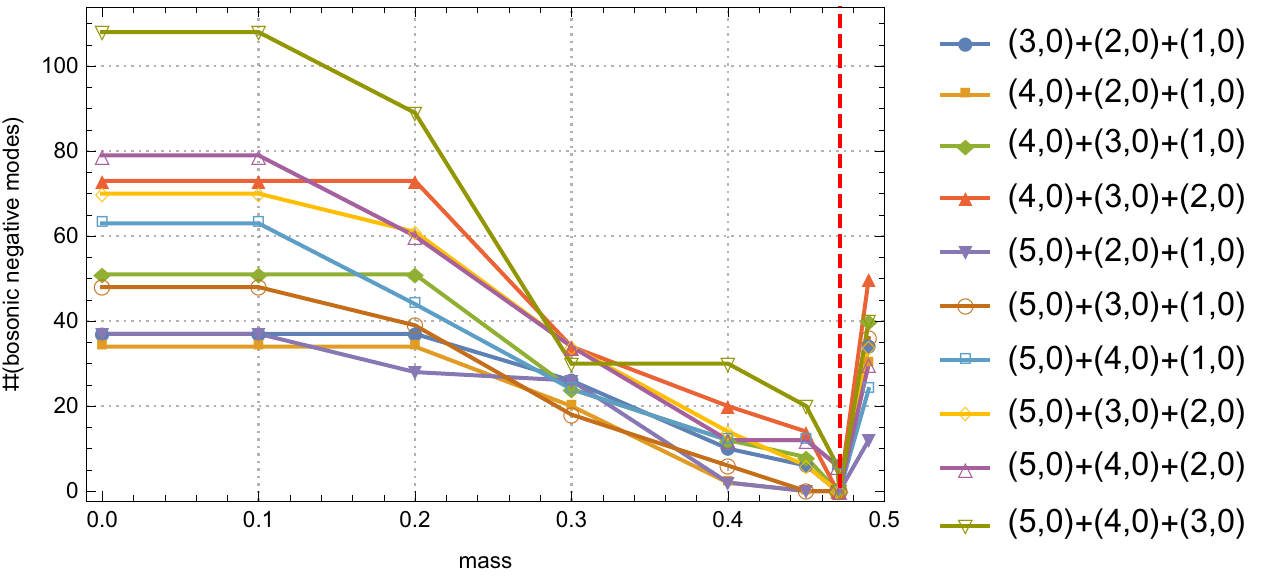}  
\caption{Configuration \ref{fig:3Branes_5}.}
\label{fig:3branes_neg_modes_x=0}
 \end{subfigure}
% 
%%%%%%%%%%%%%%%%%%%%%%%%%%%%%%%%%%%%%%%%%%%%%%%%%%%%%%%%%%%%%%%  
% 
 \begin{subfigure}{0.495\textwidth}
 \centering
\includegraphics[width=\textwidth]{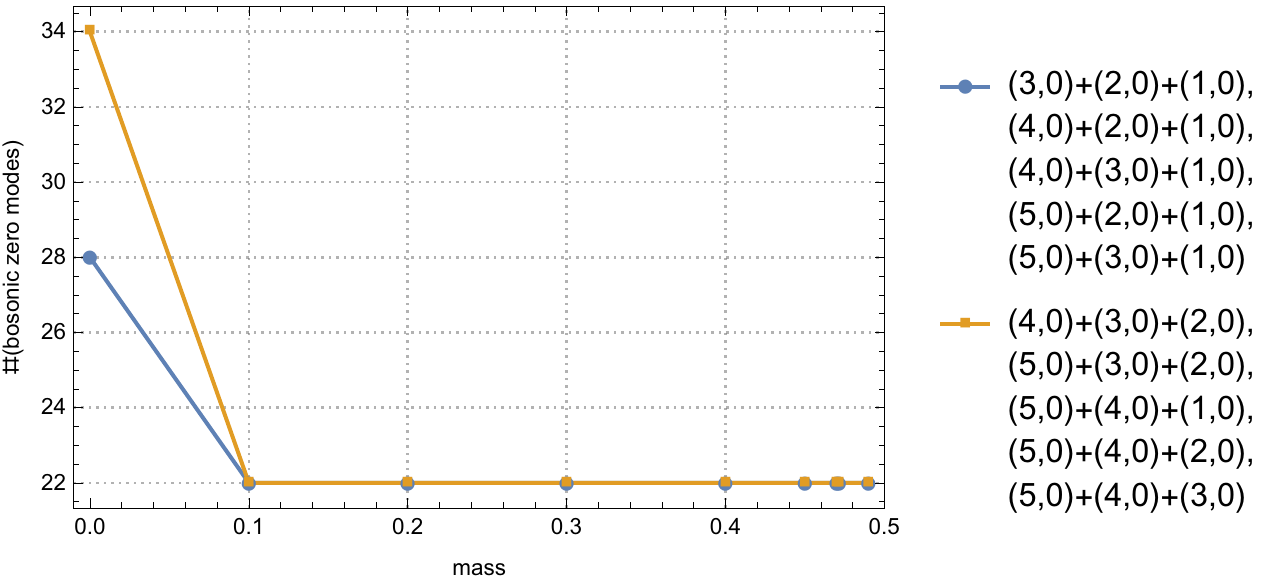}  
\caption{Configuration  \ref{fig:3Branes_1}}
\label{fig:3branes_0-modes_x=1}
 \end{subfigure}
\begin{subfigure}{0.495\textwidth}
 \centering
\includegraphics[width=\textwidth]{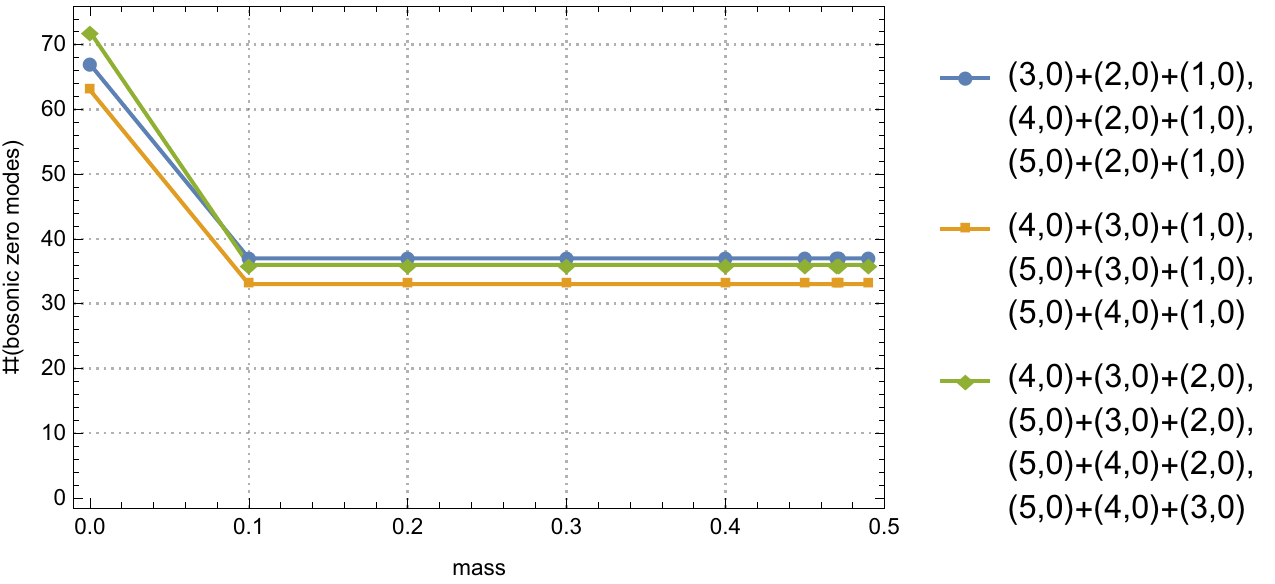}  
\caption{Configuration \ref{fig:3Branes_5}}
\label{fig:3branes_0-modes_x=0}
 \end{subfigure}
 % 
%%%%%%%%%%%%%%%%%%%%%%%%%%%%%%%%%%%%%%%%%%%%%%%%%%%%%%%%%%%%%%%  
% 
 \begin{subfigure}{0.495\textwidth}
 \centering
\includegraphics[width=\textwidth]{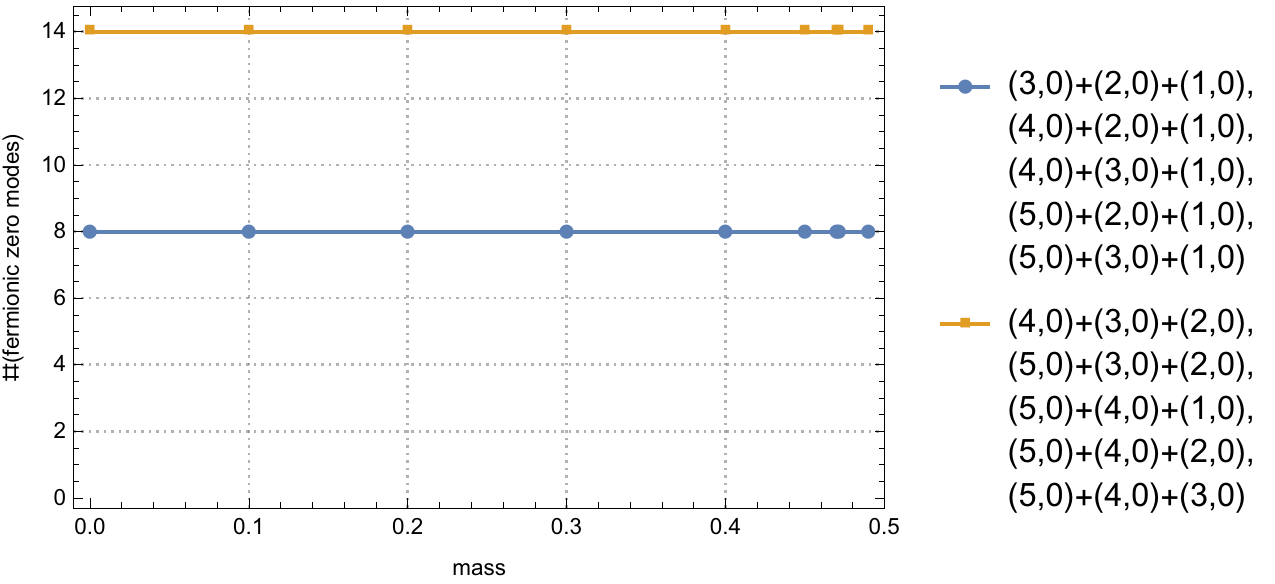}  
\caption{Configuration \ref{fig:3Branes_1}}
\label{fig:3branes_Dirac-modes_x=1}
 \end{subfigure}
\begin{subfigure}{0.495\textwidth}
 \centering
\includegraphics[width=\textwidth]{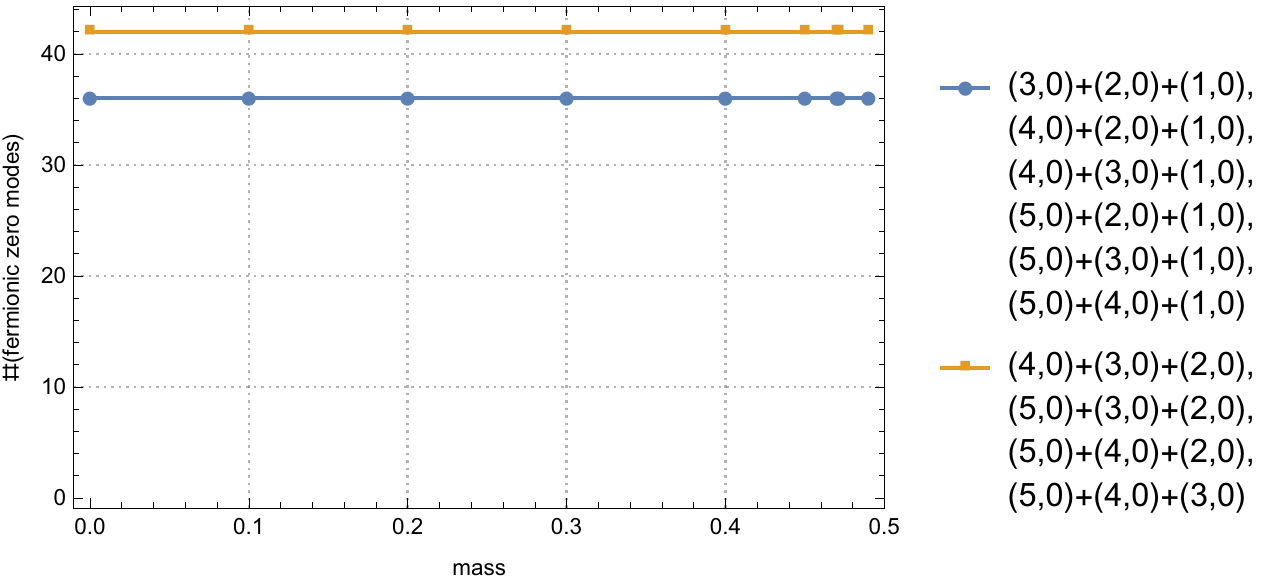}  
\caption{Configuration \ref{fig:3Branes_5}}
\label{fig:3branes_Dirac-modes_x=0}
 \end{subfigure}
% 
%%%%%%%%%%%%%%%%%%%%%%%%%%%%%%%%%%%%%%%%%%%%%%%%%%%%%%%%%%%%%%%  
%
  \caption{$\cC[(N_1,0)] + \cC[(N_2,0)]+ \cC[(N_3,0)]$ with Higgs 
configuration as in figure  
\ref{fig:3Branes_1} or \ref{fig:3Branes_5}, respectively.
  The number of 
negative modes of $\cO_V^{X+\phi}$, for each choice of 
mass value $M$, are depicted in (\subref{fig:3branes_neg_modes_x=1}) and 
(\subref{fig:3branes_neg_modes_x=0}). The red 
dashed vertical lines indicates the mass value $M^*$.
In addition, we provide the number of bosonic zero modes in  
(\subref{fig:3branes_0-modes_x=1}), (\subref{fig:3branes_0-modes_x=0}) and the 
number of fermionic zero modes in  
(\subref{fig:3branes_Dirac-modes_x=1}), (\subref{fig:3branes_Dirac-modes_x=0}).}
\label{fig:3branes_spectrum}
\end{figure}

Analogous to the other set-ups, we can evaluate the spectrum of the vector 
Laplacian and the Dirac operator around a combined background. We restrict 
ourselves to two 
configurations: (i) the solution of figure \ref{fig:3Branes_1}, and (ii) the 
maximal intra-brane solution of figure \ref{fig:3Branes_5}. The number of 
negative modes is shown in figure \ref{fig:3branes_neg_modes_x=1} and 
\ref{fig:3branes_neg_modes_x=0}, respectively. Again, we observe that for large 
enough mass values $0.47 \lesssim M \leq M^*$ all negative modes 
can be lifted consistently.
% 
%%%%%%%%%%%%%%%%%%%%%%%%%%%%%%%%%%%%%%%%%%%%%%%%%%%

For configuration \ref{fig:3Branes_5}, we observe from figure 
\ref{fig:3branes_0-modes_x=0} that a large number of zero modes disappears after 
inclusion of uniform mass values and the resulting number depends slightly on 
the system size. The fermionic zero modes exhibit that the system roughly 
corresponds to three independent copies of $\cC[(N_i,0)]$ plus maximal 
intra-brane Higgs. In detail, we observe from figure 
\ref{fig:3branes_Dirac-modes_x=0} that for the system $\cC[(N_1,0)] + \cC[(N_2,0)]+ 
\cC[(1,0)]$ there are 36 zero modes, which matches $2 \cdot 14$ modes from 
$\cC[(N_1,0)]$ and $\cC[(N_2,0)]$ (each with maximal Higgs) and 8 modes from 
the minimal brane plus maximal Higgs. Similarly, $\cC[(N_1,0)] + \cC[(N_2,0)] + 
\cC[(N_3,0)]$ with $N_i >1$ shows 42 zero modes, which are $3 \cdot 14$ 
originating from independent copies of $\cC[(N_i,0)]$ plus maximal Higgs.

The more intricate configuration \ref{fig:3Branes_1} shows a qualitatively 
similar behavior. Note, however, that the number of bosonic and fermionic zero 
modes, see figure \ref{fig:3branes_0-modes_x=1} and 
\ref{fig:3branes_Dirac-modes_x=1} respectively, is much lower compared to the 
other configuration.
% 
%%%%%%%%%%%%%%%%%%%%%%%%%%%%%%%%%%%%%%%%%%%%%%%%%%%
%%%%%%%%%%%%%%%%%%%%%%%%%%%%%%%%%%%%%%%%%%%%%%%%%%%
% 
\subsection{\texorpdfstring{$(1,0)$}{(1,0)} brane + 
\texorpdfstring{$(0,1)$}{(0,1)} brane + point brane}
\label{subsec:flipped_min_brane+point}
\subsubsection{Solution to eom}
With the notation introduced in figures 
\ref{fig:flipped_brane+point_Higgs_partI}--\ref{%
fig:flipped_brane+point_Higgs_partII} we employ the following ansatz
\begin{equation}
\begin{aligned}
 Y_j^+=  X_j^{+} 
 &+f_j \phi_j^+
 +h_j \tilde{\phi}_j^+ 
 +p_j \varphi_j^+ + q_j(\tilde{\varphi}_j^-)^\dagger
 +x_j \zeta_j^+  + y_j (\tilde{\zeta}_j^-)^\dagger \\
 &+r_j \varrho_j^+ + s_j (\tilde{\varrho_j}^-)^\dagger
 +u_j \sigma_j^+ + v_j (\tilde{\sigma}_j^-)^\dagger 
\end{aligned} 
\end{equation}
for $f_j,h_j,p_j, q_j, r_j,s_j,u_j,v_j,x_j,y_j \in \C$. Inspecting the various 
Higgs modes there are various exact solutions.
\paragraph{Maximal intra-brane Higgs.}
As seen in previous cases, the brane background together with the maximal 
regular zero modes $\phi_j$ and / or $\tilde{\phi}_j$ leads to an exact solution 
of the equations of motion.
\paragraph{$G_2$ solution, type I.}
Setting $f_j=h_j=p_j=q_j=x_j=s_j=u_j=0$ and solving for the remaining variables 
we find various exact solutions corresponding to figure 
\ref{fig:flipped_brane+pt_G2-sol1}.
\begin{subequations}
\begin{alignat}{2}
r_1 &= -r_2 = -r_3 = -v_1=v_2 =- v_3= \pm \frac{1}{\sqrt{2}} \, , \quad&
-y_1&= -y_2 = -y_3 = \frac{1}{2} \,,\\
r_1 &= r_2 = -r_3 = v_1=v_2 = v_3= \pm \frac{1}{\sqrt{2}} \, , \quad&
-y_1&= y_2 = -y_3 = \frac{1}{2} \,,\\
r_1 &= r_2 = r_3 = v_1=v_2 = -v_3= \pm \frac{1}{\sqrt{2}} \, , \quad&
y_1&= -y_2 = -y_3 = \frac{1}{2}  \,, \\
r_1 &= -r_2 = r_3 = -v_1=v_2 = v_3= \pm \frac{1}{\sqrt{2}} \, , \quad&
y_1&= y_2 = -y_3 = \frac{1}{2} \,, \\
r_1 &= -r_2 = r_3 = v_1=-v_2 = -v_3= \pm \frac{1}{\sqrt{2}} \, , \quad&
-y_1&= -y_2 = y_3 = \frac{1}{2} \,, \\
r_1 &= r_2 = r_3 = -v_1=-v_2 = v_3= \pm \frac{1}{\sqrt{2}} \, , \quad&
-y_1&= y_2 = y_3 = \frac{1}{2} \,, \\
r_1 &= r_2 = -r_3 = -v_1=-v_2 = -v_3= \pm \frac{1}{\sqrt{2}} \, , \quad&
y_1&= -y_2 = y_3 = \frac{1}{2} \,, \\
r_1 &= -r_2 = -r_3 = v_1=-v_2 = v_3= \pm \frac{1}{\sqrt{2}} \, , \quad&
y_1&= y_2 = y_3 = \frac{1}{2} \,.
\end{alignat}
\end{subequations}
Besides these solutions containing all $r_j, v_j,y_j$ non-trivial, there are 
also solutions like 
\begin{equation}
 -r_2=r_3= v_2 = v_3=y_1=1 \, , \quad r_1=v_1=y_2=y_3 =0 \; .
\end{equation}
Again, there are various different sign assignments possible.
This type of solution corresponds to a subset of the full $G_2$-type solution.
\paragraph{$G_2$ solution, type II.}
Setting $f_j=h_j=p_j=q_j=y_j=r_j=v_j=0$ and solving for the remaining variables 
we find various exact solutions corresponding to figure 
\ref{fig:flipped_brane+pt_G2-sol2}. These are as in the previous case, and we 
refrain from repeating them here.
\subsubsection{Spectrum}
\begin{figure}[t!]
 \centering
 \begin{subfigure}{0.495\textwidth}
 \centering
\includegraphics[width=\textwidth]{%
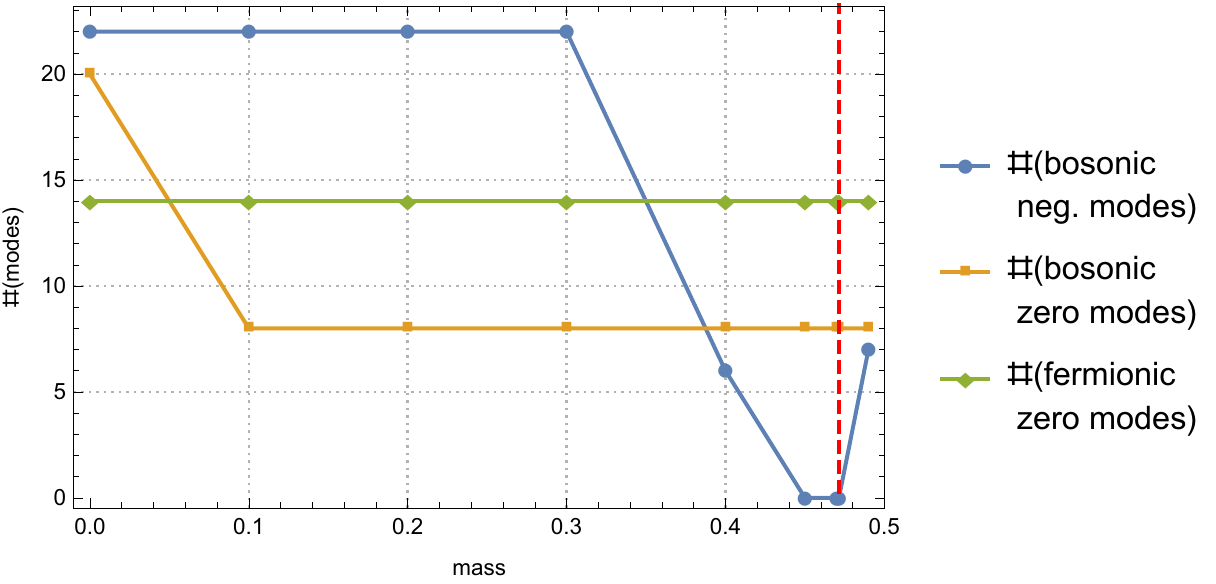}  
\caption{}
\label{fig:spectrum_flipped+pt_brane}
 \end{subfigure}
\begin{subfigure}{0.495\textwidth}
 \centering
 \includegraphics[width=\textwidth]{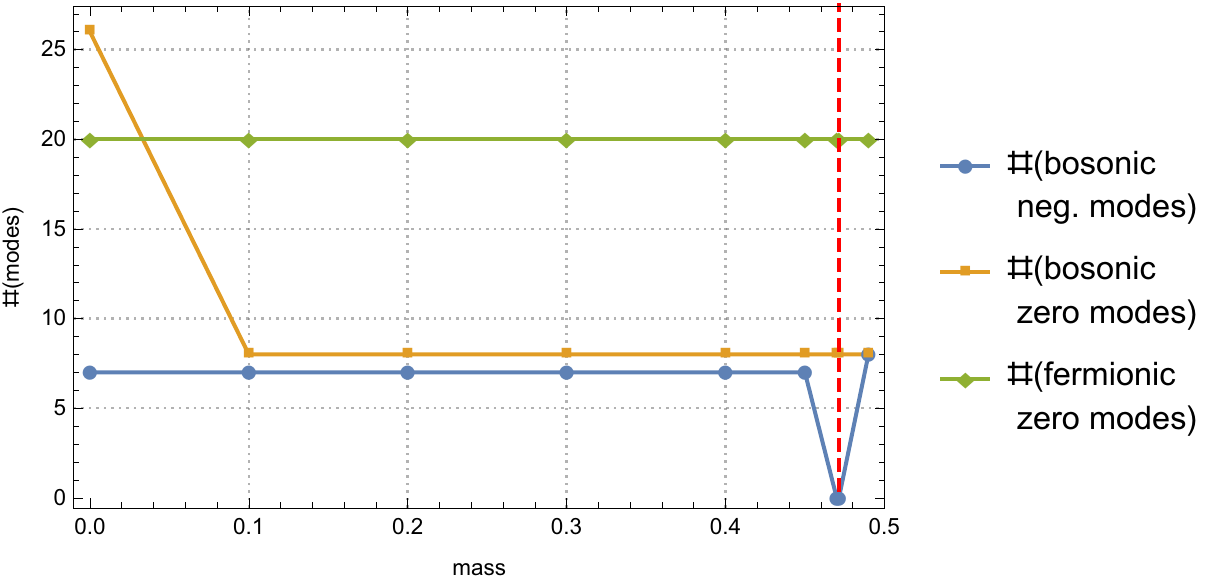}
 \caption{}
 \label{fig:spectrum_1-1_brane}
\end{subfigure}
 \caption{(\subref{fig:spectrum_flipped+pt_brane}) The number of negative and 
zero modes  of $\cO_V^X$ for the  
$\cC[(1,0)]+\cC[(0,1)]+\cD$ branes with several Higgs modes as 
background. %
(\subref{fig:spectrum_1-1_brane})
The number of negative and zero modes  of $\cO_V^X$ for the  
$\cC[(1,1)]$ brane with rank 2 Higgs modes as background. The red dashed 
vertical lines indicates $M^*$.}
\end{figure}

Around the combined solution $X+\phi$ we check the spectrum of the vector 
Laplacian and the Dirac operator. As in the other cases discussed so far, we can
lift negative modes by inclusion of uniform mass terms $M_i 
\equiv M$. We summarize the behavior of negative and zero modes in figure 
\ref{fig:spectrum_flipped+pt_brane}.
% 
%%%%%%%%%%%%%%%%%%%%%%%%%%%%%%%%%%%%%%%%%%%%%%%%%%%
%%%%%%%%%%%%%%%%%%%%%%%%%%%%%%%%%%%%%%%%%%%%%%%%%%%
% 
\subsection{\texorpdfstring{$(1,1)$}{(1,1)}-brane}
\label{subsec:1-1_brane}
Considering the $(1,1)$ brane we are faced with the maximal Higgs as displayed 
in Fig. \ref{fig:brane_1-1_rank1}; unfortunately, those three fields cannot be 
arranged to form a triangular subsystem. However, by considering the  
non-maximal Higgs of figure 
\ref{fig:brane_1-1_rank2_1}, \ref{fig:brane_1-1_rank2_2} and
\ref{fig:brane_1-1_rank2_3}, we can set up configurations that form triangles 
and give rise to exact solutions for the eom.
\subsubsection{Solution to eom}
Following the conventions of figure \ref{fig:brane_1-1_Higgs}, we employ the 
ansatz:
\begin{equation}
\begin{aligned}
 Y_j^{+}= X_{j}^{+} 
 + g_j \ \varphi_j^{+} +  h_j \ \sigma_j^{+}  \; ,
\end{aligned} 
\end{equation}
with $g_j,h_j\in \C$ and $ Y_j^{-}= (Y_j^{+})^\dagger$.
There are two classes of solutions (restricting to real coefficient does not 
exclude any non-trivial solution) which are
\begin{subequations}
\begin{align}
 (g_1,g_2,g_3) =(1,1,1) \quad \text{and} 
\quad h_j=0
\end{align}
up to phases, and
\begin{align}
 (h_1,h_2,h_3) =(-1,-1,-1) \quad  \text{and} 
\quad g_j=0 \; 
\end{align}
\end{subequations}
up to phases.
These correspond to the configurations in \ref{fig:brane_1-1_config_1} and 
\ref{fig:brane_1-1_config_2}, respectively.
\subsubsection{Spectrum}
For the configurations \ref{fig:brane_1-1_config_1} and 
\ref{fig:brane_1-1_config_2} we have computed the spectrum of the vector 
Laplacian, both gauge fixed and not, and of the Dirac operator. We find the number 
of negative modes of $\cO^{X+\phi}$ to be 
$7$, and their eigenvalues are all $-0.33548$.
Including equal masses $M_i\equiv M$ 
is sufficient to lift the negative modes for $0.47\lesssim  M < M^*$. 
The numerical results are depicted in figure 
\ref{fig:spectrum_1-1_brane}. 

Moreover, and similar to all previous cases, the number of bosonic zero modes 
is reduced by non-trivial mass values and is found to be $8$. We can understand these as 6 Goldstone bosons 
plus the two phases in the $g_j$ resp. $h_j$, as in section \ref{subsec:brane_spectrum}.
In addition, 
there are 20 fermionic zero modes.
%

%%%%%%%%%%%%%%%%%%%%%%%%%%%%%%%%%%%%%%%%%%%%%%%%%%%%%%%%%%%%%%%%%%%%%%%%%%%
%%%%%%%%%%%%%%%%%%%%%%%%%%%%%%%%%%%%%%%%%%%%%%%%%%%%%%%%%%%%%%%%%%%%%%%%%%%
\bibliographystyle{JHEP}
\bibliography{papers}
\end{document}